\documentclass[12pt]{article}
\usepackage[a4paper,pdftex]{geometry}
\usepackage[T1]{fontenc}
\usepackage[utf8x]{inputenc}
\usepackage{amsmath,amsthm,amssymb,graphicx,enumerate,booktabs,bigstrut,rotating,multirow,float,etoolbox,lmodern,comment,listings,qtree,a4wide,moresize,bbm,subcaption,tikz,pdfescape,mathtools,flafter,enumitem,setspace,ragged2e,colortbl,lineno,lscape,pdflscape,stackengine}
\usepackage[justification=centering]{caption}
\usepackage[authoryear]{natbib}
\usepackage[english]{babel}
\usepackage[small]{titlesec}

\DeclareMathOperator{\E}{\mathbb{E}}
\DeclareMathOperator{\Var}{\mathbb{V}ar}

\usetikzlibrary{snakes,decorations.pathreplacing,positioning, arrows.meta,shapes}

\usepackage{hyperref}
\hypersetup{colorlinks,linkcolor={red},citecolor={blue},urlcolor={blue}}  

\makeatletter
	\renewcommand\subsubsection{\@startsection{subsubsection}{3}{\z@}%
		{-3.25ex\@plus -1ex \@minus -.2ex}%
		{-1.5ex \@plus -.2ex}
		{\normalfont\normalsize\bfseries}
	}
	
	\def\@biblabel#1{\hspace*{-\labelsep}}
	
\makeatother

\def\sym#1{\ifmmode^{#1}\else\(^{#1}\)\fi}
\pdfpageheight\paperheight
\pdfpagewidth\paperwidth

\newcolumntype{L}[1]{>{\raggedright\let\newline\\\arraybackslash\hspace{0pt}}m{#1}}
\newcolumntype{C}[1]{>{\centering\let\newline\\\arraybackslash\hspace{0pt}}m{#1}}
\newcolumntype{R}[1]{>{\raggedleft\let\newline\\\arraybackslash\hspace{0pt}}m{#1}}
	
\begin{document}
	
	\title{\vspace{-3.7cm} Cognitive Endurance, Talent Selection, and the Labor Market Returns to Human Capital} 	
	
	
	\author{Germán Reyes\thanks{Department of Economics, Cornell University, 457 Uris Hall, Ithaca, NY 14853, United States (e-mail: gjr66@cornell.edu). I thank especially my advisor Ted O'Donoghue for invaluable guidance. For helpful discussions and comments, I thank Ned Augenblick, Michèle Belot, Nicolas Bottan, Emily Breza, Aviv Caspi, Zo\"{e} Cullen, Neel Datta, Stefano DellaVigna, Josh Dean, Christa Deneault, Rebecca Deranian, Gary Fields, Thomas Graeber, Ori Heffetz, Alex Imas, Guy Ishai, Judd Kessler, Yizhou (Kyle) Kuang, Shengwu Li, Yucheng Liang, George Loewenstein, Michael Lovenheim, Suraj Malladi, Alejandro Martínez-Marquina, Francesca Molinari, Kevin Ng, Muriel Niederle, Ricardo Perez-Truglia, Grace Phillips, Alex Rees-Jones, Evan Riehl, Seth Sanders, Paola Sapienza, Frank Schilbach, Heather Schofield, Peter Schwardmann, Dmitry Taubinsky, participants in the Cornell behavioral economics group, participants in the UC Berkeley psychology and economics group, and numerous seminar participants. I also thank Marco Pereira and other members of SEDAP for invaluable help using the secured data room. Financial support from the National Science Foundation is gratefully acknowledged.} 
	}

	
	\renewcommand{\today}{\ifcase \month \or January\or February\or March\or %
		April\or May \or June\or July\or August\or September\or October\or November\or %
		December\fi \ \number \year} 
	\date{\today \vspace{-1cm}}

	\maketitle
	
	\begin{abstract} 	
	\begin{singlespace}
	\noindent %
 	Cognitive endurance---the ability to sustain performance on a cognitively-demanding task over time---is thought to be a crucial productivity determinant. However, a lack of data on this variable has limited researchers' ability to understand its role for success in college and the labor market. This paper uses college-admission-exam records from 15 million Brazilian high school students to measure cognitive endurance based on changes in performance throughout the exam. By exploiting exogenous variation in the order of exam questions, I show that students are 7.1 percentage points more likely to correctly answer a given question when it appears at the beginning of the day versus the end (relative to a sample mean of 34.3\%). I develop a method to decompose test scores into fatigue-adjusted ability and cognitive endurance. I then merge these measures into a higher-education census and the earnings records of the universe of Brazilian formal-sector workers to quantify the association between endurance and long-run outcomes. I find that cognitive endurance has a statistically and economically significant wage return. Controlling for fatigue-adjusted ability and other student characteristics, a one-standard-deviation higher endurance predicts a 5.4\% wage increase. This wage return to endurance is sizable, equivalent to a third of the wage return to ability. I also document positive associations between endurance and college attendance, college quality, college graduation, firm quality, and other outcomes. Finally, I show how systematic differences in endurance across students interact with the exam design to determine the sorting of students to colleges. I discuss the implications of these findings for the use of cognitive assessments for talent selection and investments in interventions that build cognitive endurance.          
 	

 
	\end{singlespace}
	\end{abstract}
	
	\clearpage
	\section{Introduction} \label{sec:intro}

The human capital framework posits that individuals' skills and knowledge act as a form of capital that improves productivity and, thus, labor earnings \citep{becker1962investment}. The positive relationship between human capital and earnings is one of the most robust findings in the social sciences \citep{deming2022}, and is supported by a large body of work \citep[e.g.,][]{mincer1958investment, griliches1977estimating, card1999causal, card2001estimating}. While early studies focused on aggregate measures of human capital---like years of schooling---more recent work has focused on estimating the economic returns to specific skills, such as social skills \citep{deming_growing_2017} or cognitive skills \citep{hermo2022}. Identifying skills that foster productivity is essential for the design of effective education and labor-market policies \citep{almlund_chapter_2011, kautz2014fostering}.

In this paper, I study one dimension of human capital that may be particularly important for knowledge workers: \textit{cognitive endurance}, that is, the ability to sustain performance on a cognitively-demanding task for an extended duration. I first document that the performance of individuals on a college admission exam tends to decline, which allows me to measure cognitive endurance. Specifically, I develop a method to decompose test scores into fatigue-adjusted ability and endurance. I use the decomposition to investigate the relationship between endurance and long-run outcomes. I show that endurance has a sizable wage return in the labor market, comparable to the wage return to ability. I also show that, due to systematic differences in endurance across students, seemingly neutral exam design choices, such as the exam length, can have equity and efficiency consequences by affecting the sorting of students across colleges.


Psychologists and self-help books have long hypothesized that cognitive endurance is an important productivity determinant. 
Research on the nature of expertise---popularized in influential books like \textit{Focus} \citep{goleman_focus_2013} or \textit{Deep Work} \citep{newport2016deep}---often identifies this skill as a key driver of performance.\footnote{Psychologists have identified cognitive fatigue effects and highlighted the importance of mental endurance for high performance at least since the early 20th century \citep[e.g.,][]{james_energies_1907,dodge1917laws}.} Relatedly, biographers of extraordinary achievers often ascribe their accomplishments to unusually-high endurance.\footnote{For example, in describing Newton's accomplishments, \cite{keynes1956} noted that his greatest skill was ``the power of holding continuously in his mind a purely mental problem until he had seen straight through it.'' See \cite{lykken2005mental} for many other examples.} %
Consistent with this, researchers have documented the negative consequences of limited endurance for task performance in many settings.\footnote{Specifically, researchers have shown that individual-level job performance tends to deteriorate over relatively short time spans. For example, over the course of a day: nurses are less likely to wash their hands \citep{dai2015impact, steiny2022}; doctors make more diagnostic mistakes \citep{chan_fewer_2009, linder2014time, kim_variations_2018}; financial analysts make less accurate forecasts \citep{hirshleifer2019decision}; and umpires make more incorrect calls in baseball games \citep{archsmith_dynamics_2021}. \label{fn:decline}} %
The hypothesized link between endurance and productivity is also consistent with the large markets for endurance enhancers like coffee or nootropics (e.g., Adderall).\footnote{For example, in the US, 65\% of adults drink coffee daily \citep{lampkin_dietary_2012}, and about 20\% of college students report using nootropics without a prescription to enhance focus and cognition \citep{benson2015misuse}. Relatedly, over-the-counter focus-enhancing drugs have entire sections in chain drug stores (e.g., Appendix Figure \ref{fig:cvs}), and there is a growing variety of products marketed as endurance training (e.g., brain-training games like ``Lumosity'' or interval-based training technologies like  ``Pomodoros'').} 

These observations suggest that cognitive endurance and task performance are intimately linked. Yet, despite this popular perception, empirical economists have had little to say about the role of endurance in the labor market, possibly because of a lack of data on this variable. I address this problem by using data from the college admission exam in Brazil (called ``ENEM'') to create an individual-level measure of endurance that is based on performance declines throughout the exam \citep{borghans2018decomposing, kaur2022endurance}.


The ENEM is an ideal setting to study cognitive endurance for several reasons. First, the exam is administered under uniform conditions, and the scoring is standardized---two crucial properties for generating measures that are comparable across individuals \citep{almlund_chapter_2011}. Second, it is a high-stakes environment. Test scores largely determine the college options of the millions of high school students who take the ENEM every year. Since test-takers have incentives to exert maximal effort, limits to cognitive endurance are more likely to drive systematic declines in performance rather than low motivation \citep{duckworth2011role, gneezy_measuring_2019}. Third, the exam is grueling. The ENEM is ten hours long and is conducted over two consecutive days of testing. Thus, we might expect cognitive endurance to be an especially valuable skill in this setting and cross-person differences in endurance to be reflected in test performance. 

My analysis takes advantage of three features of the ENEM. First, the dataset contains students' responses to each exam question, which enables me to measure student performance throughout the exam. Second, students are randomly assigned different test booklets. Each booklet has the same set of questions (or ``items'') but in a different order, which enables me to study how students perform on a given question when they are relatively ``fresh'' versus mentally fatigued. Third, the ENEM can be linked to other administrative datasets to measure students' long-run outcomes. Specifically, I link the ENEM records to a census of all Brazilian college students and an employee-employer matched dataset that covers the universe of formal-sector workers in Brazil. 

I measure cognitive endurance as the impact of a one-position increase in the order of a given question on the likelihood of correctly answering the question. A potential-outcomes framework reveals that this measure captures the combined impact of two structural parameters: how cognitively fatigued an individual becomes throughout the exam and how an increase in fatigue affects test performance. These two parameters, and thus, my endurance measure, likely capture a variety of psychological mechanisms, including intrinsic motivation, grit, and attention capacity. 

Applying this framework, I first estimate \textit{mean} cognitive endurance across all students using two empirical strategies. The first research design compares average student performance \textit{on a given question} as a function of its position on each booklet, which I implement by regressing the fraction of students who correctly answer a question on its position on the exam, controlling for question fixed effects. This approach provides the more credible estimates of mean cognitive endurance; however, since each student only receives one exam booklet, it cannot be used to estimate \textit{individual-level} endurance. Thus, I also use a second research design that can be used to identify both average and individual-level endurance. The second approach consists of creating a position-adjusted measure of question difficulty, and then using this measure as a control variable instead of the question fixed effects. Both strategies deliver a similar-sized estimate of mean cognitive endurance. A one-position increase in the order of a given question decreases the chances of correctly answering the question by 0.08 percentage points. Scaled by the number of questions per testing day, this estimate implies that daily performance decreases by 7.1 percentage points due to limited endurance. 


Next, I estimate the difficulty-adjusted regression separately for each individual. This allows me to decompose an individual's test score into  a measure of cognitive endurance and a measure of fatigue-adjusted academic ability. My measure of cognitive endurance is the same as above but now estimated separately for each student. My measure of fatigue-adjusted ability is the residual of an individual's test score after subtracting from it the component explained by cognitive endurance. Using a sample of students who took the exam multiple times, I show that this measure of cognitive endurance has a test-retest reliability comparable to that of other commonly used constructs like risk aversion \citep{mata2018risk} or teacher value-added \citep{chetty2014measuring}. 

The measures generated by the decomposition enable me to investigate the importance of cognitive endurance for success in college and the labor market. I find that, holding fixed fatigue-adjusted ability and other student characteristics, individuals with more cognitive endurance are more likely to attend college, enroll in higher-quality colleges, are more likely to graduate, earn higher wages, and work for higher-paying firms. The associations are sizable. For example, controlling for ability and other variables, a one standard deviation (SD) increase in cognitive endurance predicts a 5.4\% increase in early-career wages. The corresponding prediction for a one SD increase in ability equals 15.4\%. Hence, the wage return to endurance is about a third of the size of the return to ability. Instrumental variable regressions show that the association between endurance and wages is larger after accounting for measurement error (on the order of 70\% the size of the return to ability) and also reveal that the predicted effect is not driven by a mechanical relationship between endurance and test scores.

Heterogeneity analysis reveals substantial variation in the wage return to ability and to endurance across college majors, occupations, and industries. On average, occupations and industries that pay higher wages also offer a higher wage return to ability and to endurance. This result documents a novel type of assortative matching between high-endurance workers and high-paying jobs. Furthermore, occupations and industries with a high wage return to endurance also tend to have a high wage return to ability, suggesting these two skills are complements in production. Some occupations with the highest wage return to endurance include those where lapses in sustained attention can have high costs, like facility operators in chemical plants or professionals in the aviation industry. This finding suggests that the value of endurance depends on a job's task requirements and the role of endurance in the production function of those tasks.

Finally, I use the measure of endurance to examine how differences in this skill across individuals affect the sorting of students to colleges. I focus on identifying the \textit{distributional} and \textit{informational} effects of an exam reform that decreases the exam length by half, thereby reducing the importance of endurance in determining test scores. The distributional effect asks how the exam reform would impact socioeconomic status (SES) test-score gaps. The informational effect asks how the reform would impact the exam's ``predictive validity'' (a measure of the exam's information content), as measured by the correlation between test scores and long-run outcomes. I derive formulas showing that test-score gaps and predictive validity can be written as linear functions of ability and endurance, with the weight on endurance proportional to the exam length.

The exam reform would decrease test-score gaps by 1.3--4.8 percentage points (a 26\%--29\% reduction from pre-reform gaps, depending on the measure of SES) and increase the predictive validity of the exam's test scores for long-run outcomes by as much as 95\%. Intuitively, the reform would reduce test-score gaps because, conditional on academic ability, low-SES students have lower endurance than high-SES students and, thus, perform disproportionally worse in questions toward the end of the exam. Similarly, the reform would increase the predictive validity of the exam partly because differences in performance at the beginning of the exam mainly reflect differences in ability (roughly, because most students are ``fresh''), which are highly predictive of long-run outcomes. In contrast, performance differences towards the end of the exam disproportionally reflect the noise associated with mental fatigue, which reduces the information content of test responses.

My findings yield three broad lessons. First, cognitive endurance matters for success in college and the labor market. My results provide empirical evidence on the long-standing hypothesis of endurance being a valuable skill. Thus, investing in the development of this skill, possibly at school during early ages, may have significant societal returns. Second, distinguishing between endurance and ability can improve how talent is selected and trained. Since the value of endurance varies among college majors, the student-major match may improve if majors where high endurance is required to succeed screen applicants partly based on this skill. Similarly, workers in endurance-intensive occupations may be more productive if the training necessary to enter into these occupations includes components aimed at building this skill. Third, seemingly neutral exam design decisions---the ``choice architecture'' of the exam---such as length or number of breaks, can have equity and efficiency consequences. By influencing the importance of endurance for test performance, the exam design can affect test-score gaps and predictive validity and, thus, the diversity of colleges' student bodies and the student-college match quality. 

This paper relates to the literature that studies cognitive endurance and fatigue effects in field settings. Limited cognitive endurance has been documented in a wide variety of environments (see footnote \ref{fn:decline}). Recent experimental evidence shows that cognitive endurance can be trained in children, which leads to less pronounced performance declines \citep{kaur2022endurance}. I contribute by linking individual-level endurance to long-run outcomes and establishing a novel set of associations. I do this in a high-stakes exam, which complements previous studies documenting performance declines in the low-stakes PISA test \citep[e.g.,][]{debeer_student_2014, borghans2018decomposing, zamarro_comparing_2018, balart_females_2019}. My findings provide a micro perspective to the results of \cite{balart_test_2018}, who show that the average performance decline in the PISA test among a country's test-takers has a sizable predictive power in cross-country growth regressions.

This paper also contributes to a growing literature documenting the importance of different dimensions of human capital for long-run outcomes. A large body of work shows that cognitive skills are valuable in the labor market \citep[e.g.,][]{hanushek2008role, hanushek2012better, fe2022cognitive, hermo2022}. This work often uses test scores as a measure of cognitive skills. I show that, even in a high-stakes setting, test scores partly measure cognitive endurance and provide methods to decompose test scores into fatigue-adjusted ability and endurance. Relatedly, a growing body of work shows that skills other than intelligence and technical skills (``noncognitive skills'') are also important predictors of long-run outcomes \citep{bowles_determinants_2001, heckman_effects_2006, borghans_economics_2008, almlund_chapter_2011, lindqvist_labor_2011, deming_growing_2017, jackson2018test, buser2021can, edin2022}. I document the strong predictive power of one noncognitive skill for college and labor-market outcomes.

Finally, this paper contributes to the literature on the design of college admission exams \citep{rothstein2004college, ackerman2009test, bettinger2013improving, hoxby2013expanding, bulman2015effect, goodman2016learning, goodman2020take, riehl2022}. These exams are designed to rank a large number of applicants. This requires discerning small ability differences, and as a consequence, they tend to be long and arduous. I show that performance on college admission exams measures not only academic preparedness but also the capacity to endure mental fatigue. Hence, there is a limit to how much information an exam can extract about student academic achievement. A lengthier exam may not lead to more precise measures of ability but rather to a selection mechanism that puts more weight on endurance. This may be desirable for programs where endurance is crucial to succeed, but it may come at the cost of screening out high-ability low-endurance students.

The rest of the paper is structured as follows. Section \ref{sec:context} describes the context and data. Section \ref{sec:framework} presents a statistical framework and describes my research designs. Section \ref{sec:fatigue-effects} presents estimates of average cognitive endurance. Section \ref{sec:recover} decomposes test scores into fatigue-adjusted ability and cognitive endurance. Section \ref{sec:endurance} examines the association between cognitive endurance and long-run outcomes. Section \ref{sec:equity-efficiency} studies the implication of limited endurance for the sorting of students to colleges. Section \ref{sec:conclusions} concludes.

	\section{Institutional Context and Data} \label{sec:context}

\subsection{The ENEM exam} \label{sub:educ}

The High School Assessment Exam (\textit{Exame Nacional do Ensino Médio}, or ENEM for short) is an achievement test created in 1998 by the Brazilian Ministry of Education to make high schools accountable for their students' progress. Some universities used the ENEM for college admissions; however, most institutions had university-specific admission exams. In 2009, the Ministry of Education expanded the ENEM to encourage universities to use it as their admission exam, and created a centralized admission system that uses ENEM scores to assign students to the highly-selective federal universities. Since then, many universities have started using the ENEM for admissions \citep{machado2021centralized, otero2021}.


The ENEM contains 180 multiple-choice questions equally divided across four subject tests (language arts, math, natural sciences, and social sciences) and an essay. The exam takes place over two consecutive days (two subjects per day, plus the essay on the second day). Test-takers have four and a half hours to complete the test on the first day and five and a half hours on the second day. There are no allocated breaks. To combat cheating, examinees randomly receive one of four different booklets each day. The order of the subjects and the set of questions is the same across booklets, but the order of the questions within a subject is randomized across booklets. A score for each subject is calculated based on item response theory (IRT), but most colleges ask applicants to submit their average score across all subjects. 

The exam is simultaneously taken across the country once a year at the end of the year. It costs approximately \$20 to take the exam, although this fee is waived for low-income applicants. Between 2009 and 2016, over 50 million individuals signed up to take the ENEM, making it the second-largest college admission exam globally. In Appendix \ref{app:enem-exam}, I describe the main changes in the ENEM over time, explain how ENEM scores are used in the higher-education system (other than for college admissions), and compare the ENEM to the US SAT and ACT exams.

\subsection{Data} \label{sub:data}

I combine three administrative databases from Brazil. The base dataset contains exam records from the ENEM from 2009--2016. This dataset contains both student-level and question-level information. The student-level data includes self-reported demographic and socioeconomic status (SES) measures, such as sex, race, high-school type (public/private), parental education, and family income. The question-level data includes each student's responses to each exam question, the position of the question, skill tested, etc.

To study individuals' trajectories through college and the labor market, I link the ENEM records to two other administrative datasets using individuals' national ID numbers (\textit{Cadastro de Pessoas Físicas}).\footnote{The linkage was conducted in the secured data room at the facilities of the Ministry of Education in Brasilia, Brazil.} To measure college outcomes, I use Brazil's higher-education census from 2010--2019. This dataset includes information on all college enrollees' major, university, year of enrollment, number of credits, and year of graduation. To measure labor-market outcomes, I use an administrative employee-employer matched dataset called RAIS (\textit{Relação Anual de Informações Sociais}) from 2016--2018.\footnote{The RAIS does not contain information on workers employed in the informal sector, self-employed individuals, or the unemployed.} The RAIS covers the universe of formal-sector workers in Brazil and includes information about both the worker and the firm. Workers' data include educational attainment, occupation, and earnings. Firms' data include the number of employees, industry, and geographical location.

\subsection{Samples and Summary Statistics} \label{sub:sample}

\textit{High-school-students sample.} To construct this sample, I impose several sample restrictions. First, I only consider individuals who take the ENEM during high school. This restriction excludes individuals who take the exam after dropping out or graduating from high school. Second, I only include individuals with a non-zero non-missing score on each subject test. This restriction excludes, for example, students who missed one of the days of testing. I also exclude a small fraction of students with special accommodations, usually due to a disability. After these restrictions, the high-school-students sample contains information on approximately 15 million students who took the ENEM from 2009--2016. To examine students' long-run outcomes, I focus on 1.9 million high-school seniors in the first two cohorts in my data (2009--2010), for whom I observe college and labor-market outcomes 6--9 years after taking the admission exam.

\textit{Retakers sample.} To assess the temporal stability of my measure of cognitive endurance, I identify students who take the ENEM more than once, usually as high-school juniors to practice and again in their senior year to apply for college. Approximately 16\% of test-takers in the high-school-students sample take the exam more than once.\footnote{ENEM scores are only valid for one year. Thus, students cannot use their junior-year ENEM results to apply for college. Some high-school students take the ENEM more than two times in the data, possibly because of grade repetition. I exclude a small fraction of students who take the ENEM more than three times.} I only include students with a valid exam score in all the years. The retakers sample contains information on 1.5 million students or 3.1 million student-years.

\textit{Summary Statistics.} Table \ref{tab:summ-enem} shows summary statistics on the samples. The average student in the high-school-students sample is 18.2 years old, 59.8\% are female, 47.6\% are white, and 22.2\% went to a private high school (column 1). Over half of students (53.4\%) have a high-school-educated mother, and 38.8\% live in a household that earns an income above twice the minimum wage.\footnote{Students self-report their household income and other SES measures when they enroll to take the ENEM. Household income is elicited in ranges and expressed as a multiple of the minimum wage. I divide students into those whose household earns more than five minimum wages and those whose household earns less than twice minimum wage. Using the Brazilian National Household Survey, I find that the former households are in the top 30\% of the national income distribution, while the latter households are in the bottom 30\%.} On average, students correctly respond to only 34.3\% of exam questions, which shows that the ENEM is a hard exam. High-school seniors from the 2009--2010 cohorts are slighly older, slightly more likely to be females, and white (column 2). Students in the retakers sample are slighly younger, their parents tend to have higher incomes, and they tend to perform better on the exam (column 3). Student characteristics are balanced across booklet colors (Appendix Table \ref{tab:summ-enem-book}).

\subsection{Definition of Main Outcomes} \label{sub:var-def}


\textit{Test score.} I define a student's exam score as the fraction of correct responses across all four academic subjects. The advantage of this measure is that it is intuitive and consistent with the existing literature \citep[e.g.,][]{zamarro_when_2019}. However, this measure differs from how the Brazilian testing agency calculates the ENEM score, which is based on IRT (see Appendix \ref{app-sub:irt-grading}). Reassuringly, the correlation between the fraction of correct responses and the IRT-based score is above 0.90 (Appendix Table \ref{tab:irt-pct-corr}).

\textit{College enrollment.} I define college enrollment as an indicator for appearing in the higher-education census one year after taking the ENEM. The rest of the college outcomes are defined conditional on college enrollment.

\textit{College quality.} I construct an earnings-based index of college quality. To do this, I group all college-educated workers in the RAIS (not just the workers in my sample) based on the university they attended and compute the average earnings of the graduates from each university.\footnote{This index is analogous to the college quality measure used by \cite{chetty2011does} and \cite{chetty2014impacts} to study the long-term impacts of kindergarten quality and teachers, respectively.}

\textit{College degree quality}. I create an index of college degree (or major) quality using the average earnings of the graduates of each college degree. To allow for international comparisons, I classify majors based on the International Standard Classification of Education \citep{unesco2012international}. 


\textit{Degree progress.} I calculate the ratio between the number of credits completed at the end of each year and the total number of credits required to graduate. This variable is available starting in the 2015 higher-education census. Thus, I use data from the cohort enrolled in 2015 to measure this outcome.

\textit{Likelihood of graduating.} I define an indicator for graduating one to six years after enrolling in college. Most students who ever graduate do so within the first six years (Appendix Figure \ref{fig:cdf-grad}). As robustness, I define a measure of on-time graduation based on expected degree length. The higher-education census contains information on how long a student in good standing should take to graduate from each program. I use this information to define an indicator for graduating within the expected number of years.

\textit{Formal employment.} I define formal employment as an indicator for appearing in the employee-employer matched dataset in any year in my sample. This variable is defined for all test-takers. The rest of the labor-market outcomes are defined conditional on formal employment. If an individual has multiple jobs, I use the data from the job with the highest number of hours. I use the job monthly earnings as a tiebreaker.

\textit{Monthly earnings.} This variable represents the average salary of a worker across all months in a given year. To report this variable, firms have to calculate the worker's total earnings for the year and divide them by the number of months the firm employed the worker. If a worker appears in multiple years in the RAIS, I calculate the inflation-adjusted average monthly earnings across all years. I adjust earnings for inflation using the consumer price index. 

\textit{Hourly wage.} I calculate the hourly rate of each worker as the ratio between a worker's inflation-adjusted monthly earnings and the hours worked per month.\footnote{Firms do not record the number of hours individuals actually work each week. Instead, the data on hours indicates the number of hours per week that the worker is expected to work based on her contract.} If a worker appears in multiple years in the RAIS, I calculate the average hourly wage across all years.

\textit{Firm, industry, and occupation mean wage.} I calculate the average hourly wage at each firm, industry, and occupation. I use leave-one-out measures so that an individual's own employment outcomes do not affect the mean wage. I define firms using the 14-digit CNPJ,\footnote{The CNPJ is a tax identifier for legally incorporated identities. The first eight digits identify the company. The rest of the digits identify the branch or subsidiary of the company.} industries using the Brazilian National Classification of Economic Activities (CNAE), and occupations using the Brazilian Occupational Code Classification (CBO). I calculate the wage indices separately for each year and use the average value across years.

I measure labor-market outcomes for the 2009--2010 cohort using employment data from 2016--2018. This means that, for the 2009 cohort, I measure outcomes 7--9 years after taking the ENEM, and for the 2010 cohort, 6--8 years after taking the ENEM. I account for this variation by controlling for an individual's potential years of experience throughout the analysis. I measure potential experience as the individual's age minus the years of schooling minus six.

\section{Empirical Framework} \label{sec:framework}

This section lays out a simple potential-outcomes framework. I use the framework to formally define cognitive endurance in terms of empirical estimands and to clarify the identification assumptions.

\subsection{Statistical Model} \label{sub:potential-outcomes}

Let $C_{ij}$ be the probability of individual $i$ correctly answering question $j$. I model $C_{ij}$ as a function of the student's level of cognitive fatigue, $f_{ij}$. Fatigue can affect performance by impairing cognitive functions such as attention, memory, or reasoning. The effects can be manifested in many ways, including students forgetting a crucial formula, making a computation mistake, misinterpreting or ignoring an important aspect of a question, and filling in the wrong bubble in the multiple-choice sheet.

To build intuition, first consider an environment in which fatigue is binary: individuals can be either mentally ``fresh'' ($f_{ij} = 0$) or ``fatigued'' ($f_{ij} = 1$). Let $C_{ij}(0)$ be the likelihood of individual $i$ correctly answering question $j$ if she is fresh and $C_{ij}(1)$ the likelihood if she is fatigued. These two probabilities denote potential outcomes for different fatigue levels, but only one of the two outcomes is observed. The observed performance, $C_{ij}(f_{ij})$, can be written in terms of these potential outcomes as
\begin{align} \label{eq:pot-out}
	C_{ij}(f_{ij}) = C_{ij}(0) + \underbrace{\Big( C_{ij}(1) -  C_{ij}(0) \Big)}_{\text{``Fatigue effect'' } (\kappa_i)} f_{ij},
\end{align}
where $C_{ij}(1) -  C_{ij}(0) \equiv \kappa_i$ measures the effect of fatigue on performance, or ``fatigue effect,'' for short. I allow the fatigue effect to be heterogeneous across individuals, although for simplicity I assume that it is constant across types of questions. Suppose for the moment that we observed whether the individual was fresh or fatigued when she answered each exam question. Then, one could compare $i$'s average performance in questions she answered while fatigued ($\E[C_{ij} | f_{ij} = 1]$) to her average performance in questions she answered while rested ($\E[C_{ij} | f_{ij} = 0]$). This comparison can be written as
\begin{align} \label{eq:pot-out-decomp}
	\E[C_{ij} | f_{ij} = 1] - \E[C_{ij} | f_{ij} = 0] = \underbrace{\left(\E[C_{ij}(1) | f_{ij} = 1] -  \E[C_{ij}(0) | f_{ij} = 1]\right)}_{\text{Term 1: Fatigue effect }} \notag \\ + \underbrace{\left(\E[C_{ij}(0) | f_{ij} = 1] -  \E[C_{ij}(0) | f_{ij} = 0]\right)}_{\text{Term 2: Selection bias}},
\end{align}
Equation \eqref{eq:pot-out-decomp} shows that a comparison of average performance would yield the sum of two terms. The first one is the fatigue effect for questions answered while fatigued. The second term is a selection bias that arises when comparing performance across different questions. For example, if individuals become fatigued over time, a selection bias might arise if questions become increasingly hard over the course of the exam. In this case, $i$'s average performance may deteriorate even if she had not experience fatigued. 

In practice, cognitive fatigue is not binary; rather, an individual can have different gradations of ``tiredness.'' In what follows, I assume $f_{ij}$ is continuous and interpret $\kappa_i$ as the impact of a unit change of cognitive fatigue on performance. Because cognitive fatigue cannot be directly observed, estimating $\kappa_i$ is not feasible. In the empirical analysis, I use the position of question $j$ on the version of the exam answered by $i$ ($\text{Position}_{ij}$), under the reasoning that students become increasingly fatigued over the course of the exam.\footnote{This idea is supported by research showing that time-on-task is a strong determinant of cognitive fatigue. For a review of the literature on the determinants of cognitive fatigue, see \cite{ackerman2011cognitive}.} To understand how cognitive fatigue relates to question position, consider a hypothetical linear projection of $f_{ij}$ on $\text{Position}_{ij}$:
\begin{align} \label{eq:pot-projection}
	f_{ij} = \omega_i + \pi_i \text{Position}_{ij} + \eta_{ij}.
\end{align}

The intercept of the projection, $\omega_i$, measures $i$'s cognitive fatigue at the beginning of the test. The slope of the projection, $\pi_i$, measures the change in cognitive fatigue due to a one-position increase in the order of a given question. $\eta_{ij}$ is a mean-zero projection error, uncorrelated with $\text{Position}_{ij}$ by definition. If student $i$ answers the exam in chronological order and finds the exam mentally taxing, we would expect $\pi_i > 0$. Using equation \eqref{eq:pot-projection}, it is possible to re-write equation \eqref{eq:pot-out} as a regression equation that can be estimated in observational data:
\begin{align} \label{eq:pot-out-reg}
	C_{ij} = \alpha_i + \beta_i \text{Position}_{ij} + \varepsilon_{ij}.
\end{align}

The intercept of the regression, $\alpha_i \equiv \E[C_{ij}(0)] + \kappa_i \omega_i$, measures $i$'s expected performance on the test if she were fresh ($\E[C_{ij}(0)]$), plus the impact of her initial level of fatigue on performance ($\kappa_i \omega_i$). Henceforth, I interpret $\alpha_i$ as a measure of $i$'s academic ability. The slope of the regression, $\beta_i \equiv \kappa_i \pi_i$, is the estimand of interest. This reduced-form measure is the product of two structural parameters, $\kappa_i$ and $\pi_i$, that are likely determined by several psychological mechanisms. For example, the performance of some individuals may be less impaired by cognitive fatigue (captured by $\kappa_i$) due to, for example, high intrinsic motivation or grit. Similarly, students may not become cognitively fatigued over the course of the exam (captured by $\pi_i$) due to, for example, high attention capacity. Henceforth, I interpret $\beta_i$ as $i$'s cognitive endurance.

The random part of performance, $\varepsilon_{ij} \equiv C_{ij}(0) - \E[C_{ij}(0)] +  \eta_{ij}$, measures deviations of $i$'s potential performance on question $j$ from her average potential performance. Comparing $i$'s performance across exam questions in different positions yields the sum of cognitive endurance plus a selection bias:
\begin{align*} 
	\E[C_{ij} | \text{Pos}_{ij} = p ] - \E[C_{ij} | \text{Pos}_{ij} = p - 1] = \beta_i + \underbrace{\E[C_{ij}(0) | \text{Pos}_{ij} = p] - \E[C_{ij}(0) | \text{Pos}_{ij} = p-1] }_{\text{Selection bias}}.
\end{align*}

Next, I describe the two research designs that I use to deal with the selection bias.


\subsection{Identifying Cognitive Endurance} \label{sub:identification}

In the empirical analysis, I first estimate the mean cognitive endurance across all students, $\beta \equiv \E[\beta_i]$. This parameter represents the causal effect of increasing a question's position on average student performance, $\bar{C}_j \equiv \E[C_{ij}]$. Rejecting the null hypothesis of $\beta = 0$ would demonstrate that average student performance partly depends on cognitive endurance (i.e., this would show that $\kappa_i \pi_i \neq 0$ for some $i$).

To identify $\beta$, I use two research designs. The first research design consists of assessing how average student performance \textit{on a given question} varies as a function of the question's position. This approach is enabled by the fact that a given question is located in a different position across booklets. To illustrate this approach, Appendix Figure \ref{fig:item-11898} displays student performance in a natural science question (Appendix Figure \ref{fig:nsci-ex} shows the text of the question). This question appears as early as position 46 in the gray booklet and as late as position 87 in the blue booklet. Accordingly, the fraction of correct responses declines from 40.8\% in the gray booklet to 29.9\% in the blue booklet. Comparing student performance in these two booklets reveals that an increase of 41 positions reduces performance on this question by 10.9 percentage points. Analogous pairwise comparisons can be made for any two booklets.\footnote{Not all questions appear in a different position across all booklets. Appendix Figure \ref{fig:hist-chg-pos} shows the variation in question position across all questions for every pairwise booklet combination.} I exploit this information using the following fixed effects specification:
\begin{align} \label{eq:reg-spec-fe}
	\bar{C}_{jb} = \alpha_j + \beta \text{Position}_{jb} + \xi_{jb}, 
\end{align}
where $\bar{C}_{jb}$ is the fraction of students who correctly answered question $j$ in booklet $b$ and $\alpha_j$ are question fixed effects. Appendix Figure \ref{fig:mean-corr-item} illustrates the mechanics of identification by plotting average student performance on selected questions as a function of their position on the four exam booklets and the corresponding best-fit lines. $\beta$ is identified by first estimating the effect of question position on average student performance separately for each question and then aggregating these question-specific best-fit lines (like the ones plotted in the figure) using the OLS weights. The advantage of this approach is that it relies on a weak identification assumption---the random allocation of booklets across students. However, since each student only receives one exam booklet, I cannot compare a student's performance across different booklets to identify $\beta_i$. Thus, I also use a second empirical strategy that can be used to identify both $\beta$ and $\beta_i$. 

The second empirical approach consists of controlling for question difficulty ($\text{Difficulty}_{j}$) in equation \eqref{eq:reg-spec-fe} instead of the question fixed effects. To estimate $\beta$, I assess how average student performance changes throughout the exam in regressions of the form: 
\begin{align} \label{eq:reg-spec-diff}
	\bar{C}_{jb} = \alpha + \beta \text{Position}_{jb} + \delta \text{Difficulty}_{j} + \mu_{jb}.
\end{align}

One challenge in implementing this approach is measuring question difficulty. An intuitive and often used measure of a question's difficulty is the fraction of students who correctly answered the question. However, a given question has a different fraction of correct responses depending on where it is located on the booklet. Thus, a question might appear to be more difficult simply because it is located later in the exam on average across booklets. To deal with this, I exploit the within-question position variation to construct a ``position-adjusted'' measure of a question's difficulty. This measure of question difficulty represents the fraction of correct responses we would expect to observe if question $j$ appeared in the first position of the exam (see Appendix \ref{app:difficulty} for details). To avoid a spurious correlation, I calculate question difficulty using data from test-takers outside my sample.\footnote{These are mainly individuals who took the ENEM after graduating from high school. The results are very similar if I use my sample to generate the measures of question difficulty. The correlation between the measure of question difficulty estimated with test-takers in my sample and outside my sample is 0.98.}

This strategy yields a consistent estimate of $\beta$ under the assumption that unobserved question characteristics are conditionally independent of average student performance. Below, I provide evidence in support of this assumption. Importantly, as I describe in Section \ref{sec:recover}, this second empirical strategy can also be used to identify the cognitive endurance of each individual. In this case, the identification assumption is stronger, requiring that unobserved question characteristics are conditionally independent of $i$'s performance. In the following two sections, I present estimates of mean cognitive endurance (Section \ref{sec:fatigue-effects}) and individual-level endurance (Section \ref{sec:recover}).

	
	\section{Cognitive Endurance and Test Performance} \label{sec:fatigue-effects}

This section presents estimates of average cognitive endurance using two research designs.

\subsection{Student Performance over the Course of the ENEM} \label{sub:pattern-enem}

\begin{figure}[H]
	\caption{Average student performance over the course of the ENEM}\label{fig:mean-corr}
	\centering
	\includegraphics[width=.75\linewidth]{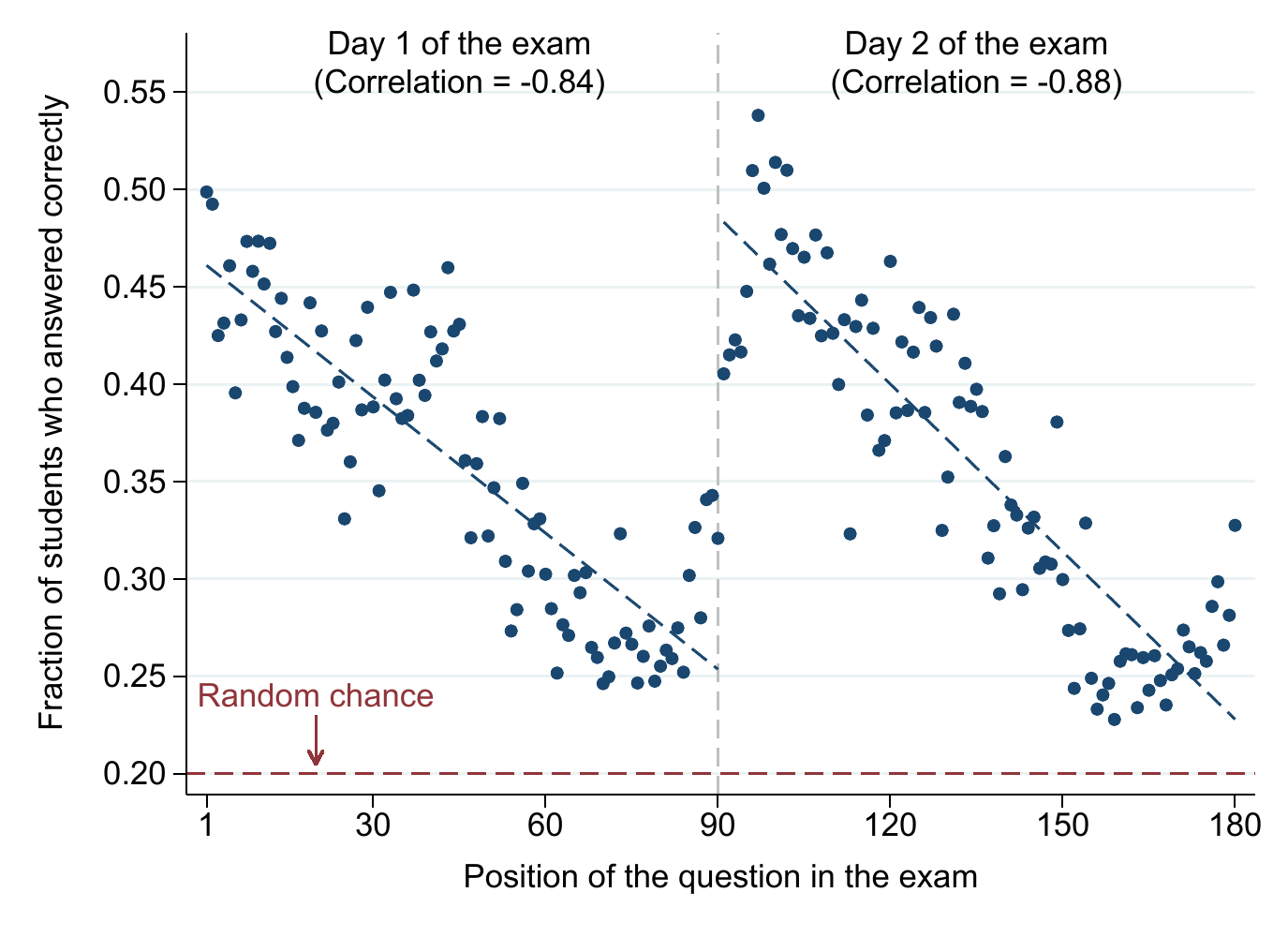}
				
	{\footnotesize
		\singlespacing \justify
		
		\textit{Notes:} This figure shows student performance over the course of each testing day in the ENEM. The $y$-axis displays the fraction of students who correctly responded to each question, averaged across all years in my sample. The $x$-axis displays the position of each question in the exam. The dashed lines are predicted values from a linear regression estimated separately for each testing day. The horizontal red dashed line shows the expected performance if students randomly guessed the answer to each question.

	}
\end{figure}

To motivate the analysis, I begin by studying student performance over the duration of the exam without controlling for question difficulty or any other performance determinant that may be changing throughout the exam. Figure \ref{fig:mean-corr}, plots the fraction of students who correctly responded to each exam question ($y$-axis) against the position of the question in the test ($x$-axis). As a benchmark, the red dashed line shows the expected performance if students randomly guessed the answer to each question. 

There is a strong negative relationship between student performance and question position. Average performance decreases from about 45\% at the beginning of the exam to about 24\% at the end of the exam. A bivariate regression of the fraction of correct responses on question position indicates that average student performance declines by 21.4 percentage points over the course of each testing day ($p < 0.01$), as shown in Table \ref{tab:perf-pos}, column 1. Interestingly, Figure \ref{fig:mean-corr} shows that average performance \textit{increases} from about 30\% at the end of the first day to about 45\% at the beginning of the second day.\footnote{Another interesting feature of Figure \ref{fig:mean-corr} is that student performance seems to \textit{increase} towards the end of each testing day. This pattern is not unique to the ENEM; a similar pattern has been found in the SAT \citep{mandinach2005impact} and the PISA test \citep{borghans2018decomposing}. One possible explanation is what \cite{mullainathan2013scarcity} refer to as the ``the focus dividend,'' that is, the notion that when a resource is scarce (in this case, the time left to finish the exam), the mind becomes better at focusing and blocking distractions. Another explanation is that some students answer the exam in reverse order.}

Limited cognitive endurance can provide a parsimonious explanation of these patterns. As students advance through the exam, their mental resources may become increasingly taxed, and thus they become more prone to committing mistakes. Cognitive resources are replenished after taking a break \citep{sievertsen2016cognitive} and overnight via sleep \citep{baumeister_yielding_2002, lim_sleep_2008}, which may explain why performance increases between the end of the first day and the beginning of the second day. Next, I implement the research designs described in Section \ref{sub:identification} to identify mean cognitive endurance.

\subsection{Estimates of Mean Cognitive Endurance} \label{sub:fatigue-empirics}

Table \ref{tab:perf-pos} presents the regression estimates from the two research designs. To facilitate the interpretation of the coefficients, I scale $\beta$ so that it can be interpreted as decrease in student performance due to limited endurance over the course of each testing day. 

Estimating the question-fixed-effects specification (equation \ref{eq:reg-spec-fe}) yields an average cognitive endurance $\beta = -0.072$ ($p < 0.01$), as shown in column 2. This estimate indicates that student daily performance decreases, on average, by 7.2 percentage points due to limited cognitive endurance. The difficulty-adjusted regression specification (equation \ref{eq:reg-spec-diff}) yields an estimate of average cognitive endurance $\beta = -0.058$ ($p < 0.01$), as shown in column 3. The similarity of this estimate relative to that obtained from the fixed effects specification suggests that controlling for question difficulty is adequate to account for differences in question characteristics. Moreover, the R-squared indicates that 97\% of the variation in $\bar{C_{j}}$ is explained by a question's position and difficulty. This high R-squared shows that there is little scope for unobservable variables to affect $\bar{C_{j}}$, further providing supporting evidence for the selection-on-observables assumption  \citep{oster2019unobservable}.

\begin{figure}[htpb!]
	\caption{The effect of an increase in the order of a given question on student performance}\label{fig:chg-pos}
	\centering
	
	\includegraphics[width=.75\linewidth]{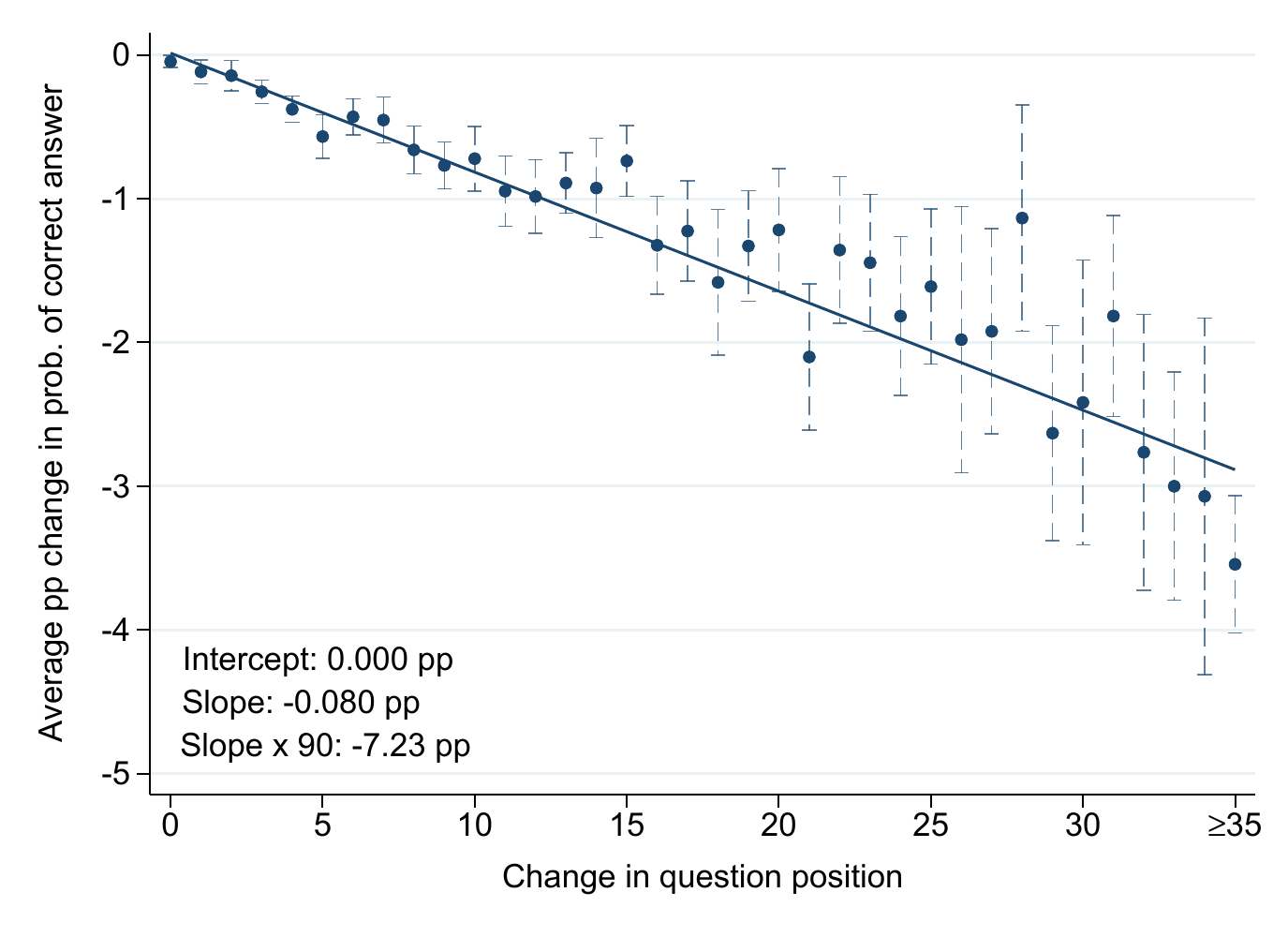}
	
		{\footnotesize
		\singlespacing \justify
		
		\textit{Notes:} This figure shows estimates of the impact of an increase in the order of a given question on the fraction of students who correcly answer the question. %
		The $y$-axis plots the average change (in percentage points) in the fraction of students who correctly respond to a question. The $x$-axis displays changes in a question position between each possible booklet pair. See Appendix Figure \ref{fig:hist-chg-pos}, Panel A for a histogram of the values in the $x$-axis. To construct this figure, I first compute the change in student performance and the distance in a question's position between each possible booklet pair. Then, I calculate the average change in performance for each observed distance.
		The solid line denotes predicted values from a linear regression estimated on the plotted points, using as weights the number of questions used to estimate each point. The vertical dashed lines denote 95\% confidence intervals, estimated with heteroskedasticity-robust standard errors. 
		
	}

\end{figure}

Figures \ref{fig:mean-corr-res} and \ref{fig:chg-pos} provide visual evidence on the effect of limited cognitive endurance on student performance. Figure \ref{fig:mean-corr-res} plots average student performance over the course of the exam after removing the influence of question difficulty on performance. To construct this figure, I first regress $\bar{C}_{jb}$, the fraction of students who correctly answered question $j$ in booklet $b$, on question difficulty, $\text{Difficulty}_{j}$, and estimate the residual from this regression, $\bar{C}^{r}_{jb} = \bar{C}_{jb} - \E[\bar{C}_{jb} | \text{Difficulty}_{j}]$. I add back the sample mean to $\bar{C}^{r}_{jb}$ to facilitate interpretation of units. Finally, I plot the mean value of $\bar{C}^{r}_{jb}$ across the exam. The figure shows that difficulty-adjusted performance tends to decline linearly throughout the exam. Daily performance decreases by about 5.2 percentage points each day, an effect consistent with the regression estimates.

Figure \ref{fig:chg-pos} plots the average percentage point change in the probability of correctly answering a question ($y$-axis) against the change in question position ($x$-axis) across all questions. The line is the predicted value from a linear regression estimated on the micro data. Its intercept is statistically equal to zero, indicating that a given question is, on average, equally likely to be answered if it appears in the same position in two different booklets. The slope indicates that, on average, a given question is 0.08 percentage points less likely to be correctly answered if it appears one position later in the test ($p < 0.01$). Thus, due to limited endurance, performance decreases by about one percentage point roughly every 12 questions (or 36 minutes if students spend the exam time uniformly across questions). The implied daily change in performance due to limited endurance equals 7.2 percentage points ($p < 0.01$), an estimate quantitatively identical to the question-fixed-effects specification.

Taken together, the evidence indicates that average student performance decreases by about 5--7 percentage points per day due to limited cognitive endurance. This effect is sizable. The fixed-effects-specification estimate represents a 16\% decrease of the estimated performance at the beginning of the exam (equal to 45\%, Table \ref{tab:perf-pos}, column 1) or about 60\% of the standard deviation of overall test score (equal to 11.6 percentage points). The effect is comparable to that of a decrease of half a standard deviation in teacher quality \citep{chetty2014measuring}, an increase in the class size of about 16 pupils \citep{angrist1999using}, or taking the exam under 66 degrees Fahrenheit hotter conditions \citep{park2022hot}.


\subsection{Limited Cognitive Endurance or Time Pressure?} \label{sub:time-pressure}

Throughout this section, I have interpreted the causal effect of an increase in question position on performance as a manifestation of limited cognitive endurance. This interpretation is in line with the framework in Section \ref{sec:framework}. However, an estimate of $\beta < 0$ could also potentially be generated by students running out of time toward the end of the exam.

In Appendix \ref{app:time}, I provide two pieces of evidence against this alternative interpretation. First, very few students leave any responses unanswered. Second, performance declines are present even when students respond to questions while they are likely not time-pressured (such as when responding to questions in the first half of each testing day). This evidence supports the interpretation of $\beta < 0$ as a consequence of limited cognitive endurance. 
	
	\section{Decomposing Test Scores into Ability and Cognitive Endurance} \label{sec:recover}

The results in Section \ref{sec:fatigue-effects} demonstrate that test scores reflect not only students' academic preparedness (``ability'') but also their capacity to endure mental fatigue (``cognitive endurance''). This section decomposes individuals' test scores into these two skills and examines the test-retest reliability of the generated measures.

 \subsection{Linear Decomposition}

To quantify the relative influence of ability and endurance on a student's test score, I estimate the difficulty-adjusted regression specification separately for each student:  
\begin{align} \label{reg:lpm-ind}
	C_{ij}  = \alpha_i + \beta_i \text{PosNorm}_{ij} + \delta_i \text{Difficulty}_j + \varepsilon_{ij} \quad \text{for } i = 1, ..., N,
\end{align}
where $C_{ij}$ equals one if student $i$ answered question $j$ correctly, $\text{PosNorm}_{ij}$ is question position normalized such that the first question of each day equals zero and the last question equals one,  and $\text{Difficulty}_j$ is the position-adjusted measure of question difficulty, normalized to have mean zero. In the baseline specification, I estimate equation \eqref{reg:lpm-ind} pooling student responses from both testing days and all academic subjects and show robustness to including day and subject fixed effects, as well as to estimating the parameters separately by day and subject.

Without further assumptions, $\hat{\alpha}_i$ and $\hat{\beta}_i$ simply describe how $i$'s performance changes throughout the test. The intercept of each regression, $\hat{\alpha}_i$, measures the predicted performance of student $i$ in the first question of the test for a question of average difficulty. Thus, $\hat{\alpha}_i$ represents $i$'s performance after accounting for the effect of a question's position and difficulty on performance. The slope of each regression, $\hat{\beta}_i$, measures the predicted performance change between the first and last question of each testing day after accounting for question difficulty.\footnote{In Appendix \ref{app:ols-endurance}, I derive the OLS estimate of $\beta_i$. The formula shows that $\hat{\beta}_i$ is calculated as a weighted average of deviations of $i$'s performance on each exam question from $i$'s average performance. Thus, $\hat{\beta}_i$ captures the intuition that a student who tends to do worse in the latter parts of the exam---relative to her average--- has low endurance.} Importantly, equation \eqref{reg:lpm-ind} can be interpreted as an observational analog of the model \eqref{eq:pot-out-reg}. Under this model,  $\hat{\alpha}_i$ measures $i$'s academic ability and $\hat{\beta}_i$ measures $i$'s cognitive endurance. 

\subsection{Limitations of Measuring Endurance using Standardized Tests}
 
This approach to measuring cognitive endurance has advantages but also important limitations. The main advantage is that it is based on observed behavior (``revealed preference''). This deals with some of the well-known biases of measures based on self-reports (``stated preferences''). Examples include social-desirability bias (i.e., respondents want to look good in front of the interviewer), reference-group bias (i.e., respondents judge their behavior using different standards), and framing effects (i.e., slightly different ways of asking the same question cause large changes in respondents' answers). %

However, there are at least three important concerns with the measure. %
First, estimating individual-level endurance requires a large number of orthogonality conditions. My research design requires any unobserved determinants of test performance to be uncorrelated with question position (conditional on question difficulty). This assumption is unlikely to hold exactly for \textit{all} students. For example, some students may happen to be unprepared for the questions that appear at the end of the exam, which would lead to biased estimates of endurance for these students. If the departures of the identification assumption are not systematic (e.g., some students are unprepared for questions at the end, but others are unprepared for questions at the beginning), then this issue is equivalent to measurement error, which would attenuate the effects documented below. Using the sample of retakers, I provide evidence consistent with this interpretation. In addition, I show that the results are similar using several alternative measures of endurance (e.g., calculated separately for each academic subject and using the average)

Second, my endurance measure hinges on the assumption that students answer the exam in chronological order. However, some students might respond in a different order (or strategically skip some questions).\footnote{For example, students with limited endurance that answer the exam in reverse chronological order will appear in the data as exhibiting a performance \textit{increase}. My estimate of endurance is biased for these students.} While I do not have data on the order in which students answered the exam, below I show that the results are robust to excluding individuals with \textit{positive} estimated endurance (i.e., those students who possibly answered the exam in reverse order).

Finally, my measure of endurance is biased in the presence of floor or ceiling effects. For example, individuals with extremely low ability or endurance may randomize their responses throughout the entire exam and show up in the data as having high endurance due to their stable performance. While this issue is not specific to my measure of endurance, it may be a concern for the empirical analysis. Below, I show that the results are robust to excluding students in the tails of the ability and the endurance distributions (i.e., students for whom floor and ceiling effects are more likely to be binding).

\subsection{Assessing the Reliability of the Cognitive Endurance Measure}  \label{sub:reliability}

Are the measures of academic ability and cognitive endurance generated by the decomposition reliable? To assess the reliability of a construct, researchers typically measure the construct multiple times and calculate the ``temporal stability'' or correlation between these measures \citep{miller_measurement_2009}. The size of the correlation is a measure of construct reliability. The higher the correlation, the more reliable the construct is said to be.\footnote{Reliability estimates vary significantly across constructs. Appendix Table \ref{tab:reliability} includes examples of reliability estimates for some well-known economic and psychological constructs. IQ is the construct with the highest known reliability, with correlations on the order of 0.80 \citep{hopkins1975ten, schuerger1989temporal}. Other commonly used constructs have lower temporal stability. For example, reliability estimates of risk aversion range 0.20--0.40 \citep{mata2018risk}; big five personality range 0.49--0.70 \citep{wooden2012stability}; and teacher value-added range 0.23--0.47 \citep{chetty2014measuring}.}

I compute two measures of test-retest reliability. First, I estimate ability and endurance separately for each testing day and calculate the correlation between consecutive days. The advantage of this approach is that it can be implemented in my main sample. The drawback is that the academic subjects tested vary each day, which could affect the reliability estimates.\footnote{For example, students who are good at natural science (a subject test on the first day) might not be as good at math (a subject test on the second day). This would lead to an imperfect between-day correlation.} Second, I estimate the temporal stability of ability and endurance between consecutive years. This analysis produces more comparable estimates, but it can only be done using the smaller sample of retakers.

\begin{figure}[H]
	\caption{The temporal stability of ability and endurance estimates}\label{fig:temp-stab}
	\centering
	\begin{subfigure}[t]{.48\textwidth}
		\caption*{Panel A. Ability in day $d$ and day $d+1$}
		\centering
		\includegraphics[width=\linewidth]{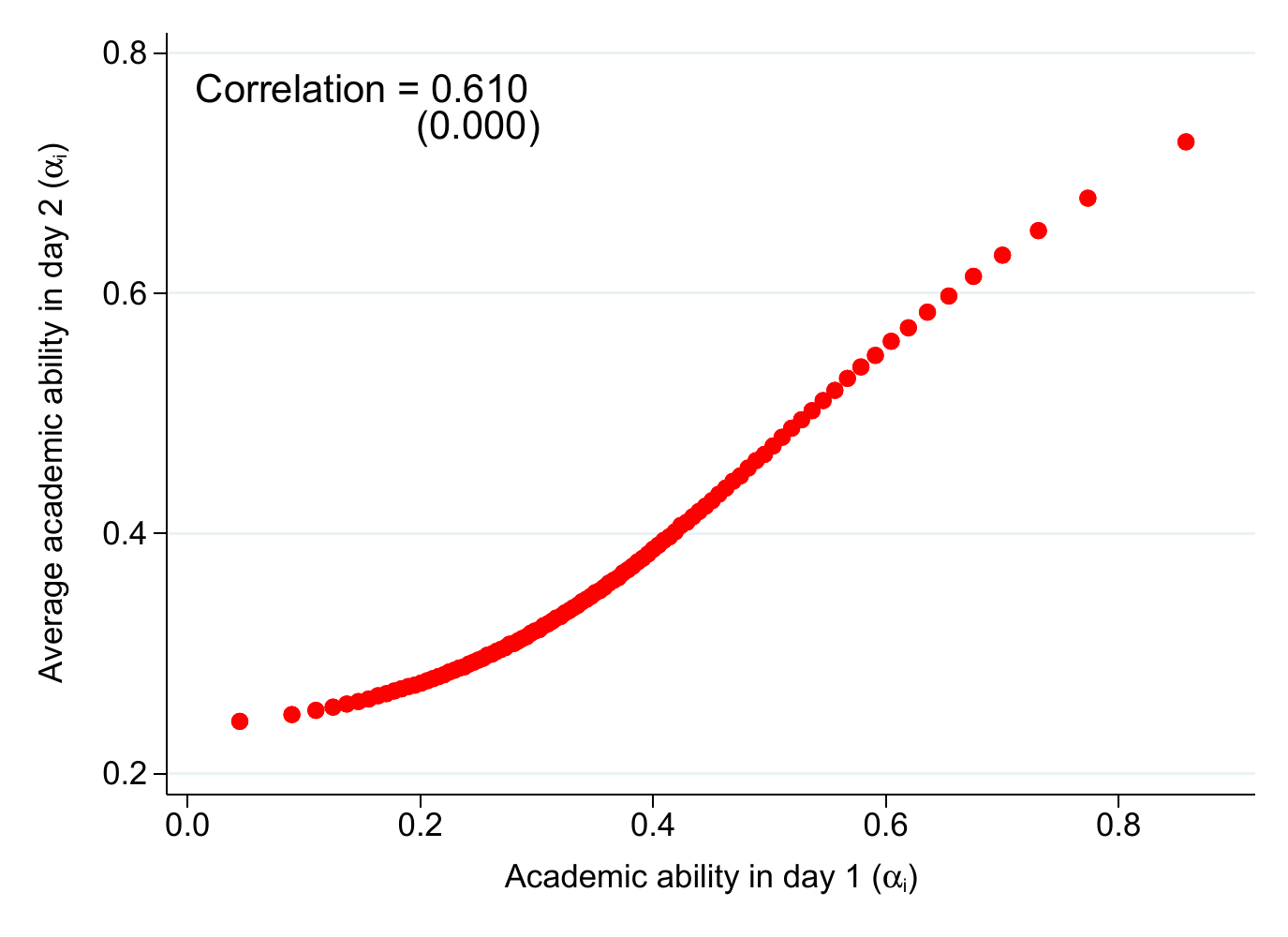}
	\end{subfigure}
	\hfill		
	\begin{subfigure}[t]{0.48\textwidth}
		\caption*{Panel B. Endurance in day $d$ and day $d+1$}
		\centering
		\includegraphics[width=\linewidth]{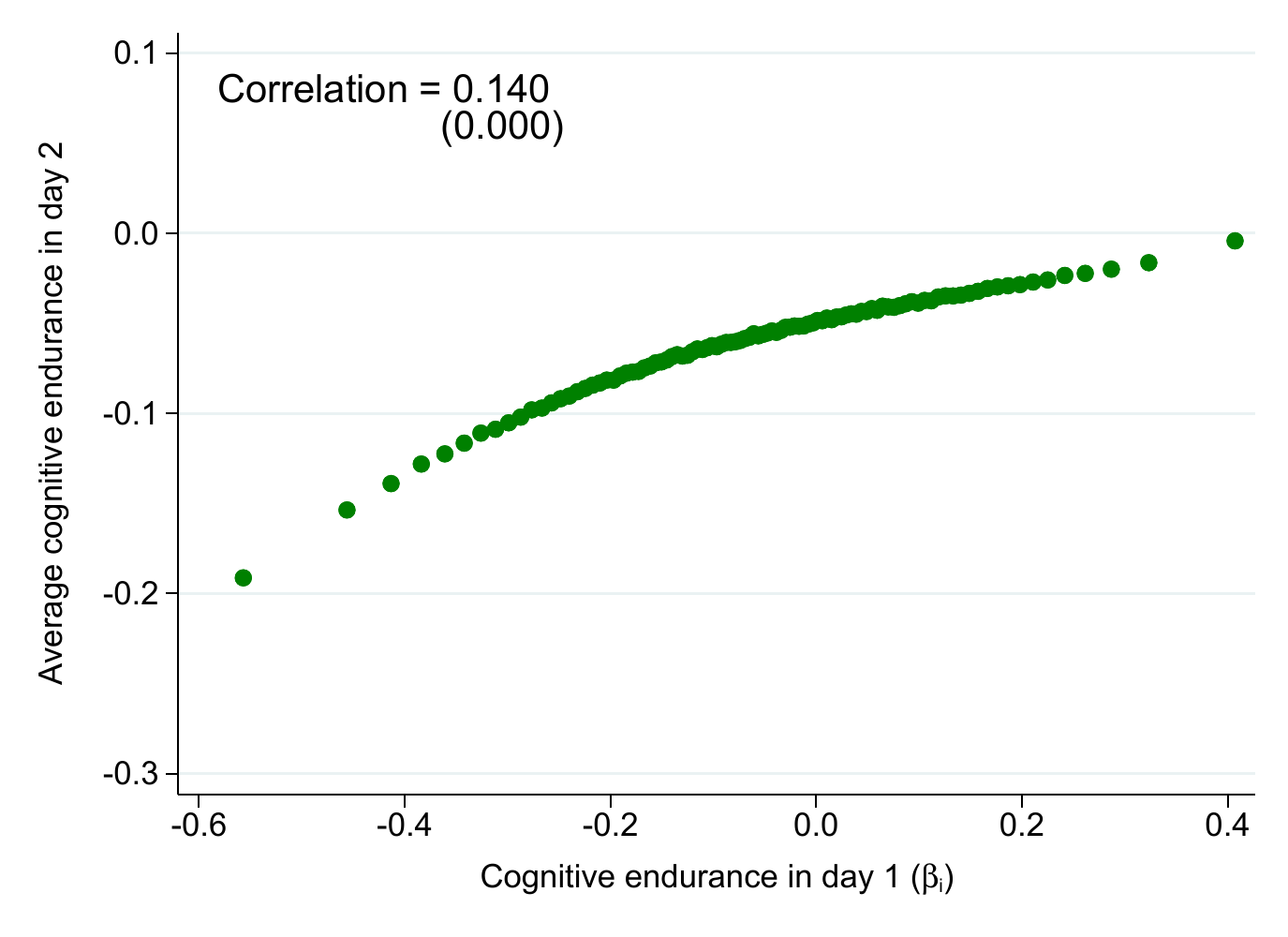}
	\end{subfigure}	
	\hfill		
	\begin{subfigure}[t]{.48\textwidth}
		\caption*{Panel C. Ability in year $t$ and year $t+1$}
		\centering
		\includegraphics[width=\linewidth]{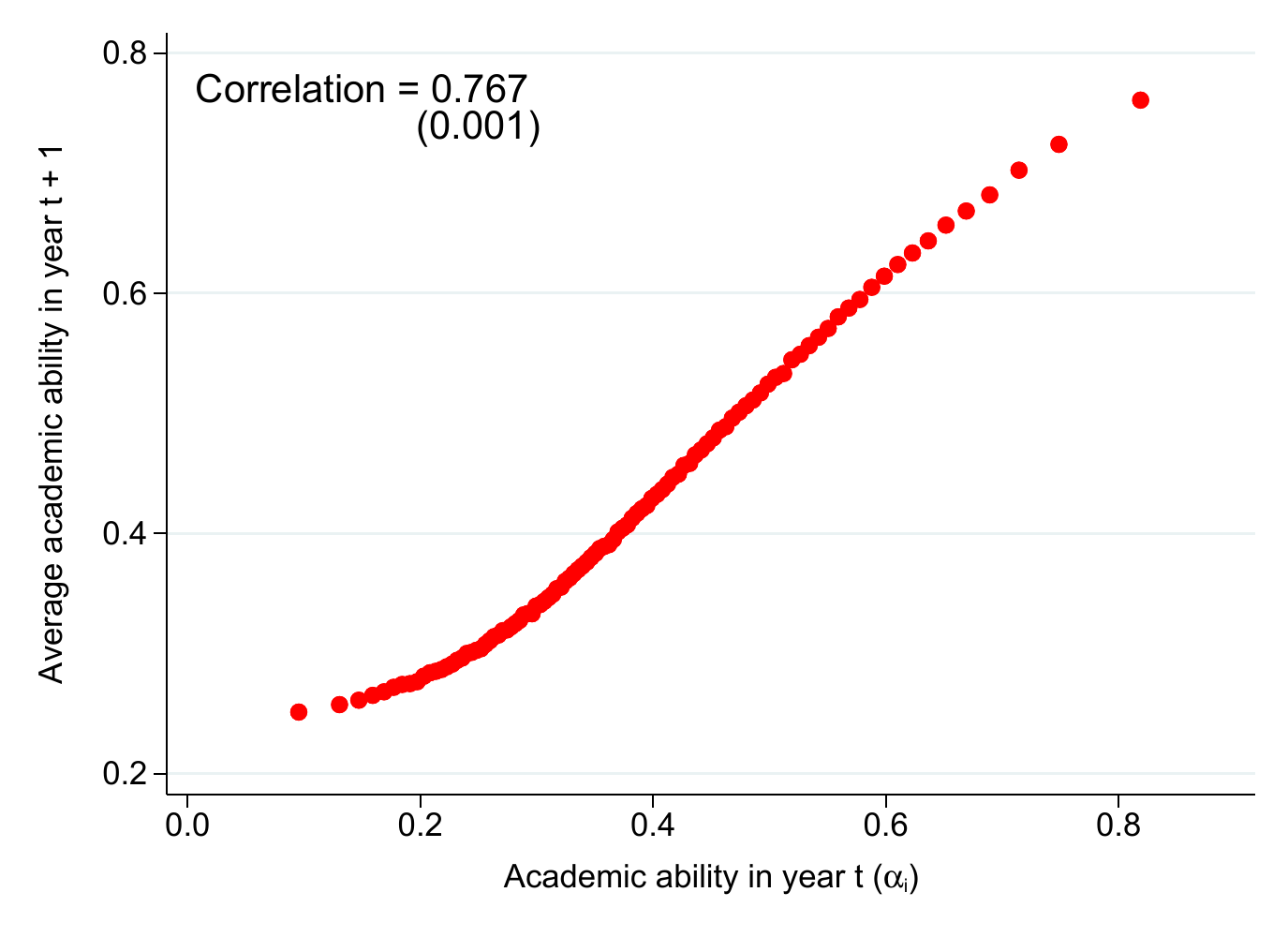}
	\end{subfigure}
	\hfill		
	\begin{subfigure}[t]{0.48\textwidth}
		\caption*{Panel D. Endurance in year $t$ and year $t+1$}
		\centering
		\includegraphics[width=\linewidth]{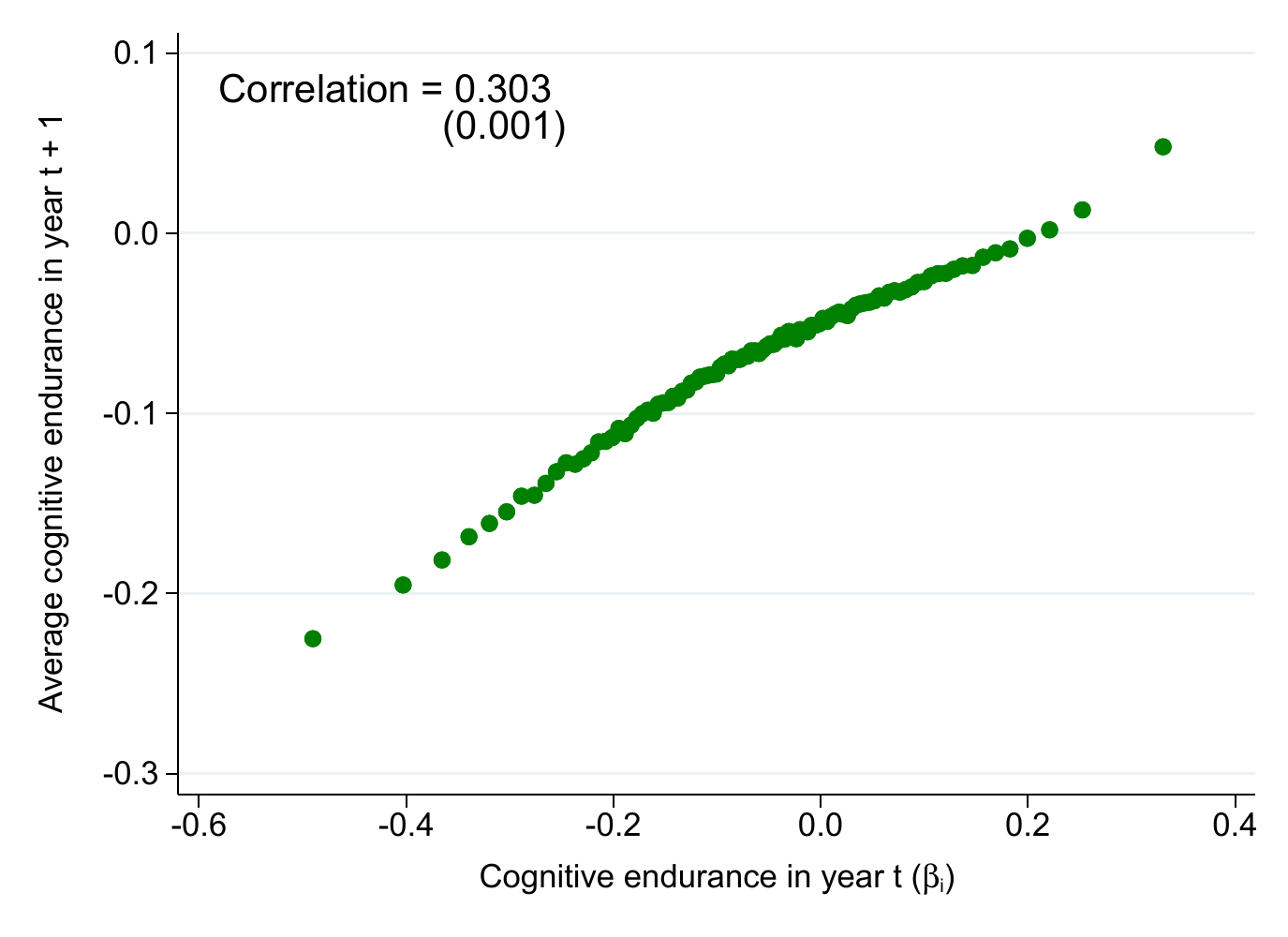}
	\end{subfigure}
	\hfill		
	\begin{subfigure}[t]{.48\textwidth}
		\caption*{Panel E. Ability in year $t$ and year $t+2$}
		\centering
		\includegraphics[width=\linewidth]{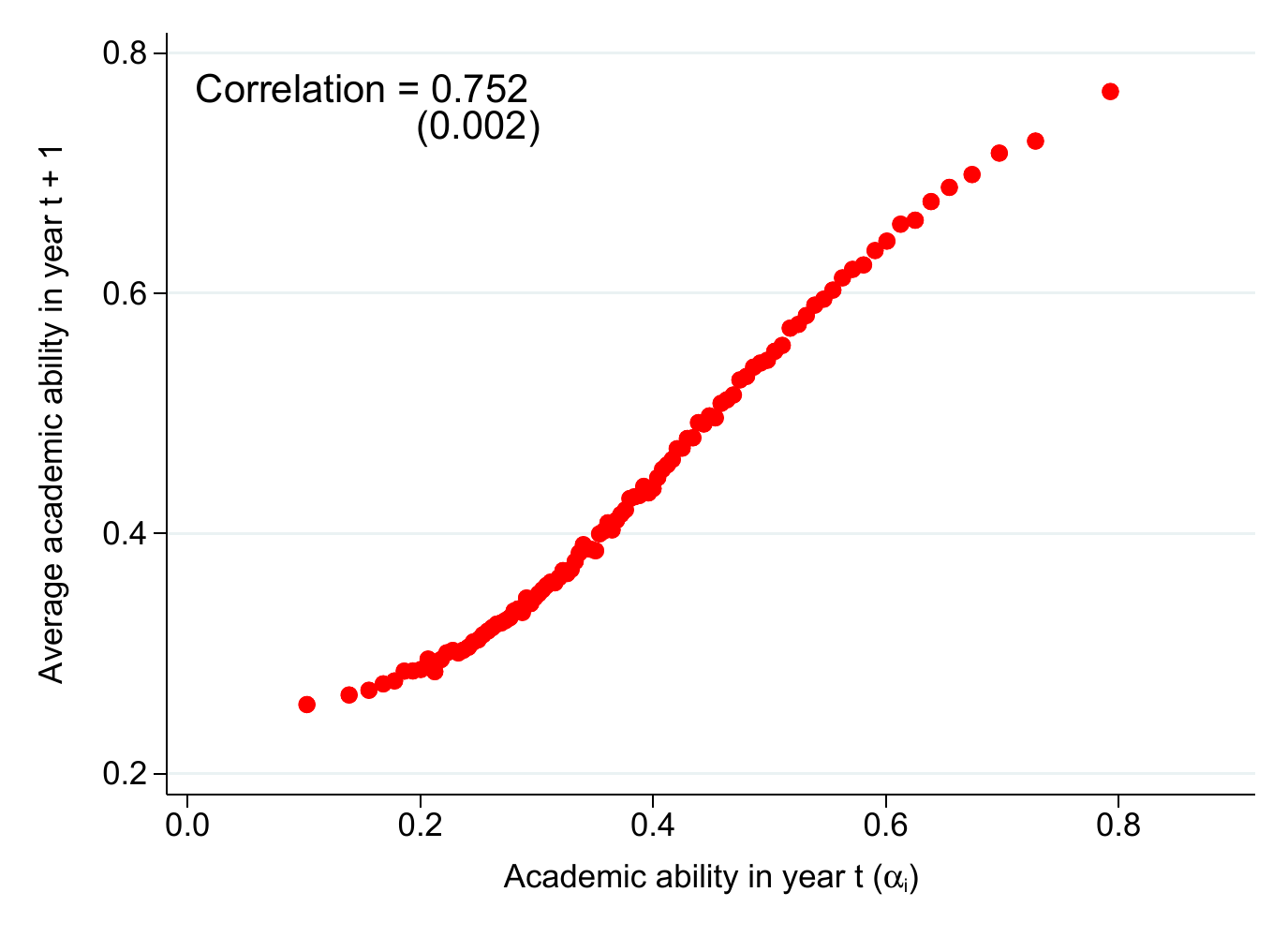}
	\end{subfigure}
	\hfill		
	\begin{subfigure}[t]{0.48\textwidth}
		\caption*{Panel F. Endurance in year $t$ and year $t+2$}
		\centering
		\includegraphics[width=\linewidth]{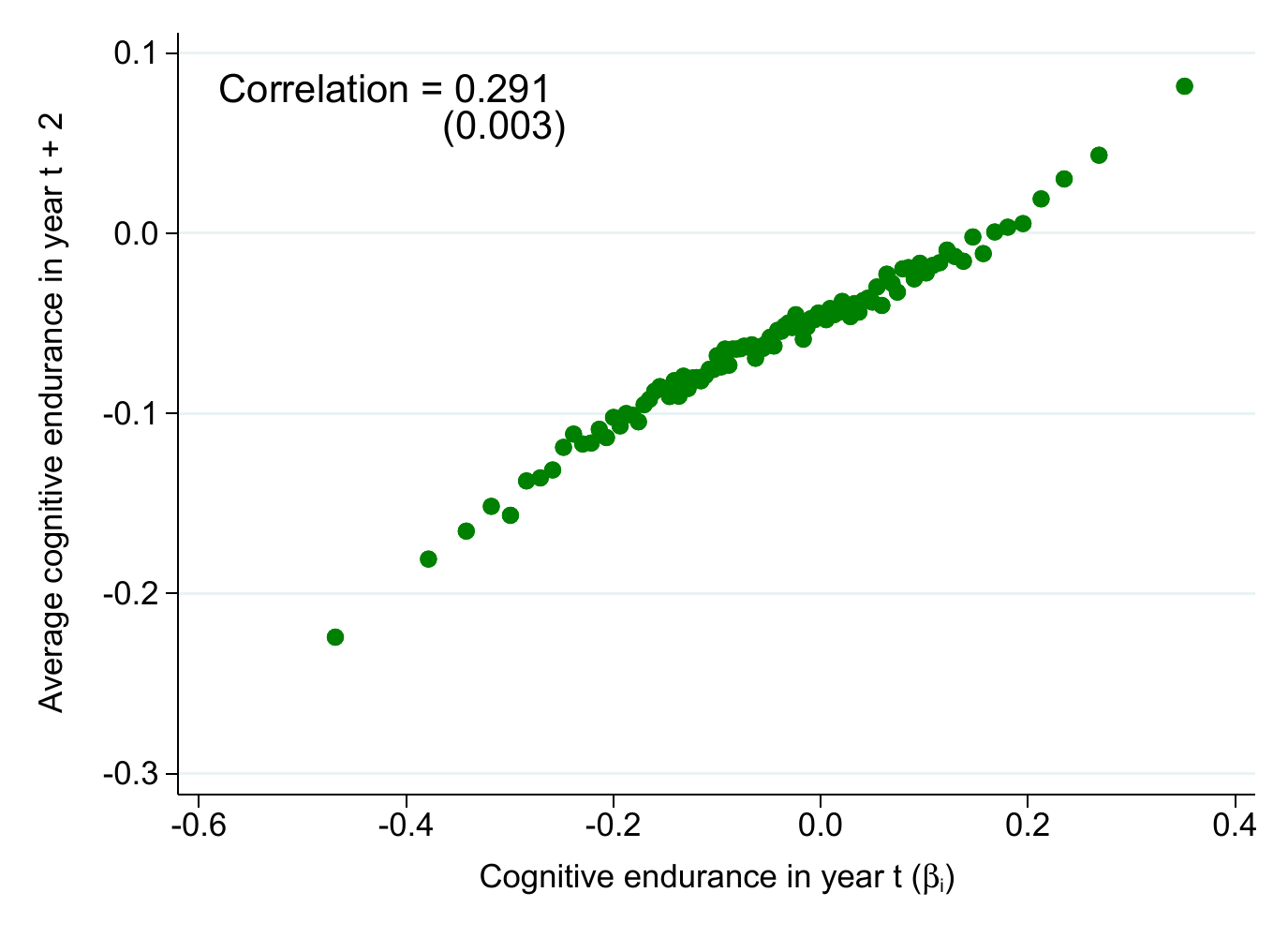}
	\end{subfigure}
	{\footnotesize
		\singlespacing \justify
		
		\textit{Notes:} This figure shows the correlation between the measures of academic ability and cognitive endurance measured at two different points in time. Each panel shows a binned scatterplot plotting the estimates of ability/endurance at two different times. To construct this figure, I first divide students into 100 equally-sized bins based on their ability/endurance at time $t$. Then, I calculate the average ability/endurance at time $t'>t$ for students in each bin. The panel title indicates the two time periods in which I measure ability and endurance.
		
	}
\end{figure}

The test-retest reliability of academic ability and cognitive endurance is comparable to that of other well-known constructs. Figure \ref{fig:temp-stab} show a series of binned scatterplots plotting the average $t+1$ estimate of ability/endurance as a function of the time $t$ estimate. The temporal stability of ability ranges from 0.61 (between consecutive days) to 0.77 (between consecutive years). The temporal stability of cognitive endurance ranges from 0.14 (between consecutive days) to 0.30 (between consecutive years). These results suggest that the ability and endurance measures are reliable for use in economic analysis.

\subsection{Summary Statistics on Ability and Cognitive Endurance} \label{sub:summ-endurance}

Average cognitive endurance is $\hat{\beta} = -0.058$, meaning that, due to limited endurance, the performance of the average student decreases by 5.8 percentage points over the course of the exam. This estimate is consistent with the quasi-experimental results shown in Section \ref{sec:fatigue-effects}. The standard deviation of $\hat{\beta}_i$ is $\sigma_{\hat{\beta}}$ = 14.4 percentage points.\footnote{Appendix Figure \ref{fig:hist-endurance} shows the distribution of estimated ability (Panel A) and endurance (Panel B).} Because of sampling error in $\hat{\beta}_i$, this raw standard deviation overstates the variability of true latent $\beta_i$, $\sigma_{\beta}$. Following \cite{angrist2017leveraging}, I estimate $\sigma^2_{\beta}$ as
\begin{align} 
	\hat{\sigma}^2_{\beta} =  \sigma^2_{\hat{\beta}} - \E[\text{SE}^2_{\hat{\beta}}],
\end{align}
where $\E[\text{SE}^2_{\hat{\beta}}]$ is the average squared standard error of $\hat{\beta}_i$. I construct an analogous estimate for the standard deviation of latent ability, $\hat{\sigma}_{\alpha}$ (see Appendix \ref{app:std-dev} for details).

The standard deviation (SD) of $\beta_i$ is $\hat{\sigma}_{\beta} = 0.088$. This means that an increase of one SD in cognitive endurance predicts a 8.8 percentage point increase in test score. The corresponding estimate for ability is $\hat{\sigma}_{\alpha} = 0.132$. Hence, $\hat{\sigma}_{\beta}$ is about two-thirds the magnitude of  $\hat{\sigma}_{\alpha}$, meaning that ability is more dispersed than endurance across students. These estimates can be translated into percentage effects by dividing by the average test score of 0.344 (Table \ref{tab:summ-enem}, Panel D). Under this rescaling, the estimates imply that a one SD increase in endurance leads to a 25.6\% increase in test score. The corresponding impact of ability equals 38.3\%.

\begin{figure}[H]
	\caption{Joint distribution of ability and endurance estimates}\label{fig:corr-alpha-beta}
	\centering
	\includegraphics[width=.75\linewidth]{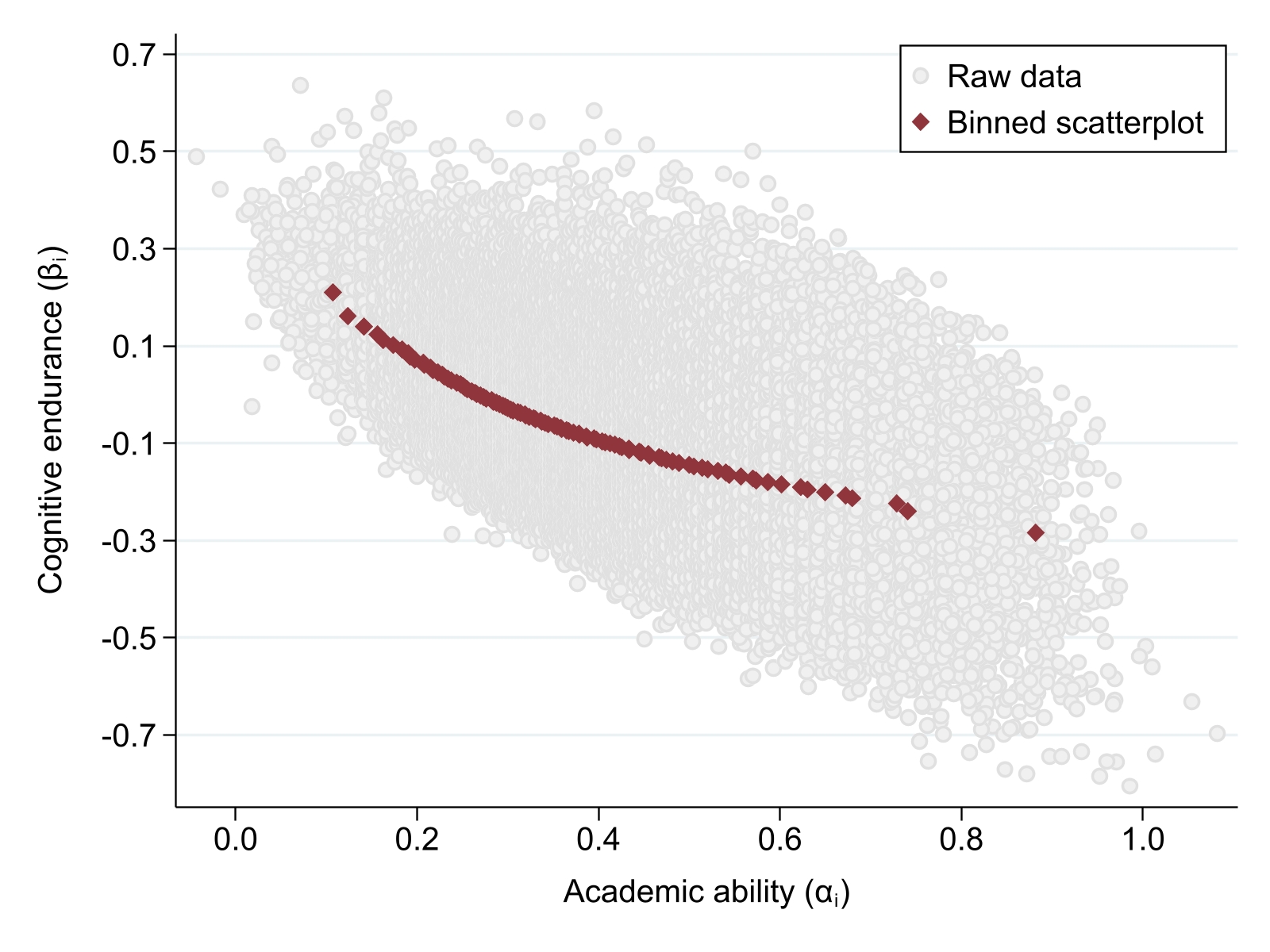}
	
	{\footnotesize
		\singlespacing \justify
		
		\textit{Notes:} This figure shows estimates of the relationship between academic ability and cognitive endurance. Gray circles display a scatterplot of $\hat{\beta}_i$ against $\hat{\alpha}_i$ for a randomly-selected one percent of my sample. The red diamonds show a binned scatterplot of average endurance as a function of ability. To construct the binned scatterplot, I first divide students into 100 equally-sized bins based on their ability. Then, I calculate the average endurance for students in each bin. Finally, I plot average endurance against ability in each bin.
		
	}
\end{figure}

Figure \ref{fig:corr-alpha-beta} shows the joint distribution of estimated ability and endurance. The red diamonds show a binned scatterplot of mean endurance as a function of ability, calculated by dividing students into 100 equally-sized ability bins. The gray circles display a scatterplot of $\hat{\beta}_i$ against $\hat{\alpha}_i$ for a randomly-selected one percent of my sample. 

Figure \ref{fig:corr-alpha-beta} reveals two important patterns. First, there is substantial variation in individuals' ability-endurance combination.\footnote{For example, for individuals with $\hat{\alpha} \simeq 0.50$, their estimates of endurance ranges from $\hat{\beta} = -0.50$ (a value roughly in the bottom one percent of the endurance distribution) to $\hat{\beta} = 0.50$ (a value in the top one percent).} Second, there is a negative relationship between $\hat{\alpha}$ and $\hat{\beta}$. On average, individuals with low values of $\hat{\alpha}$ tend to have higher values of $\hat{\beta}$. This relationship is largely mechanical and it is driven by floor and ceiling effects. Low-ability individuals have a limited margin to decrease their performance throughout the exam because test scores are bounded. A similar argument holds for high-ability individuals. This generates a ``missing mass'' of individuals with low-ability low-endurance and high-ability high-endurance, inducing a negative correlation between the two variables.\footnote{For individuals with intermediate values of ability (for whom ceiling and floor effects are less likely to be binding), the correlation is negligible. For example, the correlation between the two measures is -0.08 for individuals with estimated ability between 0.50 and 0.60.} In the analysis below, I always control for both variables to account for their mechanical relationship and show robustness to excluding individuals in the tails of the ability/endurance distribution.

In the following two sections, I use the estimates of ability and endurance to (i) revisit the association between test scores and long-run outcomes through the lens of the ability-endurance decomposition and (ii) characterize how systematic differences in endurance across students affect test-score gaps and the information contained by test scores.
	
\section{Cognitive Endurance and Student Outcomes in Adulthood}  \label{sec:endurance}


In this section, I use the decomposition to separately quantify the contribution of ability and endurance to the well-known association between test scores and long-run outcomes \citep[e.g.,][]{bishop1989test, hanushek2008role, hanushek2012better}. 

\subsection{Estimating the Return to Academic Ability and Cognitive Endurance} 

To assess how test scores and their component skills relate to college and labor-market outcomes, I estimate regressions of the form:
\begin{align} 
	Y_{i} &= \phi         + \lambda X_i  + \psi_T \text{TestScore}_i  + \nu_{i} \label{reg:score-outcomes} \\
	Y_{i} &= \tilde{\phi} + \tilde{\lambda} X_i + \psi_A \text{Ability}_i  + \psi_E \text{Endurance}_{i} + \tilde{\nu}_{i}, \label{reg:endurance-outcomes}
\end{align}
where $Y_i$ is an outcome of student $i$ (e.g., earnings); $\text{Ability}_i$  and $\text{Endurance}_{i}$ are the measures of academic ability and cognitive endurance estimated in Section \ref{sec:recover}; and $X_i$ is a vector that contains demographic variables and socioeconomic status.\footnote{For students with a missing value for a control variable, I define the missing value as equal to the sample mean value and include a dummy for missing student characteristics in the regressions.} For labor-market outcomes, I additionally control for educational attainment and potential years of experience. Because students can enroll in multiple college degrees, each observation denotes a student--degree combination. I account for the fact that an individual can appear multiple times in the dataset by clustering the standard errors at the individual level.

 
To compare the magnitude of the predicted effect of endurance on a given outcome with the corresponding effect of academic ability, I normalize both variables such that their coefficients represent the effect of a one standard-deviation (SD) increase on a given outcome.


\subsection{Baseline Estimates} 

I first discuss college outcomes and then turn to labor-market outcomes.

\subsubsection{College outcomes.}

Table \ref{tab:reg-coll-lmkt}, Panel A displays estimates of the relationship between test scores, its component skills (ability and endurance), and college outcomes. The first row shows estimates of equation \eqref{reg:score-outcomes}. Consistent with a sizable literature on the strong predictive power of test scores, I find that students with higher test scores tend to have better college outcomes. Students with a one SD higher test score are 8.8 percentage points more likely to enroll in college (relative to a mean of 24.4\%, column 1). Conditional on enrolling in college, the quality of their institution and college major---as measured by the average earnings of previous graduates---is 8.2\%--11.7\% higher (columns 2--3), the share of total credits they complete by the end of their first year is 1.4 percentage points higher (an 8.8\% increase relative to the mean of 15.8\%, column 4), and they are 6.0 percentage points more likely to graduate (column 5). Conditional on graduating, they take 0.12 fewer years to graduate (a 3.1\% decrease relative to the mean of 3.4 years, column 6). These estimates are comparable to those in the literature.\footnote{For example, \cite{chetty2014impacts} estimates that a one-standard-deviation increase in test scores is associated with a 5.5 percentage point increase in college enrollment at age 20, a 7.8\% increase in college quality as measured by the earnings of previous graduates, and an 11.9\% increase in earnings at age 28 (see their Appendix Table 3, row 2).} 

The second and third rows in Panel A show estimates of equation \ref{reg:endurance-outcomes}. There are two things to notice. First, controlling for endurance produces associations between ability and outcomes that are stronger than the associations between test scores and outcomes. Second, endurance has an economically and statistically significant effect on college outcomes. A one SD increase in endurance predicts a 2.9 percentage points increase in the likelihood of enrolling in college ($p < 0.01$); a 8.2\% increase in the college quality  ($p < 0.01$), and a 6.0 percentage point increase in the six-year graduation rate ($p < 0.01$). To benchmark the size of these associations, I compute the ratio between the predicted effect of endurance on an outcome and the predicted effect ability ($\hat{\psi}_E/\hat{\psi}_A$). This estimate is shown in the third-to-last row in Panel A. The effect of endurance as a percent of the effect of ability ranges from 31.6\%--36.2\%, depending on the outcome.

Figure \ref{fig:binsc-coll-lmkt}, Panels A--C present binned scatterplots of selected college outcomes against cognitive endurance. To construct each panel, I first regress $Y_i$ and $\text{Endurance}_{i}$ on student-level characteristics and ability, and estimate the residuals from these regressions, $Y^r_i$ and $\text{Endurance}^r_{i}$ (adding back the unconditional sample mean to facilitate the interpretation of units). Then, I group individuals into 10 equally-sized bins (deciles) based on $\text{Endurance}^r_{i}$. Finally, I plot the mean value of $Y^r_i$ for each bin. Consistent with the regression results, there is a strong relationship between cognitive endurance and college enrollment (Panel A), college quality (Panel B), and the six-year graduation rate (Panel C).
 
These results suggest that both ability and endurance are crucial for college success. While the importance of academic ability had been widely documented, the results show that traditional estimates of ability---usually measured through test scores---are confounded with the effect of cognitive endurance due to fatigue effects. More importantly, the results suggest that endurance plays a commensurate role in college success.\footnote{The positive association between endurance and college outcomes is consistent with \cite{kaur2022endurance}, who show that a cognitive-endurance-enhancing intervention improves student performance in elementary school.}

\subsubsection{Labor-market outcomes.}

Table \ref{tab:reg-coll-lmkt}, Panel B displays the results for labor-market outcomes. On average, students with higher test scores face better prospects in the labor market. Students with a one SD higher test scores are 0.1 percentage points more likely to have a formal-sector job (column 1), have a 12.7\% higher hourly wage (column 2), earn a 10.9\% higher monthly salary (column 3), work in firms that pay 9.1\% higher wages (column 4), choose occupations that pay 4.1\% higher wages (column 5), and work in industries that pay 1.3\% higher wages (column 6).

\begin{figure}[H]	
	\caption{The relationship between cognitive endurance and long-run outcomes}\label{fig:binsc-coll-lmkt}
	\begin{subfigure}[t]{.45\textwidth}
		\caption*{Panel A. College enrollment}
		\centering
		\includegraphics[width=\textwidth]{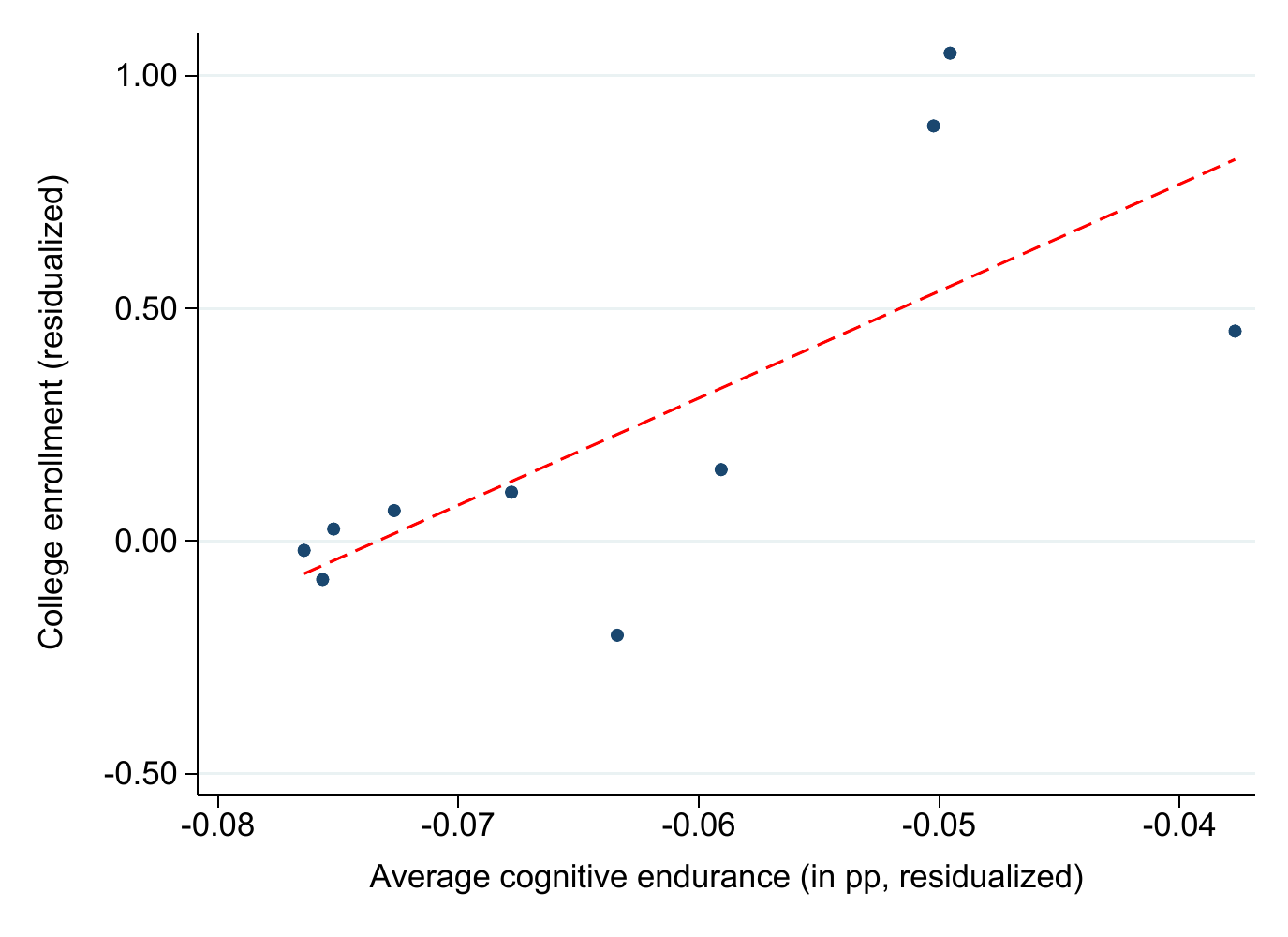}
	\end{subfigure}
	\hfill     
	\begin{subfigure}[t]{.45\textwidth}
		\caption*{Panel B. College quality}
		\centering
		\includegraphics[width=\textwidth]{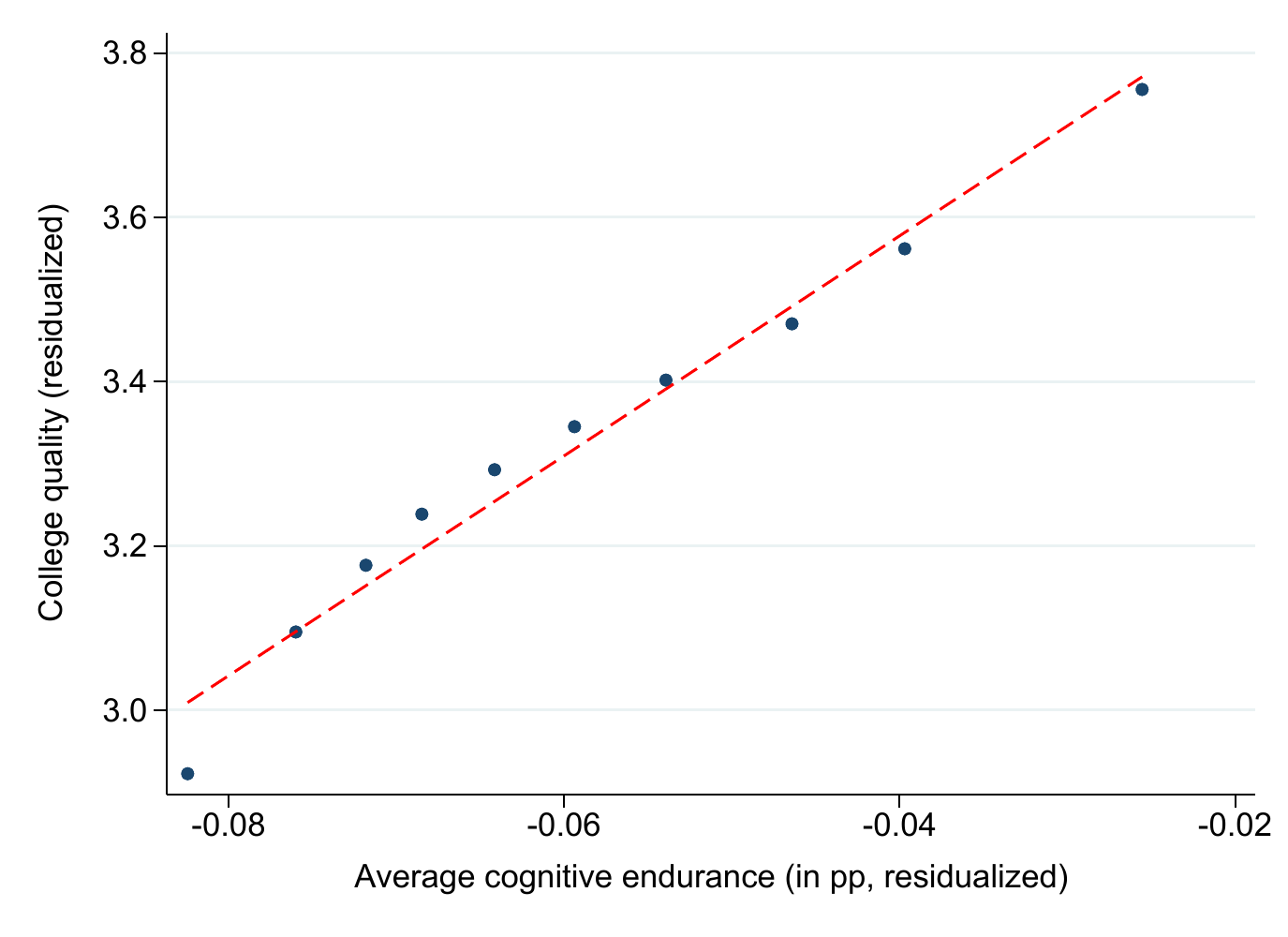}
	\end{subfigure} 
	\hfill       	
	\begin{subfigure}[t]{0.45\textwidth}
		\caption*{Panel C. Six-year graduation rate}
		\centering
		\includegraphics[width=\textwidth]{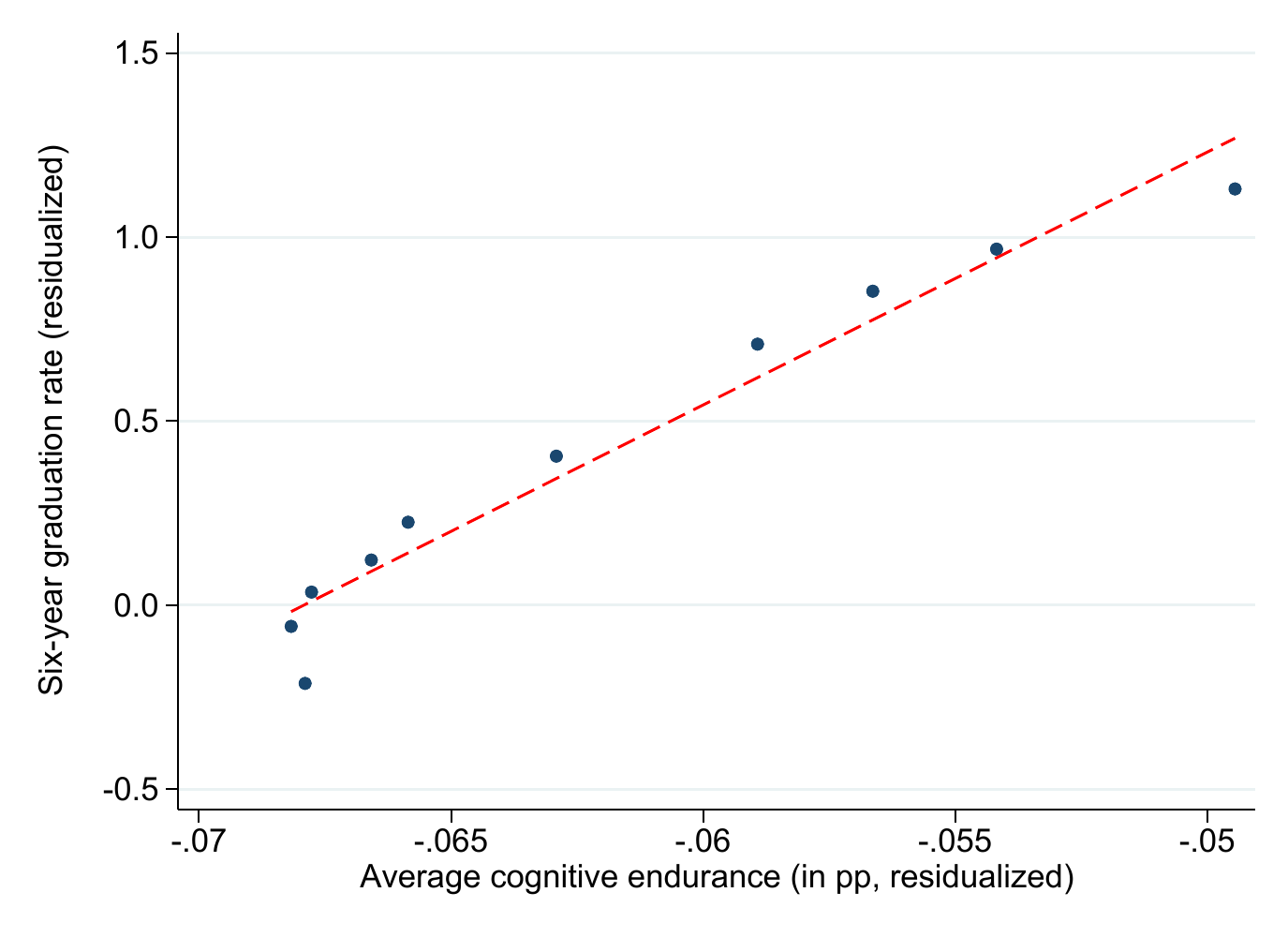}
	\end{subfigure}		
	\hfill        
	\begin{subfigure}[t]{.45\textwidth}
		\caption*{Panel D. Log hourly wage}
		\centering
		\includegraphics[width=\textwidth]{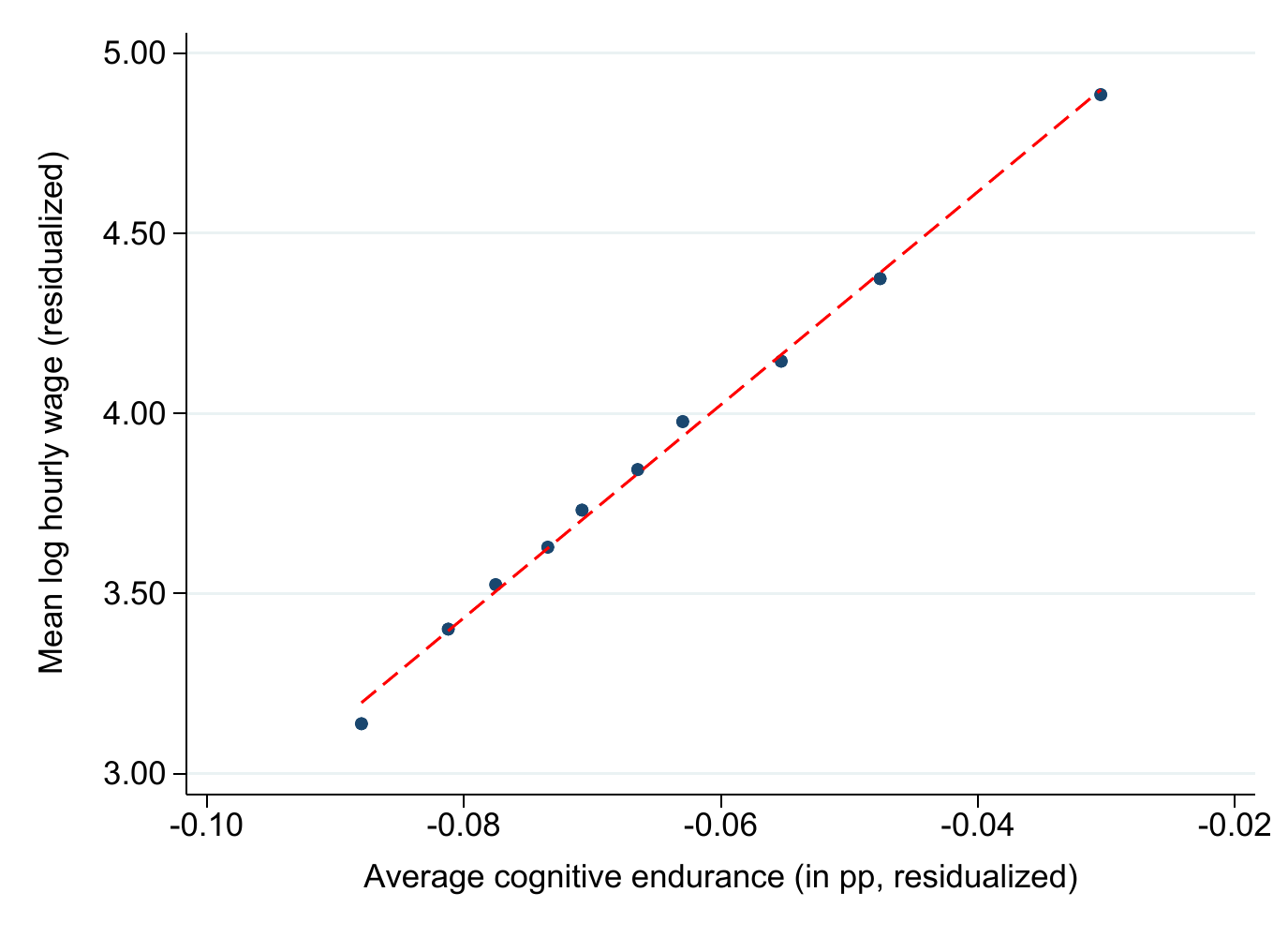}
	\end{subfigure}
	\hfill       	
	\begin{subfigure}[t]{0.45\textwidth}
		\caption*{Panel E. Log monthly earnings}
		\centering
		\includegraphics[width=\textwidth]{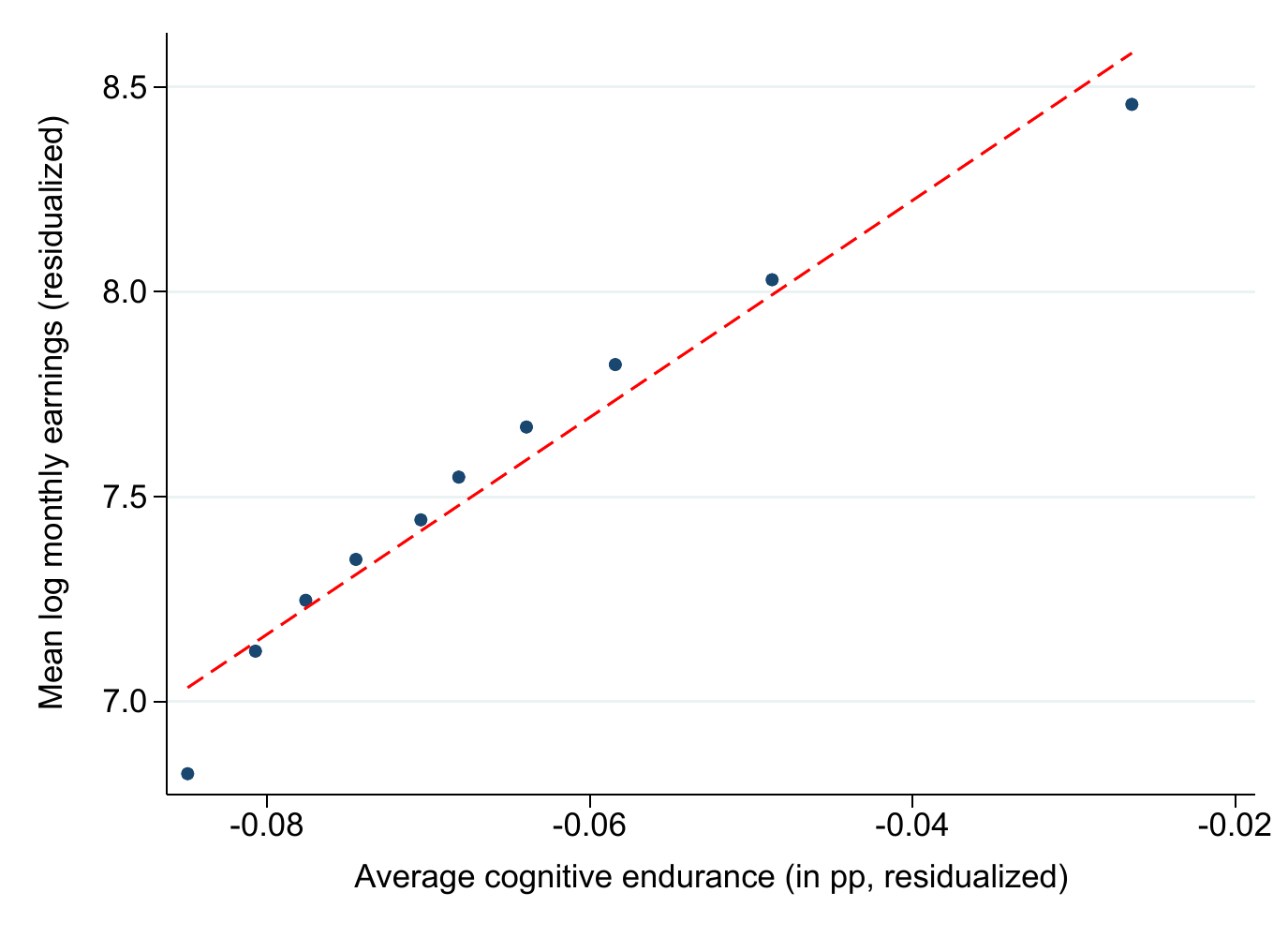}
	\end{subfigure}		
	\hfill        
	\begin{subfigure}[t]{0.45\textwidth}
		\caption*{Panel F. Firm mean wage (leave-one-out)}
		\centering
		\includegraphics[width=\textwidth]{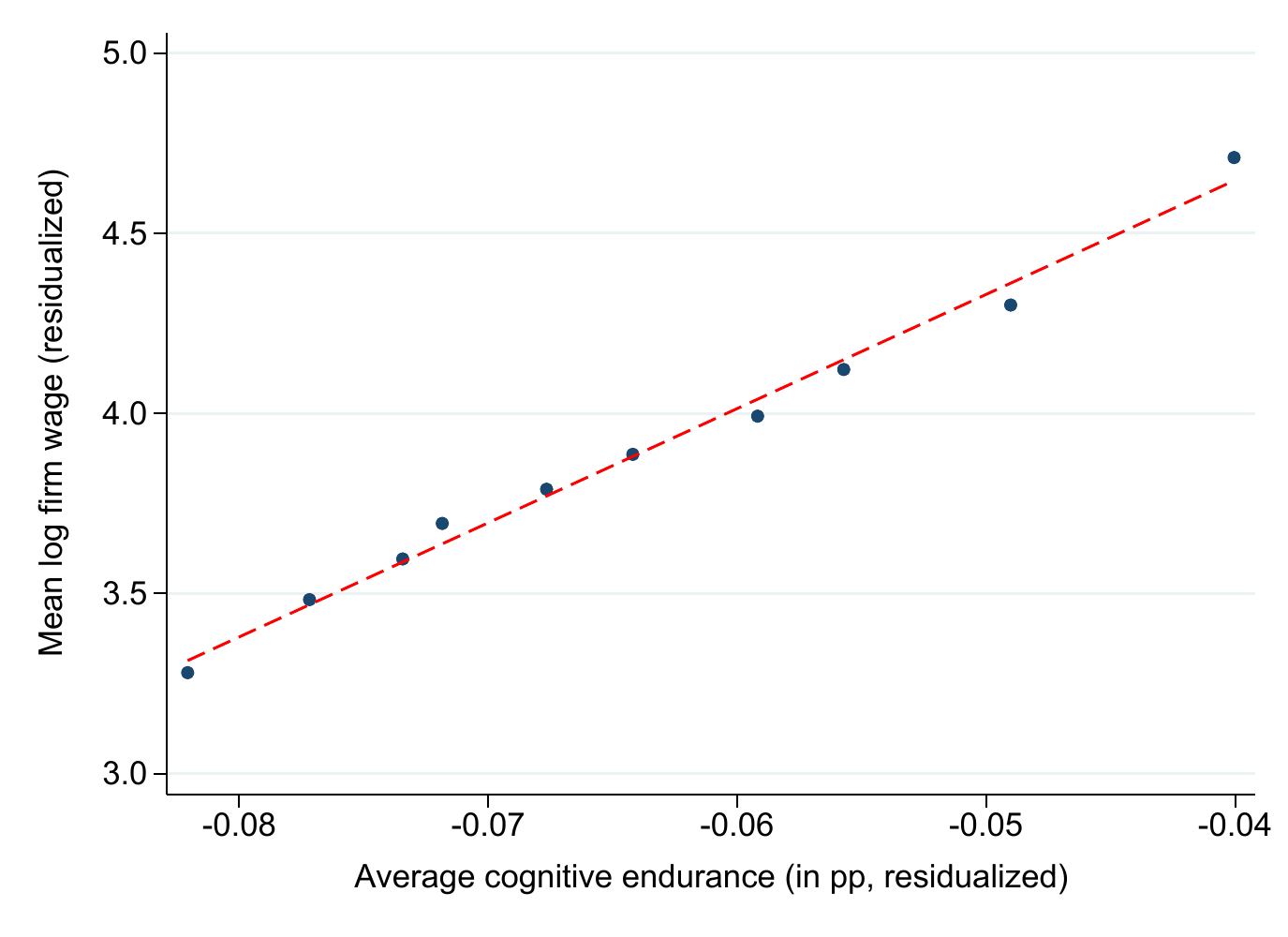}
	\end{subfigure}
	
	{\footnotesize
		\singlespacing \justify
		
		\textit{Notes:} This figure shows the relationship between cognitive endurance and selected college and labor-market outcomes. Each panel shows a binned scatterplot plotting the average value of the outcome ($y$-axis) against cognitive endurance ($x$-axis). To construct this figure, I first residualize cognitive endurance and each outcome on student-level characteristics and academic ability. I add back the unconditional sample mean to facilitate interpretation. Then, I divide students into 10 equally-sized bins (deciles) based on their residualized endurance and plot the average outcome for students of each bin. The red dashed lines are predicted values from a linear regression on the plotted points. Each panel shows the results for the outcome listed in the panel title. See Section \ref{sub:var-def} for variable definitions.

	}
	
\end{figure}

These associations reflect both the influence of academic ability and cognitive endurance, both of which have statistically and economically significant effects on all labor-market outcomes. For example, a one SD increase in endurance predicts a 5.4\% increase in hourly wages ($p < 0.01$), a 5.2\% increase in monthly earnings ($p < 0.01$), and a 3.6\% increase in the average firm wage ($p < 0.01$). The strong relationship between cognitive endurance and these three labor-market outcomes is illustrated in binned scatterplots in Figure \ref{fig:binsc-coll-lmkt},  Panels D--F. These figures show that mean wages and earnings increase roughly linearly with endurance. For labor-market outcomes, the predicted effect of cognitive endurance as a percent of the predicted effect of ability ranges from 25.5\%--38.7\%, depending on the outcome.

These results indicate that endurance has a sizable wage return in the labor market. Under complete information and frictionless markets, the price of a skill equals the present value of the future returns generated by the skill \citep{abraham2022}. Thus, the sizable wage return to endurance suggests that this skill is a key productivity determinant. The positive wage return to ability and endurance is consistent with models in which firms pay workers according to their productivity, and output is generated by combining ability with cognitive effort. Cognitive endurance enables workers to sustain effort for a longer time, allowing them to produce a higher total output. The results also reveal a novel type of assortative matching in the labor market: workers with high cognitive endurance are more likely to work for high-paying firms. This is relevant given that the sorting between workers and firms is an important driver of labor-market outcomes \citep{card2018firms}.

\subsection{Robustness and IV Estimates} 

Appendix \ref{app:robustness} presents a series of robustness and specification checks. The baseline results are robust to computing the effects nonparametrically, estimating ability and endurance with alternative specifications (e.g., with day or subject fixed effects), and imposing several sample restrictions (e.g., excluding the tails of the ability or endurance distribution).

Appendix Tables \ref{tab:reg-rob-coll-iv}--\ref{tab:reg-rob-lmkt-iv} display instrumental variables (IV) estimates of the effect of ability and endurance on long-run outcomes, estimated on the retakers sample. I instrument the year $t$ measure of ability/endurance with the year $t-1$ measure. Using repeated measures of a skill as an instrument is a common strategy to deal with measurement error in the literature \citep[e.g.,][]{gronqvist_intergenerational_2017,edin2022}. In addition, by using the year $t-1$ measures as instruments, this strategy eliminates the mechanical relationship between year $t$ ability/endurance and year $t$ test scores. 

Panel A report OLS estimates on the retakers sample. The effects are comparable to those estimated on the main sample. Panel B reports the IV estimates. In general, these tend to be larger than the OLS estimates. For example, the OLS estimate of the effect of a one SD increase of endurance [ability] on wages is 12.1\% [23.1\%], while the IV estimate is 18.8\% [25.0\%]. Relative to the OLS estimates, the IV estimates of the endurance effects tend to increase more than the ability effects, consistent with the endurance measure containing more measurement error than ability. Hence, the IV estimates suggest that the wage return to endurance---as a fraction of the return ability---is significantly higher, on the order of 75\%.

\subsection{The Value of Endurance across Degrees, Occupations, and Industries} 

Does the value of cognitive endurance vary across degrees, occupations, and industries? The task-based approach to labor markets highlights that workers produce output by performing job tasks, and tasks differ in their skill requirements \citep{acemoglu2011skills}. Consequently, the value of endurance should vary according to the tasks individuals have to accomplish in a given job and the importance of endurance in the production function of those tasks. For example, endurance may be particularly important for some jobs because mistakes due to attentional lapses can dramatically reduce the output value, as in ``O-ring'' production functions \citep{kremer1993ring}.
 
To assess this, I estimate the wage return to endurance separately for each degree, occupation, and industry. Intuitively, the wage return to endurance should reflect the increase in productivity due to an increase in this skill. Thus, a high wage return to endurance in a given occupation may indicate that this skill is particularly valuable in the production function of the tasks required by such an occupation.\footnote{There are two important caveats with this approach to measuring the value of endurance. The first one is that individuals may select into degrees, occupations, and industries partly based on their endurance. The second one is that an increase in productivity may not lead to a corresponding increase in wages in some occupations or industries due to institutional factors (e.g., collective bargaining).}

\begin{figure}[H]	
	\caption{Heterogeneity in the wage return to ability and cognitive endurance}\label{fig:endurance-het}
	\begin{subfigure}[t]{.45\textwidth}
		\caption*{Panel A. Distribution of wage returns across college degrees}
		\centering
		\includegraphics[width=\textwidth]{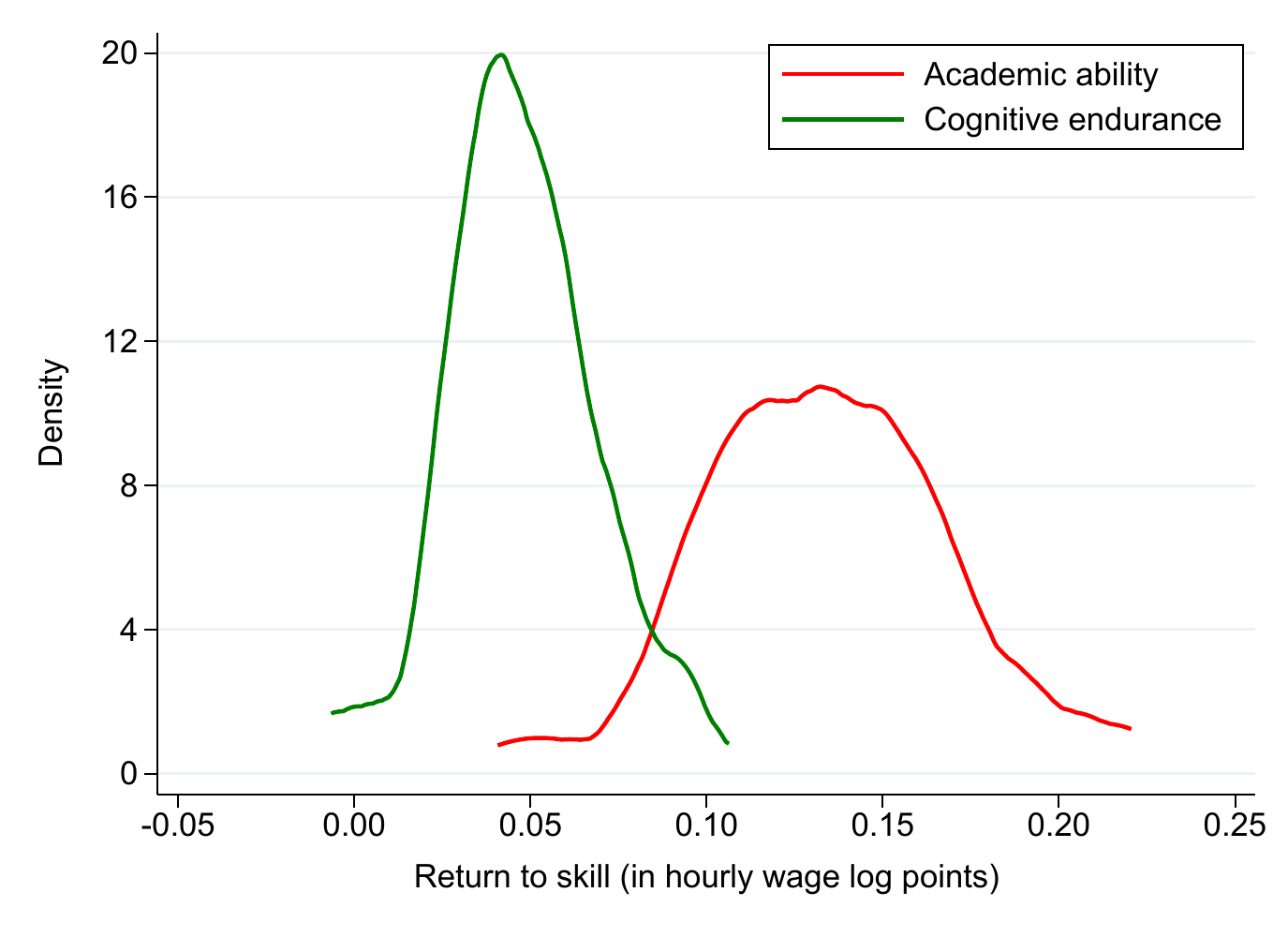}
	\end{subfigure}
	\hfill     
	\begin{subfigure}[t]{.45\textwidth}
		\caption*{Panel B. Return to ability/endurance vs. average wage across college degrees}
		\centering
		\includegraphics[width=\textwidth]{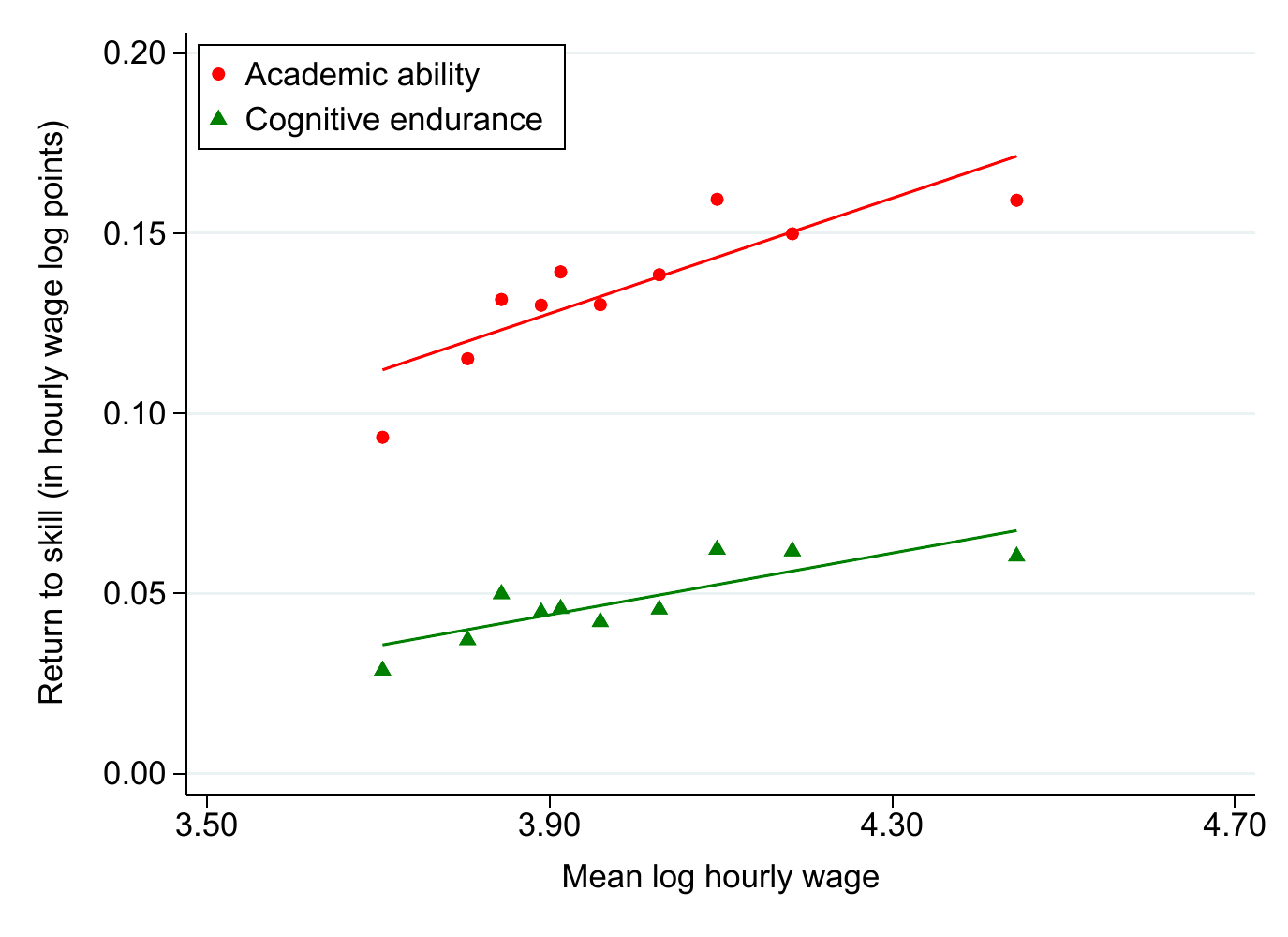}
	\end{subfigure} 
	\hfill       	
	\begin{subfigure}[t]{0.45\textwidth}
		\caption*{Panel C. Distribution of wage returns  across occupations}
		\centering
		\includegraphics[width=\textwidth]{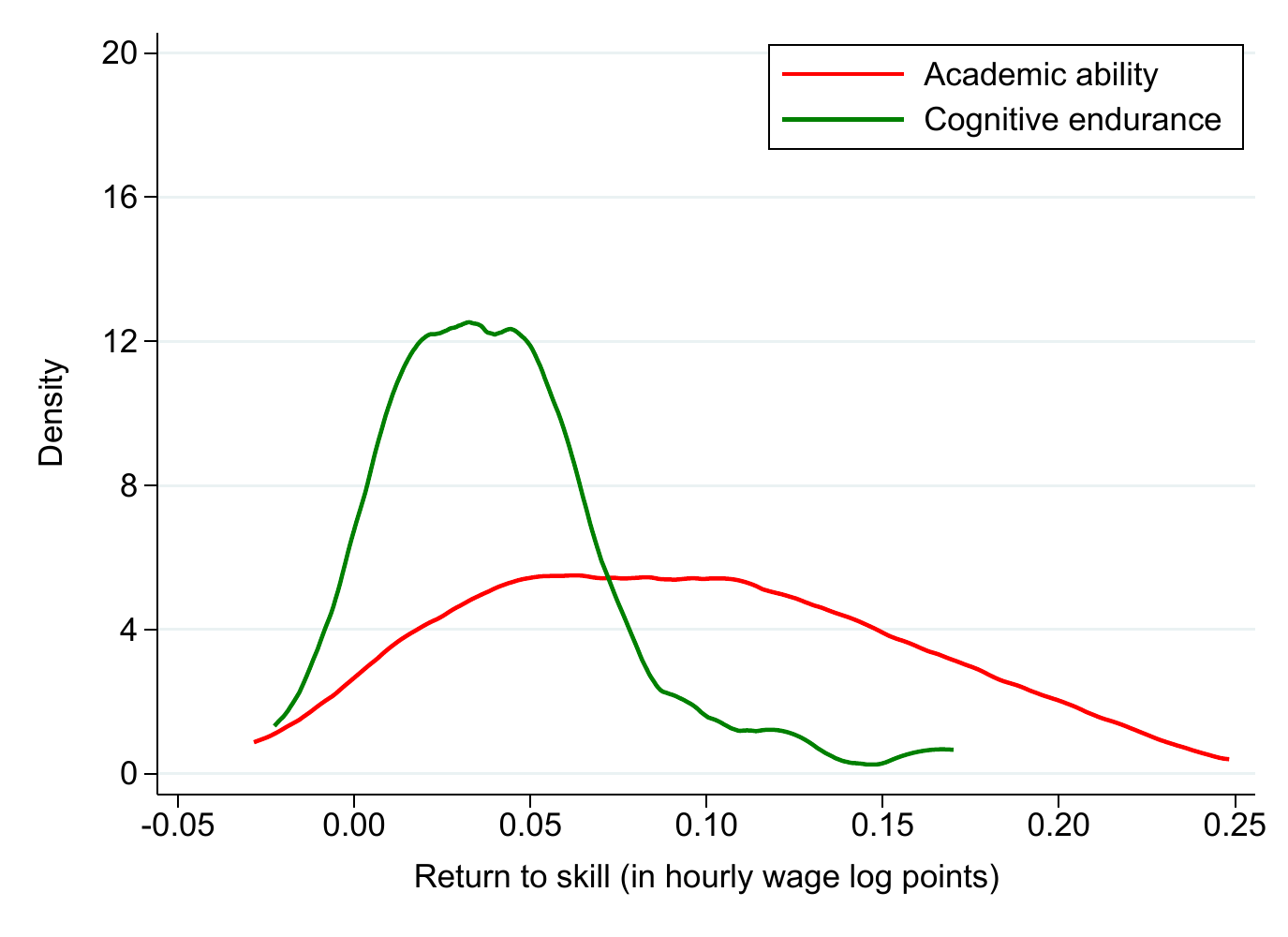}
	\end{subfigure}		
	\hfill        
	\begin{subfigure}[t]{.45\textwidth}
		\caption*{Panel D. Return to ability/endurance vs. average wage across occupations}
		\centering
		\includegraphics[width=\textwidth]{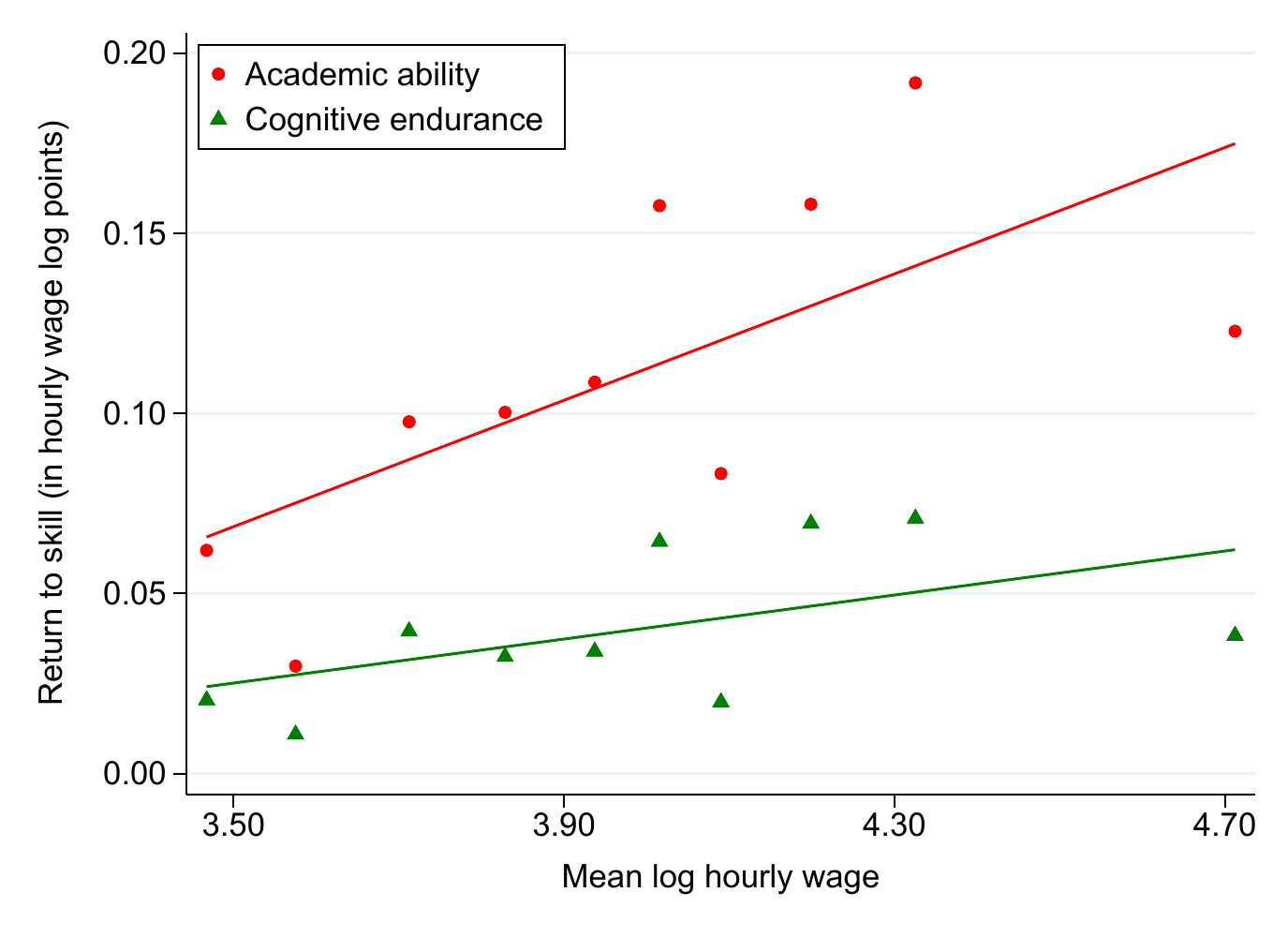}
	\end{subfigure}
	\hfill       	
	\begin{subfigure}[t]{0.45\textwidth}
		\caption*{Panel E. Distribution of wage returns across industries}
		\centering
		\includegraphics[width=\textwidth]{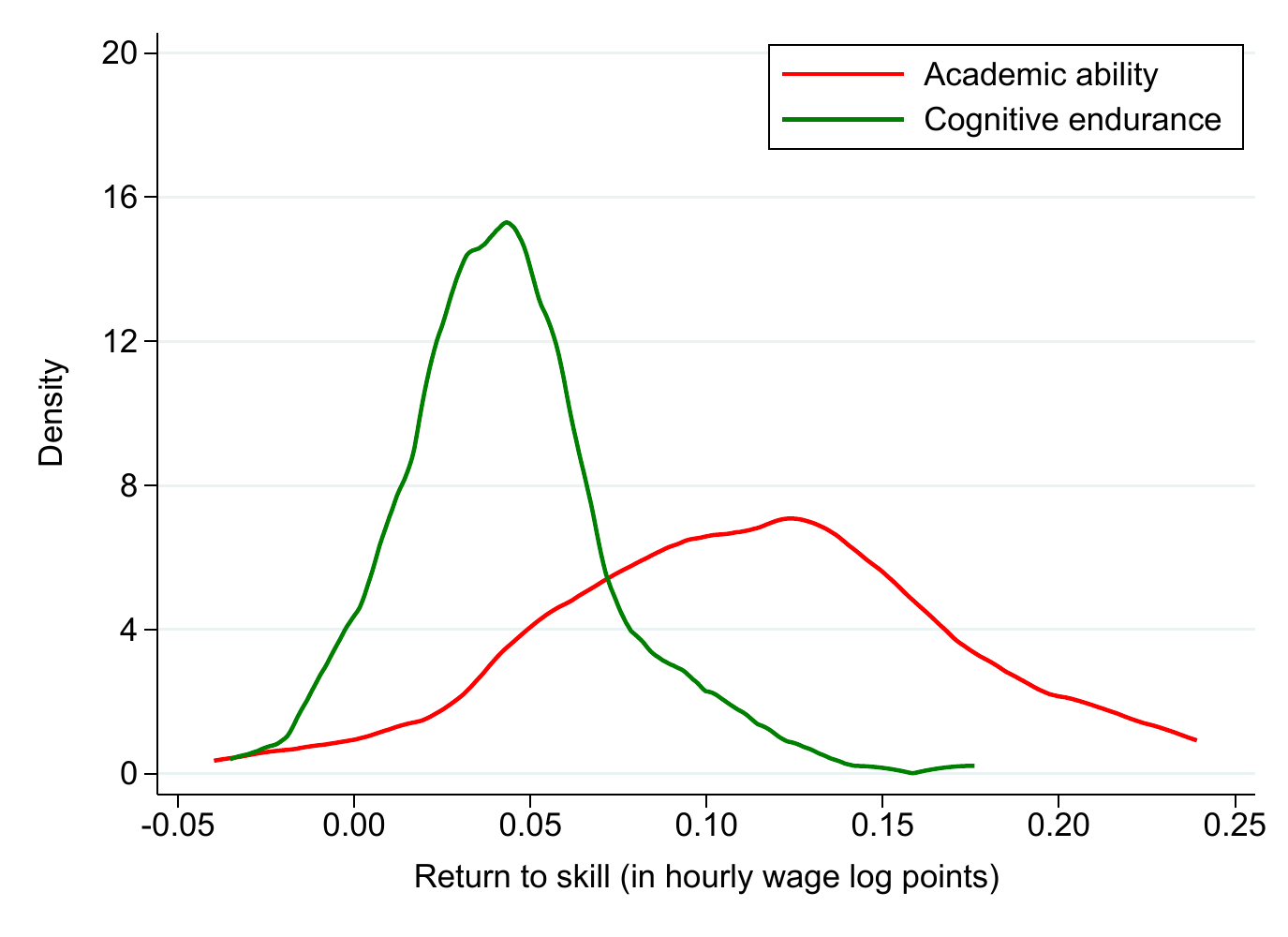}
	\end{subfigure}		
	\hfill        
	\begin{subfigure}[t]{0.45\textwidth}
		\caption*{Panel F. Return to ability/endurance vs. average wage across across industries}
		\centering
		\includegraphics[width=\textwidth]{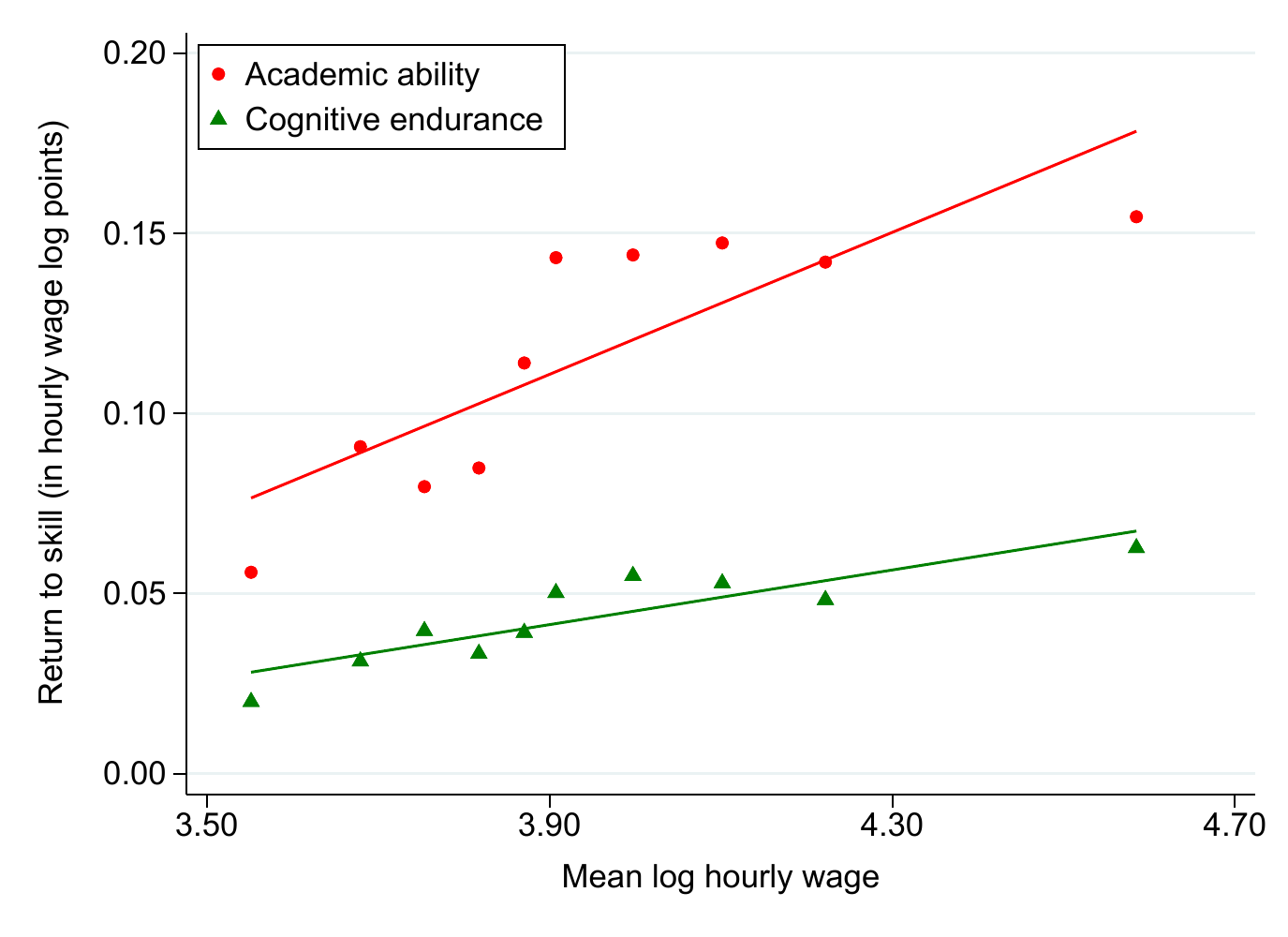}
	\end{subfigure}
	
	{\footnotesize
		\singlespacing \justify
		
		\textit{Notes:} Panels A, C, and E show nonparametric estimates of the distribution of the wage return to ability (red line) and the wage return to endurance (green line) across degrees, occupations, and industries. The wage return to ability and endurance are the coefficients $\psi_A$ and $\psi_E$ in equation \eqref{reg:endurance-outcomes} using log hourly wage as outcome, estimated separately for each degree, occupation, and industry. The figure excludes outliers (i.e., estimates of the returns below -0.05 or above 0.25).
		
		Panels B, D, and F display a series of binned scatterplots plotting the wage return to ability/endurance ($y$-axis) against the mean hourly wage in bins ($x$-axis). To construct this figure, I first divide degrees, occupations, and industries into 10 equally-sized bins based on their mean wage. Then, I estimate the average return to ability/endurance in each bin. Finally, I plot the average return to ability/endurance against the mean wage in each bin.
		
	}
	
\end{figure}

Figure \ref{fig:endurance-het} plots the distribution of wage returns across degrees (Panel A), occupations (Panel C), and industries (Panel E). There is substantial heterogeneity in the wage return to ability and to endurance. For example, while the average return to endurance across degrees is 4.9\%, the return across degrees in the bottom decile of the return distribution is 0.1\% and in the top decile is 9.8\%. This suggests that cognitive endurance is more valuable for success in some college degrees. Occupations and industries also exhibit substantial heterogeneity in wage returns.

Figure \ref{fig:endurance-het} also show that degrees, occupations, and industries that tend to pay higher average wages tend to offer higher returns to ability and endurance (Panels B, D, and F). For example, the return to endurance among the top ten percent highest-paying occupations is about three times higher than the return to endurance among the bottom-ten-percent-paying occupations (4.9\% vs. 1.6\%, respectively). This finding is consistent with high-paying jobs requiring high-endurance workers, suggesting that the value of this skill is higher in high-paying jobs. 

\begin{figure}[H]
	\caption{The relationship between the wage return to ability and endurance}\label{fig:ability-endurance-return}
	\begin{subfigure}[t]{.45\textwidth}
		\caption*{Panel A. Across college degrees}
		\centering
		\includegraphics[width=\textwidth]{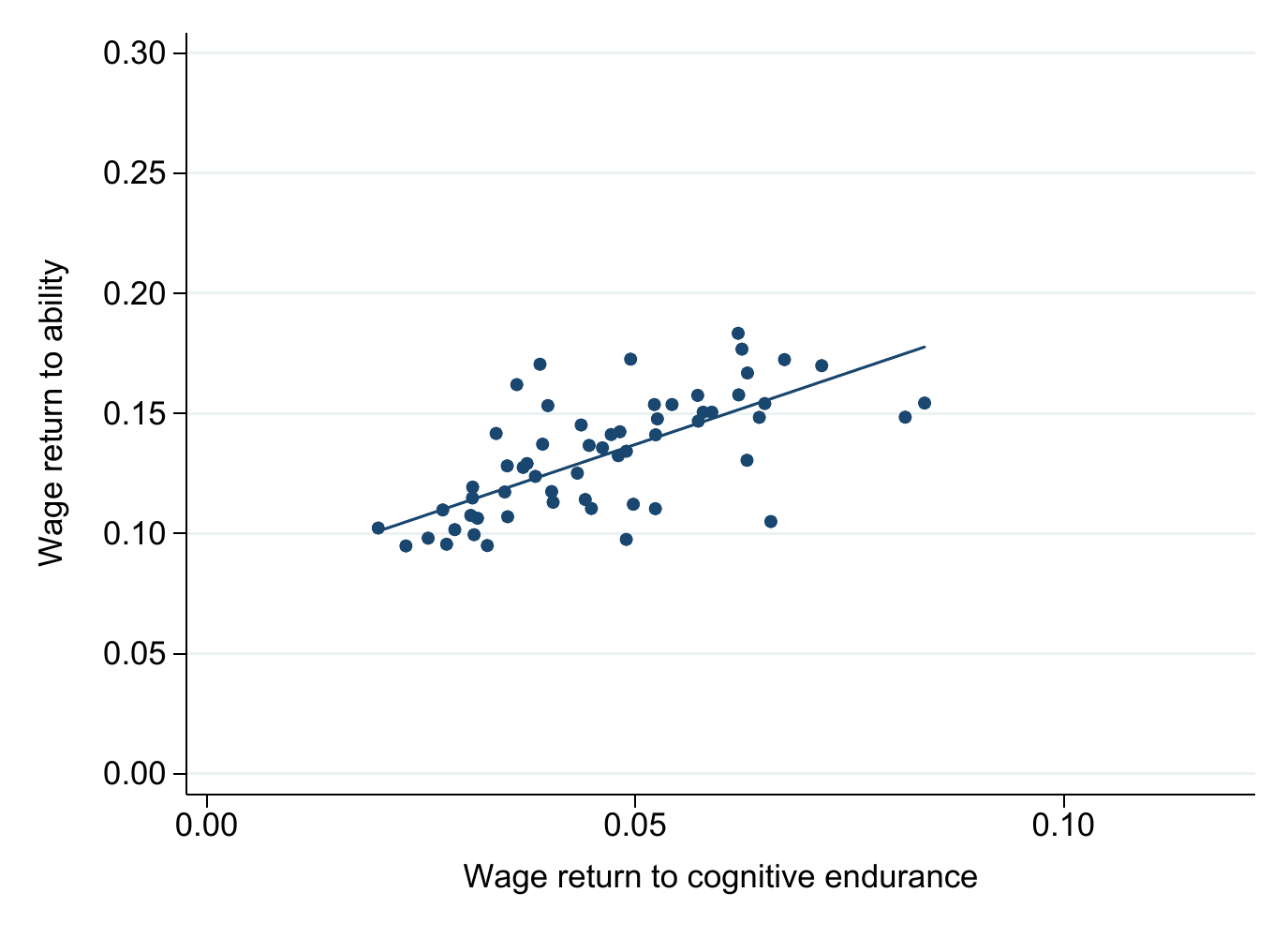}
	\end{subfigure}
	\hfill     
	\begin{subfigure}[t]{.45\textwidth}
		\caption*{Panel B. Across occupations}
		\centering
		\includegraphics[width=\textwidth]{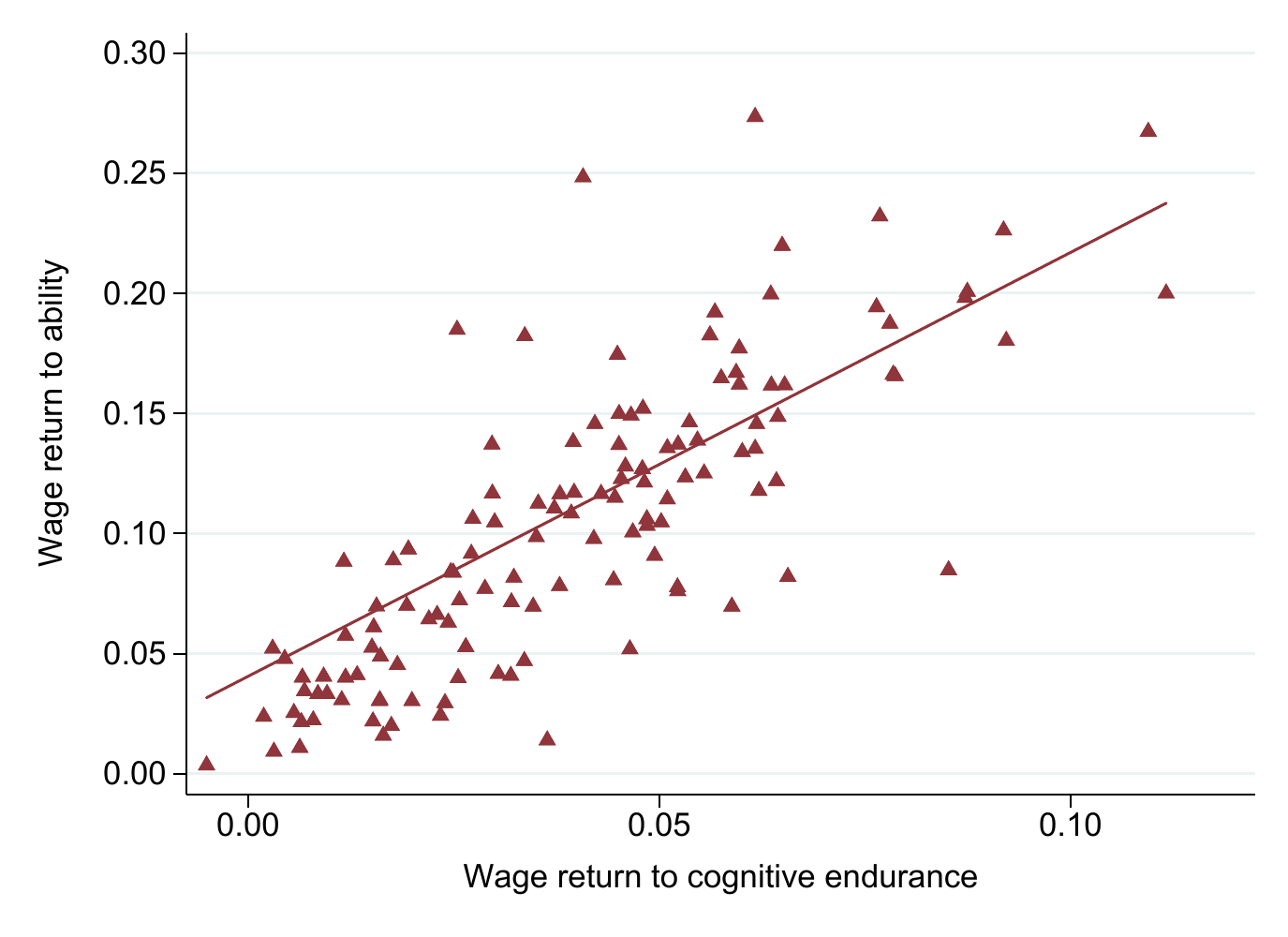}
	\end{subfigure}
	\hfill       	
	\begin{subfigure}[t]{0.45\textwidth}
		\caption*{Panel C. Across industries}
		\centering
		\includegraphics[width=\textwidth]{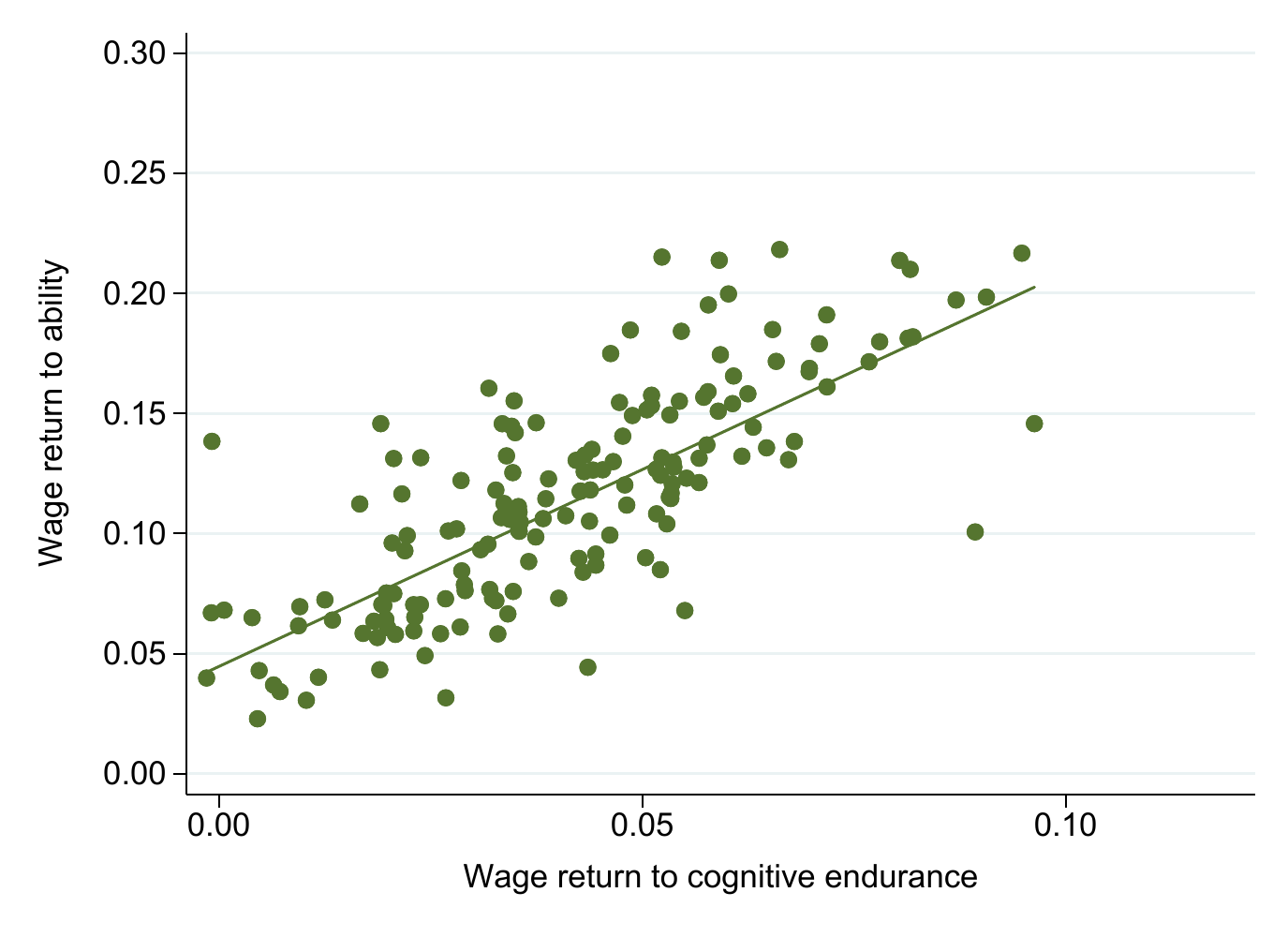}
	\end{subfigure}			
	\hfill        
	\begin{subfigure}[t]{0.45\textwidth}
		\caption*{Panel D. Binned scatterplot}
		\centering
		\includegraphics[width=\textwidth]{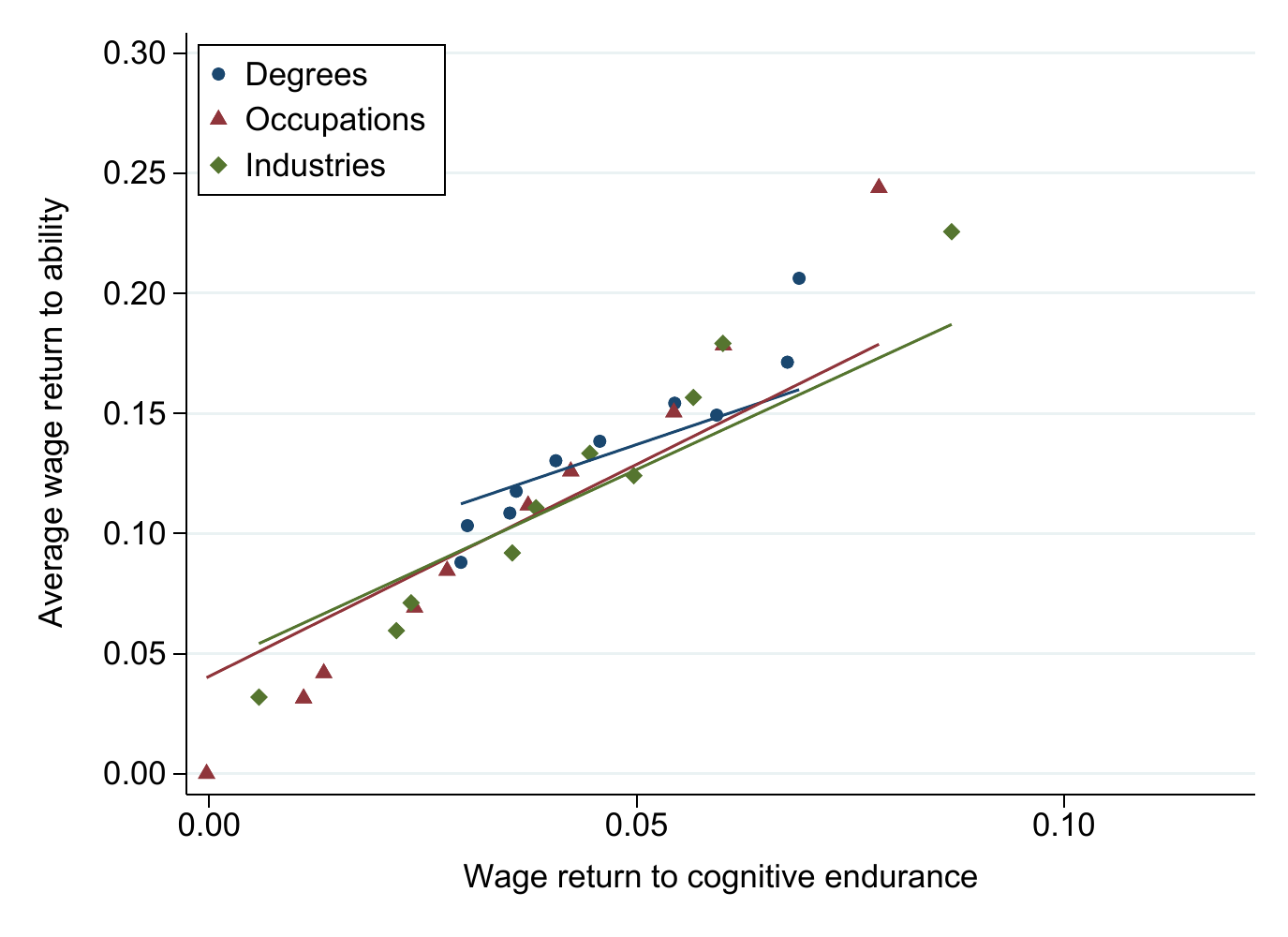}
	\end{subfigure}		
	\hfill        
	
	{\footnotesize
		\singlespacing \justify
		
		\textit{Notes:} This figure shows the relationship between the wage return to ability ($y$-axis) against the wage return to endurance ($x$-axis). Panels A--C show scatterplots of the wage return to ability in a given college degree (Panel A), occupation (Panel B), and industry (Panel C), against the wage return to endurance. The scatterplots exclude outliers (wage returns in the bottom 5\% or top 5\% of the distribution). The solid lines denote predicted values from linear regressions estimated on the microdata (including all observations).
		
		Panel D shows a binned scatterplot plotting the mean wage return to ability against the wage return to endurance. To construct this figure, I first divide degrees (blue circles), occupations (red triangles), and industries (green diamonds) into 10 equally-sized bins based on their wage return to endurance. Then, I calculate the average wage return to ability in each bin, using the number of individuals in each bin as weights. The solid lines denote predicted values from linear regressions estimated on the plotted points.		
		
	}
\end{figure}

Figure \ref{fig:ability-endurance-return} shows the joint distribution of the wage return to ability and the wage return to endurance across college degrees (Panel A), occupations (Panel B), and industries (Panel C). The figure reveals a strong association between the wage return to ability and the wage return to endurance across degrees, occupations, and industries. For example, on average, a 10\%-increase in the wage return to endurance across occupations predicts a 22.1\% increase in the wage return to ability ($p < 0.01$). This finding suggests that ability and endurance are complementary skills in production.


To make tangible some of the real-world tasks for which endurance may be particularly valuable, Table \ref{tab:reg-rank-return} list the top-five degrees, occupations, and industries with the highest wage return to endurance. The list includes occupations where attentional lapses may be extremely costly---such as facility operators in petrochemical plants or air navigation professionals---but also academically-oriented occupations, like mathematicians and statisticians (Panel B). The list also includes degrees conducive to these occupations (e.g., aeronautics, Panel A) and related industries (e.g., oil extraction, Panel C). While suggestive, this list is consistent with the proposition that the value of endurance depends on the type of tasks required by a job and the importance of endurance in the production function of those tasks. 


	\section{Endurance and the Sorting of Students to Colleges} \label{sec:equity-efficiency}

The sorting of students to colleges has important implications for the education and labor-market outcomes of these students \citep{macleod2017big}. In Brazil, as in many other countries, this sorting is largely mediated by admission exam scores. My results indicate that test scores reflect two valuable but distinct skills: ability and endurance. An important question is how these two skills contribute to the sorting of students to colleges. 

In this section, I estimate the effects of an exam reform that reduces the exam length by half. Such a reform would decrease the relative weight of endurance for determining test scores and increase the relative weight of ability. I focus on two channels through which the reform could affect the sorting of students to colleges. First, I study the \textit{distributional} effects of the reform, that is, how the reform would affect test-score gaps due to systematic differences in average endurance across types of students. Second, I study the \textit{informational} effects of the reform, that is, how the reform would affect the information on the quality of each applicant contained in test scores due to a change in the skills being measured by the exam.

\subsection{Cognitive Endurance and Test-Score Gaps} \label{sub:correlates}

Standardized tests often exhibit large racial and income {test-score gaps} \citep[e.g.,][]{fryer2006black, card2007racial, riehl2022}. In the context of college admission exams, these gaps lead to inequitable college access and amplify earnings disparities \citep{chetty2020income}. An important question is what explains those test-score gaps. Next, I examine the contribution of differences in cognitive endurance to these gaps. 

\subsubsection{Decomposing Test-Score Gaps.}

To begin with, notice that the linear decomposition \eqref{reg:lpm-ind} can be used to parsimoniously summarize an individual's test score, $\text{TestScore}_i \equiv \E[C_{ij}]$, as a linear combination of her academic ability and endurance:

\begin{align} \label{eq:test-score}
	\text{TestScore}_i  = \hat{\alpha}_i + \hat{\beta}_i \overline{\text{Position}}.
\end{align}

Let $X \in \{0, 1\}$ be a student observable characteristic. For example, $X = 1$ may denote high-income students and $X = 0$ low-income students. The average test score of students with characteristic $x$ can be written as
\begin{align} \label{eq:test-score-grp}
	\E[\text{TestScore}_i | X_i = x]  = \E[\hat{\alpha}_i | X_i = x]  +  \E[\hat{\beta}_i | X_i = x] \overline{\text{Position}}.
\end{align}

Using this expression, the test-score gap, $\text{ScoreGap}$, can be decomposed into differences in average academic ability and differences in average cognitive endurance as follows:			
\begin{align} \label{eq:oaxaca}
	\text{ScoreGap} =  
	\underbrace{{\alpha}_{1} - {\alpha}_{0}}_{\substack{\text{Difference in average} \\ \text{academic ability} \\ \text{between groups}}} + 
	\underbrace{({\beta}_{1} - {\beta}_{0}) \overline{\text{Position}}}_{\substack{\text{Difference in average} \\ \text{cognitive endurance} \\ \text{between groups}}},
\end{align}
where $\alpha_x \equiv \E[\hat{\alpha}_i | X_i = x]$ and $\beta_x \equiv \E[\hat{\beta}_i | X_i = x]$. Equation \eqref{eq:oaxaca} shows that, in the absence of systematic differences in limited endurance ($\beta_1 = \beta_0$), test scores gaps would be purely a reflection of gaps in academic ability. Thus, exam design features that put a higher or lower weight on endurance, such as the length of the exam or the number of breaks, should not affect test-score gaps. This is no longer true in the presence of systematic differences in endurance. If student-level characteristics are associated with endurance, then an exam design that puts more weight on endurance will affect test-score gaps.  

I focus on estimating the impact of an exam reform that decreases the length of the test by half on test-score gaps.\footnote{Such a reform would be equivalent to changing the ENEM from its current length to roughly the length of the ACT exam.} This reform would decrease the average question position (from $\overline{\text{Position}}$ to $\overline{\text{Position}}/2$), thereby decreasing the influence of endurance gaps on test-score gaps.\footnote{An important concern is that, by reducing the number of questions, the exam would determine the place in the score distribution of any one student with less precision. However, the reform could be achieved without sacrificing much precision by using an adaptive exam that selects questions based on the student's ability level.} While I focus on test length, other exam features can also affect the influence of endurance for test-score gaps. For example, Figures \ref{fig:mean-corr}--\ref{fig:mean-corr-res} show that student performance starkly increases between the end of the first day and the beginning of the second day, suggesting that giving students more breaks would decrease the importance of endurance for test-scores gaps. Similarly, in Appendix \ref{app:het-question}, I show that the \textit{type} of exam questions impacts cognitive fatigue, thereby influencing the importance of endurance for test-score gaps. Thus, the reform can also be interpreted as, for example, introducing a long break in the middle of each testing day. 


Using equation \eqref{eq:test-score}, I estimate the effects of the reform on test-score gaps between:

\begin{enumerate}
	\item Male and female students,
	\item White and non-white (Black, Brown, and Indigenous) students,
	\item Students in households in the top 30\% and bottom 30\% of the income distribution,
	\item Students with a college-educated mother and non-college-educated mother,
	\item Students enrolled in a private high school and public high school.
\end{enumerate}

\subsubsection{The Impact of an Exam Reform on Test-Score Gaps.}

Table \ref{tab:corr-endurance} shows estimates of the contribution of gaps in ability and endurance to test-score gaps. Column 1 shows the difference in average test scores between the groups of students listed in the row header, column 2 shows the difference in average academic ability (in a regression that controls for endurance), and column 3 shows the difference in average cognitive endurance (controlling for ability).

By reducing the contribution of endurance gaps to test-score gaps by half, the reform would: (i) Reduce the gender test-score gap by 0.85 percentage points (a 32\% decrease from the pre-reform gap of 2.6 percentage points); (ii) Reduce the racial test-score gap by 0.08 percentage points (a 14\% decrease from the pre-reform gap of 5.7 percentage points); and (iii) Reduce the SES test-score gap by 1.3--3.1 percentage points (a 13\%--16\% decrease from pre-reform gaps), depending on the SES measure.

The predicted impact of the exam reform is robust to (i) measuring the gaps in percentiles (Appendix Table \ref{tab:corr-rob-pctil}); (ii) Estimating ability and endurance with alternative specifications (Appendix Table \ref{tab:corr-rob-meas}); (iii) Using alternative measures of question difficulty when estimating ability and endurance (Appendix Table \ref{tab:corr-rob-diff}); (iv) Excluding individuals in the tails of the ability or endurance distributions (Appendix Table \ref{tab:corr-rob-samp}); and (v) Using precision-weighted estimates (Appendix Table \ref{tab:corr-rob-prec}).

\subsection{Cognitive Endurance and the Predictive Validity of Test Scores} \label{sub:fatigue-informational}

Admission officers use test scores to screen applicants because they are {informative} about which applicants will succeed in college. The standard approach to assess the informative content of an exam test is to calculate the cross-individual correlation between test scores and a long-run outcome that colleges want to screen their applicants based on (such as first-year college GPA or on-time graduation). This correlation is known as the \textit{predictive validity} of an exam \citep{rothstein2004college}. Next, I study how an exam's predictive validity depends on cognitive endurance.

\subsubsection{Decomposing Predictive Validity.} 

In the presence of limited endurance, features of the exam that affect the weight of endurance will affect the exam's predictive validity. To see this, notice that the predictive validity of test scores for outcome $Y$, $\rho^Y$, can be written as:
\begin{align} \label{eq:pred-val}
	\rho^Y  &\equiv \text{Corr}(Y_i, \text{TestScore}_i)   \notag \\
	        &= \text{Corr}(Y_i,  \alpha_i + \beta_i \overline{\text{Position}}) \notag \\
	        &= \frac{\sigma_{\alpha}}{\sigma_T} \text{Corr}(Y_i,\alpha_i) +  \frac{\sigma_{\beta}}{\sigma_T}\text{Corr}(Y_i,\beta_i)\overline{\text{Position}},
\end{align}
where $\sigma_{\alpha}, \sigma_{\beta}, \sigma_T$ are the standard deviations of ability, endurance, and test scores. Equation \eqref{eq:pred-val} shows that the predictive validity of an exam can be expressed as a linear combination of the correlation between the outcome and the skills measured by the exam. The weight of each skill depends on its dispersion and the exam's length. Unfortunately, equation \eqref{eq:pred-val} cannot be directly used to predict how an exam reform that changes the test length would affect its predictive validity, i.e.,  $\partial \rho^Y/\partial\overline{\text{Position}}$. This is because, as shown in Section \ref{sub:correlates}, such a reform would change the ranking of students, thereby affecting $\text{Corr}(Y_i,\alpha_i)$ and $\text{Corr}(Y_i,\beta_i)$.

I sidetrack this problem by studying how predictive validity varies throughout the exam. In particular, I ask how a given question's predictive validity $\rho^Y_j \equiv \text{Corr}(Y_i, C_{ij})$, changes when its position changes. Notice that the predictive validity of the overall exam can be written as a weighted average of the predictive validity of each exam question $j \in \{1, ..., J\}$:
\begin{align} \label{eq:pred-val-decomp}
	\rho^Y = \frac{1}{J} \sum_{j=1}^J  \frac{\sigma_{C_{ij}}}{\sigma_T} \rho^Y_j.
\end{align}
Equation \eqref{eq:pred-val-decomp} is helpful because it allows me to exploit the random variation in whether a given question is presented when students are relatively fresh or cognitively fatigued. To build intuition on the mechanics of this analysis, notice that $\rho^Y_j$ is given by the difference in average outcomes between students who correctly and incorrectly responded to question $j$, multiplied by the ratio of standard deviations:
\begin{align} \label{eq:pred-val-exp}
	\rho^Y_j = \Big(\E[Y_i | C_{ij} = 1] - \E[Y_i | C_{ij} = 0] \Big) \frac{\sigma_{C_{ij}}}{\sigma_{Y_{i}}}.
\end{align}

Equation \eqref{eq:pred-val-exp} indicates that limited endurance affects the predictive validity of a question by changing the composition of students who correctly answer the question. Loosely speaking, correct responses at the beginning of the exam are driven by high-ability students (regardless of their endurance level since all students are ``fresh'') and low-ability students who guessed the answer. Toward the end of the exam, correct responses are driven by students with high ability \textit{and} high cognitive endurance and students with either low ability \textit{or} low endurance who guessed the answer. 

How this compositional change affects a question's predictive power for an outcome is theoretically ambiguous. It depends on how the outcome correlates to academic ability relative to cognitive endurance, the distribution of ability and endurance in the population, and how difficult the question is. Thus, I empirically assess this by estimating regressions of the form:
\begin{align} \label{reg:validity-pos}
	\rho^Y_{jb} &= \alpha_j + \gamma^Y \text{Position}_{jb} + \eta_{jb}.
\end{align}
where $\rho^Y_{jb}$ is the predictive validity of question $j$ in booklet $b$, $\alpha_j$ are question fixed effects, and $\text{Position}_{jb}$ is the position of question $j$ in booklet $b$. The coefficient of interest is $\gamma^Y$, which measures the impact of a one-position increase in the order of a given question on the question's predictive validity for outcome $Y$. I scale $\gamma^Y$ so that it represents the change in predictive validity due to a reform that decreases the average question position by half.

I estimate the effect of the reform for eight main outcomes: test score (calculated without the contribution of question $j$ to avoid mechanical effects), college enrollment, college quality, degree progress, six-year graduation rate, hourly wage, monthly earnings, and firm leave-individual-out mean earnings. Since the dependent variable is an estimate, I weight each observation using the inverse square of its standard error. I cluster standard errors at the question position level.


\subsubsection{The Impact of an Exam Reform on the Test's Predictive Validity.} 

Table \ref{tab:val-pos} presents the results. Panel A shows the average predictive validity across all test questions. I find that individual questions are predictive of student long-run outcomes, although the size of the correlations tends to be small. For example, on average across all questions, correctly responding to a question has a 0.05 positive correlation with enrolling in college (column 2, $p < 0.01$), 0.10 correlation with college quality (column 3, $p < 0.01$), and 0.10 correlation with wages and earnings (columns 6--7, $p < 0.01$).

Panel B reports the estimates of equation \eqref{reg:validity-pos}. The exam reform would generate a sizable increase in the predictive validity of the exam for the majority of outcomes. For example, the exam reform would increase the average predictive validity of test responses for college enrollment by 0.05 points (a 95.2\% increase relative to the pre-reform mean, $p < 0.01$), for college quality by 0.09 points (a 91.1\% increase relative to the pre-reform mean), and for earnings by 0.07--0.08 points (a 75.7\%--79.6\% increase relative to the pre-reform mean, $p < 0.01$). The predicted effect for the six-year graduation rate is also positive but not statistically different from zero. The reform would decrease the predictive validity for degree progress.

These estimated effects of the reform are driven by the fact that a given question tends to be less predictive of long-run outcomes if it appears later in the exam. This can be seen visually in Appendix Figure \ref{fig:pred-val-chg-pos}, which shows binned scatterplots plotting the change in the predictive ability of a question ($y$-axis) against the change in the question position ($x$-axis) for selected outcomes. In all cases, the average predictive validity of test questions tends to decrease if the question appears later in the test.\footnote{The negative association can also be seen in Appendix Figure \ref{fig:pred-val-pos}, which plots the predictive validity of a question ($y$-axis) against its position on the exam ($x$-axis). There is an evident decline in predictive validity between the exam's beginning and end.}


These results can help explain puzzling empirical findings in the literature. \cite{kobrin2008validity} study how the predictive validity of the SAT changed after the exam increased the number of questions in 2005. Intuitively, more test questions should lead to more precise student ability estimates and thus to more predictive test scores. Yet, the predictive validity of the exam did not change. This finding can be explained by cognitive fatigue eroding the predictive power of test responses at the end of the exam. \cite{bettinger2013improving} show that performance on the English and Math sections of the ACT predict college outcomes, but not performance on the Science and Reading sections. Based on this finding, the authors propose eliminating the Science and Reading sections. Notably, the Science and Reading questions are the last to appear in the ACT. Hence, the null predictive power of these two subjects may be driven by students being cognitively fatigued by the time they reach those sections---and not by the skills assessed by Science and Reading questions being irrelevant for long-run outcomes.\footnote{If the goal is to maximize the exam's predictive power, an alternative reform that might be more desirable is to reduce the number of questions in all sections (instead of eliminating some sections). Cognitive fatigue plays a smaller role in shorter tests, allowing students to reveal their academic preparedness across all questions. By not removing any subjects, colleges would have a measure of academic preparedness for all topics. I discuss the implications of my findings for exam design in Section \ref{sec:conclusions}.} 

\subsection{Discussion} 

In summary, this section shows that, due to heterogeneity in cognitive endurance, the design of college admission exams can have equity and efficiency consequences. I estimate that a reform that halves the exam length would reduce SES gaps by 26\%--29\%, possibly leading to a more diverse college student body. In addition, the shorter exam would be more informative about the quality of each applicant (as measured by its predictive validity), possibly leading to a better allocation of students to colleges. The first result is driven by the fact that, conditional on academic ability, low-SES students have lower endurance than high-SES students; thus, their performance declines at a steeper rate throughout the exam. The second result is driven by the fact that differences in student performance at the beginning of the exam mainly reflect differences in ability (roughly, because most students are ``fresh''), whereas performance differences towards the end of the exam increasingly reflect differences in endurance. Since ability is a stronger predictor of long-run outcomes than endurance, the predictive validity of a given question decreases if it appears later on the exam.
 
	\section{Conclusion} \label{sec:conclusions}


Cognitive effort underlies most, if not all, productive activities. Just like individuals differ in preferences and personality traits, they also differ in their ability to endure mental fatigue. This paper shows that cognitive endurance affects student performance in college admission exams and has a substantial earnings return in the labor market.

My findings have implications for investments in different types of human capital. I find that endurance has a substantial return in the labor market. Yet, a typical school curriculum does not include any material directly aimed at building this dimension of human capital. Policymakers should consider investing in the development of cognitive endurance, possibly during early ages when neuroplasticity is higher. While research in this area is still in its infancy, some examples of protocols that build cognitive endurance include mindfulness meditation \citep{levy2012effects, goleman2017altered}, noninvasive brain stimulation \citep{rubia2018cognitive}, and the restriction of smartphones in learning environments \citep{thornton2014mere}.\footnote{Some of these protocols are already being implemented in the private sector. For example, meditation practices are commonly used among tech companies in Silicon Valley to enhance worker productivity \citep{shachtman2013silicon}.} An important caveat is the lack of exogenous variation in cognitive endurance in my analysis. While my findings provide evidence of a positive link between endurance and earnings, these estimates may be misleading if the available control variables are inadequate to provide meaningful estimates of the return to this skill. Future work should establish if this link is causal. 




The findings also have implications for the design of standardized tests. In a typical test, all questions contribute the same to an individual's score. However, the results show that questions that appear early in the exam are more predictive of long-run outcomes. An aggregation mechanism of individuals' test responses that takes into account students' varying fatigue levels throughout the exam may lead to more informative test scores. For example, testing agencies can weight each question based on the position in which it was answered, assigning more weight to questions students responded to earlier. 

The ability-endurance score decomposition developed in this paper generates directions for future research. Test scores are commonly used in economics research, for example, as measures of cognitive skills \citep{hanushek2008role, hanushek2012better}; as a ``surrogate'' variable to measure the impact of an intervention on long-run outcomes \citep{athey2019surrogate}; and to measure the effectiveness of educational inputs \citep{chetty2011does, dobbie2015medium, angrist2016stand}. The decomposition allows researchers to explore the role of cognitive endurance in these and other applications. For example, researchers can use conventional value-added methods to identify teachers who might be particularly effective at building cognitive endurance. 

	
    \clearpage
\section*{Figures and Tables}

\clearpage 
\begin{figure}[H]
	\caption{Performance residuals after controlling for question difficulty}\label{fig:mean-corr-res}
	\centering
	\includegraphics[width=.75\linewidth]{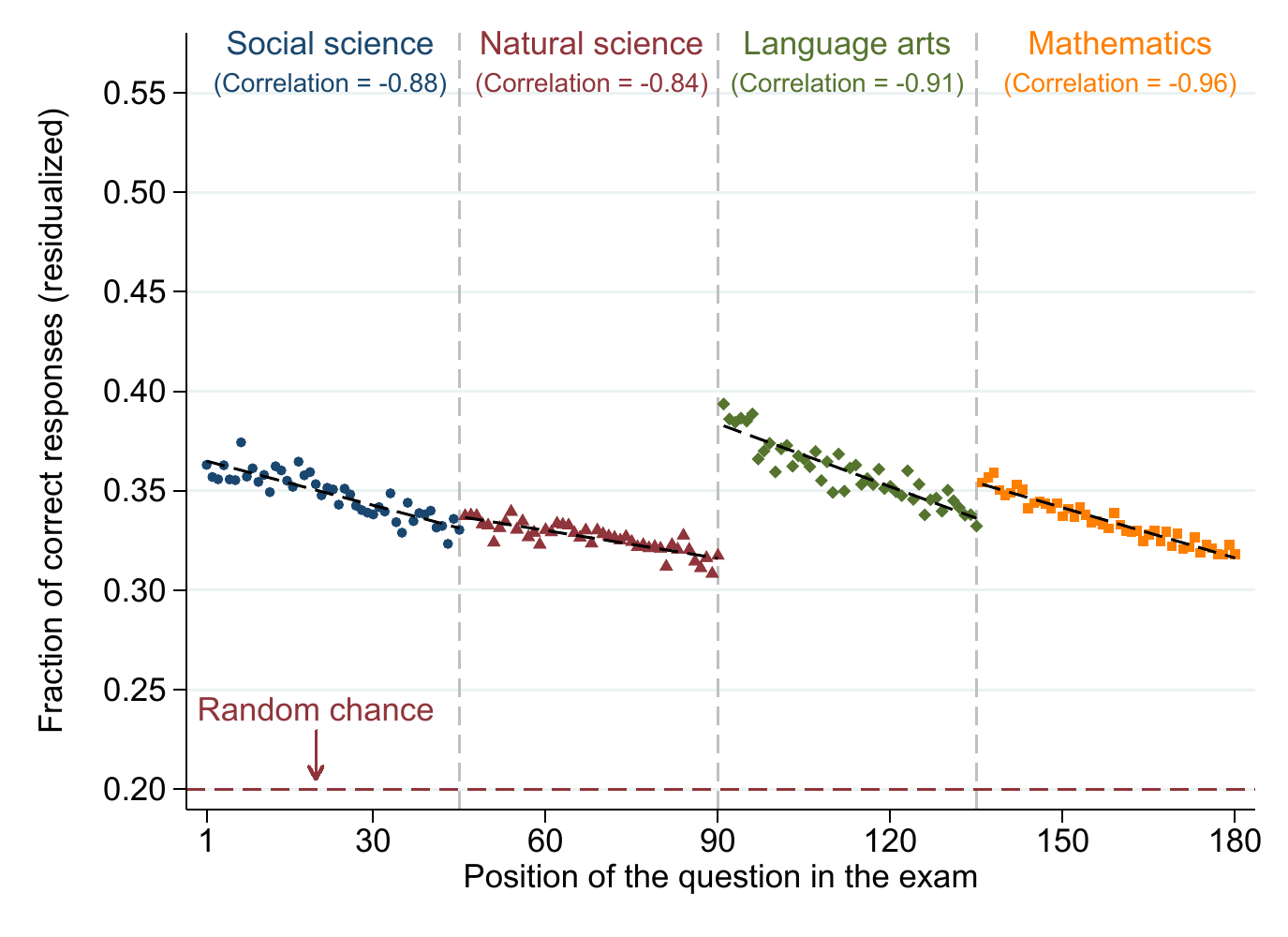}

	{\footnotesize
		\singlespacing \justify
		
		\textit{Notes:} This figure shows student performance over the course of each testing day after removing the influence of question difficulty on performance. The $y$-axis displays the residuals of a regression of (i) $\bar{C}_{jb}$, the fraction of students who correctly answered question $j$ in booklet $b$ on (ii) $\text{Difficulty}_{j}$, a  position-adjusted measure of question difficulty (adding back the sample mean to facilitate interpretation of units). The $x$-axis displays the position of each question in the exam. Marker colors denote each academic subject tested. Appendix \ref{app:difficulty} describes how I construct the measure of question difficulty. The dashed lines are predicted values from a linear regression estimated separately for each academic subject. The horizontal red dashed line shows the expected performance if students randomly guessed the answer to each question.
		
	}
\end{figure}

\clearpage

\clearpage

\begin{figure}[H]
	\caption{Change in a question's position and change in predictive validity}\label{fig:pred-val-chg-pos}
	\begin{subfigure}[t]{.45\textwidth}
		\caption*{Panel A. Test score (leave-question-out)}
		\centering
		\includegraphics[width=\textwidth]{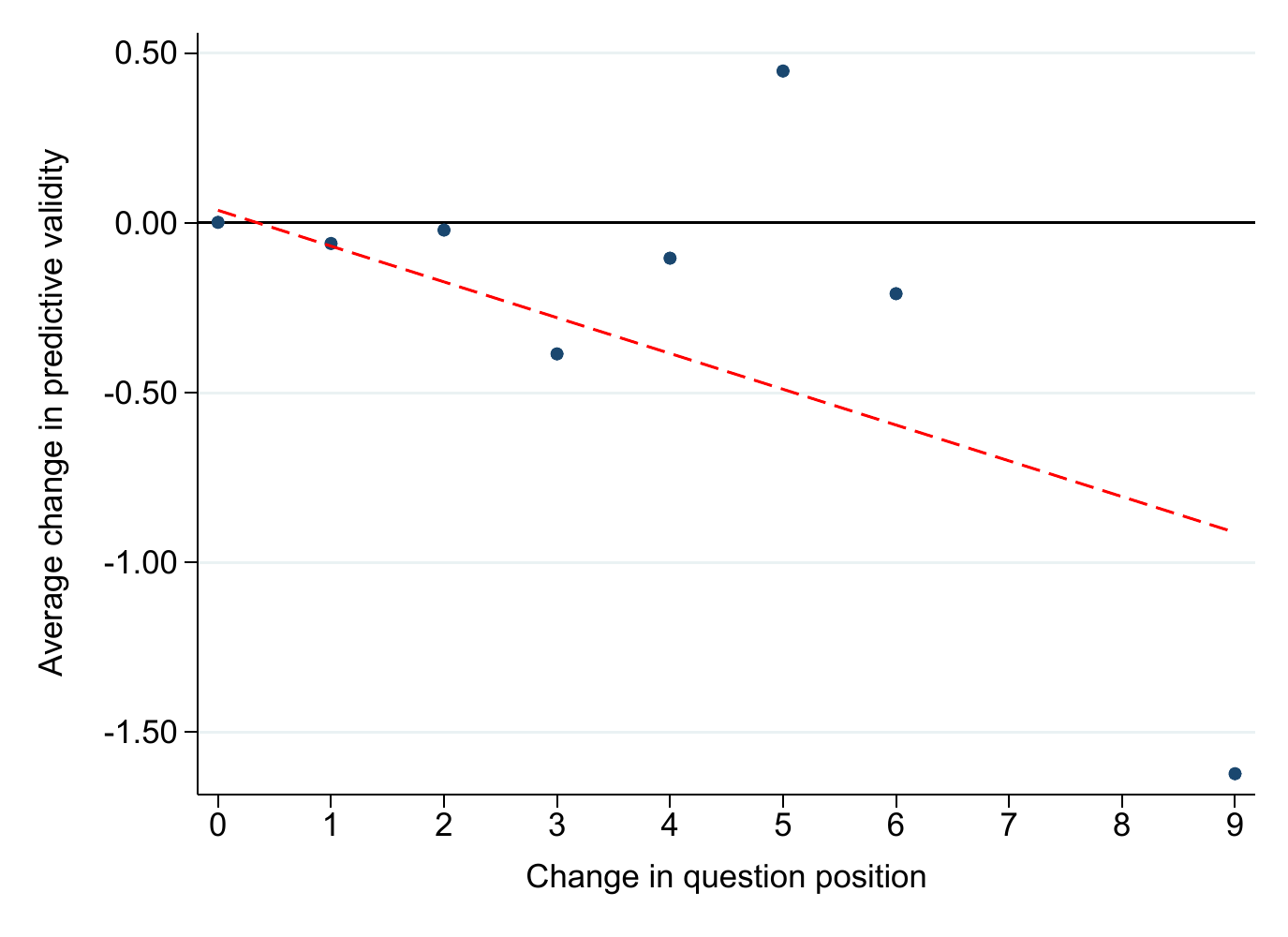}
	\end{subfigure}
	\hfill     
	\begin{subfigure}[t]{.45\textwidth}
		\caption*{Panel B. College enrollment}
		\centering
		\includegraphics[width=\textwidth]{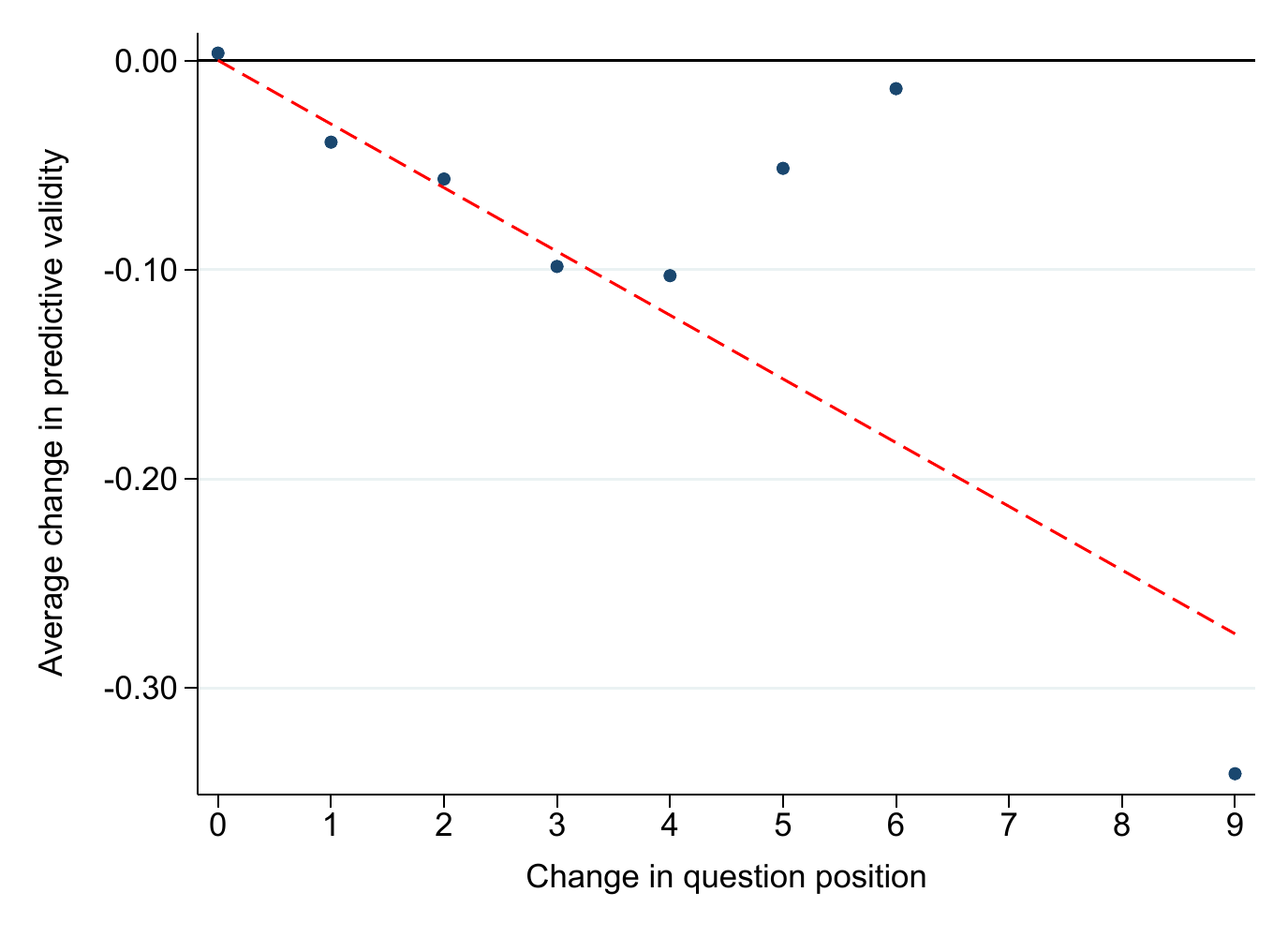}
	\end{subfigure}
	\hfill       	
	\begin{subfigure}[t]{0.45\textwidth}
		\caption*{Panel C. College quality}
		\centering
		\includegraphics[width=\textwidth]{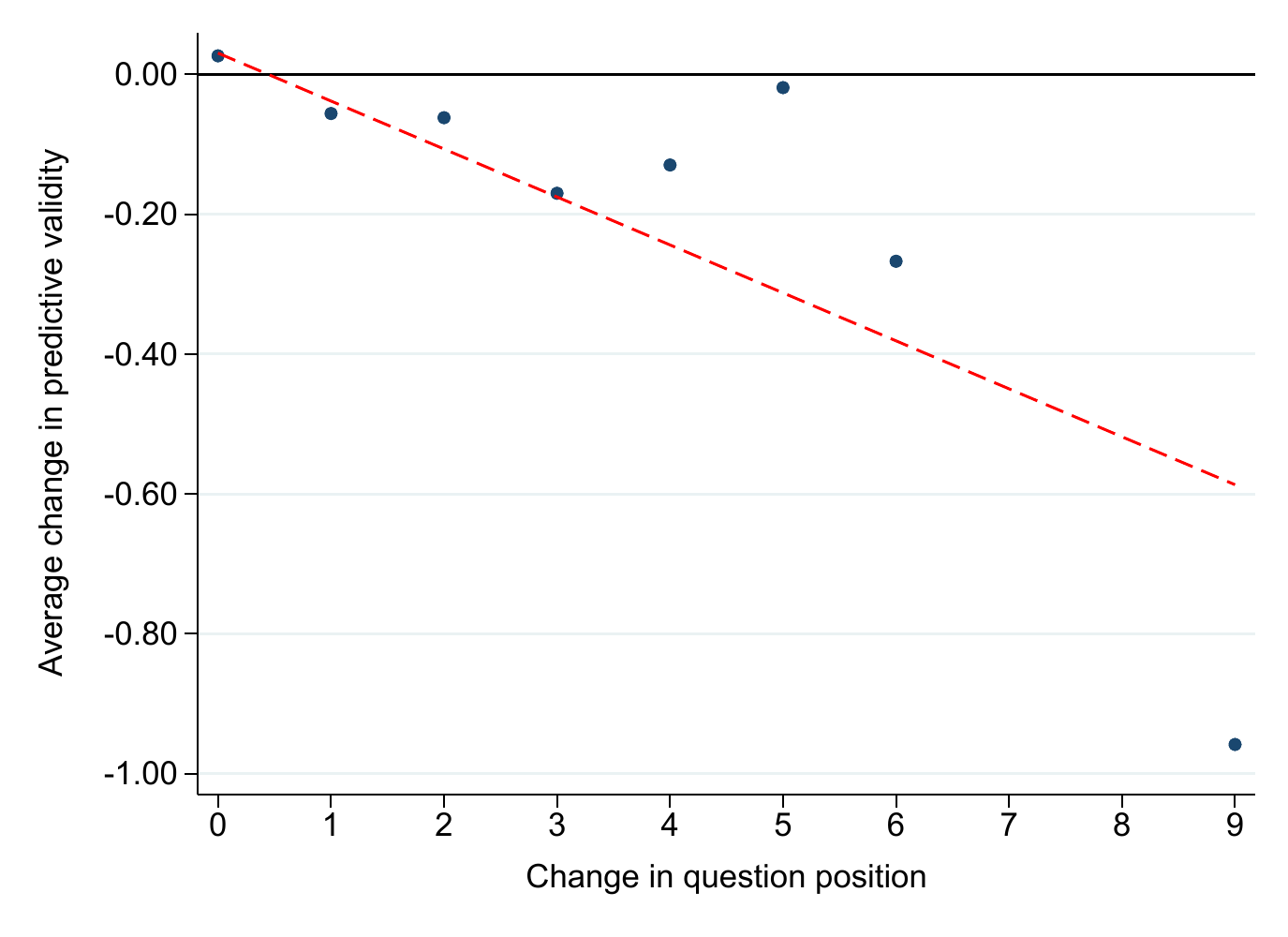}
	\end{subfigure}			
	\hfill        
	\begin{subfigure}[t]{0.45\textwidth}
		\caption*{Panel D. Hourly wage}
		\centering
		\includegraphics[width=\textwidth]{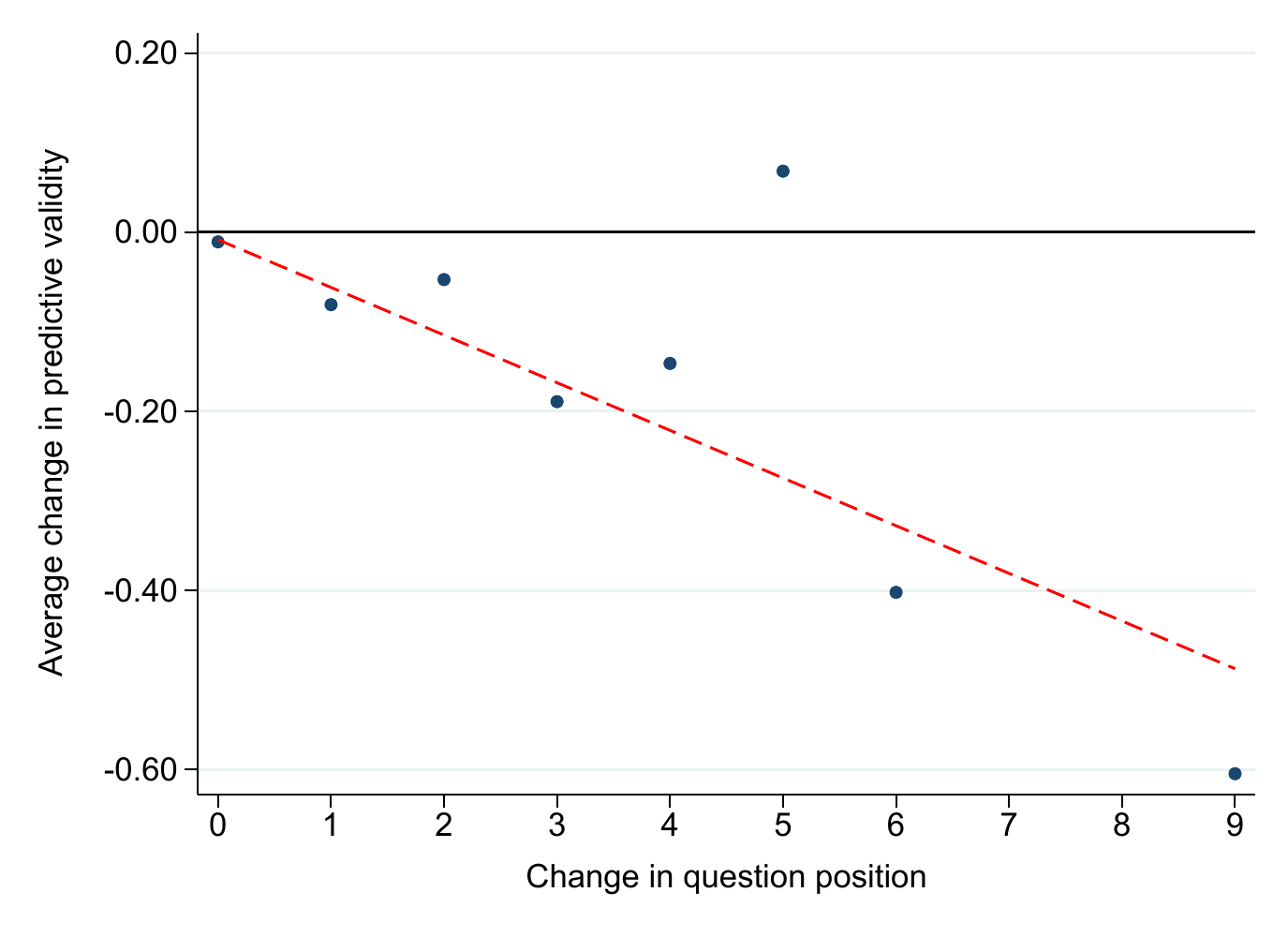}
	\end{subfigure}		
	\hfill        
	\begin{subfigure}[t]{.45\textwidth}
		\caption*{Panel E. Monthly earnings}
		\centering
		\includegraphics[width=\textwidth]{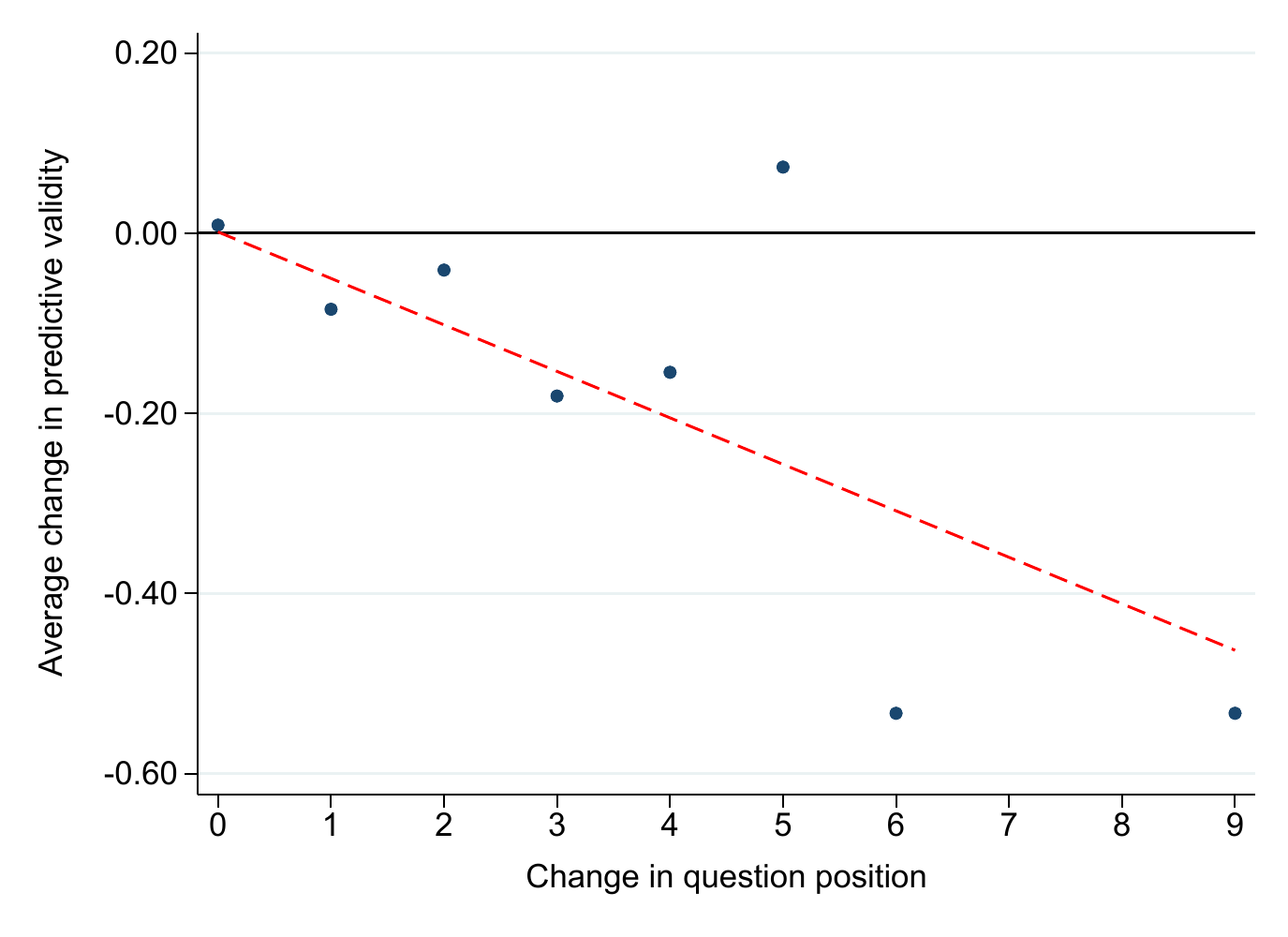}
	\end{subfigure}
	\hfill       
	\begin{subfigure}[t]{0.45\textwidth}
		\caption*{Panel F. Firm mean wage (leave-one-out)}
		\centering 
		\includegraphics[width=\textwidth]{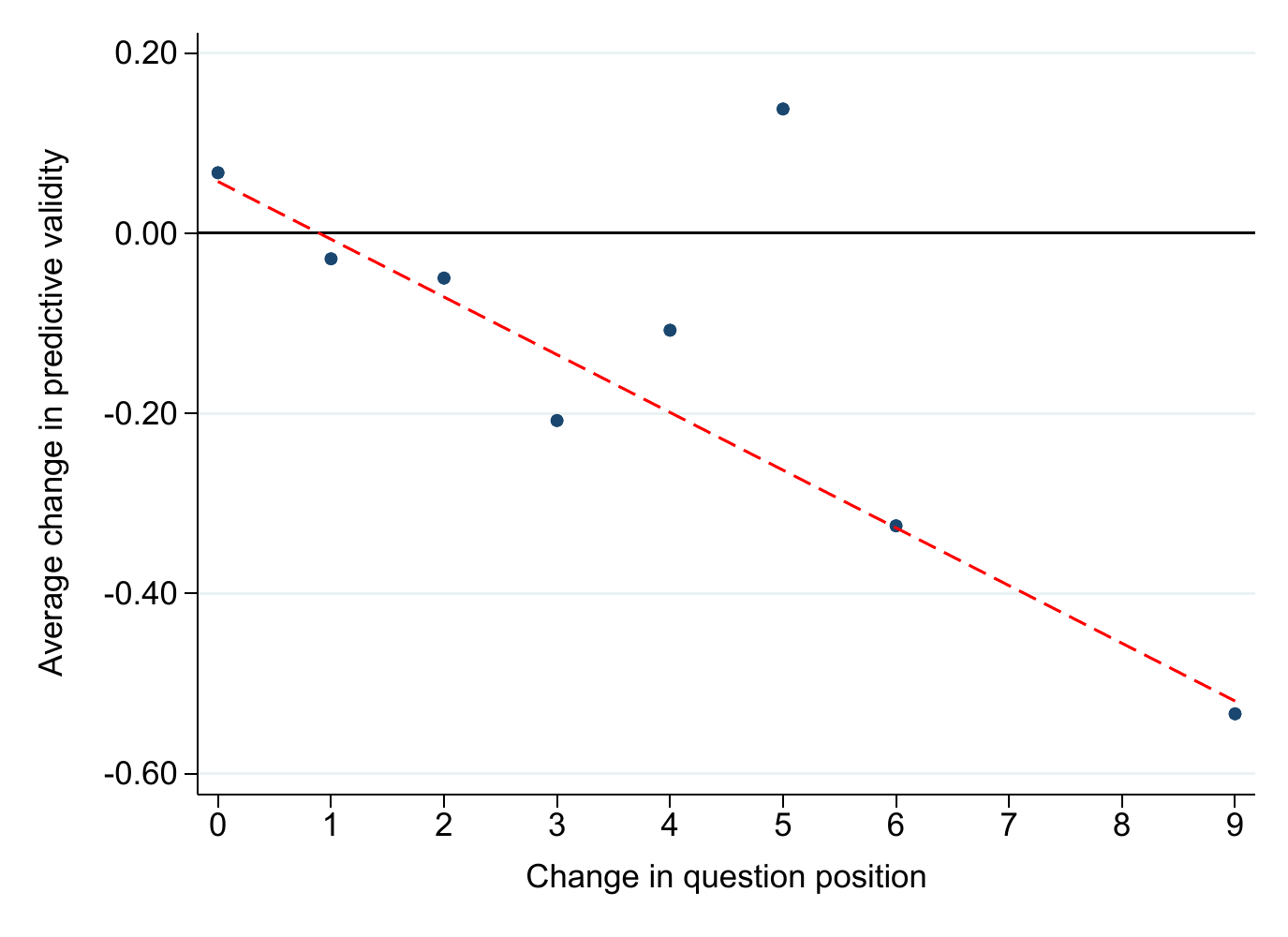}
	\end{subfigure}
	
	{\footnotesize
		\singlespacing \justify
		
		\textit{Notes:} This figure displays estimates of the effect of an increase in the order of a given question on the question's predictive validity. Each panel shows a binned scatterplot plotting the average change in the predictive validity of a test question on a given outcome ($y$-axis) against the change in the position of the question on the exam ($x$-axis). Each panel shows the results for the outcome listed in the panel title. See Section \ref{sub:var-def} for variable definitions. The red dashed lines are predicted values from a linear regression on the microdata. See Appendix Figure \ref{fig:hist-chg-pos}, Panel B for a histogram of the values in the $x$-axis.
		
	}
	
\end{figure}


\clearpage
\begin{table}[htpb!]\caption{Summary statistics of the samples} \label{tab:summ-enem}
	{\footnotesize
		\begin{centering} 
			\protect
			\begin{tabular}{lcccc}
				\addlinespace \addlinespace \midrule			
				&  \multicolumn{2}{c}{High-school-students sample} && Retakers sample \\ \cmidrule{2-3} \cmidrule{5-5}
				& All & 2009-2010  && All \\
				& (1) & (2)        && (3)  \\
				\midrule 	
				\multicolumn{5}{l}{\hspace{-1em} \textbf{Panel A. Demographic characteristics and race}} \\ 
				Age &18.204 &19.151 & &18.062 \\
Female &0.598 &0.611 & &0.618 \\
White &0.476 &0.510 & &0.504 \\
Black/Brown &0.505 &0.450 & &0.483 \\
 \midrule
				
				\multicolumn{5}{l}{\hspace{-1em} \textbf{Panel B. SES and household characteristics}} \\ 
				Attends a private HS &0.222 &0.222 & &0.342 \\
Mom completed high school &0.534 &0.506 & &0.606 \\
Mom completed college &0.205 &0.186 & &0.270 \\
Family earns above 2x M.W. &0.388 &0.379 & &0.432 \\
Family earns above 5x M.W. &0.062 &0.071 & &0.087 \\
 \midrule												
				
				\multicolumn{5}{l}{\hspace{-1em} \textbf{Panel C. Exam preparation}} \\ 
				Took a foreign lang. course &0.241 &0.269 & &0.263 \\
Took a test prep course &0.119 &0.167 & &0.160 \\
 \midrule
				
				\multicolumn{5}{l}{\hspace{-1em} \textbf{Panel D. Fraction of correct responses}} \\ 
				Natural Science &0.283 &0.333 & &0.317 \\
Social Science &0.398 &0.388 & &0.446 \\
Language &0.408 &0.449 & &0.468 \\
Math &0.283 &0.287 & &0.320 \\
Average &0.343 &0.364 & &0.388 \\
 \midrule

				\multicolumn{5}{l}{\hspace{-1em} \textbf{Panel E. Geographical location}} \\ 
				Lives in the North &0.089 &0.081 & &0.082 \\
Lives in the Northeast &0.305 &0.261 & &0.354 \\
Lives in the Southeast &0.389 &0.426 & &0.365 \\
Lives in the South &0.131 &0.150 & &0.113 \\
Lives in the Midwest &0.086 &0.081 & &0.085 \\
 \midrule
				
				Number of test-takers &14,941,156 &1,910,502 & &1,519,842 \\
 \midrule \addlinespace \addlinespace

			\end{tabular}
			\par\end{centering}

		\begin{singlespace} \vspace{-.5cm}
			\noindent \justify  \textit{Notes}: This table shows summary statistics on all test-takers in the high-school-students sample (column 1), those who took the exam in 2009--2010 as high-school seniors (column 2), and students in the retakers sample (column 3). For students who took the exam multiple times, I compute the summary statistics using data from the last year in which I observe them in my sample. See Section \ref{sub:sample} for sample definitions.
		\end{singlespace}	
		
	}
\end{table}

\clearpage
\begin{table}[H]{\footnotesize
		\begin{center}
			\caption{The effect of question position on test performance} \label{tab:perf-pos}
			\newcommand\w{2.2}
			\begin{tabular}{l@{}lR{\w cm}@{}L{0.5cm}R{\w cm}@{}L{0.5cm}R{\w cm}@{}L{0.5cm}}
				\midrule
				&& \multicolumn{6}{c}{Outcome: Correctly responded the question} \\ \cmidrule{3-8}
				&& (1) && (2) && (3) \\
				\midrule
				Question position (normalized)&&$-$0.214&\sym{***}&$-$0.071&\sym{***}&$-$0.058&\sym{***}\\
&&(0.013&)&(0.004&)&(0.002&)\\
Constant&&0.450&\sym{***}&&&&\\
&&(0.008&)&&&&\\
 \midrule
				\(N\) (Item$-$Booklets)&&5,896&&5,896&&5,896&\\

				\(N\) (Students)&&14,940,464&&14,940,464&&14,940,464&\\
    
				\(N\) (Question responses)&&2,689,345,707&&2,689,345,707&&2,689,345,707&\\
R$-$squared&&0.85&&0.99&&0.97&\\
  \midrule  
				Question fixed effects           && No && Yes && No \\ 
				Controls for question difficulty && No && No  && Yes \\ \midrule
			\end{tabular}
		\end{center}
		\begin{singlespace}  \vspace{-.5cm}
			\footnotesize \noindent \justify \textit{Notes:} This table displays estimates of the effect of a question position on the likelihood of correctly answering the question. 
			
			Each column displays an estimate from a different specification. Column 1 presents estimates from a bivariate regression of average student performance on question position. Column 2 presents estimates from equation \eqref{eq:reg-spec-fe}, which includes question fixed effects. Column 3 presents estimates from equation \eqref{eq:reg-spec-diff}, which controls for question difficulty. I normalize question position such that the first question in each testing day is equal to zero and the last question is equal to one.

			Heteroskedasticity-robust standard errors clustered at the question level in parentheses. $^{***}$, $^{**}$ and $^*$ denote significance at 10\%, 5\% and 1\% levels, respectively.
			
		\end{singlespace} 	
	}
\end{table}

\begin{table}[H]{\footnotesize
		\begin{center}
			\caption{The effect of academic ability and cognitive endurance on long-run outcomes} \label{tab:reg-coll-lmkt}
			\newcommand\w{1.45}
			\begin{tabular}{l@{}lR{\w cm}@{}L{0.45cm}R{\w cm}@{}L{0.45cm}R{\w cm}@{}L{0.45cm}R{\w cm}@{}L{0.45cm}R{\w cm}@{}L{0.45cm}R{\w cm}@{}L{0.45cm}}
			
				\addlinespace
				\multicolumn{12}{l}{\hspace{-1em} \textbf{Panel A. College outcomes}}  \\ \midrule
				
				& \multicolumn{12}{c}{Dependent variable} \\ \cmidrule{3-14} 
				&& Enrolled    && College && Degree   && 1st-year && Grad.     && Time to    \\
				&& college     && quality && quality  && credits  && rate      && grad.      \\
				&& (1)         && (2)     && (3)      && (4)      && (5)       &&  (6)       \\ \midrule
				Test score&&0.088&\sym{***}&0.082&\sym{***}&0.117&\sym{***}&0.014&\sym{***}&0.060&\sym{***}&$-$0.119&\sym{***}\\
&&(0.000&)&(0.000&)&(0.000&)&(0.000&)&(0.001&)&(0.002&)\\
 \midrule
				Endurance&&0.032&\sym{***}&0.030&\sym{***}&0.051&\sym{***}&0.006&\sym{***}&0.026&\sym{***}&$-$0.048&\sym{***}\\
&&(0.000&)&(0.000&)&(0.000&)&(0.000&)&(0.000&)&(0.001&)\\
Ability&&0.102&\sym{***}&0.095&\sym{***}&0.140&\sym{***}&0.016&\sym{***}&0.072&\sym{***}&$-$0.140&\sym{***}\\
&&(0.000&)&(0.000&)&(0.000&)&(0.000&)&(0.001&)&(0.002&)\\
 \midrule
				Ratio coef.&&0.310&\sym{***}&0.319&\sym{***}&0.365&\sym{***}&0.358&\sym{***}&0.361&\sym{***}&0.342&\sym{***}\\
&&(0.002&)&(0.001&)&(0.001&)&(0.004&)&(0.003&)&(0.005&)\\

				Mean DV&&0.244&&3.326&&3.244&&0.158&&0.418&&3.817&\\
\(N\)&&2,501,519&&1,800,546&&1,768,707&&1,124,972&&1,472,916&&793,822&\\
 \midrule	\addlinespace \addlinespace \addlinespace
				
				\multicolumn{12}{l}{\hspace{-1em} \textbf{Panel B. Labor-market outcomes}}  \\ \midrule
							
				& \multicolumn{12}{c}{Dependent variable} \\ \cmidrule{3-14} 
				&& Formal        && Hourly && Monthly   && Firm  && Occup.     && Industry  \\
				&& sector        && wage   && earnings  && wage  && wage       && wage      \\
				&& (1)           && (2)     && (3)      && (4)   && (5)        && (6)       \\	\midrule
				Test score&&0.001&\sym{***}&0.129&\sym{***}&0.111&\sym{***}&0.092&\sym{***}&0.041&\sym{***}&0.013&\sym{***}\\
&&(0.000&)&(0.001&)&(0.001&)&(0.001&)&(0.001&)&(0.000&)\\
 \midrule
				Endurance&&0.000&\sym{***}&0.054&\sym{***}&0.052&\sym{***}&0.036&\sym{***}&0.017&\sym{***}&0.004&\sym{***}\\
&&(0.000&)&(0.001&)&(0.001&)&(0.001&)&(0.000&)&(0.000&)\\
Ability&&0.002&\sym{***}&0.154&\sym{***}&0.135&\sym{***}&0.108&\sym{***}&0.049&\sym{***}&0.014&\sym{***}\\
&&(0.000&)&(0.001&)&(0.001&)&(0.001&)&(0.001&)&(0.000&)\\
 \midrule
				Ratio coef.&&0.276&\sym{***}&0.351&\sym{***}&0.387&\sym{***}&0.330&\sym{***}&0.346&\sym{***}&0.255&\sym{***}\\
&&(0.015&)&(0.003&)&(0.003&)&(0.004&)&(0.006&)&(0.010&)\\

				Mean DV&&0.326&&3.865&&7.551&&3.885&&3.886&&3.858&\\
\(N\)&&2,523,029&&818,590&&818,590&&692,880&&818,374&&818,590&\\
 \midrule

			\end{tabular}%
		\end{center}
		\begin{singlespace}  \vspace{-.5cm}
			\noindent \justify \textit{Notes:} This table displays estimates of the relationship between ability/endurance and college outcomes (Panel A) and labor market-outcomes (Panel B).
			
			The first row of each panel shows estimates of the association between test scores and the outcome listed in the column header (coefficient $\psi_T$ in equation \eqref{reg:score-outcomes}). The following rows show estimates of the association between ability and cognitive endurance and a given outcome (coefficients $\psi_A$ and $\psi_E$ in equation \eqref{reg:endurance-outcomes}). All regressions control for age, gender, race, high school type, parental income, cohort fixed effects, and municipality fixed effects. In addition to the baseline controls, the regressions in Panel A, columns 4--6, include college-degree fixed effects to remove the influence of a student's program choice, while the regressions in Panel B control for potential years of experience and years of education. Heteroskedasticity-robust standard errors clustered at the individual level in parentheses. See Section \ref{sub:var-def} for outcome definitions.
			
			The third-to-last row in each panel shows the ratio between the predicted effect of academic ability and the effect of cognitive endurance on a given outcome. Standard errors estimated through the delta method in parentheses. $^{***}$, $^{**}$ and $^*$ denote significance at 10\%, 5\% and 1\% levels.			
			
		\end{singlespace} 	
	}
\end{table}%

\clearpage 
\begin{table}[H]{\footnotesize
		\begin{center}
			\caption{Degrees, occupations, and industries with the largest return to endurance} \label{tab:reg-rank-return}
			\newcommand\w{1.15}
			\begin{tabular}{l@{}lR{\w cm}@{}L{0.45cm}R{\w cm}@{}L{0.45cm}R{\w cm}@{}L{0.45cm}R{\w cm}@{}L{0.45cm}R{\w cm}@{}L{0.45cm}}
				\midrule
				&& Return    && Return   && Ratio   && Wage  && Sample   \\
			    && ability   && endur.   && returns && pctil. && size \\
				&& (1)       && (2)      && (3)     && (4) && (5)  \\ \midrule
				
				\multicolumn{8}{l}{\hspace{-1em} \textbf{Panel A. Top five degrees}}  \\ \addlinespace
				1.  Aeronautics and related degrees&&0.173&&0.106&&0.613&& 69.7 &&      670 &\\
&&(0.026&)&(0.015&)&(0.077&)& && &\\

				2. Music and performing arts&&0.221&&0.093&&0.420&& 59.3 &&      502 &\\
&&(0.060&)&(0.035&)&(0.110&)& && &\\

				3. Religion&&0.186&&0.088&&0.471&& 49.0 &&      356 &\\
&&(0.047&)&(0.031&)&(0.097&)& && &\\

				4. History and archeology&&0.211&&0.087&&0.414&& 60.0 &&      988 &\\
&&(0.043&)&(0.022&)&(0.064&)& && &\\

				5. Forestry engineering&&0.154&&0.084&&0.543&& 52.3 &&      397 &\\
&&(0.045&)&(0.024&)&(0.115&)& && &\\
 \midrule
				
				\multicolumn{8}{l}{\hspace{-1em} \textbf{Panel B. Top five occupations}}  \\ \addlinespace
				1. Public tax auditors&&0.454&&0.168&&0.369&& 49.2 &&      255 &\\
 &&(0.106&)&(0.057&)&(0.084&)& && &\\

				2. Professionals in air&&0.278&&0.114&&0.412&& 88.3 &&      347 &\\
\hspace{.3cm}~navigation,~sea~and~fluvial &&(0.049&)&(0.028&)&(0.072&)& && &\\

				3. Technicians in operation of radio,&&0.200&&0.112&&0.558&& 66.3 &&      658 &\\
\hspace{.3cm}~TV~systems~and~video~producers &&(0.047&)&(0.027&)&(0.091&)& && &\\

				4. Plant operators in chemical,&&0.267&&0.109&&0.409&& 62.8 &&    1,221 &\\
\hspace{.3cm}~petrochemical~and~related~occup. &&(0.031&)&(0.020&)&(0.057&)& && &\\

				5. Instrument and precision&&0.180&&0.092&&0.511&& 70.8 &&      203 &\\
\hspace{.3cm}~equipment~repairers &&(0.068&)&(0.040&)&(0.194&)& && &\\
 \midrule
	
				\multicolumn{8}{l}{\hspace{-1em} \textbf{Panel C. Top five industries}}  \\ \addlinespace
				1. Oil extraction and&&0.276&&0.135&&0.488&& 91.5 &&      948 &\\
\hspace{.3cm}~related~services &&(0.034&)&(0.021&)&(0.052&)& && &\\

				2. Financial intermediation and&&0.215&&0.087&&0.402&& 54.7 &&    6,177 &\\
\hspace{.3cm}~and~insurance~(aux.~services) &&(0.013&)&(0.007&)&(0.024&)& && &\\

				3. Research and development&&0.198&&0.084&&0.426&& 73.7 &&    1,323 &\\
 &&(0.024&)&(0.015&)&(0.052&)& && &\\

				4. Electricity, gas and hot water&&0.223&&0.083&&0.373&& 78.1 &&    1,966 &\\
 &&(0.020&)&(0.011&)&(0.036&)& && &\\

				5. Manufacture of office machinery&&0.132&&0.073&&0.553&& 54.7 &&      920 &\\
 &&(0.033&)&(0.019&)&(0.095&)& && &\\
 \midrule
			\end{tabular}%
		\end{center}

		\begin{singlespace}  \vspace{-.5cm}
			\noindent \justify \textit{Notes:} This table lists the top five 3-digit academic degrees (Panel A), 3-digit occupations (Panel B), and 2-digit industries (Panel C) with the highest wage return to cognitive endurance (column 2). 
			
			Column 1 shows the wage return to ability. Column 3 shows the ratio between the wage return to endurance and the wage return to ability. Column 4 shows the average wage percentile of workers in each degree, occupation, or industry. Column 5 shows the sample size used to estimate each wage return.
			
			The wage return to ability and endurance are the coefficients $\psi_A$ and $\psi_E$ in equation \eqref{reg:endurance-outcomes} using as outcome log hourly wage, estimated separately for each degree, occupation, and industry. 
			
			Heteroskedasticity-robust standard errors clustered at the individual level in parentheses.
			
		\end{singlespace} 	
		
	}
\end{table}%

\begin{table}[H]{\footnotesize
		\begin{center}
			\caption{The contribution of gaps in ability and endurance to test-score gaps} \label{tab:corr-endurance}
			\newcommand\w{1.75}
			\begin{tabular}{l@{}lR{\w cm}@{}L{0.5cm}R{\w cm}@{}L{0.5cm}R{\w cm}@{}L{0.5cm}R{\w cm}@{}L{0.5cm}R{\w cm}@{}L{0.5cm}}
				\midrule
				&&  \multicolumn{10}{c}{Gap between} \\ \cmidrule{2-12}
				&&    Male /  && White /   &&  Priv HS /   && Mom coll / && High-inc /  \\
				&&    Female  && Non-white &&  Public HS      && No coll  && Low-inc  \\
				&&     (1)    && (2)       && (3)             && (4)          && (5)       \\ \midrule \addlinespace 
							
				\multicolumn{12}{l}{\hspace{-1em} \textbf{Panel A. Difference in average test score}}  \\ \addlinespace   
				Test-score gap&&0.026&\sym{***}&0.057&\sym{***}&0.130&\sym{***}&0.098&\sym{***}&0.192&\sym{***}\\
&&(0.000&)&(0.000&)&(0.000&)&(0.000&)&(0.000&)\\
 \midrule \addlinespace  

				\multicolumn{12}{l}{\hspace{-1em} \textbf{Panel B. Contribution of gaps in ability and endurance to test-score gaps}}  \\ \addlinespace   
				Ability gap&&0.030&\sym{***}&0.056&\sym{***}&0.127&\sym{***}&0.095&\sym{***}&0.188&\sym{***}\\
&&(0.000&)&(0.000&)&(0.000&)&(0.000&)&(0.000&)\\

				Endurance gap&&0.017&\sym{***}&0.016&\sym{***}&0.038&\sym{***}&0.026&\sym{***}&0.063&\sym{***}\\
&&(0.000&)&(0.000&)&(0.000&)&(0.000&)&(0.000&)\\
  \midrule \addlinespace

				\multicolumn{12}{l}{\hspace{-1em} \textbf{Panel C. Impact of a reform that halves the exam length on test-score gaps}}  \\ \addlinespace   
				P.p. change gap&&-0.008&\sym{***}&-0.008&\sym{***}&-0.019&\sym{***}&-0.013&\sym{***}&-0.031&\sym{***}\\
&&(0.000&)&(0.000&)&(0.000&)&(0.000&)&(0.000&)\\

				Pct. change gap&&-0.322&\sym{***}&-0.137&\sym{***}&-0.144&\sym{***}&-0.130&\sym{***}&-0.163&\sym{***}\\
&&(0.001&)&(0.000&)&(0.000&)&(0.000&)&(0.000&)\\
\midrule \addlinespace 

				\(N\) (Students)&&14,941,097&&14,565,550&&9,924,652&&14,290,759&&9,996,959&\\
 \addlinespace

			\end{tabular}
		\end{center}
		\begin{singlespace}  \vspace{-.5cm}
			\footnotesize \noindent \justify \textit{Notes:} This table shows test-score gaps in the ENEM and the contribution of differences in ability and endurance to those gaps.
			
			Each column shows the result for a different test-score gap. Column 1 shows gaps between male and female students. Column 2 shows gaps between  white and non-white (Black, Brown, and Indigenous) students. Column 3 shows gaps between students enrolled in a private high school and public high school. Column 4 shows gaps between students with a college-educated mother and non-college-educated mother. Column 5 shows gaps between students in households in the top 30\% and bottom 30\% of the income distribution.
			
			Panel A shows the average test score difference between the two groups displayed in the column header, $\E[\text{TestScore}_i | X_i = 1] - \E[\text{TestScore}_i | X_i = 0]$.
			
			Panel B shows the contribution of differences in ability and differences in endurance to the test-score gap. The ability gap is the average difference in ability, controlling for endurance,  $\E[\hat{\alpha}_i | X_i = 1, \hat{\beta}_i] - \E[\hat{\alpha}_i | X_i = 0, \hat{\beta}_i]$. The endurance gap is the average difference in endurance, controlling for ability and scaled by the average question position, $\Big(\E[\hat{\beta}_i | X_i = 1, \hat{\alpha}_i] - \E[\hat{\beta}_i | X_i = 0, \hat{\alpha}_i]\Big) \times \overline{\text{Position}}.$
	
			Panel C shows estimates of the impact of a reform that changes the length of the exam from $\overline{\text{Position}}$ to $\overline{\text{Position}}/2$. The first row shows the percentage point change in the test-score gap due to the reform, which is equal to $-\Big(\E[\hat{\beta}_i | X_i = 1, \hat{\alpha}_i] - \E[\hat{\beta}_i | X_i = 0, \hat{\alpha}_i]\Big) \times \overline{\text{Position}}/2$. The second row shows the percentage change in the test-score gap, which equals the percentage point change in the gap divided by the pre-reform test-score gap (shown in Panel A). Standard errors estimated through the delta method in parentheses. $^{***}$, $^{**}$ and $^*$ denote significance at 10\%, 5\% and 1\% levels, respectively.

		\end{singlespace} 	
	}
\end{table}

\begin{landscape}
	\begin{table}[H]{\footnotesize
			\begin{center}
				\caption{The effect of an exam reform that halves the exam length on its predictive validity} \label{tab:val-pos}
				\newcommand\w{1.3}
				\begin{tabular}{l@{}lR{\w cm}@{}L{0.5cm}R{\w cm}@{}L{0.5cm}R{\w cm}@{}L{0.5cm}R{\w cm}@{}L{0.5cm}R{\w cm}@{}L{0.5cm}R{\w cm}@{}L{0.5cm}R{\w cm}@{}L{0.5cm}R{\w cm}@{}L{0.5cm}}
					\midrule
					&& \multicolumn{16}{c}{Outcome: Predictive validity of question $j$ for} \\ \cmidrule{3-18}
					&& Test       && College   && College && Degree   && Grad. && Hourly && Monthly && Firm   \\
					&& score      && enrol.    && quality && progress && rate    && wage   && earnings && wage \\
					&& (1)        && (2)        && (3)    && (4)       && (5)     && (6)    && (7)    && (8)\\							
					\midrule \addlinespace
					\multicolumn{14}{l}{\hspace{-1em} \textbf{Panel A. Average predictive validity}}  \\ \midrule \addlinespace
					Constant&&0.285&\sym{***}&0.057&\sym{***}&0.109&\sym{***}&0.003&\sym{***}&0.016&\sym{***}&0.106&\sym{***}&0.101&\sym{***}&0.084&\sym{***}\\
&&(0.007&)&(0.002&)&(0.003&)&(0.001&)&(0.001&)&(0.003&)&(0.003&)&(0.002&)\\
 \addlinespace 
						
					\multicolumn{14}{l}{\hspace{-1em} \textbf{Panel B. Effect of the exam reform}}  \\ \midrule \addlinespace
					Change in Pred. Val.&&0.115&\sym{*}&0.055&\sym{***}&0.099&\sym{***}&$-$0.005&\sym{**}&0.003&&0.084&\sym{**}&0.077&\sym{**}&0.072&\sym{**}\\
&&(0.069&)&(0.018&)&(0.032&)&(0.002&)&(0.013&)&(0.034&)&(0.032&)&(0.030&)\\
 \addlinespace \midrule 
					Chg. Val./Mean&&0.404&\sym{***}&0.952&\sym{***}&0.911&\sym{***}&$-$1.500&\sym{***}&0.201&&0.796&\sym{***}&0.757&\sym{***}&0.855&\sym{***}\\
&&(0.110&)&(0.191&)&(0.144&)&(0.494&)&(0.662&)&(0.193&)&(0.194&)&(0.239&)\\
 \addlinespace 
										
					\(N\) (Item$-$Booklets)&&1,416&&1,416&&1,416&&700&&1,416&&1,416&&1,416&&1,416&\\
	\midrule
					
				\end{tabular}
			\end{center}
			\begin{singlespace}  \vspace{-.5cm}
				\noindent \justify \textit{Notes:} This table displays the estimated effect of an exam reform that changes the exam length from $\overline{\text{Position}}$ to $\overline{\text{Position}}/2$ on the predictive validity of the exam questions for long-run outcomes. 
				
				Each column displays the estimates of equation \eqref{reg:validity-pos} for a different outcome. In Panel A, the regression only includes a constant. In Panel B, the regression includes question fixed effects. I the coefficients so that they can be interpreted as the effect of decreasing the exam length by half. See Section \ref{sub:var-def} for outcome definitions.
				
				Heteroskedasticity-robust standard errors clustered at the question level in parentheses. $^{***}$, $^{**}$ and $^*$ denote significance at 10\%, 5\% and 1\% levels, respectively.
				
			\end{singlespace} 	
		}
	\end{table}
\end{landscape}

	\clearpage
	\begin{singlespace}
		\bibliographystyle{apa}
		\bibliography{0-endurance}
	\end{singlespace}

\clearpage 
\appendix
\begin{center}\noindent {\LARGE \textbf{Appendix}}\end{center}
\label{app:figs}

\setcounter{table}{0}
\setcounter{figure}{0}
\setcounter{equation}{0}	
\renewcommand{\thetable}{A\arabic{table}}
\renewcommand{\thefigure}{A\arabic{figure}}
\renewcommand{\theequation}{A\arabic{equation}}

\section{Appendix Figures and Tables}

\begin{figure}[H]
	\caption{Examples of ``focus support'' products in a local CVS}\label{fig:cvs}
	\begin{subfigure}[t]{.42\textwidth}
		\caption*{ }
		\centering
		\includegraphics[width=\textwidth]{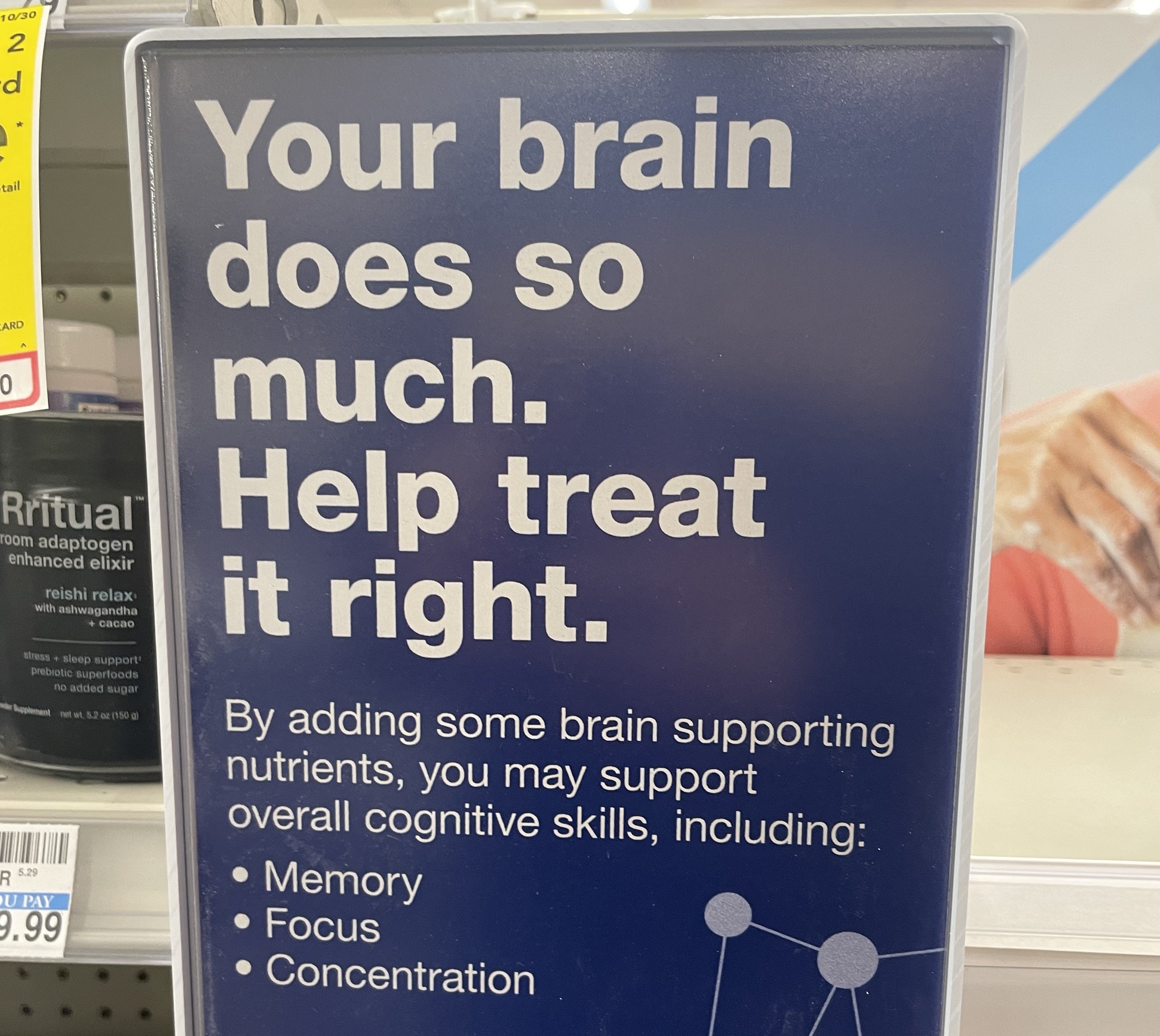}
	\end{subfigure}
	\hfill        
	\begin{subfigure}[t]{.48\textwidth}
		\caption*{ }
		\centering
		\includegraphics[width=\textwidth]{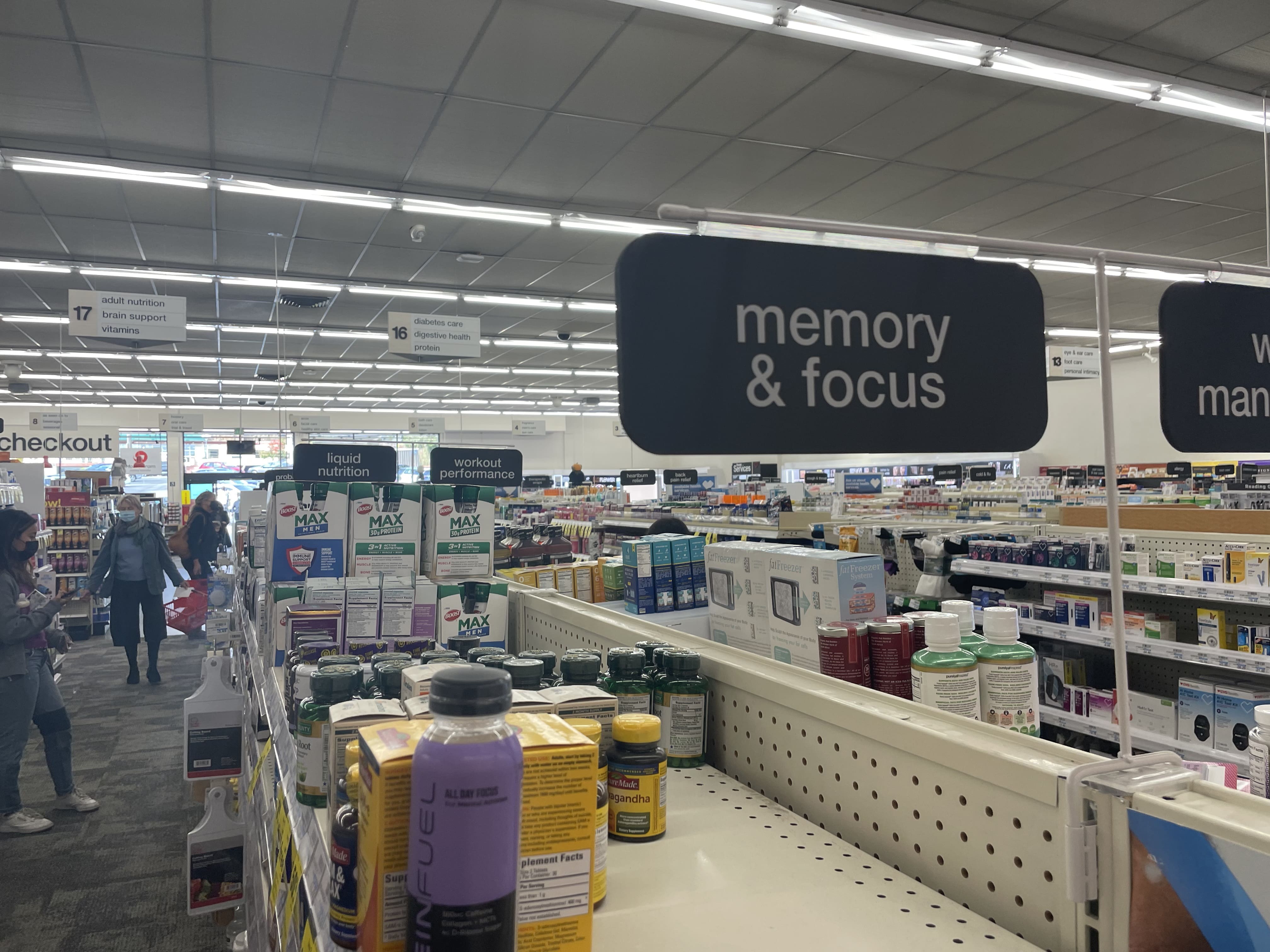}
	\end{subfigure}
	\hfill       	
	\begin{subfigure}[t]{0.48\textwidth}
		\caption*{ }
		\centering
		\includegraphics[width=\textwidth]{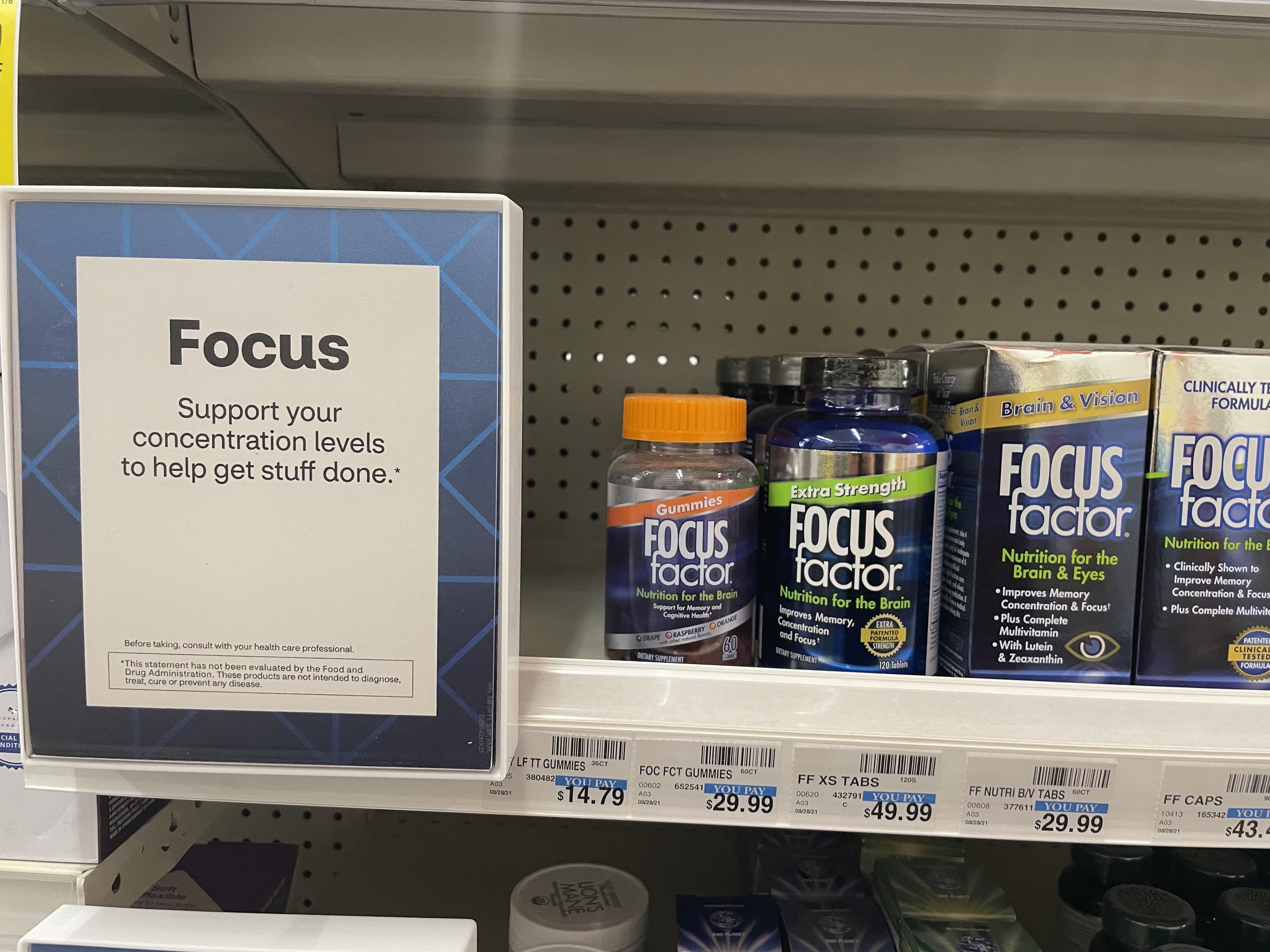}
	\end{subfigure}		
	\hfill        
	\begin{subfigure}[t]{.48\textwidth}
		\caption*{ }
		\centering
		\includegraphics[width=\textwidth]{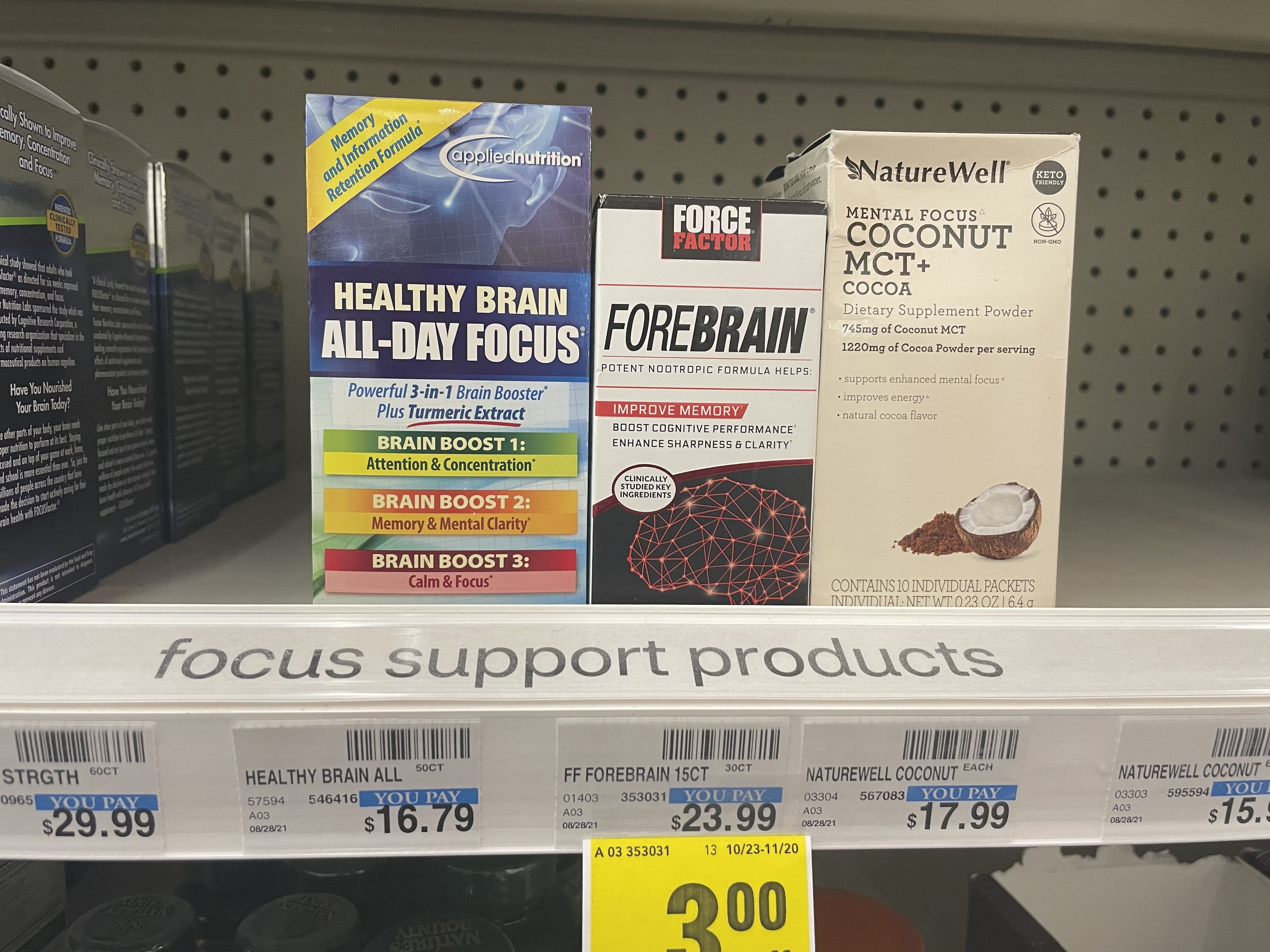}
	\end{subfigure}	
	{\footnotesize
		\singlespacing \justify
		
		\textit{Notes:} These pictures show examples of over-the-counter products aimed at enhancing focus and cognition. The pictures were taken at a local pharmacy by the author.
	}
\end{figure}

\begin{figure}[H]
	\caption{Fraction of students who graduate from college by years since enrollment}\label{fig:cdf-grad}
	\centering
	\includegraphics[width=.75\linewidth]{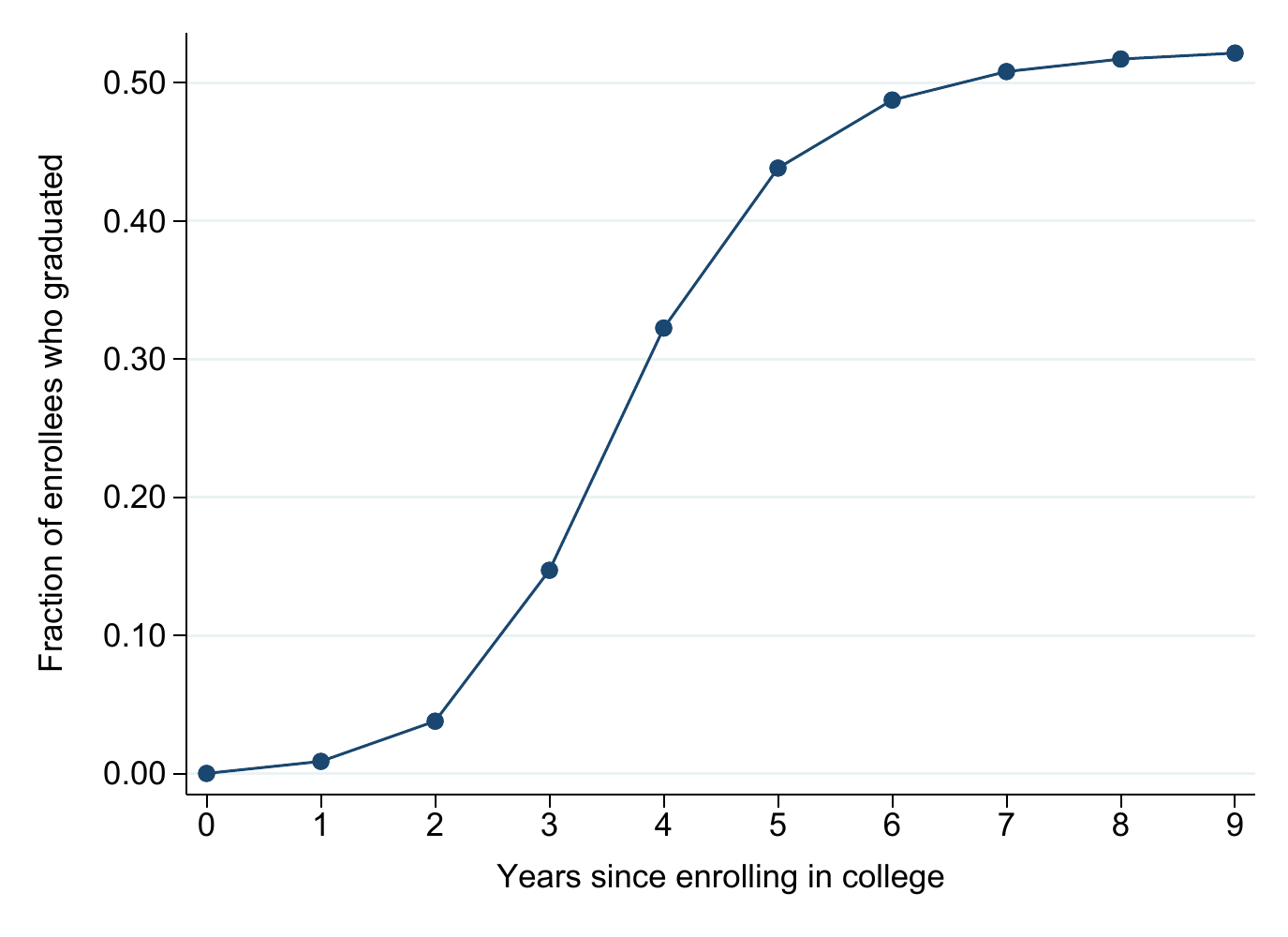}
	
	\hfill						
	{\footnotesize
		\singlespacing \justify
		
		\textit{Notes:} This figure shows the empirical cumulative distribution function of the graduation rate of individuals in the high-school-students sample.
		
	}
	
\end{figure}

\clearpage
\begin{figure}[H]
	\caption{Histogram of the change in a question's position across exam booklets}\label{fig:hist-chg-pos}
	\centering
	\begin{subfigure}[t]{.45\textwidth}
		\caption*{Panel A. All years (2009--2016)}
		\centering
		\includegraphics[width=\textwidth]{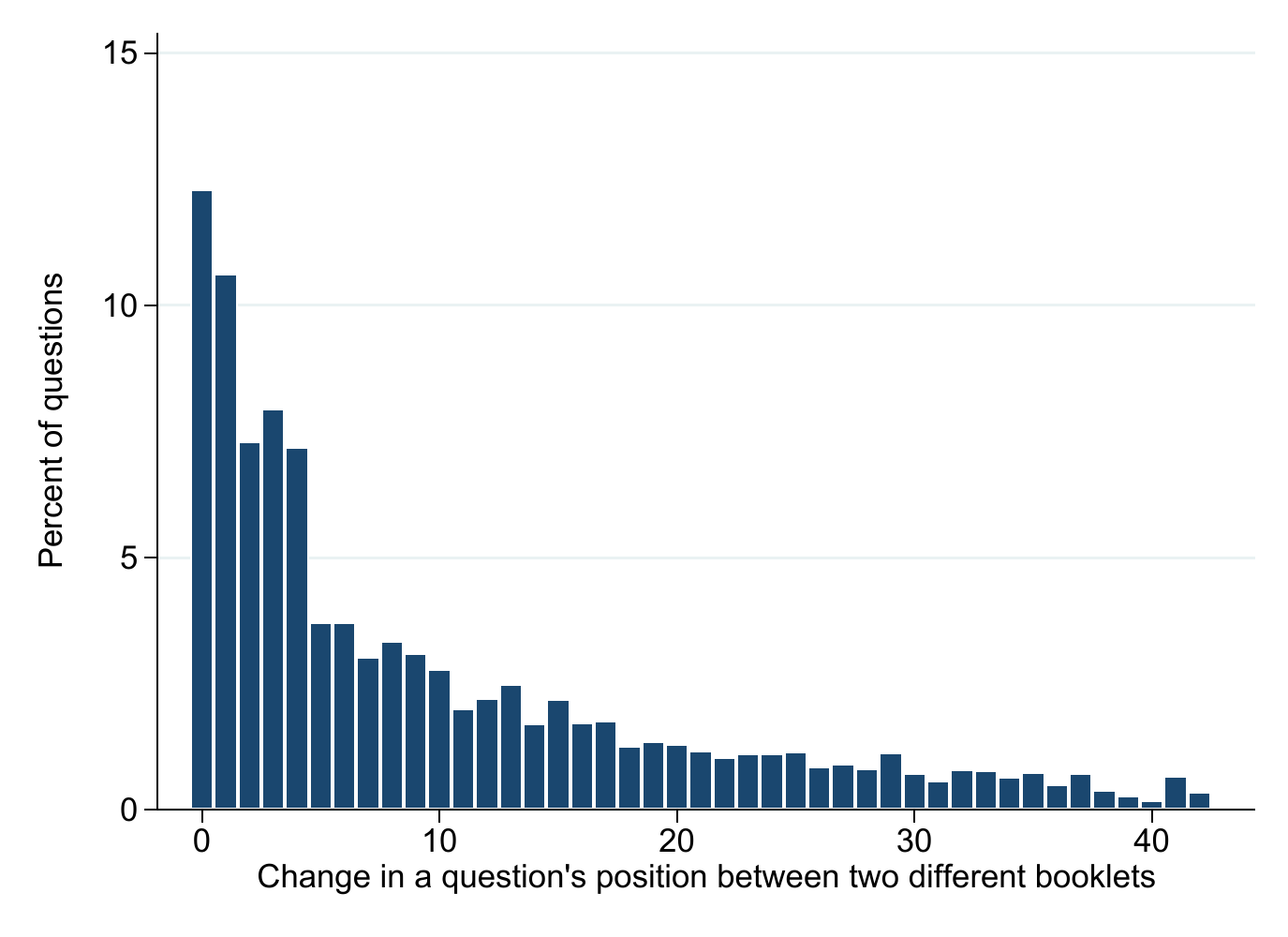}
	\end{subfigure}
	\hfill     
	\begin{subfigure}[t]{.45\textwidth}
		\caption*{Panel B. First two cohorts (2009--2010)}
		\centering
		\includegraphics[width=\textwidth]{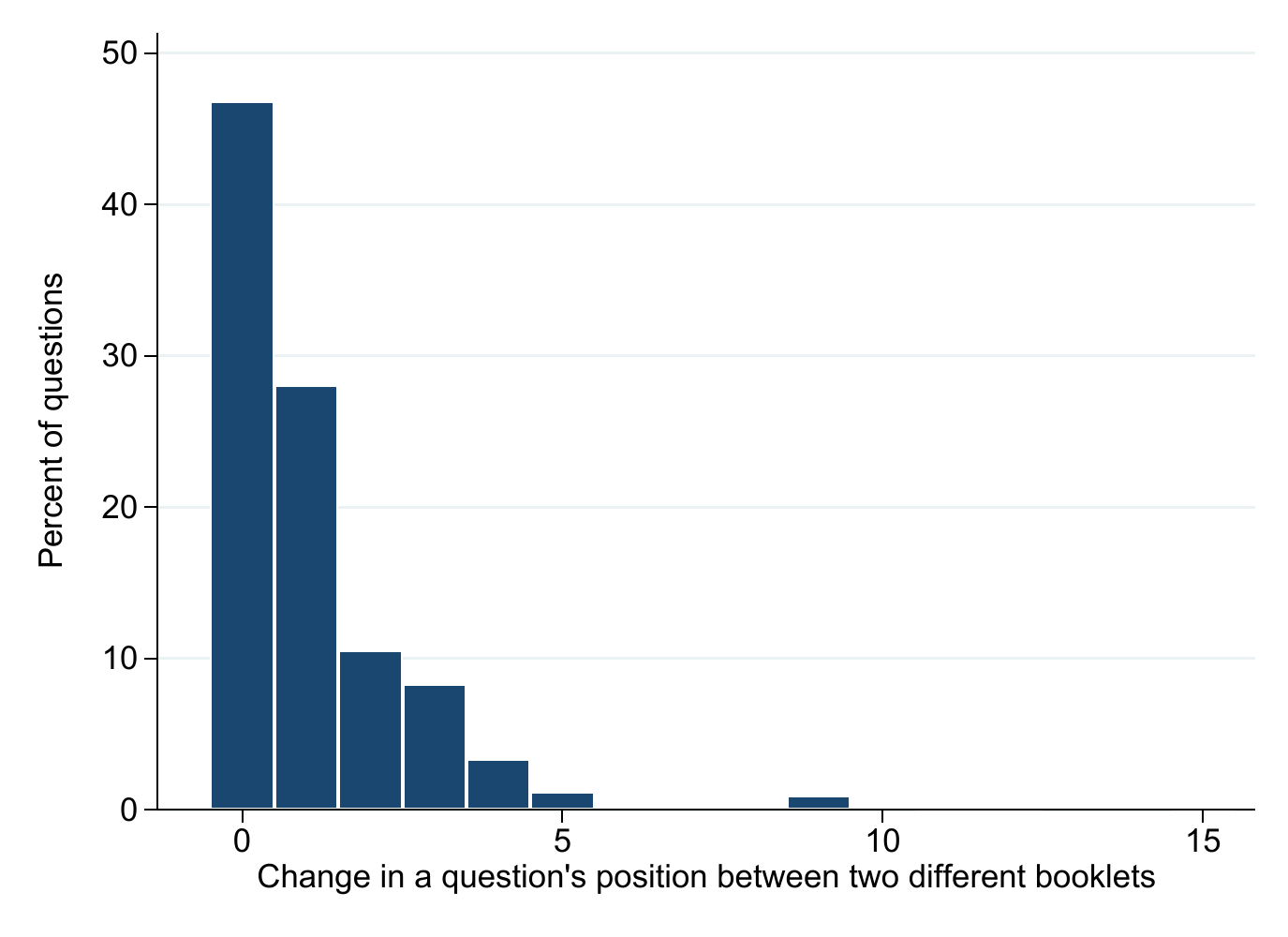}
	\end{subfigure} 
	\hfill       
	
	{\footnotesize
		\singlespacing \justify
		
		\textit{Notes:} This figure shows the amount of variation available in a given question's position between different exam booklets. To construct this figure, I first calculate the difference (in absolute value) in a question's position in two exam booklets. This difference ranges from zero (if a question is in the same position in two different booklets) to 44 (if a question is in the first position of a section in one booklet and the last position of a section in another booklet). I repeat this process for each question and each possible booklet pair. The figure plots the resulting histogram of position differences.
		
	}	
\end{figure}

\clearpage
\begin{figure}[H]
	\caption{Average student performance on selected questions by question position}\label{fig:mean-corr-item}
	\centering
	\includegraphics[width=.75\linewidth]{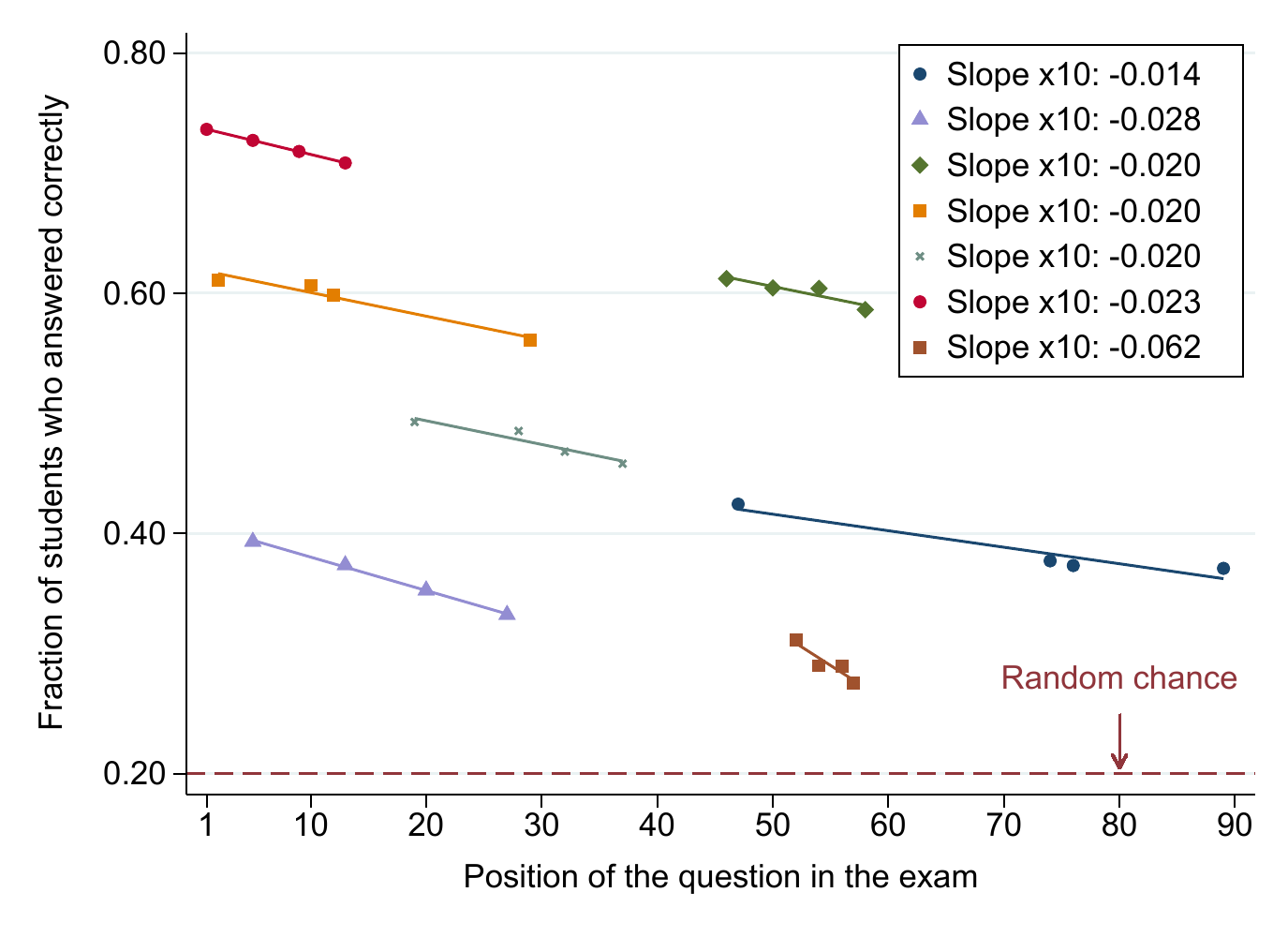}
	
	{\footnotesize
		\singlespacing \justify
		
		\textit{Notes:} This figure plots the fraction of correct responses on seven selected exam questions as a function of their position on the four different exam booklets. Solid lines denote predicted values from linear regressions estimated on the plotted points. 
		
	}
	
\end{figure}

\clearpage
\begin{figure}[H]
	\caption{Histogram of question-level position effects}\label{fig:hist-item-effect}
	\centering
	\includegraphics[width=.75\linewidth]{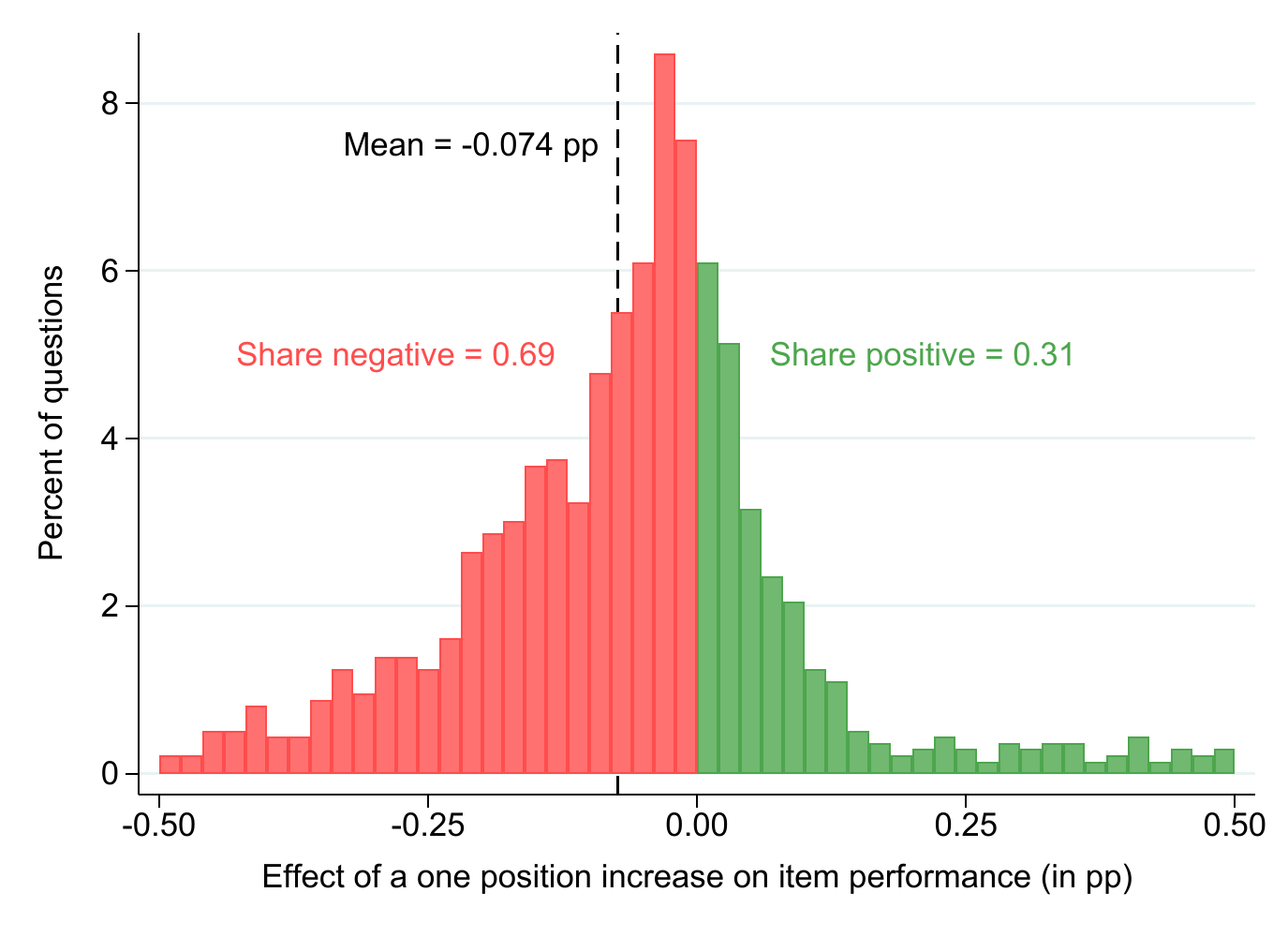}
	
	{\footnotesize
		\singlespacing \justify
		
		\textit{Notes:} This figure plots the distribution of item-level position effects. To construct this figure, I estimate the impact of an increase in the position of a given question on student performance separately for each question. The figure displays the distribution of estimated $\beta$'s (one for each item). The figure excludes outliers (i.e., questions for which the effect is below -0.50 or above 0.50 percentage points).
		
	}
	
\end{figure}

\clearpage
\begin{figure}[H]
	\caption{Distribution of academic ability and cognitive endurance}\label{fig:hist-endurance}
	\begin{subfigure}[t]{.45\textwidth}
		\caption*{Panel A. Academic ability ($\alpha_i$)}
		\centering
		\includegraphics[width=\textwidth]{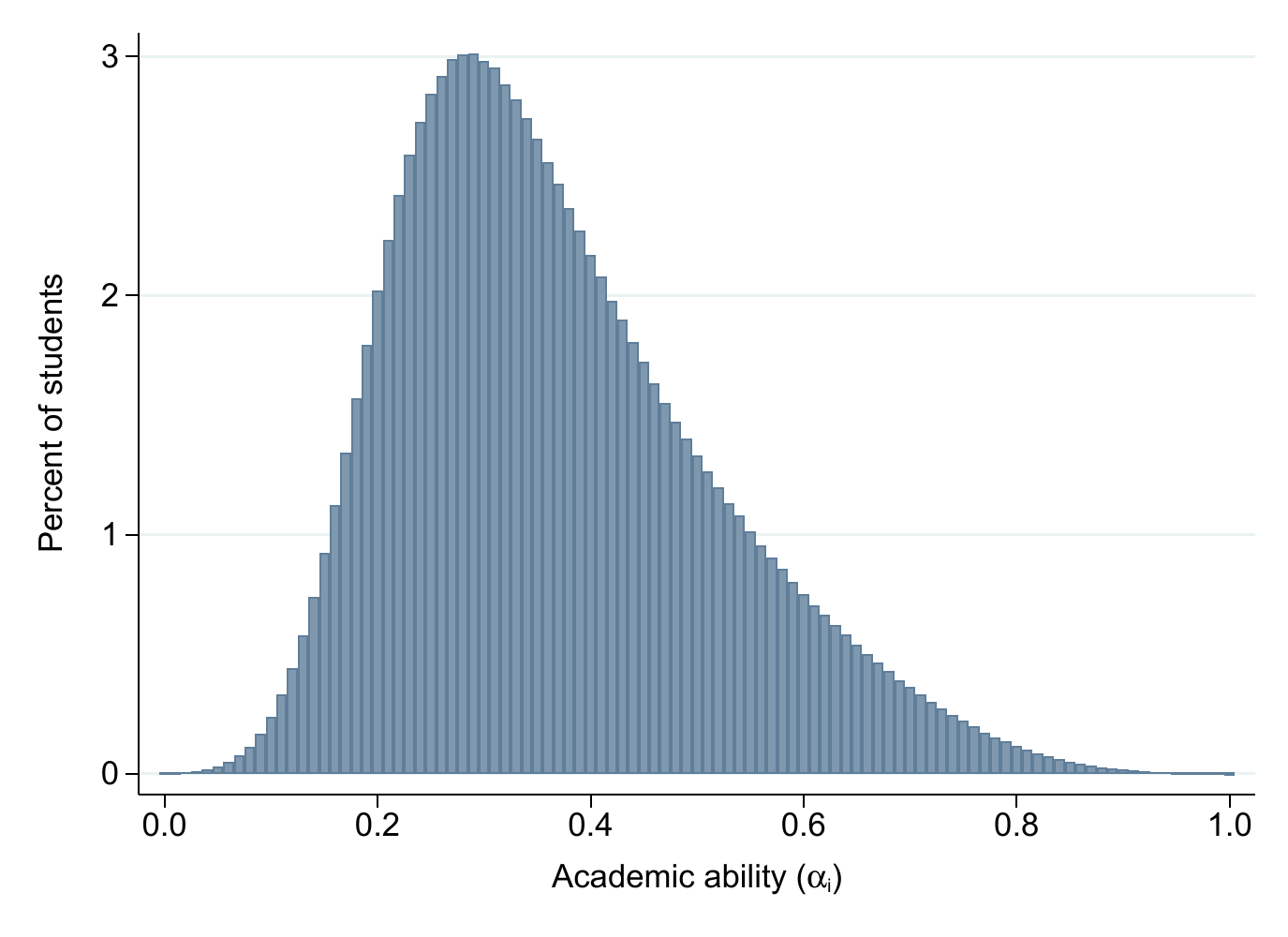}
	\end{subfigure}
	\hfill     
	\begin{subfigure}[t]{.45\textwidth}
		\caption*{Panel B. Cognitive endurance ($\beta_i$)}
		\centering
		\includegraphics[width=\textwidth]{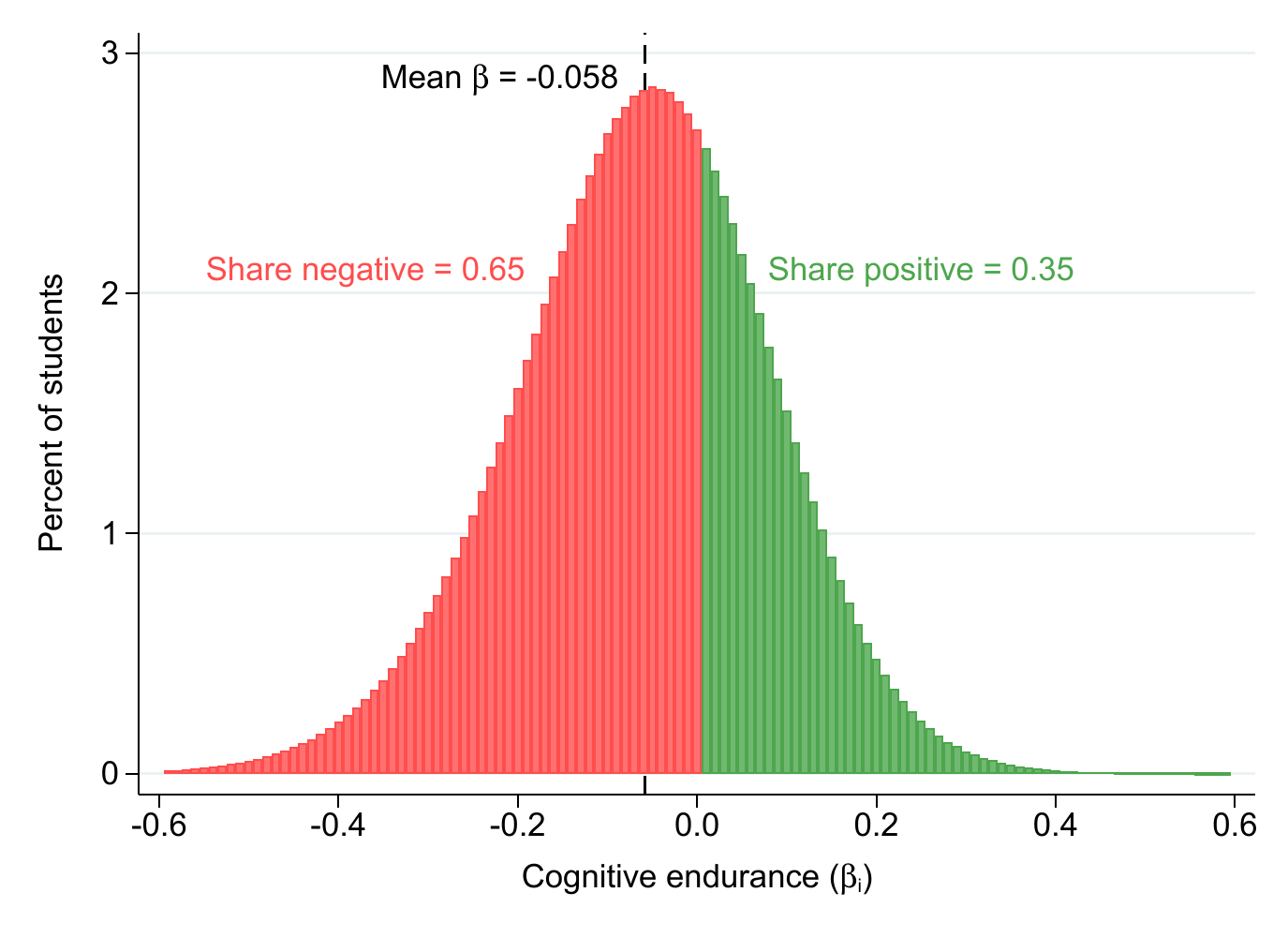}
	\end{subfigure} 
	\hfill    		
	
	{\footnotesize\singlespacing \justify
		
		\textit{Notes:} This figure shows the distribution of my estimates of academic ability (Panel A) and cognitive endurance (Panel B) among individuals in the high-school-students sample. The measure of an individual's ability is the estimated intercept ($\alpha_i$) in equation \eqref{reg:lpm-ind}. The measure of an individual's cognitive endurance is the estimated slope ($\beta_i$) in equation \eqref{reg:lpm-ind}. 
		
	}
\end{figure}

\begin{figure}[H]
	\caption{The relationship between a question's predictive validity and its position}\label{fig:pred-val-pos}
	\begin{subfigure}[t]{.45\textwidth}
	\caption*{Panel A. Test score (leave-question-out)}
	\centering
	\includegraphics[width=\textwidth]{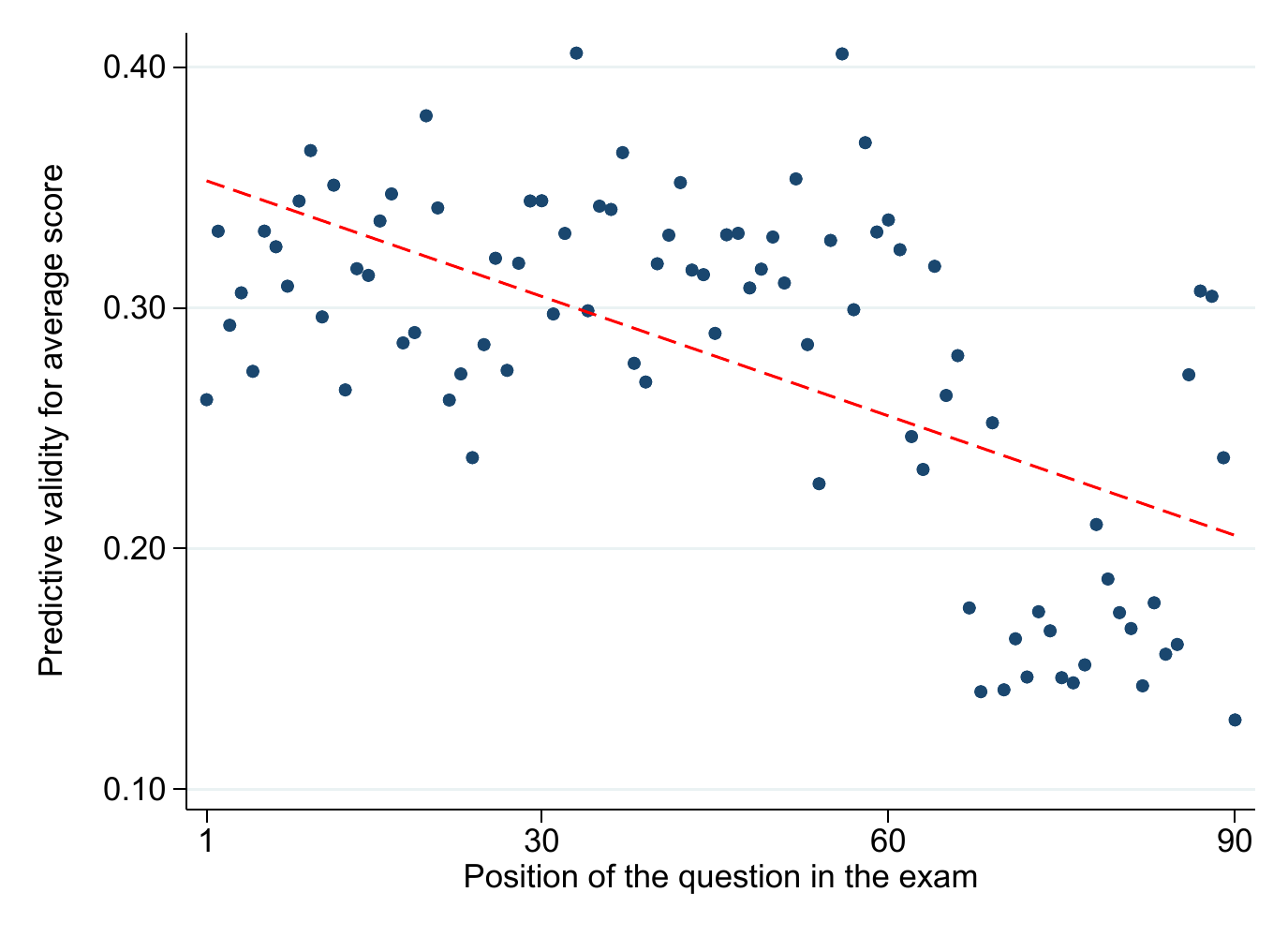}
	\end{subfigure}
	\hfill     
	\begin{subfigure}[t]{.45\textwidth}
		\caption*{Panel B. College enrollment}
		\centering
		\includegraphics[width=\textwidth]{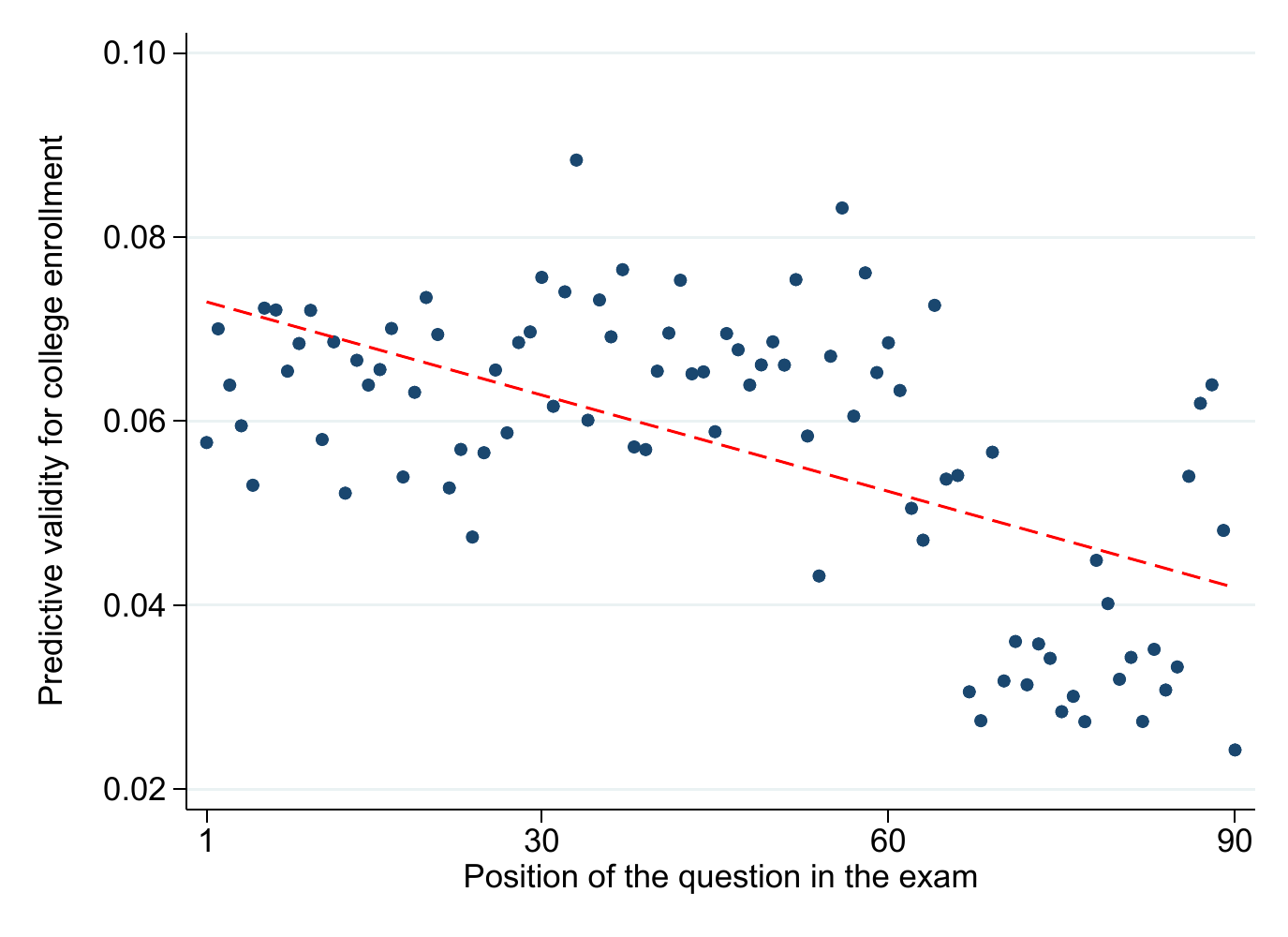}
	\end{subfigure}
	\hfill       	
	\begin{subfigure}[t]{0.45\textwidth}
		\caption*{Panel C. College quality}
		\centering
		\includegraphics[width=\textwidth]{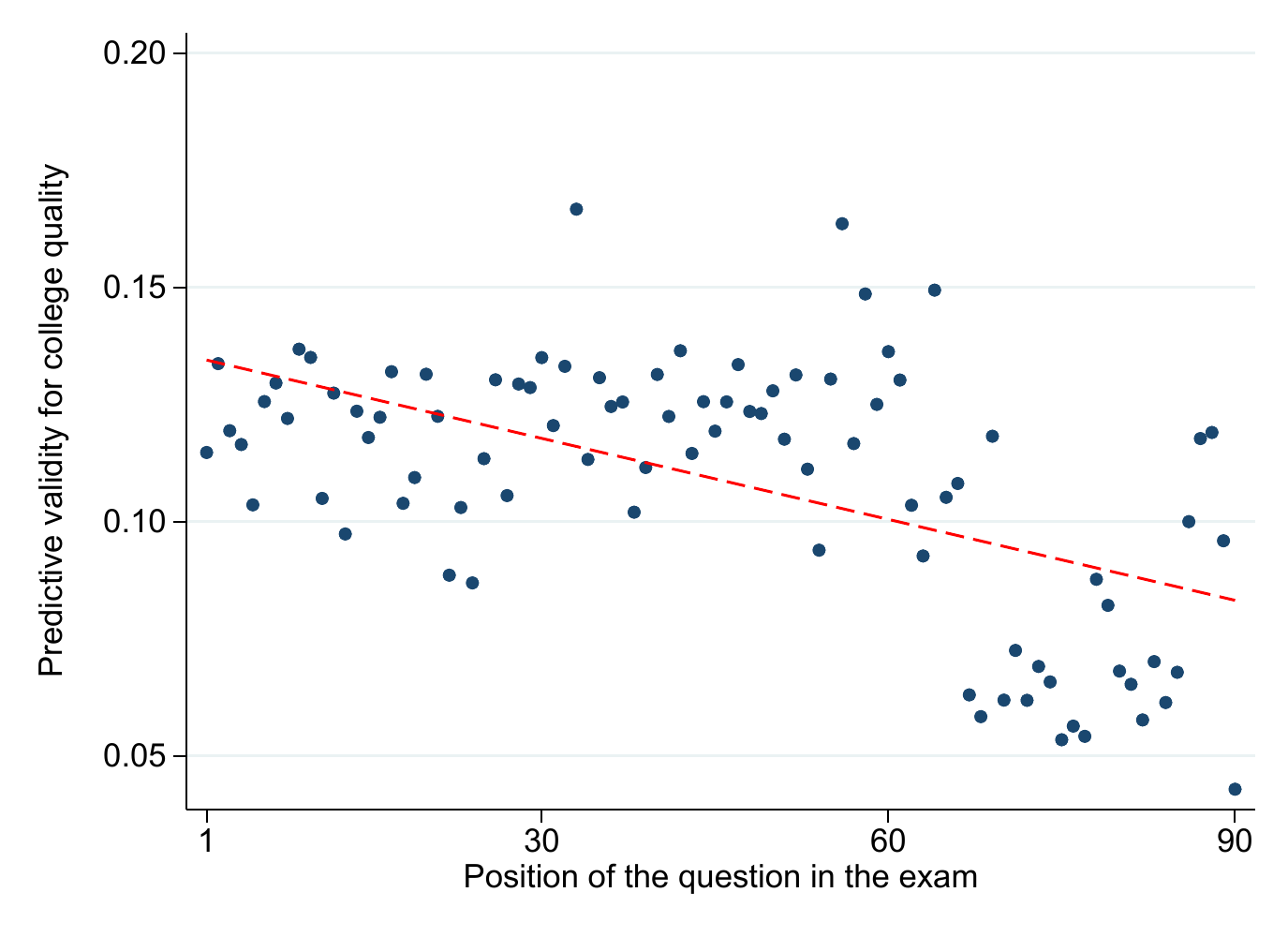}
	\end{subfigure}			
	\hfill        
	\begin{subfigure}[t]{0.45\textwidth}
		\caption*{Panel D. Hourly wage}
		\centering
		\includegraphics[width=\textwidth]{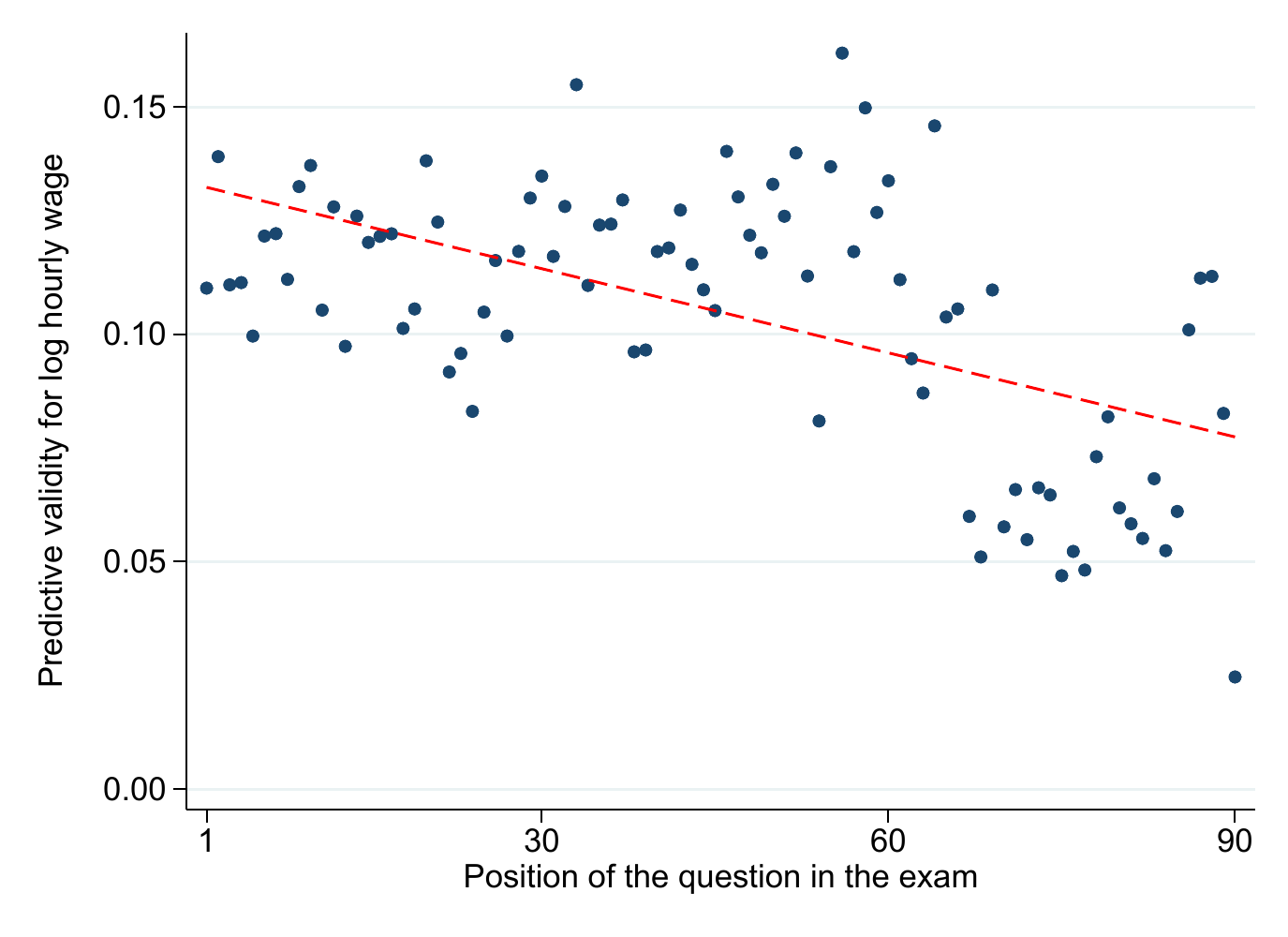}
	\end{subfigure}		
	\hfill        
	\begin{subfigure}[t]{.45\textwidth}
		\caption*{Panel E. Monthly earnings}
		\centering
		\includegraphics[width=\textwidth]{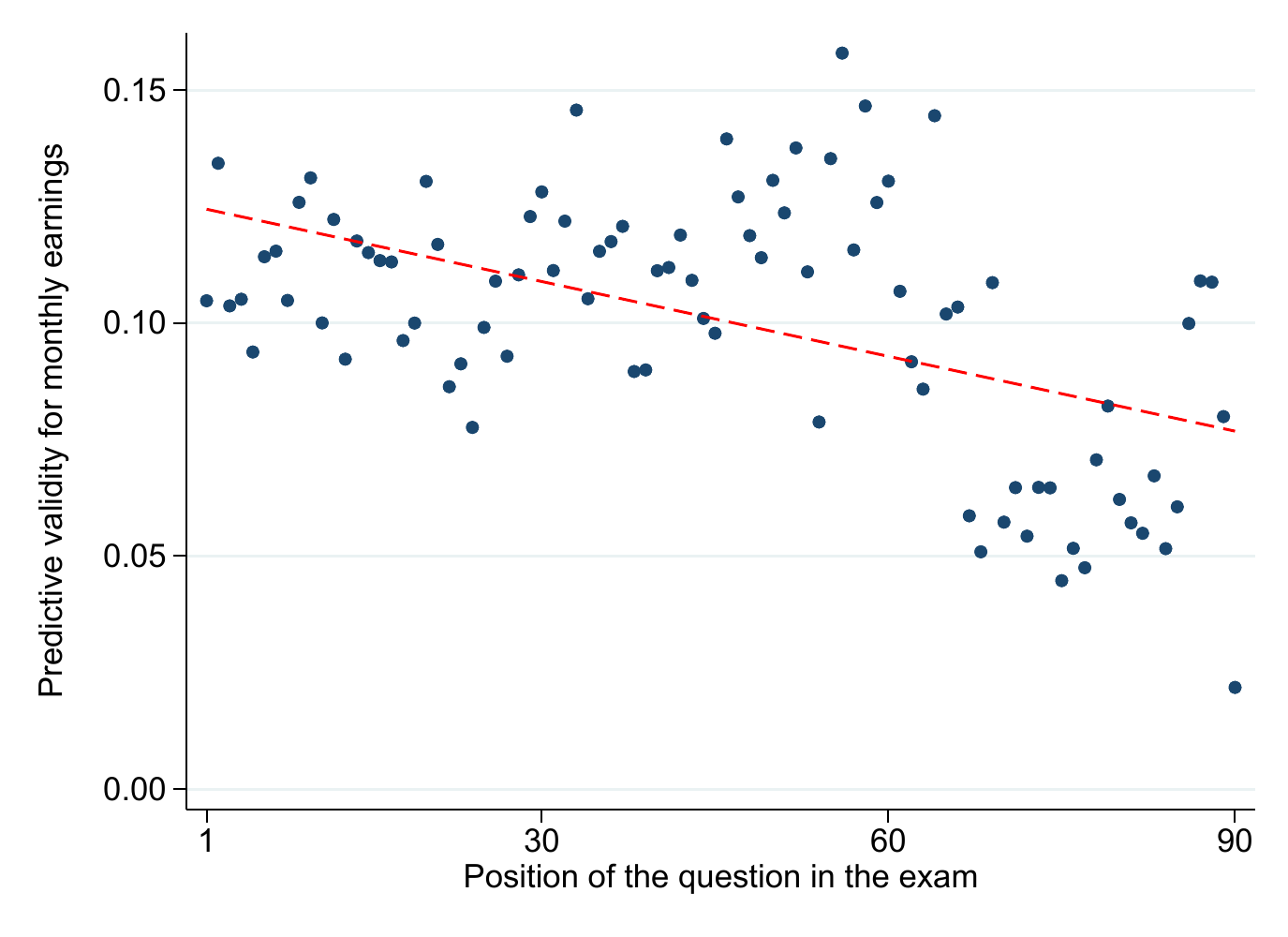}
	\end{subfigure}
	\hfill       
	\begin{subfigure}[t]{0.45\textwidth}
		\caption*{Panel F. Firm leave-one-out mean wage}
		\centering
		\includegraphics[width=\textwidth]{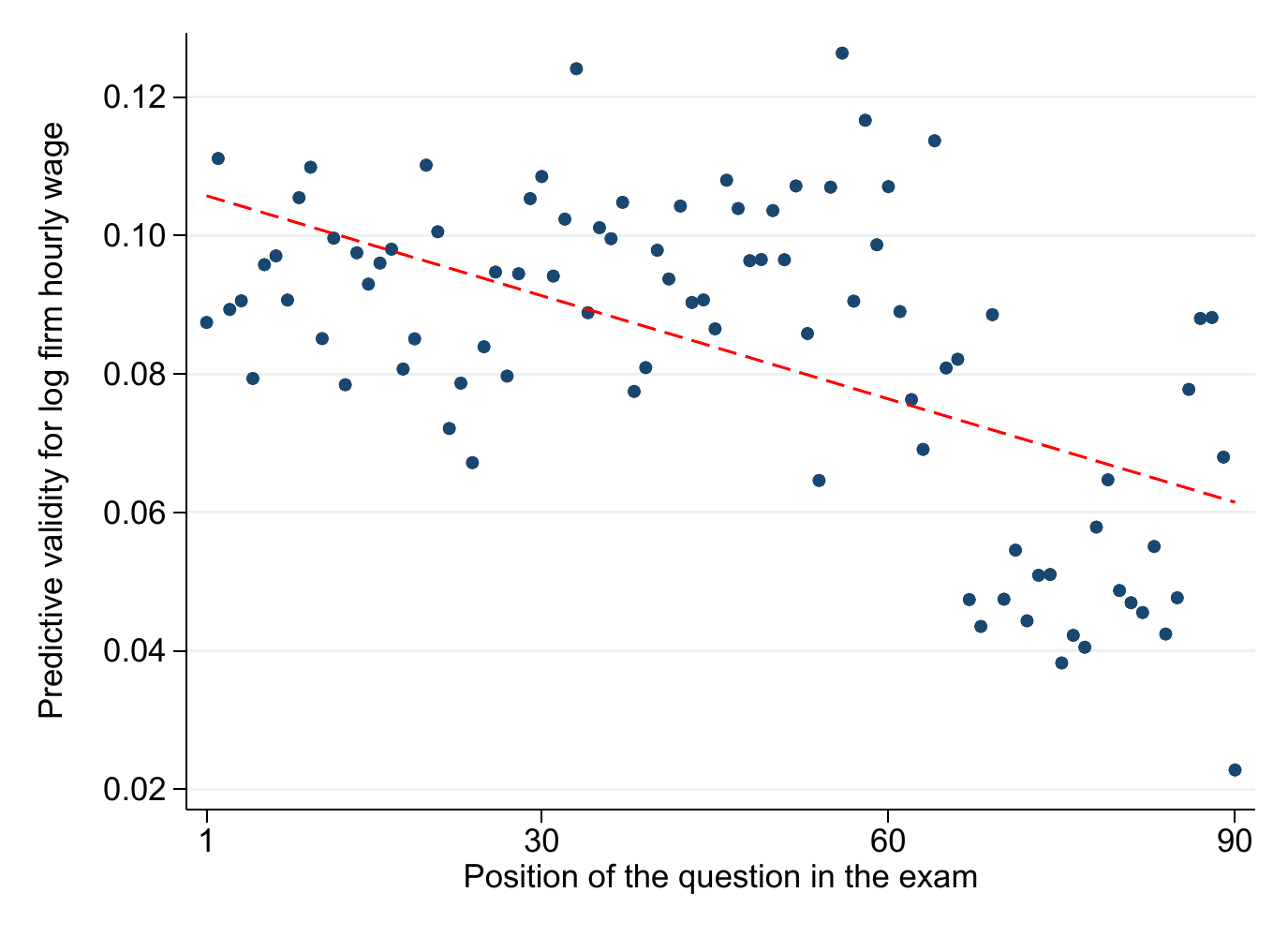}
	\end{subfigure}

	{\footnotesize
		\singlespacing \justify
		
		\textit{Notes:} This figure shows the relationship between (i) the predictive validity of an exam question for a given outcome and (ii) the position of the question on the exam. The $y$-axis plots the correlation between correctly responding to the question in position $j$ and a given outcome $Y$. The $x$-axis show the position of the question on the exam. Each plot shows the results for the outcome listed in the panel title. See Section \ref{sub:var-def} for outcome definitions. The red dashed lines are predicted values from a linear regression on the plotted points.
		
		
	}
\end{figure}

\clearpage
\begin{table}[htpb!]\caption{Summary statistics of the high-school-student sample by booklet color} \label{tab:summ-enem-book}
	{\footnotesize
		\begin{centering} 
			\protect
			\begin{tabular}{lcccccc}
				\addlinespace \addlinespace \midrule			
				&     && \multicolumn{4}{c}{Day 1 booklet color}     \\ \cmidrule{3-7} 
				& All && Yellow  & Blue & Pink & White              \\
				& (1) && (2)     & (3)  & (4)  & (5)                 \\
				\midrule 	
				\multicolumn{6}{l}{\hspace{-1em} \textbf{Panel A. Demographic characteristics and race}} \\ 
				Age &18.204 & &18.201 &18.210 &18.209 &18.196 \\
Female &0.598 & &0.595 &0.600 &0.595 &0.599 \\
White &0.476 & &0.478 &0.476 &0.476 &0.474 \\
Black/Brown &0.505 & &0.503 &0.505 &0.505 &0.507 \\
 \midrule
				
				\multicolumn{6}{l}{\hspace{-1em} \textbf{Panel B. Household characteristics}} \\ 
				Attends a private HS &0.222 & &0.225 &0.220 &0.223 &0.220 \\
Mom completed high school &0.534 & &0.538 &0.532 &0.535 &0.531 \\
Mom completed college &0.205 & &0.208 &0.203 &0.207 &0.203 \\
Family earns above 2x M.W. &0.388 & &0.392 &0.386 &0.390 &0.385 \\
Family earns above 5x M.W. &0.062 & &0.064 &0.061 &0.063 &0.061 \\
 \midrule												
				
				\multicolumn{6}{l}{\hspace{-1em} \textbf{Panel C. Exam preparation}} \\ 
				Took a foreign lang. course &0.241 & &0.241 &0.241 &0.240 &0.241 \\
Took a test prep course &0.119 & &0.121 &0.119 &0.119 &0.118 \\
 \midrule
				
				\multicolumn{6}{l}{\hspace{-1em} \textbf{Panel D. Fraction of correct responses}} \\ 
				Natural Science &0.283 & &0.284 &0.283 &0.283 &0.283 \\
Social Science &0.398 & &0.398 &0.398 &0.398 &0.398 \\
Language &0.408 & &0.410 &0.408 &0.408 &0.407 \\
Math &0.283 & &0.284 &0.283 &0.283 &0.282 \\
Average &0.343 & &0.344 &0.343 &0.343 &0.343 \\
 \midrule
				
				\multicolumn{7}{l}{\hspace{-1em} \textbf{Panel E. Geographical location}} \\ 
				Lives in the North &0.089 & &0.089 &0.089 &0.090 &0.089 \\
Lives in the Northeast &0.305 & &0.305 &0.303 &0.306 &0.305 \\
Lives in the Southeast &0.389 & &0.388 &0.390 &0.388 &0.389 \\
Lives in the South &0.131 & &0.131 &0.132 &0.130 &0.131 \\
Lives in the Midwest &0.086 & &0.086 &0.086 &0.086 &0.086 \\
 \midrule
				
				F-statistic &-- & &0.875 &1.051 &0.857 &0.887 \\
p-value F-statistic &-- & &0.599 &0.452 &0.614 &0.588 \\
Number of test-takers &14,941,156 & &3,655,807 &3,903,653 &3,590,977 &3,790,719 \\
 \midrule \addlinespace \addlinespace
				
			\end{tabular}
			\par\end{centering}
		
		\begin{singlespace} \vspace{-.5cm}
			\noindent \justify  \textit{Notes:} This table shows summary statistics on all test-takers in the high-school-students sample (column 1) and based on the booklet color they received on the first day of testing (columns 2--5). The last panel reports the $F$-statistics and $p$-values from $F$-tests that the coefficients on all pre-determined covariates (Panels A, B, C, and E) are jointly equal across booklet colors.

		\end{singlespace}	
	}
\end{table}

\clearpage		
\begin{table}[H]{\footnotesize
		\begin{center}
			\caption{\centering Examples of reliability estimates in economics and psychology} \label{tab:reliability}
			\begin{tabular}{ccc}
				\midrule
				Construct                & Reliability estimate & Reference \\ 
				(1)                      &    (2)      &    (3)   \\ \midrule
				IQ                       & 0.80        & \cite{schuerger1989temporal} \\
				Risk aversion            & 0.20--0.40  & \cite{mata2018risk} \\ 
				Big 5 personality traits & 0.60--0.73  & \cite{wooden2012stability} \\
				Present bias			 & 0.36        & \cite{meier2015temporal} \\
				Loss aversion            & 0.88        & \cite{stango2020behavioral} \\
				Teacher value added      & 0.23--0.47  & \cite*{chetty2014measuring} \\
				Life satisfaction        & 0.67        & \cite{anusic2016stability} \\
				Self-esteem              & 0.71        & \cite{anusic2016stability} \\
				Academic ability         & 0.61--0.77  & This paper \\
				Cognitive endurance      & 0.14--0.30  & This paper \\
				\midrule
			\end{tabular}%
		\end{center}
		\begin{singlespace} \vspace{-.5cm}
			\noindent \justify \textit{Notes:} This table displays examples of reliability estimates from the economics and psychology literature. The last two rows show the test-retest reliability of the measures of academic ability and cognitive endurance estimated in Section \ref{sec:recover}.
		\end{singlespace}		
	}
\end{table}

\begin{table}[H]{\footnotesize
		\begin{center}
			\caption{IV estimates of the relationship between ability/endurance and college outcomes} \label{tab:reg-rob-coll-iv}
			\newcommand\w{1.45}
			\begin{tabular}{l@{}lR{\w cm}@{}L{0.45cm}R{\w cm}@{}L{0.45cm}R{\w cm}@{}L{0.45cm}R{\w cm}@{}L{0.45cm}R{\w cm}@{}L{0.45cm}R{\w cm}@{}L{0.45cm}}
				
				\midrule
				& \multicolumn{12}{c}{Dependent variable} \\ \cmidrule{3-14} 
				&& Enrolled    && College && Degree   && 1st-year && Grad.      && Time to    \\
				&& college     && quality && quality  && credits  && on time    && grad.      \\
				&& (1)         && (2)     && (3)      && (4)      && (5)        &&  (6)       \\ \midrule \addlinespace  \addlinespace 
				
				\multicolumn{12}{l}{\hspace{-1em} \textbf{Panel A. OLS estimates on retakers sample}}  \\ \midrule				
				Endurance&&0.048&\sym{***}&0.057&\sym{***}&0.110&\sym{***}&0.010&\sym{***}&0.026&\sym{***}&$-$0.082&\sym{***}\\
&&(0.001&)&(0.001&)&(0.002&)&(0.000&)&(0.002&)&(0.006&)\\
Ability&&0.110&\sym{***}&0.129&\sym{***}&0.217&\sym{***}&0.018&\sym{***}&0.049&\sym{***}&$-$0.154&\sym{***}\\
&&(0.002&)&(0.001&)&(0.002&)&(0.000&)&(0.003&)&(0.009&)\\
 \midrule
				Ratio coef.&&0.441&\sym{***}&0.443&\sym{***}&0.509&\sym{***}&0.571&\sym{***}&0.543&\sym{***}&0.533&\sym{***}\\
&&(0.011&)&(0.006&)&(0.007&)&(0.008&)&(0.037&)&(0.032&)\\
 \midrule 
				Mean DV&&0.367&&3.420&&3.390&&0.146&&0.808&&4.191&\\
\(N\)&&132,634&&111,409&&109,390&&339,727&&51,066&&51,066&\\
 \midrule \addlinespace \addlinespace \addlinespace
				
				\multicolumn{12}{l}{\hspace{-1em} \textbf{Panel B. IV estimates on retakers sample}}  \\ \midrule				
				Endurance&&0.046&\sym{***}&0.067&\sym{***}&0.141&\sym{***}&0.016&\sym{***}&0.041&\sym{***}&$-$0.120&\sym{***}\\
&&(0.006&)&(0.004&)&(0.007&)&(0.001&)&(0.010&)&(0.028&)\\
Ability&&0.107&\sym{***}&0.140&\sym{***}&0.239&\sym{***}&0.022&\sym{***}&0.057&\sym{***}&$-$0.169&\sym{***}\\
&&(0.002&)&(0.001&)&(0.003&)&(0.000&)&(0.005&)&(0.013&)\\
 \midrule
				Ratio coef.&&0.430&\sym{***}&0.479&\sym{***}&0.590&\sym{***}&0.738&\sym{***}&0.722&\sym{***}&0.711&\sym{***}\\
&&(0.053&)&(0.025&)&(0.028&)&(0.026&)&(0.156&)&(0.145&)\\
 \midrule 
				Mean DV&&0.367&&3.420&&3.390&&0.146&&0.808&&4.191&\\
\(N\)&&132,614&&111,394&&109,375&&339,725&&51,056&&51,056&\\
 \midrule	\addlinespace

			\end{tabular}%
		\end{center}
		\begin{singlespace}  \vspace{-.5cm}
			\noindent \justify \textit{Notes:} This table displays OLS and IV estimates of the relationship between ability/endurance and college outcomes. 
			
			The OLS estimates are analogous to Table \ref{tab:reg-coll-lmkt} but estimated on the sample of retakers. See notes to Table \ref{tab:reg-coll-lmkt} for details. The IV estimates instrument the year $t$ measure of ability and cognitive endurance with the $t-1$ measures of these skills.
			
			Heteroskedasticity-robust standard errors clustered at the individual level in parentheses. $^{***}$, $^{**}$ and $^*$ denote significance at 10\%, 5\% and 1\% levels, respectively.		
			
		\end{singlespace} 	
	}
\end{table}%

\begin{table}[H]{\footnotesize
		\begin{center}
			\caption{IV estimates of the relationship between ability/endurance and labor-market outcomes} \label{tab:reg-rob-lmkt-iv}
			\newcommand\w{1.45}
			\begin{tabular}{l@{}lR{\w cm}@{}L{0.45cm}R{\w cm}@{}L{0.45cm}R{\w cm}@{}L{0.45cm}R{\w cm}@{}L{0.45cm}R{\w cm}@{}L{0.45cm}R{\w cm}@{}L{0.45cm}}		
				\midrule		
				& \multicolumn{12}{c}{Dependent variable} \\ \cmidrule{3-14} 
				&& Formal        && Hourly && Monthly   && Firm  && Occup.     && Industry  \\
				&& sector        && wage   && earnings  && wage  && wage       && wage      \\
				&& (1)           && (2)     && (3)      && (4)   && (5)        && (6)       \\	\midrule \addlinespace  \addlinespace 
				
				\multicolumn{12}{l}{\hspace{-1em} \textbf{Panel A. OLS estimates on retakers sample}}  \\ \midrule				
				Endurance&&0.001&\sym{***}&0.121&\sym{***}&0.124&\sym{***}&0.081&\sym{***}&0.030&\sym{***}&0.008&\sym{***}\\
&&(0.000&)&(0.005&)&(0.005&)&(0.005&)&(0.003&)&(0.001&)\\
Ability&&0.002&\sym{***}&0.231&\sym{***}&0.201&\sym{***}&0.163&\sym{***}&0.057&\sym{***}&0.018&\sym{***}\\
&&(0.000&)&(0.006&)&(0.006&)&(0.005&)&(0.003&)&(0.002&)\\
 \midrule
				Ratio coef.&&0.518&\sym{***}&0.525&\sym{***}&0.615&\sym{***}&0.496&\sym{***}&0.518&\sym{***}&0.413&\sym{***}\\
&&(0.072&)&(0.018&)&(0.020&)&(0.024&)&(0.040&)&(0.057&)\\
 \midrule 
				Mean DV&&0.286&&4.049&&7.702&&4.014&&3.992&&3.875&\\
\(N\)&&133,904&&37,814&&37,814&&32,908&&37,798&&37,814&\\
	\midrule \addlinespace  \addlinespace  \addlinespace 
				
				\multicolumn{12}{l}{\hspace{-1em} \textbf{Panel B. IV estimates on retakers sample}}  \\ \midrule											
				Endurance&&0.003&\sym{***}&0.188&\sym{***}&0.232&\sym{***}&0.120&\sym{***}&0.052&\sym{***}&0.001&\\
&&(0.001&)&(0.018&)&(0.017&)&(0.015&)&(0.009&)&(0.004&)\\
Ability&&0.002&\sym{***}&0.250&\sym{***}&0.215&\sym{***}&0.180&\sym{***}&0.061&\sym{***}&0.018&\sym{***}\\
&&(0.000&)&(0.008&)&(0.007&)&(0.007&)&(0.004&)&(0.002&)\\
 \midrule
				Ratio coef.&&1.171&\sym{***}&0.753&\sym{***}&1.077&\sym{***}&0.670&\sym{***}&0.845&\sym{***}&0.050&\\
&&(0.323&)&(0.069&)&(0.079&)&(0.081&)&(0.150&)&(0.241&)\\
 \midrule 
				Mean DV&&0.286&&4.049&&7.702&&4.014&&3.992&&3.875&\\
\(N\)&&133,884&&37,801&&37,801&&32,902&&37,785&&37,801&\\
 \midrule

			\end{tabular}%
		\end{center}
		\begin{singlespace}  \vspace{-.5cm}
			\noindent \justify \textit{Notes:} This table displays  OLS and IV estimates of the relationship between ability/endurance and labor-market outcomes. 
			
			The OLS estimates are analogous to Table \ref{tab:reg-coll-lmkt} but estimated on the sample of retakers. See notes to Table \ref{tab:reg-coll-lmkt} for details. The IV estimates instrument the year $t$ measure of ability and cognitive endurance with the $t-1$ measures of these skills.
			
			Heteroskedasticity-robust standard errors clustered at the individual level in parentheses. $^{***}$, $^{**}$ and $^*$ denote significance at 10\%, 5\% and 1\% levels, respectively.		
			
		\end{singlespace} 	
	}
\end{table}%

\begin{table}[H]{\footnotesize
		\begin{center}
			\caption{Robustness of baseline test-score-gaps decomposition to measuring variables in percentiles} \label{tab:corr-rob-pctil}
			\newcommand\w{1.75}
			\begin{tabular}{l@{}lR{\w cm}@{}L{0.5cm}R{\w cm}@{}L{0.5cm}R{\w cm}@{}L{0.5cm}R{\w cm}@{}L{0.5cm}R{\w cm}@{}L{0.5cm}}
				\midrule
				&&  \multicolumn{10}{c}{Gap between} \\ \cmidrule{2-12}
				&&    Male /  && White /   &&  Priv HS /   && Mom coll / && High-inc /  \\
				&&    Female  && Non-white &&  Public HS      && No coll  && Low-inc  \\
				&&     (1)    && (2)       && (3)             && (4)          && (5)       \\ \midrule \addlinespace 
				
				\multicolumn{12}{l}{\hspace{-1em} \textbf{Panel A. Difference in average test-score percentile}}  \\ \addlinespace   
				Score pctil gap&&5.871&\sym{***}&14.127&\sym{***}&27.826&\sym{***}&21.609&\sym{***}&39.838&\sym{***}\\
&&(0.015&)&(0.015&)&(0.019&)&(0.018&)&(0.023&)\\
 \midrule \addlinespace  
				
				\multicolumn{12}{l}{\hspace{-1em} \textbf{Panel B. Contribution of gaps in ability and endurance percentiles to test-score gaps}}  \\ \addlinespace   
				Ability pctil gap&&5.599&\sym{***}&10.618&\sym{***}&21.952&\sym{***}&16.728&\sym{***}&31.775&\sym{***}\\
&&(0.012&)&(0.011&)&(0.017&)&(0.015&)&(0.024&)\\

				Endurance pctil gap&&3.205&\sym{***}&3.097&\sym{***}&6.628&\sym{***}&4.596&\sym{***}&10.381&\sym{***}\\
&&(0.006&)&(0.006&)&(0.010&)&(0.008&)&(0.015&)\\
  \midrule \addlinespace

				\multicolumn{12}{l}{\hspace{-1em} \textbf{Panel C. Impact of a reform that halves the exam length on test-score percentile gaps}}  \\ \addlinespace   
				Pctil. change gap&&-1.603&\sym{***}&-1.548&\sym{***}&-3.314&\sym{***}&-2.298&\sym{***}&-5.190&\sym{***}\\
&&(0.003&)&(0.003&)&(0.005&)&(0.004&)&(0.008&)\\

				Pct. change gap&&-0.546&\sym{***}&-0.219&\sym{***}&-0.238&\sym{***}&-0.213&\sym{***}&-0.261&\sym{***}\\
&&(0.001&)&(0.000&)&(0.000&)&(0.000&)&(0.000&)\\
\midrule \addlinespace				
				
				\(N\) (Students)&&14,941,097&&14,565,550&&9,924,652&&14,290,759&&9,996,959&\\
 \midrule
			\end{tabular}
		\end{center}
		\begin{singlespace}  \vspace{-.5cm}
			\footnotesize \noindent \justify \textit{Notes:} This table is analogous to Table \ref{tab:corr-endurance}, but the variables and effects are measured in percentiles. I construct the percentiles separately for each cohort. See notes to Table \ref{tab:corr-endurance} for details.

		\end{singlespace} 	
	}
\end{table}

\begin{table}[H]{\footnotesize
		\begin{center}
			\caption{Robustness of baseline test-score-gaps decomposition to alternative ways of measuring ability and endurance} \label{tab:corr-rob-meas}
			\newcommand\w{1.8}
			\begin{tabular}{l@{}lR{\w cm}@{}L{0.5cm}R{\w cm}@{}L{0.5cm}R{\w cm}@{}L{0.5cm}R{\w cm}@{}L{0.5cm}R{\w cm}@{}L{0.5cm}}
				\midrule
				&&  \multicolumn{10}{c}{Gap between} \\ \cmidrule{2-12}
				&&    Male /  && White /   &&  Priv HS /   && Mom coll / && High-inc /  \\
				&&    Female  && Non-white &&  Public HS      && No coll  && Low-inc  \\
				&&     (1)    && (2)       && (3)             && (4)          && (5)       \\ \midrule \addlinespace 
				
				\multicolumn{12}{l}{\hspace{-1em} \textbf{Panel A. Estimating ability/endurance separately by day and using the average}}  \\ \addlinespace
				Ability gap&&0.030&\sym{***}&0.056&\sym{***}&0.127&\sym{***}&0.095&\sym{***}&0.188&\sym{***}\\
&&(0.000&)&(0.000&)&(0.000&)&(0.000&)&(0.000&)\\
			
				Endurance gap&&0.034&\sym{***}&0.032&\sym{***}&0.075&\sym{***}&0.051&\sym{***}&0.126&\sym{***}\\
&&(0.000&)&(0.000&)&(0.000&)&(0.000&)&(0.000&)\\
 \midrule				
				\multicolumn{12}{l}{\hspace{-1em} \textbf{Panel B. Estimating ability/endurance separately by subject and using the average}}  \\ \addlinespace
				Ability gap&&0.027&\sym{***}&0.061&\sym{***}&0.140&\sym{***}&0.104&\sym{***}&0.206&\sym{***}\\
&&(0.000&)&(0.000&)&(0.000&)&(0.000&)&(0.000&)\\
				Endurance gap&&0.004&\sym{***}&0.042&\sym{***}&0.099&\sym{***}&0.069&\sym{***}&0.163&\sym{***}\\
&&(0.000&)&(0.000&)&(0.000&)&(0.000&)&(0.000&)\\
 \midrule		
				\multicolumn{12}{l}{\hspace{-1em} \textbf{Panel C. Including day fixed effects}}  \\ \addlinespace
				Ability gap&&0.030&\sym{***}&0.056&\sym{***}&0.128&\sym{***}&0.096&\sym{***}&0.189&\sym{***}\\
&&(0.000&)&(0.000&)&(0.000&)&(0.000&)&(0.000&)\\
		
				Endurance gap&&0.034&\sym{***}&0.031&\sym{***}&0.075&\sym{***}&0.051&\sym{***}&0.125&\sym{***}\\
&&(0.000&)&(0.000&)&(0.000&)&(0.000&)&(0.000&)\\
 \midrule		
				\multicolumn{12}{l}{\hspace{-1em} \textbf{Panel D. Including subject fixed effects}}  \\ \addlinespace
				Ability gap&&0.026&\sym{***}&0.061&\sym{***}&0.141&\sym{***}&0.104&\sym{***}&0.207&\sym{***}\\
&&(0.000&)&(0.000&)&(0.000&)&(0.000&)&(0.000&)\\
		
				Endurance gap&&0.002&\sym{***}&0.039&\sym{***}&0.095&\sym{***}&0.066&\sym{***}&0.158&\sym{***}\\
&&(0.000&)&(0.000&)&(0.000&)&(0.000&)&(0.000&)\\
 \midrule	
				\multicolumn{12}{l}{\hspace{-1em} \textbf{Panel E. Using linear correlation as an alternative measure of endurance}}  \\ \addlinespace
				Ability gap&&0.030&\sym{***}&0.057&\sym{***}&0.129&\sym{***}&0.097&\sym{***}&0.191&\sym{***}\\
&&(0.000&)&(0.000&)&(0.000&)&(0.000&)&(0.000&)\\

				Endurance gap&&0.021&\sym{***}&0.020&\sym{***}&0.047&\sym{***}&0.032&\sym{***}&0.079&\sym{***}\\
&&(0.000&)&(0.000&)&(0.000&)&(0.000&)&(0.000&)\\
 \midrule	
				
				\(N\) (Students)&&14,941,097&&14,565,550&&9,924,652&&14,290,759&&9,996,959&\\
 \addlinespace

			\end{tabular}
		\end{center}
		\begin{singlespace}  \vspace{-.5cm}
			\footnotesize \noindent \justify \textit{Notes:} This table shows estimates of the contribution of gaps in ability and endurance to test-score gaps using alternative specifications to estimate ability and endurance.
						
			Each column shows the result for a different test-score gap. Each panel shows the result from estimating ability and endurance with a different specification. In Panels A--B, I estimate a student's ability/endurance separately for each testing day (Panel A) and academic subject (Panel B) and then average the estimates across days or subjects. In Panels C--D, I estimate endurance in a regression that controls for day fixed effects (Panel C) or subject fixed effects (Panel D). Finally, in Panel E, I use the correlation between question position and a dummy for correctly answering a question as an alternative measure of endurance.

			Heteroskedasticity-robust standard errors clustered at the question level in parentheses. $^{***}$, $^{**}$ and $^*$ denote significance at 10\%, 5\% and 1\% levels, respectively.

		\end{singlespace} 	
	}
\end{table}

\begin{table}[H]{\footnotesize
		\begin{center}
			\caption{Robustness of baseline test-score-gaps decomposition to alternative ways of controlling for question difficulty when estimating ability/endurance} \label{tab:corr-rob-diff}
			\newcommand\w{1.8}
			\begin{tabular}{l@{}lR{\w cm}@{}L{0.5cm}R{\w cm}@{}L{0.5cm}R{\w cm}@{}L{0.5cm}R{\w cm}@{}L{0.5cm}R{\w cm}@{}L{0.5cm}}
				\midrule
				&&  \multicolumn{10}{c}{Gap between} \\ \cmidrule{2-12}
				&&    Male /  && White /   &&  Priv HS /   && Mom coll / && High-inc /  \\
				&&    Female  && Non-white &&  Public HS      && No coll  && Low-inc  \\
				&&     (1)    && (2)       && (3)             && (4)          && (5)       \\ \midrule \addlinespace 
				
				\multicolumn{12}{l}{\hspace{-1em} \textbf{Panel A. Not controlling for question difficulty}}  \\ \addlinespace
				Ability gap&&0.032&\sym{***}&0.049&\sym{***}&0.118&\sym{***}&0.087&\sym{***}&0.175&\sym{***}\\
&&(0.000&)&(0.000&)&(0.000&)&(0.000&)&(0.000&)\\
 				
				Endurance gap&&0.033&\sym{***}&0.026&\sym{***}&0.074&\sym{***}&0.050&\sym{***}&0.128&\sym{***}\\
&&(0.000&)&(0.000&)&(0.000&)&(0.000&)&(0.000&)\\
 \midrule
				\multicolumn{12}{l}{\hspace{-1em} \textbf{Panel B. Estimating difficulty without adjusting for average position}}  \\ \addlinespace
				Ability gap&&0.028&\sym{***}&0.057&\sym{***}&0.130&\sym{***}&0.097&\sym{***}&0.191&\sym{***}\\
&&(0.000&)&(0.000&)&(0.000&)&(0.000&)&(0.000&)\\

				Endurance gap&&0.034&\sym{***}&0.036&\sym{***}&0.080&\sym{***}&0.055&\sym{***}&0.131&\sym{***}\\
&&(0.000&)&(0.000&)&(0.000&)&(0.000&)&(0.000&)\\
 \midrule		
				\multicolumn{12}{l}{\hspace{-1em} \textbf{Panel C. Estimating difficulty using question-specific position effects}}  \\ \addlinespace
				Ability gap&&0.032&\sym{***}&0.055&\sym{***}&0.126&\sym{***}&0.095&\sym{***}&0.186&\sym{***}\\
&&(0.000&)&(0.000&)&(0.000&)&(0.000&)&(0.000&)\\

				Endurance gap&&0.036&\sym{***}&0.031&\sym{***}&0.082&\sym{***}&0.058&\sym{***}&0.135&\sym{***}\\
&&(0.000&)&(0.000&)&(0.000&)&(0.000&)&(0.000&)\\
 \midrule		
				\multicolumn{12}{l}{\hspace{-1em} \textbf{Panel D. Estimating difficulty using shrunk question-specific position effects}}  \\ \addlinespace
				Ability gap&&0.031&\sym{***}&0.056&\sym{***}&0.127&\sym{***}&0.095&\sym{***}&0.187&\sym{***}\\
&&(0.000&)&(0.000&)&(0.000&)&(0.000&)&(0.000&)\\
 
				Endurance gap&&0.036&\sym{***}&0.032&\sym{***}&0.080&\sym{***}&0.057&\sym{***}&0.131&\sym{***}\\
&&(0.000&)&(0.000&)&(0.000&)&(0.000&)&(0.000&)\\
 \midrule		
				\multicolumn{12}{l}{\hspace{-1em} \textbf{Panel E. Estimating position effects separately by fraction of correct responses}}  \\ \addlinespace
				Ability gap&&0.030&\sym{***}&0.056&\sym{***}&0.128&\sym{***}&0.096&\sym{***}&0.189&\sym{***}\\
&&(0.000&)&(0.000&)&(0.000&)&(0.000&)&(0.000&)\\
 
				Endurance gap&&0.035&\sym{***}&0.032&\sym{***}&0.078&\sym{***}&0.053&\sym{***}&0.130&\sym{***}\\
&&(0.000&)&(0.000&)&(0.000&)&(0.000&)&(0.000&)\\
 \midrule		
				\multicolumn{12}{l}{\hspace{-1em} \textbf{Panel F.  Estimating position effects separately by subject}}  \\ \addlinespace
				Ability gap&&0.029&\sym{***}&0.057&\sym{***}&0.128&\sym{***}&0.096&\sym{***}&0.190&\sym{***}\\
&&(0.000&)&(0.000&)&(0.000&)&(0.000&)&(0.000&)\\

				Endurance gap&&0.034&\sym{***}&0.034&\sym{***}&0.077&\sym{***}&0.053&\sym{***}&0.128&\sym{***}\\
&&(0.000&)&(0.000&)&(0.000&)&(0.000&)&(0.000&)\\
 \midrule
				
				\(N\) (Students)&&14,941,097&&14,565,550&&9,924,652&&14,290,759&&9,996,959&\\
 \addlinespace
				
			\end{tabular}
		\end{center}
		\begin{singlespace}  \vspace{-.5cm}
			\footnotesize \noindent \justify \textit{Notes:} This table shows estimates of the contribution of gaps in ability and endurance to test-score gaps using alternative measures of difficulty in the specification used to estimate ability and endurance.
						
			Each column shows the result for a different test-score gap. Each panel shows the result from a different way of controlling for question difficulty in equation \eqref{reg:lpm-ind}. In Panel A, I compute the estimate equation \eqref{reg:lpm-ind} without controlling for question difficulty. In Panel B, I measure question difficulty as the fraction of students who incorrectly answer to the question across all booklets. In Panels C--F, I adjust for average question position by estimating the positon effects with alternative specifications. In column C, I compute question-specific position effects. In Panel D, I compute a shrinkage estimator of the position effects. In Panel E, I compute the position effects separately for questions with a below/above fraction of correct responses. In Panel F, I compute the position effects separately by subject. See Appendix \ref{app:difficulty} for details on each measure of question difficulty.
			
			Heteroskedasticity-robust standard errors clustered at the question level in parentheses. $^{***}$, $^{**}$ and $^*$ denote significance at 10\%, 5\% and 1\% levels, respectively.

		\end{singlespace} 	
	}
\end{table}

\begin{table}[H]{\footnotesize
		\begin{center}
			\caption{Robustness of baseline test-score-gaps decomposition to alternative sample restrictions} \label{tab:corr-rob-samp}
			\newcommand\w{1.8}
			\begin{tabular}{l@{}lR{\w cm}@{}L{0.5cm}R{\w cm}@{}L{0.5cm}R{\w cm}@{}L{0.5cm}R{\w cm}@{}L{0.5cm}R{\w cm}@{}L{0.5cm}}
				\midrule
				&&  \multicolumn{10}{c}{Gap between} \\ \cmidrule{2-12}
				&&    Male /  && White /   &&  Priv HS /   && Mom coll / && High-inc /  \\
				&&    Female  && Non-white &&  Public HS      && No coll  && Low-inc  \\
				&&     (1)    && (2)       && (3)             && (4)          && (5)       \\ \midrule \addlinespace 
				\multicolumn{12}{l}{\hspace{-1em} \textbf{Panel A. Excluding students in the bottom or top 10\% of the ability distribution}}  \\ \addlinespace
				Ability gap&&0.018&\sym{***}&0.036&\sym{***}&0.076&\sym{***}&0.055&\sym{***}&0.116&\sym{***}\\
&&(0.000&)&(0.000&)&(0.000&)&(0.000&)&(0.000&)\\
	
				Endurance gap&&0.032&\sym{***}&0.031&\sym{***}&0.067&\sym{***}&0.045&\sym{***}&0.107&\sym{***}\\
&&(0.000&)&(0.000&)&(0.000&)&(0.000&)&(0.000&)\\
 \midrule				
				\multicolumn{12}{l}{\hspace{-1em} \textbf{Panel B. Excluding students in the bottom or top 10\% of the endurance distribution}}  \\ \addlinespace
				Ability gap&&0.027&\sym{***}&0.054&\sym{***}&0.125&\sym{***}&0.094&\sym{***}&0.190&\sym{***}\\
&&(0.000&)&(0.000&)&(0.000&)&(0.000&)&(0.000&)\\
 
				Endurance gap&&0.017&\sym{***}&0.017&\sym{***}&0.041&\sym{***}&0.028&\sym{***}&0.078&\sym{***}\\
&&(0.000&)&(0.000&)&(0.000&)&(0.000&)&(0.000&)\\
 \midrule			
				\multicolumn{12}{l}{\hspace{-1em} \textbf{Panel C. Excluding students in the bottom or top 10\% of either distribution}}  \\ \addlinespace
				Ability gap&&0.016&\sym{***}&0.036&\sym{***}&0.073&\sym{***}&0.053&\sym{***}&0.113&\sym{***}\\
&&(0.000&)&(0.000&)&(0.000&)&(0.000&)&(0.000&)\\
 		
				Endurance gap&&0.017&\sym{***}&0.018&\sym{***}&0.037&\sym{***}&0.025&\sym{***}&0.062&\sym{***}\\
&&(0.000&)&(0.000&)&(0.000&)&(0.000&)&(0.000&)\\
 \midrule		
				\multicolumn{12}{l}{\hspace{-1em} \textbf{Panel D. Excluding students in the bottom or top 20\% of either distribution}}  \\ \addlinespace
				Ability gap&&0.009&\sym{***}&0.023&\sym{***}&0.042&\sym{***}&0.030&\sym{***}&0.066&\sym{***}\\
&&(0.000&)&(0.000&)&(0.000&)&(0.000&)&(0.000&)\\
 
				Endurance gap&&0.009&\sym{***}&0.011&\sym{***}&0.019&\sym{***}&0.013&\sym{***}&0.032&\sym{***}\\
&&(0.000&)&(0.000&)&(0.000&)&(0.000&)&(0.000&)\\
 \midrule		
				\multicolumn{12}{l}{\hspace{-1em} \textbf{Panel E.  Excluding individuals with positive estimated endurance}}  \\ \addlinespace
				Ability gap&&0.021&\sym{***}&0.050&\sym{***}&0.112&\sym{***}&0.084&\sym{***}&0.165&\sym{***}\\
&&(0.000&)&(0.000&)&(0.000&)&(0.000&)&(0.000&)\\

				Endurance gap&&0.013&\sym{***}&0.017&\sym{***}&0.037&\sym{***}&0.025&\sym{***}&0.063&\sym{***}\\
&&(0.000&)&(0.000&)&(0.000&)&(0.000&)&(0.000&)\\
 \midrule					
				
			\end{tabular}
		\end{center}
		\begin{singlespace}  \vspace{-.5cm}
			\footnotesize \noindent \justify \textit{Notes:} This table shows estimates of the contribution of gaps in ability and endurance to test-score gaps using alternative sample restrictions.
			
			Each column shows the result for a different test-score gap. Each panel shows the result for a different sample of students. In Panel A, I exclude students in the bottom and top deciles of the ability distribution. In Panel B, I exclude students in the bottom and top deciles of the endurance distribution. In Panel C, I exclude students in the bottom and top deciles of the distribution of either skill. In Panel D, I exclude students in the bottom and top quintiles of the distribution of either skill. In Panel E, I exclude students with positive estimated endurance. I construct the deciles and quintiles using all the students in the high-school-students sample.
			
			Heteroskedasticity-robust standard errors clustered at the question level in parentheses. $^{***}$, $^{**}$ and $^*$ denote significance at 10\%, 5\% and 1\% levels, respectively.

		\end{singlespace} 	
	}
\end{table}

\begin{table}[H]{\footnotesize
		\begin{center}
			\caption{Robustness of baseline test-score-gaps decomposition to accounting for measurement error} \label{tab:corr-rob-prec}
			\newcommand\w{1.8}
			\begin{tabular}{l@{}lR{\w cm}@{}L{0.5cm}R{\w cm}@{}L{0.5cm}R{\w cm}@{}L{0.5cm}R{\w cm}@{}L{0.5cm}R{\w cm}@{}L{0.5cm}}
				\midrule
				&&  \multicolumn{10}{c}{Gap between} \\ \cmidrule{2-12}
				&&    Male /  && White /   &&  Priv HS /   && Mom coll / && High-inc /  \\
				&&    Female  && Non-white &&  Public HS      && No coll  && Low-inc  \\
				&&     (1)    && (2)       && (3)             && (4)          && (5)       \\ \midrule \addlinespace 
				
				\multicolumn{12}{l}{\hspace{-1em} \textbf{Panel A. Weighting each observation by its precision}}  \\ \addlinespace				
				Ability gap&&0.030&\sym{***}&0.056&\sym{***}&0.127&\sym{***}&0.095&\sym{***}&0.188&\sym{***}\\
&&(0.000&)&(0.000&)&(0.000&)&(0.000&)&(0.000&)\\
	
				Endurance gap&&0.034&\sym{***}&0.031&\sym{***}&0.075&\sym{***}&0.051&\sym{***}&0.126&\sym{***}\\
&&(0.000&)&(0.000&)&(0.000&)&(0.000&)&(0.000&)\\
 \midrule				
				\multicolumn{12}{l}{\hspace{-1em} \textbf{Panel B. Shrunk estimator of ability and endurance}}  \\ \addlinespace
				Ability gap&&0.021&\sym{***}&0.040&\sym{***}&0.091&\sym{***}&0.067&\sym{***}&0.136&\sym{***}\\
&&(0.000&)&(0.000&)&(0.000&)&(0.000&)&(0.000&)\\
 
				Endurance gap&&0.009&\sym{***}&0.010&\sym{***}&0.023&\sym{***}&0.016&\sym{***}&0.040&\sym{***}\\
&&(0.000&)&(0.000&)&(0.000&)&(0.000&)&(0.000&)\\
 \midrule			
				\(N\) (Students)&&14,941,097&&14,565,550&&9,924,652&&14,290,759&&9,996,959&\\
 \addlinespace				
			\end{tabular}
		\end{center}
		\begin{singlespace}  \vspace{-.5cm}
			\footnotesize \noindent \justify \textit{Notes:} This table shows estimates of the contribution of gaps in ability and endurance to test-score gaps accounting for measurement error in the estimates of ability and endurance.
			
			Each column shows the result for a different test-score gap. In Panel A, I weight each observation by the inverse of the standard error of the ability and endurance estimates. Specifically, the weight of each observation is $w = 1/(\text{SE}_{\hat{\alpha}_i}^2 + \text{SE}_{\hat{\beta}_i}^2)$, where $\text{SE}_{\hat{\alpha}_i}$ and $\text{SE}_{\hat{\beta}_i}^2$ are the standard errors of $\hat{\alpha}_i$ and $\hat{\beta}_i$. In Panel B, I estimate the baseline regression using a shrunk estimator of ability and endurance. I compute the shrunk estimator of endurance as $\beta^s_i = \omega_i  \hat{\beta}_i + (1 - \omega_j)\bar{\beta},$ where $\bar{\beta}$ is the average cognitive endurance in my sample. The individual-specific weight is $\omega_i = \frac{\Var[\beta_i] - \E[\text{SE}_{\hat{\beta}_i}^2]}{\Var[\beta_i] - \E[\text{SE}_{\hat{\beta}_i}^2] + \text{SE}_{\hat{\beta}_i}^2}.$ The shrunk estimator, $\beta^s_i$, puts more weight on estimates of $\beta_i$ that are more precisely estimated, as measured by a low standard error. I compute the shrunk estimator of ability analgously.
			
			Heteroskedasticity-robust standard errors clustered at the question level in parentheses. $^{***}$, $^{**}$ and $^*$ denote significance at 10\%, 5\% and 1\% levels, respectively.

		\end{singlespace} 	
	}
\end{table}

	\clearpage 
\section{Empirical Appendix} \label{app:empirics}

\setcounter{table}{0}
\setcounter{figure}{0}
\setcounter{equation}{0}	
\renewcommand{\thetable}{B\arabic{table}}
\renewcommand{\thefigure}{B\arabic{figure}}
\renewcommand{\theequation}{B\arabic{equation}}

\subsection{Limited Cognitive Endurance and Time Pressure}  \label{app:time}

Is the causal effect of an increase in the order of a given question on student performance a manifestation of limited cognitive endurance or is it driven by students running out of time? Two pieces of evidence suggest that time pressure does not explain the estimated $\beta < 0$.

\begin{figure}[H]
	\caption{Fraction of question left unanswered throughout the ENEM}\label{fig:miss-quest}
	\centering
	\includegraphics[width=.65\linewidth]{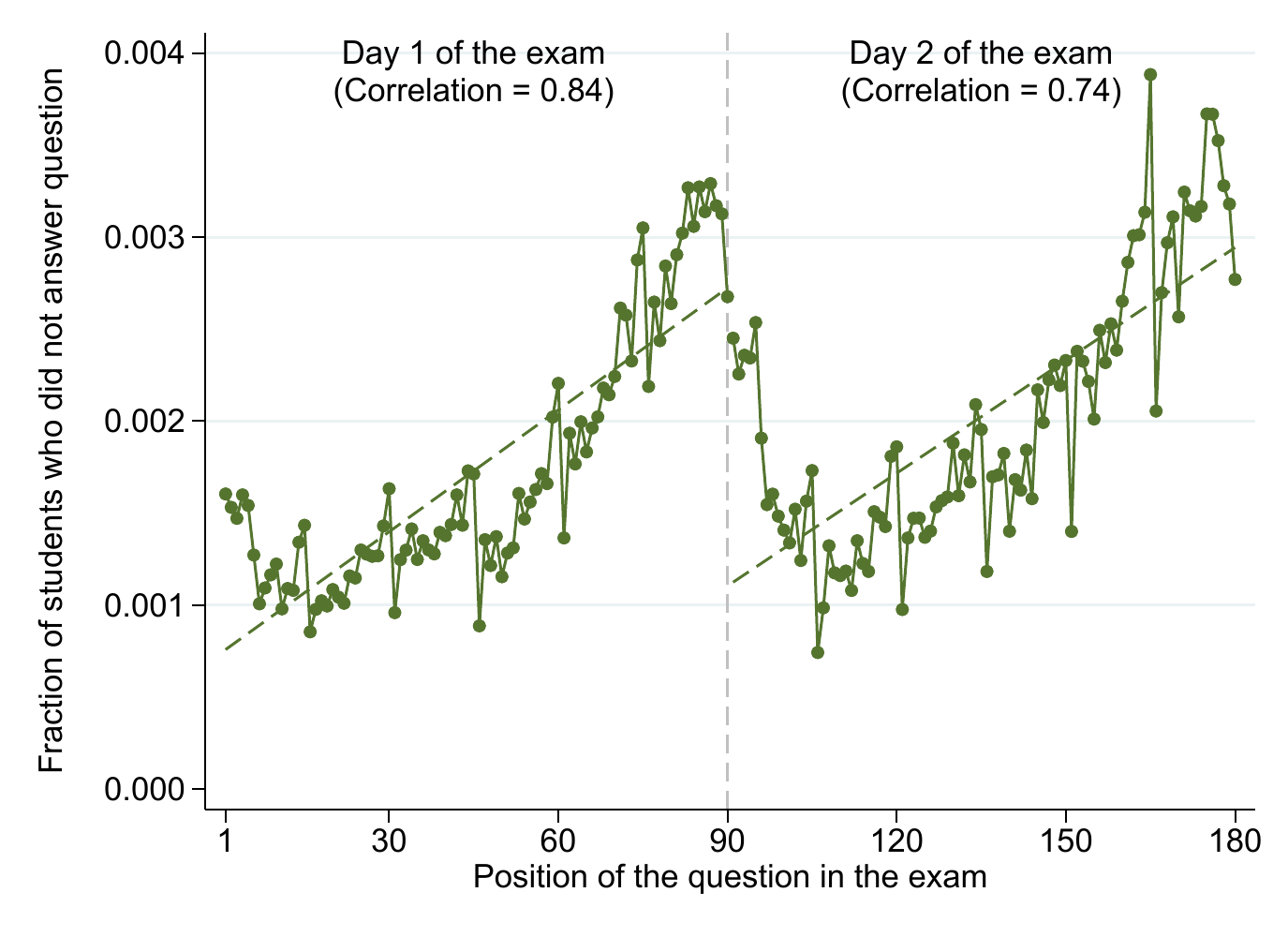}
	
	{\footnotesize
		\singlespacing \justify
		
		\textit{Notes:} This figure shows the fraction of questions left unanswered over the course of each testing day. The $y$-axis displays the fraction of students who did not select any of the multiple-choice answers to a given question. The $x$-axis displays the position of each question in the exam. The dashed lines are predicted values from a linear regression estimated separately for each testing day.
		
	}
	
\end{figure}

First, very few students leave responses unanswered. Appendix Figure \ref{fig:miss-quest} plots the fraction of students who left a question unanswered (possibly, because they ran out of time) against the question position. Questions that appear later in the test are more likely to be left unanswered. However, only a small fraction of students leave \textit{any} questions unanswered. Thus, missing responses cannot account for the large change in performance observed throughout the exam.\footnote{There is no penalty for incorrectly answering a question. Therefore, this evidence is only suggestive since leaving a question unanswered is a weakly dominated strategy.}

\begin{figure}[H]
	\caption{The heterogeneous effect of fatigue on performance by question position}\label{fig:chg-pos-firsthalf}
	\centering
	\begin{subfigure}[t]{.48\textwidth}
		\caption*{Panel A. First half of each testing day}
		\centering
		\includegraphics[width=\linewidth]{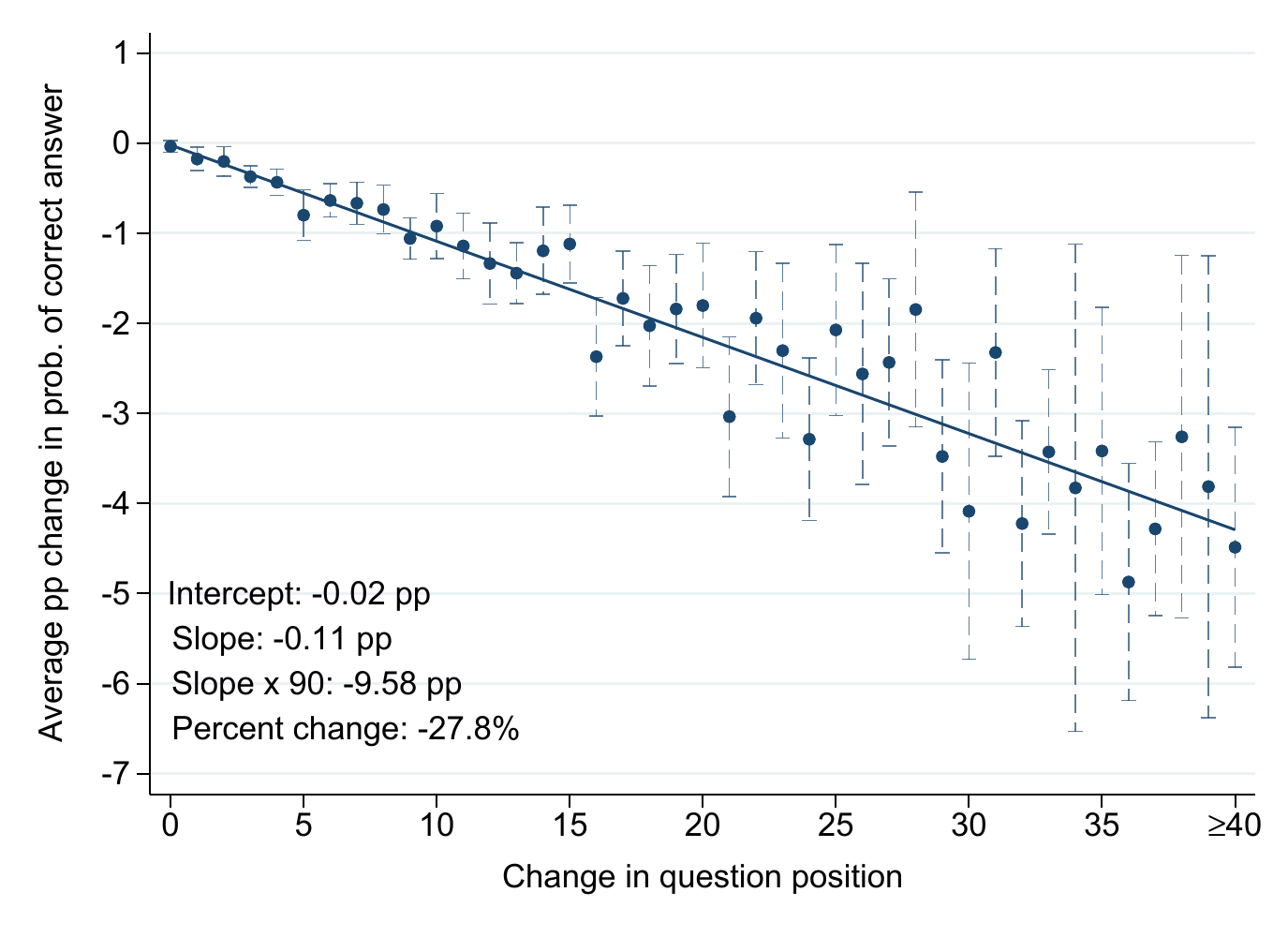}
	\end{subfigure}
	\hfill		
	\begin{subfigure}[t]{0.48\textwidth}
		\caption*{Panel B. Second half of each testing day}
		\centering
		\includegraphics[width=\linewidth]{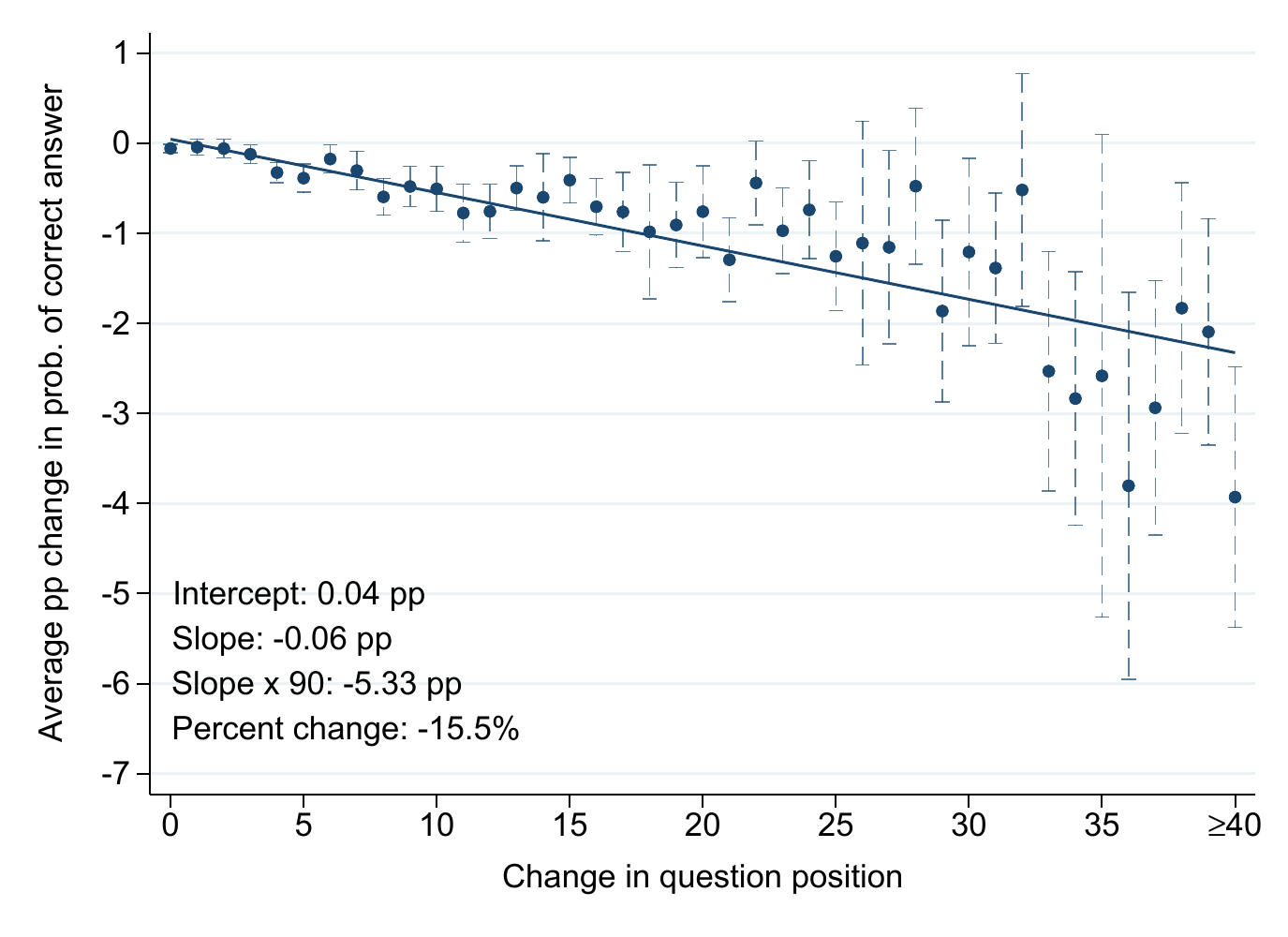}
	\end{subfigure}	
	\hfill						
	{\footnotesize
		\singlespacing \justify
		
		\textit{Notes:} This figure shows heterogeneity in the effect of limited endurance on performance by question position. Panels A and B are analogous to Figure \ref{fig:chg-pos}, but the effect is estimated separately for questions that appear on the first half of each testing day (Panel A) or the second half of each testing day (Panel B). The $y$-axis shows the average change (in percentage points) in the fraction of students correctly responding to a question. The $x$-axis plots the difference in the question position between each possible booklet pair. The dashed line denotes predicted values from a linear regression estimated on the plotted points, using the number of questions used to estimate each point as weights.
		
	}
\end{figure}

Second, student performance declines even in questions that students answer when they are likely not time-pressured. Appendix Table \ref{fig:chg-pos-char} estimates the fatigue effect separately for questions that appear in the first half (column 1) and the second half of each testing day (column 2). Presumably, students should have plenty of time to answer the first half of the exam. Yet, I still find fatigue effects that are quantitatively similar---or even larger---to those estimated on the second half of each day or with all questions (see also Appendix Figure \ref{fig:chg-pos-firsthalf}). This result is consistent with visual evidence in Figure \ref{fig:mean-corr-res}, which shows that student performance tends to decline shortly after the exam starts and with the declines in performance exhibited by the example questions that appear at the beginning of the exam in Appendix Figure \ref{fig:mean-corr-item}.

In summary, the evidence indicates that the effect of a question position on student performance is not driven by students running out of time.

\clearpage
\subsection{OLS Formulas of Academic Ability and Cognitive Endurance} \label{app:ols-endurance}

My measure of cognitive endurance is $\beta_i$ in equation \eqref{reg:lpm-ind}. Ignoring controls for question difficulty, the OLS estimator of $\beta_i$ is
\begin{align} \label{eq:lpm-slope}
	\hat{\beta}_i &= \frac{\sum_{j}(\text{Pos}_{ij} - \overline{\text{Pos}})(C_{ij}-\bar{C_i})}{\sum_{j}(\text{Pos}_{ij} - \overline{\text{Pos}})^2} \notag \\
	&= \sum_{j} \underbrace{w_{j}}_{\substack{\text{Weight of} \\ \text{question $j$}}} \times \underbrace{(C_{ij}-\bar{C_i}),}_{\substack{\text{Performance on} \\ \text{question $j$ relative to} \\ \text{$i$'s average performance}}}
\end{align}
where $\bar{C}_i$ is the fraction of questions correctly answered by student $i$, $\overline{\text{Pos}}$ is the average question position (which is constant across test-takers), and $w_{j} \equiv \frac{\text{Pos}_{j} - \overline{\text{Pos}}}{\sum_{j}(\text{Pos}_{j} - \overline{\text{Pos}})^2}$ is the weight of question $j$.

\begin{figure}[H]
	\caption{Weight of each question in a test with 90 questions}\label{fig:ols-wt}
	\centering
	\includegraphics[width=.60\linewidth]{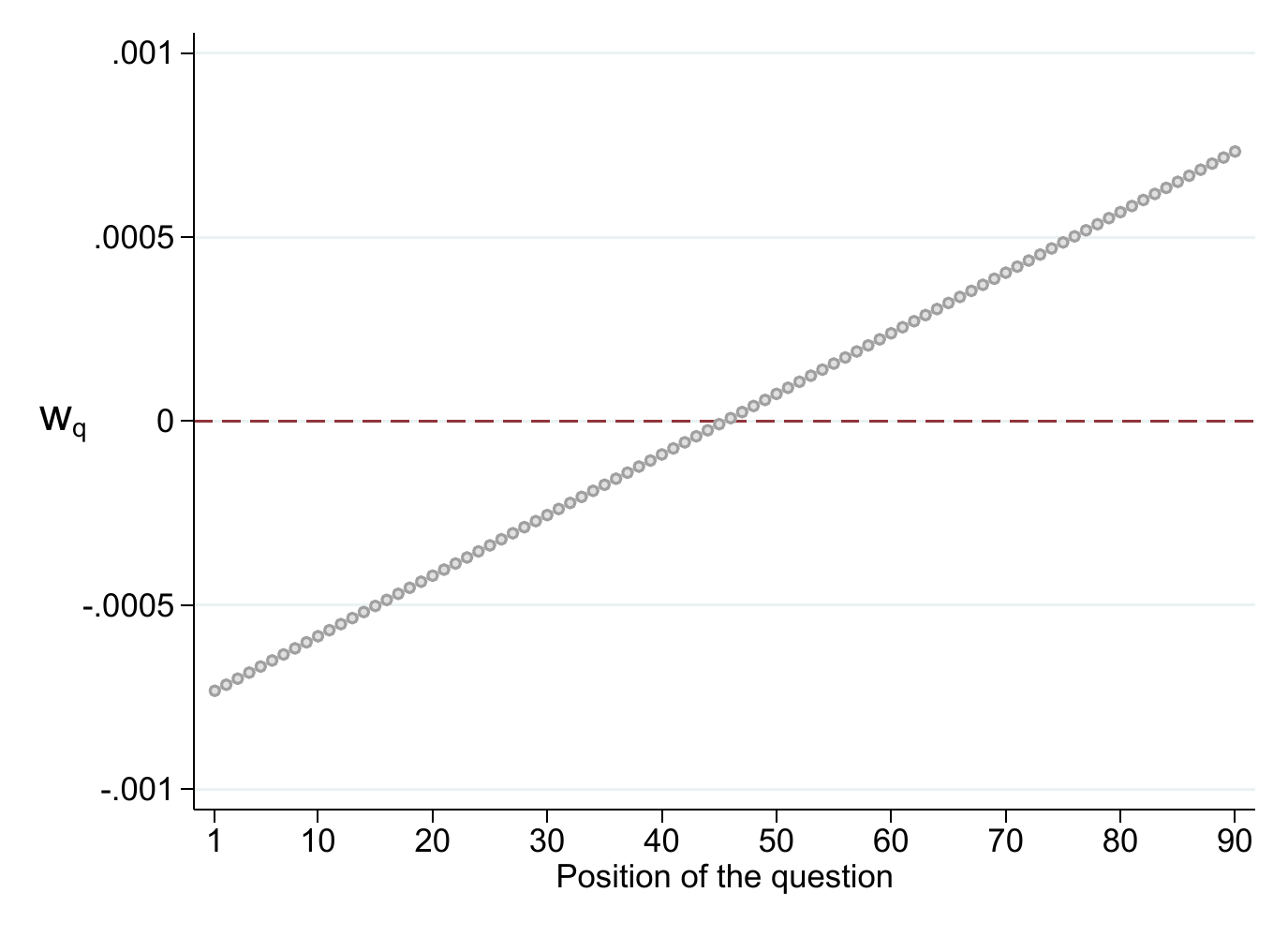}
	
	{\footnotesize
		\singlespacing \justify
		
		\textit{Notes:} This figure displays the weight put by the ordinary least squares (OLS) estimator of $\beta_i$ (equation \eqref{reg:lpm-ind}) on each question of the test.
		
	}
	
\end{figure}

Equation \eqref{eq:lpm-slope} shows that $\hat{\beta}_i$ is a weighted average of deviations from $i$'s average score. The weight of each question depends on the location of the question on the test. Appendix Figure \ref{fig:ols-wt} plots the weight OLS places on each question. The questions with the largest weights (in absolute value) are the ones at the beginning and the end of the test.

My measure of academic ability is $\alpha_i$ in equation \eqref{reg:lpm-ind}. The OLS estimator of $\alpha_i$ is
\begin{align} \label{eq:lpm-cons}
	\hat{\alpha}_i &= \bar{C_i} - \hat{\beta_i}\overline{\text{Pos}} 
\end{align}

Equation \eqref{eq:lpm-cons} shows that $\alpha_i$ can be estimated by the difference between $i$'s test score ($\bar{C_i}$) and the part of her test score that is explained by limited endurance, $\hat{\beta_i}\overline{\text{Pos}}$.

\subsection{Estimating the Standard Deviation of Ability and Endurance} \label{app:std-dev}

The estimate of cognitive endurance, $\hat{\beta}_i$, can be decomposed into latent cognitive endurance, $\beta_i$, and a sampling error $e_i$ independent of $\beta_i$ and with variance $\sigma^2_e$:
\begin{align} \label{eq:latent-error}
	\hat{\beta}_i = \beta_i + e_i
\end{align}

Calculating the variance on each side of equation \eqref{eq:latent-error} yields:
\begin{align} \label{eq:latent-variance}
	\sigma^2_{\hat{\beta}} = \sigma^2_\beta + \sigma^2_e,
\end{align}
where $\sigma^2_{\hat{\beta}}$ and $\sigma^2_\beta$ are the variances of $\hat{\beta}$ and $\beta$, respectively. Equation \eqref{eq:latent-variance} shows that the raw standard deviation of $\hat{\beta}$ overstates the variability of $\beta$ since it includes variability in the sampling error. Let $\text{SE}_{\hat{\beta}}$ be the standard error of $\hat{\beta}$. The variance of the sampling error can be estimated as $\sigma^2_e = \E[\text{SE}^2_{\hat{\beta}}]$. Thus, an estimate of the variance of $\beta$ is given by
\begin{align} \label{eq:latent-variance-beta}
	\hat{\sigma}^2_\beta = \sigma^2_{\hat{\beta}} - \E[\text{SE}^2_{\hat{\beta}}].
\end{align}

I use an analogous derivation to estimate the variance of latent ability, $\sigma^2_\alpha$.

\subsection{Estimating the Predictive Validity of a Question} \label{app:predictive-val}

Let $\pi_j$ be the fraction of students who correctly responded to question $j$. Note that the standard deviation of $C_{ij}$ is $\sigma_{C_{ij}} = \sqrt{\pi_j(1-\pi_j)}$. The predictive ability of question $j$ for outcome $Y$ is given by
\begin{align} \label{eq:pred-val-deriv}
	\rho^Y_j \equiv Corr(Y_i,C_{ij}) &= \frac{Cov(Y_i,C_{ij})}{\sigma_Y \sigma_{C_{ij}}} \notag \\
	&= \frac{\big( \E[Y_i| C_{ij} = 1] - \E[Y_i| C_{ij} = 0]\big)\pi_j(1-\pi_j)}{\sigma_Y \sigma_{C_{ij}}} \notag \\ 
	&= \big( \E[Y_i| C_{ij} = 1] - \E[Y_i| C_{ij} = 0]\big) \frac{\sigma_{C_{ij}}}{\sigma_Y}.
\end{align}

Equation \eqref{eq:pred-val-deriv} shows that the predictive validity of a question partly depends on the difference between the average outcome of students who correctly responded to the question and the average outcome of students who did not, $\E[Y_i | C_{ij} = 1] - \E[Y_i| C_{ij} = 0]$. The predictive validity also depends on the variability of correct responses relative to variability in the outcome, $\sigma_{C_{ij}}/\sigma_Y$. Thus, holding the rest of the variables constant, the more dispersion there is in the distribution of correct responses, the more predictive the question will be for future outcomes.\footnote{For example, for a question with two possible responses, the variance is maximized when $\pi_j = 0.50$, that is, when half students correctly answer the question.}

\subsection{Non-parametric Estimates} \label{app:nonpara}

I assess nonparametrically the predicted effects of endurance on long-run outcomes by estimating how a movement from the bottom decile to the top decile in the endurance distribution affects a given outcome:
\begin{align}\label{eq:pctil-chg}
	\E[Y_i | i \in \text{Top decile Endurance}] - \E[Y_i | i \in \text{Bottom decile Endurance}].
\end{align}

As a benchmark, I compare the size of a decile movement in the endurance distribution to an equivalent decile movement in the ability distribution. I compute these effects in a regression framework by estimating equations of the form:
\begin{align}
	Y_{i} &= \phi + \lambda X_i + \sum_{d=2}^{10} \mathbbm{1}\{i \in \text{TestScore decile $d$}\} +  \zeta_i \label{reg:score-pctil} \\
	Y_{i} &=  \tilde{\phi}_1 + \tilde{\lambda}_1 X_i  + \sum_{d=2}^{10} \mathbbm{1}\{i \in \text{Ability decile $d$}\} + \sum_{d=2}^{10} \mathbbm{1}\{i \in \text{Endurance decile $d$}\} + \tilde{\zeta}_i, \label{reg:endurance-pctil}
\end{align}
where the omitted category is the bottom decile.

\subsection{Robustness of the Relationship between Endurance and Long-Run Outcomes} \label{app:robustness}

Appendix Table \ref{tab:pctil-coll-lmkt} shows non-parametric estimates of the effect of ability and endurance on each outcome based on the slope of percentile changes on outcomes \citep{heckman_effects_2006}. Specifically, I estimate how a movement from the bottom decile to the top decile in the endurance distribution affects a given outcome (see Appendix \ref{app:nonpara} for details). The first row of each panel shows that moving higher in the distribution of test scores tends to improve college and labor-market outcomes. Subsequent rows show that both cognitive endurance and ability contribute to this effect. Depending on the outcome, the predicted effect of a movement from decile 1 to decile 10 in the endurance distribution represents  32.6\%--53.0\% of the corresponding effect of a movement in the ability distribution.

Appendix Tables \ref{tab:rob-meas-coll}--\ref{tab:rob-meas-lmkt} show that the results are robust to estimating ability and endurance with alternative specifications. First, I compute the estimates of ability/endurance separately for each testing day and for each academic subject, and use the average estimate across days/subjects as regressors in equation \eqref{reg:endurance-outcomes}. Second, I compute the estimates of endurance controlling for day fixed effects and subject fixed effects; thus accounting for possible differences in preparation across subjects. Finally, I use the correlation between question position and a dummy for correctly answering a question as an alternative measure of endurance. Across specifications, I find effects that are quantitatively similar and qualitatively identical to those of the baseline specification.

Appendix Tables \ref{tab:rob-diff-coll}--\ref{tab:rob-diff-lmkt} show that the results are robust to controlling for question difficulty in alternative ways when estimating ability and endurance. First, I compute the estimates of ability and endurance in equation \eqref{reg:lpm-ind} without controlling for question difficulty. Second, I calculate question difficulty without adjusting for the average position of the question across booklets. Finally, I compute question difficulty adjusting for average question position in several alternative ways (see Appendix \ref{app:difficulty}). Consistent with the baseline results, I find that the estimates are remarkably robust across specifications.

Appendix Tables \ref{tab:rob-samp-coll}--\ref{tab:rob-samp-lmkt} shows that the results are robust to different sample restrictions. Specifically, I estimate the baseline specification excluding students in the tails of the ability and the endurance distributions. These are students for whom floor and ceiling effects may be binding and, thus, for whom estimates may be biased. I also exclude students with a positive estimate of endurance. These are students who, for example, may answer the exam in reverse order. I find little impact of these sample restrictions on the estimates.

Appendix Tables \ref{tab:rob-prec-coll}--\ref{tab:rob-prec-lmkt} show robustness of the baseline regressions to accounting for measurement error. First, I weight each observation by the inverse of the standard error of the ability and endurance estimates, thus giving more weight to students for which I estimate more precise measures. Second, I estimate the baseline regressions using shrunk estimates of ability and endurance. The shrunk estimators of ability and endurance put more weight on measures estimated with more precision, as measured by a low standard error. The results are very similar to the baseline results.

\begin{table}[H]{\footnotesize
		\begin{center}
			\caption{The effect of a movement from decile 1 to decile 10 in the ability/endurance distribution on long-run outcomes} \label{tab:pctil-coll-lmkt}
			\newcommand\w{1.45}
			\begin{tabular}{l@{}lR{\w cm}@{}L{0.45cm}R{\w cm}@{}L{0.45cm}R{\w cm}@{}L{0.45cm}R{\w cm}@{}L{0.45cm}R{\w cm}@{}L{0.45cm}R{\w cm}@{}L{0.45cm}}
				
				\addlinespace
				\multicolumn{12}{l}{\hspace{-1em} \textbf{Panel A. College outcomes}}  \\ \midrule
				
				& \multicolumn{12}{c}{Dependent variable} \\ \cmidrule{3-14} 
				&& Enrolled    && College && Degree   && 1st-year && Grad.      && Time to    \\
				&& college     && quality && quality  && credits  && on time    && grad.      \\
				&& (1)         && (2)     && (3)      && (4)      && (5)        &&  (6)       \\ \midrule
				Test score&&0.300&\sym{***}&0.275&\sym{***}&0.389&\sym{***}&0.046&\sym{***}&0.215&\sym{***}&$-$0.397&\sym{***}\\
&&(0.001&)&(0.001&)&(0.002&)&(0.001&)&(0.002&)&(0.008&)\\
 \midrule
				Endurance&&0.136&\sym{***}&0.139&\sym{***}&0.232&\sym{***}&0.030&\sym{***}&0.132&\sym{***}&$-$0.224&\sym{***}\\
&&(0.002&)&(0.001&)&(0.002&)&(0.001&)&(0.002&)&(0.007&)\\
Ability&&0.319&\sym{***}&0.338&\sym{***}&0.488&\sym{***}&0.057&\sym{***}&0.272&\sym{***}&$-$0.486&\sym{***}\\
&&(0.002&)&(0.001&)&(0.002&)&(0.001&)&(0.003&)&(0.009&)\\
 \midrule
				Ratio coef.&&0.426&\sym{***}&0.412&\sym{***}&0.474&\sym{***}&0.530&\sym{***}&0.485&\sym{***}&0.462&\sym{***}\\
&&(0.005&)&(0.003&)&(0.003&)&(0.009&)&(0.007&)&(0.012&)\\

				Mean DV&&0.329&&3.327&&3.244&&0.158&&0.418&&3.842&\\
\(N\)&&1,850,938&&1,711,475&&1,681,214&&1,124,972&&1,471,569&&786,391&\\
 \midrule	\addlinespace \addlinespace \addlinespace
				
				\multicolumn{12}{l}{\hspace{-1em} \textbf{Panel B. Labor-market outcomes}}  \\ \midrule
				
				& \multicolumn{12}{c}{Dependent variable} \\ \cmidrule{3-14} 
				&& Formal        && Hourly && Monthly   && Firm  && Occup.     && Industry  \\
				&& sector        && wage   && earnings  && wage  && wage       && wage      \\
				&& (1)           && (2)     && (3)      && (4)   && (5)        && (6)       \\	\midrule
				Test score&&0.005&\sym{***}&0.457&\sym{***}&0.395&\sym{***}&0.320&\sym{***}&0.141&\sym{***}&0.046&\sym{***}\\
&&(0.000&)&(0.004&)&(0.004&)&(0.004&)&(0.002&)&(0.001&)\\
 \midrule
				Endurance&&0.002&\sym{***}&0.241&\sym{***}&0.240&\sym{***}&0.159&\sym{***}&0.084&\sym{***}&0.018&\sym{***}\\
&&(0.000&)&(0.005&)&(0.004&)&(0.004&)&(0.003&)&(0.001&)\\
Ability&&0.006&\sym{***}&0.543&\sym{***}&0.475&\sym{***}&0.387&\sym{***}&0.177&\sym{***}&0.055&\sym{***}\\
&&(0.000&)&(0.005&)&(0.005&)&(0.005&)&(0.003&)&(0.001&)\\
 \midrule
				Ratio coef.&&0.337&\sym{***}&0.443&\sym{***}&0.506&\sym{***}&0.410&\sym{***}&0.472&\sym{***}&0.326&\sym{***}\\
&&(0.034&)&(0.007&)&(0.007&)&(0.009&)&(0.012&)&(0.020&)\\

				Mean DV&&0.326&&3.865&&7.551&&3.885&&3.886&&3.858&\\
\(N\)&&2,523,032&&818,590&&818,590&&692,880&&818,374&&818,590&\\
 \midrule						
				
			\end{tabular}%
		\end{center}
		\begin{singlespace}  \vspace{-.5cm}
			\noindent \justify \textit{Notes:} This table displays estimates of the relationship between ability/endurance and college outcomes (Panel A) and labor-market outcomes (Panel B).
			
			The first row of each panel shows estimates of the mean outcome difference between individuals in the tenth and first decile of the test score distribution (the coefficient on the decile ten dummy in equation \eqref{reg:score-pctil}). The following rows show estimates of the mean outcome difference between individuals in the tenth and first decile of the ability/endurance distribution (the coefficients on the decile ten dummies in equation \eqref{reg:endurance-pctil}). See Section \ref{sec:recover} for a description of the measures of ability and endurance. See Section \ref{sub:var-def} for outcome definitions. Heteroskedasticity-robust standard errors clustered at the individual level in parentheses.
			
			The third-to-last row in each panel shows the ratio between the effect of ability and endurance on a given outcome.  Standard errors estimated through the delta method in parentheses. $^{***}$, $^{**}$ and $^*$ denote significance at 10\%, 5\% and 1\% levels, respectively.			
			
		\end{singlespace} 	
	}
\end{table}%

\begin{table}[H]{\footnotesize
		\begin{center}
			\caption{Robustness of baseline regressions to alternative ways of measuring ability and endurance: College outcomes} \label{tab:rob-meas-coll}
			\newcommand\w{1.45}
			\begin{tabular}{l@{}lR{\w cm}@{}L{0.45cm}R{\w cm}@{}L{0.45cm}R{\w cm}@{}L{0.45cm}R{\w cm}@{}L{0.45cm}R{\w cm}@{}L{0.45cm}R{\w cm}@{}L{0.45cm}}
				\midrule
				& \multicolumn{12}{c}{Dependent variable:} \\ \cmidrule{3-14} 
				&& Enrolled    && College && Degree   && 1st-year && Grad.      && Time to    \\
				&& college     && quality && quality  && credits  && on time    && grad.      \\
				&& (1)         && (2)     && (3)      && (4)      && (5)        &&  (6)       \\ \midrule		
				\multicolumn{14}{l}{\hspace{-1em} \textbf{Panel A. Estimating ability/endurance separately by day and using the average}}  \\ \addlinespace
				Endurance&&0.053&\sym{***}&0.050&\sym{***}&0.084&\sym{***}&0.010&\sym{***}&0.043&\sym{***}&$-$0.079&\sym{***}\\
&&(0.000&)&(0.000&)&(0.000&)&(0.000&)&(0.001&)&(0.002&)\\
Ability&&0.114&\sym{***}&0.107&\sym{***}&0.157&\sym{***}&0.018&\sym{***}&0.081&\sym{***}&$-$0.158&\sym{***}\\
&&(0.000&)&(0.000&)&(0.001&)&(0.000&)&(0.001&)&(0.002&)\\
 \midrule				
				\multicolumn{14}{l}{\hspace{-1em} \textbf{Panel B. Estimating ability/endurance separately by subject and using the average}}  \\ \addlinespace
				Endurance&&0.060&\sym{***}&0.055&\sym{***}&0.080&\sym{***}&0.009&\sym{***}&0.043&\sym{***}&$-$0.079&\sym{***}\\
&&(0.000&)&(0.000&)&(0.000&)&(0.000&)&(0.001&)&(0.002&)\\
Ability&&0.103&\sym{***}&0.097&\sym{***}&0.140&\sym{***}&0.016&\sym{***}&0.067&\sym{***}&$-$0.130&\sym{***}\\
&&(0.000&)&(0.000&)&(0.001&)&(0.000&)&(0.001&)&(0.002&)\\
 \midrule		
				\multicolumn{14}{l}{\hspace{-1em} \textbf{Panel C. Including day fixed effects}}  \\ \addlinespace
				Endurance&&0.031&\sym{***}&0.030&\sym{***}&0.050&\sym{***}&0.006&\sym{***}&0.025&\sym{***}&$-$0.047&\sym{***}\\
&&(0.000&)&(0.000&)&(0.000&)&(0.000&)&(0.000&)&(0.001&)\\
Ability&&0.110&\sym{***}&0.104&\sym{***}&0.152&\sym{***}&0.018&\sym{***}&0.077&\sym{***}&$-$0.151&\sym{***}\\
&&(0.000&)&(0.000&)&(0.001&)&(0.000&)&(0.001&)&(0.002&)\\
 \midrule		
				\multicolumn{14}{l}{\hspace{-1em} \textbf{Panel D. Including subject fixed effects}}  \\ \addlinespace
				Endurance&&0.019&\sym{***}&0.018&\sym{***}&0.026&\sym{***}&0.003&\sym{***}&0.014&\sym{***}&$-$0.025&\sym{***}\\
&&(0.000&)&(0.000&)&(0.000&)&(0.000&)&(0.000&)&(0.001&)\\
Ability&&0.097&\sym{***}&0.092&\sym{***}&0.133&\sym{***}&0.015&\sym{***}&0.062&\sym{***}&$-$0.119&\sym{***}\\
&&(0.000&)&(0.000&)&(0.001&)&(0.000&)&(0.001&)&(0.002&)\\
 \midrule		
				\multicolumn{14}{l}{\hspace{-1em} \textbf{Panel E. Using linear correlation as an alternative measure of endurance}}  \\ \addlinespace
				Endurance&&0.030&\sym{***}&0.030&\sym{***}&0.049&\sym{***}&0.006&\sym{***}&0.025&\sym{***}&$-$0.047&\sym{***}\\
&&(0.000&)&(0.000&)&(0.000&)&(0.000&)&(0.000&)&(0.001&)\\
Ability&&0.101&\sym{***}&0.095&\sym{***}&0.138&\sym{***}&0.016&\sym{***}&0.072&\sym{***}&$-$0.139&\sym{***}\\
&&(0.000&)&(0.000&)&(0.000&)&(0.000&)&(0.001&)&(0.002&)\\
 \midrule		
				Mean DV&&0.244&&3.326&&3.244&&0.158&&0.418&&3.817&\\
\(N\)&&2,501,519&&1,800,546&&1,768,707&&1,124,972&&1,472,916&&793,822&\\
 \midrule	\addlinespace \addlinespace 
			\end{tabular}%
		\end{center}
		\begin{singlespace}  \vspace{-.5cm}
			\noindent \justify \textit{Notes:} This table shows estimates of the relationship between ability/endurance and college outcomes using alternative specifications to estimate ability and endurance.
			
			Each column shows the result for a different dependent variable. Each panel shows the result from estimating ability and endurance with a different specification. In Panels A--B, I estimate a student's ability/endurance separately for each testing day (Panel A) and academic subject (Panel B) and then average the estimates across days or subjects. In Panels C--D, I estimate endurance in a regression that controls for day fixed effects (Panel C) or subject fixed effects (Panel D). Finally, in Panel E, I use the correlation between question position and a dummy for correctly answering a question as an alternative measure of endurance. 
			
			Heteroskedasticity-robust standard errors clustered at the individual level in parentheses.$^{***}$, $^{**}$ and $^*$ denote significance at 10\%, 5\% and 1\% levels, respectively.

		\end{singlespace} 	
	}
\end{table}%

\begin{table}[H]{\footnotesize
		\begin{center}
			\caption{Robustness of baseline regressions to alternative ways of measuring ability and endurance: Labor-market outcomes} \label{tab:rob-meas-lmkt}
			\newcommand\w{1.45}
			\begin{tabular}{l@{}lR{\w cm}@{}L{0.45cm}R{\w cm}@{}L{0.45cm}R{\w cm}@{}L{0.45cm}R{\w cm}@{}L{0.45cm}R{\w cm}@{}L{0.45cm}R{\w cm}@{}L{0.45cm}} \midrule
				& \multicolumn{12}{c}{Dependent variable:} \\ \cmidrule{3-14} 
				&& Formal        && Hourly && Monthly   && Firm  && Occup.     && Industry  \\
				&& sector        && wage   && earnings  && wage  && wage       && wage      \\
				&& (1)           && (2)     && (3)      && (4)   && (5)        && (6)       \\	\midrule	
				\multicolumn{14}{l}{\hspace{-1em} \textbf{Panel A. Estimating ability/endurance separately by day and using the average}}  \\ \addlinespace
				Endurance&&0.001&\sym{***}&0.088&\sym{***}&0.085&\sym{***}&0.058&\sym{***}&0.028&\sym{***}&0.006&\sym{***}\\
&&(0.000&)&(0.001&)&(0.001&)&(0.001&)&(0.001&)&(0.000&)\\
Ability&&0.002&\sym{***}&0.172&\sym{***}&0.152&\sym{***}&0.121&\sym{***}&0.055&\sym{***}&0.016&\sym{***}\\
&&(0.000&)&(0.001&)&(0.001&)&(0.001&)&(0.001&)&(0.000&)\\
 \midrule				
				\multicolumn{14}{l}{\hspace{-1em} \textbf{Panel B. Estimating ability/endurance separately by subject and using the average}}  \\ \addlinespace
				Endurance&&0.001&\sym{***}&0.088&\sym{***}&0.076&\sym{***}&0.061&\sym{***}&0.026&\sym{***}&0.008&\sym{***}\\
&&(0.000&)&(0.001&)&(0.001&)&(0.001&)&(0.001&)&(0.000&)\\
Ability&&0.002&\sym{***}&0.156&\sym{***}&0.135&\sym{***}&0.110&\sym{***}&0.048&\sym{***}&0.014&\sym{***}\\
&&(0.000&)&(0.001&)&(0.001&)&(0.001&)&(0.001&)&(0.000&)\\
 \midrule		
				\multicolumn{14}{l}{\hspace{-1em} \textbf{Panel C. Including day fixed effects}}  \\ \addlinespace
				Endurance&&0.000&\sym{***}&0.053&\sym{***}&0.051&\sym{***}&0.035&\sym{***}&0.017&\sym{***}&0.004&\sym{***}\\
&&(0.000&)&(0.001&)&(0.001&)&(0.001&)&(0.000&)&(0.000&)\\
Ability&&0.002&\sym{***}&0.167&\sym{***}&0.147&\sym{***}&0.117&\sym{***}&0.053&\sym{***}&0.016&\sym{***}\\
&&(0.000&)&(0.001&)&(0.001&)&(0.001&)&(0.001&)&(0.000&)\\
 \midrule		
				\multicolumn{14}{l}{\hspace{-1em} \textbf{Panel D. Including subject fixed effects}}  \\ \addlinespace
				Endurance&&0.000&\sym{***}&0.029&\sym{***}&0.025&\sym{***}&0.020&\sym{***}&0.008&\sym{***}&0.003&\sym{***}\\
&&(0.000&)&(0.000&)&(0.000&)&(0.000&)&(0.000&)&(0.000&)\\
Ability&&0.002&\sym{***}&0.147&\sym{***}&0.127&\sym{***}&0.104&\sym{***}&0.045&\sym{***}&0.013&\sym{***}\\
&&(0.000&)&(0.001&)&(0.001&)&(0.001&)&(0.001&)&(0.000&)\\
 \midrule		
				\multicolumn{14}{l}{\hspace{-1em} \textbf{Panel E. Using linear correlation as an alternative measure of endurance}}  \\ \addlinespace
				Endurance&&0.000&\sym{***}&0.052&\sym{***}&0.050&\sym{***}&0.035&\sym{***}&0.016&\sym{***}&0.004&\sym{***}\\
&&(0.000&)&(0.001&)&(0.001&)&(0.001&)&(0.000&)&(0.000&)\\
Ability&&0.002&\sym{***}&0.151&\sym{***}&0.132&\sym{***}&0.107&\sym{***}&0.048&\sym{***}&0.014&\sym{***}\\
&&(0.000&)&(0.001&)&(0.001&)&(0.001&)&(0.001&)&(0.000&)\\
 \midrule						
				Mean DV&&0.326&&3.865&&7.551&&3.885&&3.886&&3.858&\\
\(N\)&&2,523,029&&818,590&&818,590&&692,880&&818,374&&818,590&\\
 \midrule	\addlinespace \addlinespace 
			\end{tabular}%
		\end{center}
		\begin{singlespace}  \vspace{-.5cm}
			\noindent \justify \textit{Notes:} This table shows estimates of the relationship between ability/endurance and labor-market outcomes using alternative specifications to estimate ability and endurance.
			
			Each column shows the result for a different dependent variable. Each panel shows the result from estimating ability and endurance with a different specification. In Panels A--B, I estimate a student's ability/endurance separately for each testing day (Panel A) and academic subject (Panel B) and then average the estimates across days or subjects. In Panels C--D, I estimate endurance in a regression that controls for day fixed effects (Panel C) or subject fixed effects (Panel D). Finally, in Panel E, I use the correlation between question position and a dummy for correctly answering a question as an alternative measure of endurance. 
			
			Heteroskedasticity-robust standard errors clustered at the individual level in parentheses. $^{***}$, $^{**}$ and $^*$ denote significance at 10\%, 5\% and 1\% levels, respectively.

		\end{singlespace} 	
	}
\end{table}%

\begin{table}[H]{\footnotesize
		\begin{center}
			\caption{Robustness of the baseline regressions to alternative ways of controlling for question difficulty when estimating ability/endurance: College outcomes} \label{tab:rob-diff-coll}
			\newcommand\w{1.45}
			\begin{tabular}{l@{}lR{\w cm}@{}L{0.45cm}R{\w cm}@{}L{0.45cm}R{\w cm}@{}L{0.45cm}R{\w cm}@{}L{0.45cm}R{\w cm}@{}L{0.45cm}R{\w cm}@{}L{0.45cm}}
				\midrule
				& \multicolumn{12}{c}{Dependent variable:} \\ \cmidrule{3-14} 
				&& Enrolled    && College && Degree   && 1st-year && Grad.      && Time to    \\
				&& college     && quality && quality  && credits  && on time    && grad.      \\
				&& (1)         && (2)     && (3)      && (4)      && (5)        &&  (6)       \\ \midrule		
				\multicolumn{14}{l}{\hspace{-1em} \textbf{Panel A. Not controlling for question difficulty}}  \\ \addlinespace
				Endurance&&0.040&\sym{***}&0.045&\sym{***}&0.080&\sym{***}&0.007&\sym{***}&0.028&\sym{***}&$-$0.061&\sym{***}\\
&&(0.000&)&(0.000&)&(0.000&)&(0.000&)&(0.000&)&(0.002&)\\
Ability&&0.114&\sym{***}&0.112&\sym{***}&0.169&\sym{***}&0.018&\sym{***}&0.079&\sym{***}&$-$0.159&\sym{***}\\
&&(0.000&)&(0.000&)&(0.001&)&(0.000&)&(0.001&)&(0.002&)\\
 \midrule				
				\multicolumn{14}{l}{\hspace{-1em} \textbf{Panel B. Estimating difficulty without adjusting for average position}}  \\ \addlinespace
				Endurance&&0.030&\sym{***}&0.028&\sym{***}&0.045&\sym{***}&0.006&\sym{***}&0.026&\sym{***}&$-$0.046&\sym{***}\\
&&(0.000&)&(0.000&)&(0.000&)&(0.000&)&(0.000&)&(0.001&)\\
Ability&&0.097&\sym{***}&0.090&\sym{***}&0.130&\sym{***}&0.016&\sym{***}&0.068&\sym{***}&$-$0.133&\sym{***}\\
&&(0.000&)&(0.000&)&(0.000&)&(0.000&)&(0.001&)&(0.002&)\\
 \midrule		
				\multicolumn{14}{l}{\hspace{-1em} \textbf{Panel C. Estimating difficulty using question-specific position effects}}  \\ \addlinespace
				Endurance&&0.035&\sym{***}&0.038&\sym{***}&0.065&\sym{***}&0.007&\sym{***}&0.027&\sym{***}&$-$0.054&\sym{***}\\
&&(0.000&)&(0.000&)&(0.000&)&(0.000&)&(0.000&)&(0.001&)\\
Ability&&0.106&\sym{***}&0.102&\sym{***}&0.152&\sym{***}&0.017&\sym{***}&0.074&\sym{***}&$-$0.147&\sym{***}\\
&&(0.000&)&(0.000&)&(0.001&)&(0.000&)&(0.001&)&(0.002&)\\
 \midrule		
				\multicolumn{14}{l}{\hspace{-1em} \textbf{Panel D. Estimating difficulty using shrunk question-specific position effects}}  \\ \addlinespace
				Endurance&&0.034&\sym{***}&0.035&\sym{***}&0.059&\sym{***}&0.006&\sym{***}&0.026&\sym{***}&$-$0.052&\sym{***}\\
&&(0.000&)&(0.000&)&(0.000&)&(0.000&)&(0.000&)&(0.001&)\\
Ability&&0.104&\sym{***}&0.098&\sym{***}&0.146&\sym{***}&0.017&\sym{***}&0.073&\sym{***}&$-$0.144&\sym{***}\\
&&(0.000&)&(0.000&)&(0.001&)&(0.000&)&(0.001&)&(0.002&)\\
 \midrule		
				\multicolumn{14}{l}{\hspace{-1em} \textbf{Panel E. Estimating position effects separately by fraction of correct responses}}  \\ \addlinespace
				Endurance&&0.031&\sym{***}&0.030&\sym{***}&0.049&\sym{***}&0.006&\sym{***}&0.026&\sym{***}&$-$0.047&\sym{***}\\
&&(0.000&)&(0.000&)&(0.000&)&(0.000&)&(0.000&)&(0.001&)\\
Ability&&0.101&\sym{***}&0.094&\sym{***}&0.138&\sym{***}&0.016&\sym{***}&0.071&\sym{***}&$-$0.139&\sym{***}\\
&&(0.000&)&(0.000&)&(0.000&)&(0.000&)&(0.001&)&(0.002&)\\
 \midrule		
				\multicolumn{14}{l}{\hspace{-1em} \textbf{Panel F.  Estimating position effects separately by subject}}  \\ \addlinespace
				Endurance&&0.031&\sym{***}&0.029&\sym{***}&0.048&\sym{***}&0.006&\sym{***}&0.026&\sym{***}&$-$0.047&\sym{***}\\
&&(0.000&)&(0.000&)&(0.000&)&(0.000&)&(0.000&)&(0.001&)\\
Ability&&0.099&\sym{***}&0.093&\sym{***}&0.135&\sym{***}&0.016&\sym{***}&0.071&\sym{***}&$-$0.137&\sym{***}\\
&&(0.000&)&(0.000&)&(0.000&)&(0.000&)&(0.001&)&(0.002&)\\
 \midrule
				Mean DV&&0.244&&3.326&&3.244&&0.158&&0.418&&3.817&\\
\(N\)&&2,501,519&&1,800,546&&1,768,707&&1,124,972&&1,472,916&&793,822&\\
 \midrule	\addlinespace \addlinespace 		
			\end{tabular}%
		\end{center}
		\begin{singlespace}  \vspace{-.5cm}
			\noindent \justify \textit{Notes:} This table shows estimates of the relationship between ability/endurance and college outcomes using alternative measures of difficulty in the specification used to estimate ability and endurance.
			
			Each panel shows the result using a different measure of question difficulty in equation \eqref{reg:lpm-ind}. In Panel A, I estimate equation \eqref{reg:lpm-ind} without controlling for question difficulty. In Panel B, I measure question difficulty as the fraction of students who incorrectly answer the question across all booklets. In Panels C--F, I adjust for differences in average position across questions by estimating the position effects with alternative specifications. In column C, I compute question-specific position effects. In Panel D, I compute a shrinkage estimator of the position effects. In Panel E, I compute the position effects separately for questions with a below/above fraction of correct responses. In Panel F, I compute the position effects separately by subject. See Appendix \ref{app:difficulty} for details on each measure of question difficulty. Heteroskedasticity-robust standard errors clustered at the individual level in parentheses.$^{***}$, $^{**}$ and $^*$ denote significance at 10\%, 5\% and 1\% levels, respectively.

		\end{singlespace} 	
	}
\end{table}%

\begin{table}[H]{\footnotesize
		\begin{center}
			\caption{Robustness of the baseline regressions to alternative ways of controlling for question difficulty when estimating ability/endurance: Labor-market outcomes} \label{tab:rob-diff-lmkt}
			\newcommand\w{1.45}
			\begin{tabular}{l@{}lR{\w cm}@{}L{0.45cm}R{\w cm}@{}L{0.45cm}R{\w cm}@{}L{0.45cm}R{\w cm}@{}L{0.45cm}R{\w cm}@{}L{0.45cm}R{\w cm}@{}L{0.45cm}}
				\midrule
				& \multicolumn{12}{c}{Dependent variable:} \\ \cmidrule{3-14} 
				&& Formal        && Hourly && Monthly   && Firm  && Occup.     && Industry  \\
				&& sector        && wage   && earnings  && wage  && wage       && wage      \\
				&& (1)           && (2)     && (3)      && (4)   && (5)        && (6)       \\	\midrule	
				\multicolumn{14}{l}{\hspace{-1em} \textbf{Panel A. Not controlling for question difficulty}}  \\ \addlinespace
				Endurance&&0.001&\sym{***}&0.080&\sym{***}&0.077&\sym{***}&0.053&\sym{***}&0.022&\sym{***}&0.004&\sym{***}\\
&&(0.000&)&(0.001&)&(0.001&)&(0.001&)&(0.001&)&(0.000&)\\
Ability&&0.002&\sym{***}&0.183&\sym{***}&0.163&\sym{***}&0.127&\sym{***}&0.056&\sym{***}&0.015&\sym{***}\\
&&(0.000&)&(0.002&)&(0.001&)&(0.001&)&(0.001&)&(0.000&)\\
 \midrule				
				\multicolumn{14}{l}{\hspace{-1em} \textbf{Panel B. Estimating difficulty without adjusting for average position}}  \\ \addlinespace
				Endurance&&0.000&\sym{***}&0.048&\sym{***}&0.047&\sym{***}&0.032&\sym{***}&0.016&\sym{***}&0.004&\sym{***}\\
&&(0.000&)&(0.001&)&(0.001&)&(0.001&)&(0.000&)&(0.000&)\\
Ability&&0.001&\sym{***}&0.143&\sym{***}&0.125&\sym{***}&0.101&\sym{***}&0.046&\sym{***}&0.014&\sym{***}\\
&&(0.000&)&(0.001&)&(0.001&)&(0.001&)&(0.001&)&(0.000&)\\
 \midrule		
				\multicolumn{14}{l}{\hspace{-1em} \textbf{Panel C. Estimating difficulty using question-specific position effects}}  \\ \addlinespace
				Endurance&&0.001&\sym{***}&0.066&\sym{***}&0.064&\sym{***}&0.044&\sym{***}&0.019&\sym{***}&0.004&\sym{***}\\
&&(0.000&)&(0.001&)&(0.001&)&(0.001&)&(0.000&)&(0.000&)\\
Ability&&0.002&\sym{***}&0.165&\sym{***}&0.146&\sym{***}&0.115&\sym{***}&0.051&\sym{***}&0.015&\sym{***}\\
&&(0.000&)&(0.001&)&(0.001&)&(0.001&)&(0.001&)&(0.000&)\\
 \midrule		
				\multicolumn{14}{l}{\hspace{-1em} \textbf{Panel D. Estimating difficulty using shrunk question-specific position effects}}  \\ \addlinespace
				Endurance&&0.000&\sym{***}&0.061&\sym{***}&0.059&\sym{***}&0.040&\sym{***}&0.018&\sym{***}&0.004&\sym{***}\\
&&(0.000&)&(0.001&)&(0.001&)&(0.001&)&(0.000&)&(0.000&)\\
Ability&&0.002&\sym{***}&0.159&\sym{***}&0.140&\sym{***}&0.111&\sym{***}&0.050&\sym{***}&0.014&\sym{***}\\
&&(0.000&)&(0.001&)&(0.001&)&(0.001&)&(0.001&)&(0.000&)\\
 \midrule		
				\multicolumn{14}{l}{\hspace{-1em} \textbf{Panel E. Estimating position effects separately by fraction of correct responses}}  \\ \addlinespace
				Endurance&&0.000&\sym{***}&0.053&\sym{***}&0.051&\sym{***}&0.035&\sym{***}&0.017&\sym{***}&0.004&\sym{***}\\
&&(0.000&)&(0.001&)&(0.001&)&(0.001&)&(0.000&)&(0.000&)\\
Ability&&0.002&\sym{***}&0.152&\sym{***}&0.133&\sym{***}&0.107&\sym{***}&0.048&\sym{***}&0.014&\sym{***}\\
&&(0.000&)&(0.001&)&(0.001&)&(0.001&)&(0.001&)&(0.000&)\\
 \midrule		
				\multicolumn{14}{l}{\hspace{-1em} \textbf{Panel F.  Estimating position effects separately by subject}}  \\ \addlinespace
				Endurance&&0.000&\sym{***}&0.051&\sym{***}&0.049&\sym{***}&0.034&\sym{***}&0.016&\sym{***}&0.004&\sym{***}\\
&&(0.000&)&(0.001&)&(0.001&)&(0.001&)&(0.000&)&(0.000&)\\
Ability&&0.002&\sym{***}&0.149&\sym{***}&0.130&\sym{***}&0.105&\sym{***}&0.048&\sym{***}&0.014&\sym{***}\\
&&(0.000&)&(0.001&)&(0.001&)&(0.001&)&(0.001&)&(0.000&)\\
 \midrule
				Mean DV&&0.326&&3.865&&7.551&&3.885&&3.886&&3.858&\\
\(N\)&&2,523,029&&818,590&&818,590&&692,880&&818,374&&818,590&\\
 \midrule	\addlinespace \addlinespace 			
			\end{tabular}%
		\end{center}
		\begin{singlespace}  \vspace{-.5cm}
			\noindent \justify \textit{Notes:} This table shows estimates of the relationship between ability/endurance and labor-market outcomes using alternative measures of difficulty in the specification used to estimate ability and endurance.
			
			Each panel shows the result using a different measure of question difficulty in equation \eqref{reg:lpm-ind}. In Panel A, I estimate equation \eqref{reg:lpm-ind} without controlling for question difficulty. In Panel B, I measure question difficulty as the fraction of students who incorrectly answer the question across all booklets. In Panels C--F, I adjust for differences in average position across questions by estimating the position effects with alternative specifications. In column C, I compute question-specific position effects. In Panel D, I compute a shrinkage estimator of the position effects. In Panel E, I compute the position effects separately for questions with a below/above fraction of correct responses. In Panel F, I compute the position effects separately by subject. See Appendix \ref{app:difficulty} for details on each measure of question difficulty. Heteroskedasticity-robust standard errors clustered at the individual level in parentheses.$^{***}$, $^{**}$ and $^*$ denote significance at 10\%, 5\% and 1\% levels, respectively.

		\end{singlespace} 	
	}
\end{table}%

\begin{table}[H]{\footnotesize
		\begin{center}
			\caption{Robustness of the baseline regressions to sample selection: College outcomes} \label{tab:rob-samp-coll}
			\newcommand\w{1.45}
			\begin{tabular}{l@{}lR{\w cm}@{}L{0.45cm}R{\w cm}@{}L{0.45cm}R{\w cm}@{}L{0.45cm}R{\w cm}@{}L{0.45cm}R{\w cm}@{}L{0.45cm}R{\w cm}@{}L{0.45cm}}
				\midrule
				& \multicolumn{12}{c}{Dependent variable:} \\ \cmidrule{3-14} 
				&& Enrolled    && College && Degree   && 1st-year && Grad.      && Time to    \\
				&& college     && quality && quality  && credits  && on time    && grad.      \\
				&& (1)         && (2)     && (3)      && (4)      && (5)        &&  (6)       \\ \midrule		
				\multicolumn{14}{l}{\hspace{-1em} \textbf{Panel A. Excluding students in the bottom or top 10\% of the ability distribution}}  \\ \addlinespace
				Endurance&&0.032&\sym{***}&0.027&\sym{***}&0.039&\sym{***}&0.007&\sym{***}&0.030&\sym{***}&$-$0.047&\sym{***}\\
&&(0.000&)&(0.000&)&(0.000&)&(0.000&)&(0.000&)&(0.001&)\\
Ability&&0.102&\sym{***}&0.085&\sym{***}&0.107&\sym{***}&0.019&\sym{***}&0.087&\sym{***}&$-$0.146&\sym{***}\\
&&(0.000&)&(0.000&)&(0.001&)&(0.000&)&(0.001&)&(0.003&)\\
 \midrule				
				\multicolumn{14}{l}{\hspace{-1em} \textbf{Panel B. Excluding students in the bottom or top 10\% of the endurance distribution}}  \\ \addlinespace
				Endurance&&0.030&\sym{***}&0.030&\sym{***}&0.050&\sym{***}&0.006&\sym{***}&0.025&\sym{***}&$-$0.044&\sym{***}\\
&&(0.000&)&(0.000&)&(0.000&)&(0.000&)&(0.000&)&(0.002&)\\
Ability&&0.100&\sym{***}&0.093&\sym{***}&0.135&\sym{***}&0.016&\sym{***}&0.075&\sym{***}&$-$0.138&\sym{***}\\
&&(0.000&)&(0.000&)&(0.001&)&(0.000&)&(0.001&)&(0.002&)\\
 \midrule		
				\multicolumn{14}{l}{\hspace{-1em} \textbf{Panel C. Excluding students in the bottom or top 10\% of either distribution}}  \\ \addlinespace
				Endurance&&0.031&\sym{***}&0.026&\sym{***}&0.037&\sym{***}&0.007&\sym{***}&0.030&\sym{***}&$-$0.046&\sym{***}\\
&&(0.000&)&(0.000&)&(0.000&)&(0.000&)&(0.001&)&(0.002&)\\
Ability&&0.102&\sym{***}&0.083&\sym{***}&0.102&\sym{***}&0.020&\sym{***}&0.089&\sym{***}&$-$0.146&\sym{***}\\
&&(0.001&)&(0.000&)&(0.001&)&(0.000&)&(0.001&)&(0.003&)\\
 \midrule		
				\multicolumn{14}{l}{\hspace{-1em} \textbf{Panel D. Excluding students in the bottom or top 20\% of either distribution}}  \\ \addlinespace
				Endurance&&0.030&\sym{***}&0.023&\sym{***}&0.030&\sym{***}&0.008&\sym{***}&0.034&\sym{***}&$-$0.044&\sym{***}\\
&&(0.001&)&(0.000&)&(0.001&)&(0.000&)&(0.001&)&(0.003&)\\
Ability&&0.099&\sym{***}&0.074&\sym{***}&0.086&\sym{***}&0.022&\sym{***}&0.100&\sym{***}&$-$0.152&\sym{***}\\
&&(0.001&)&(0.001&)&(0.001&)&(0.000&)&(0.001&)&(0.004&)\\
 \midrule		
				\multicolumn{14}{l}{\hspace{-1em} \textbf{Panel E.  Excluding individuals with positive estimated endurance}}  \\ \addlinespace
				Endurance&&0.033&\sym{***}&0.028&\sym{***}&0.048&\sym{***}&0.005&\sym{***}&0.026&\sym{***}&$-$0.047&\sym{***}\\
&&(0.000&)&(0.000&)&(0.001&)&(0.000&)&(0.001&)&(0.002&)\\
Ability&&0.107&\sym{***}&0.097&\sym{***}&0.143&\sym{***}&0.016&\sym{***}&0.065&\sym{***}&$-$0.133&\sym{***}\\
&&(0.001&)&(0.000&)&(0.001&)&(0.000&)&(0.001&)&(0.003&)\\
 \midrule		
				
			\end{tabular}%
		\end{center}
		\begin{singlespace}  \vspace{-.5cm}
			\noindent \justify \textit{Notes:} This table shows estimates of the relationship between ability/endurance and college outcomes using alternative sample restrictions. 
			
			Each column shows the result for a different dependent variable. Each panel shows the result for a different sample of students. In Panel A, I exclude students in the bottom and top deciles of the ability distribution. In Panel B, I exclude students in the bottom and top deciles of the endurance distribution. In Panel C, I exclude students in the bottom and top deciles of the distribution of either skill. In Panel D, I exclude students in the bottom and top quintiles of the distribution of either skill. In Panel E, I exclude students with a positive estimate of endurance ($\hat{\beta}>0$). I construct the deciles and quintiles using all the students in the high-school-students sample.
			
			Heteroskedasticity-robust standard errors clustered at the individual level in parentheses.$^{***}$, $^{**}$ and $^*$ denote significance at 10\%, 5\% and 1\% levels, respectively.

		\end{singlespace} 	
	}
\end{table}%

\begin{table}[H]{\footnotesize
		\begin{center}
			\caption{Robustness of the baseline regressions to sample selection: Labor-market outcomes} \label{tab:rob-samp-lmkt}
			\newcommand\w{1.45}
			\begin{tabular}{l@{}lR{\w cm}@{}L{0.45cm}R{\w cm}@{}L{0.45cm}R{\w cm}@{}L{0.45cm}R{\w cm}@{}L{0.45cm}R{\w cm}@{}L{0.45cm}R{\w cm}@{}L{0.45cm}}
				\midrule
				& \multicolumn{12}{c}{Dependent variable:} \\ \cmidrule{3-14} 
				&& Formal        && Hourly && Monthly   && Firm  && Occup.     && Industry  \\
				&& sector        && wage   && earnings  && wage  && wage       && wage      \\
				&& (1)           && (2)     && (3)      && (4)   && (5)        && (6)       \\	\midrule		
				\multicolumn{14}{l}{\hspace{-1em} \textbf{Panel A. Excluding students in the bottom or top 10\% of the ability distribution}}  \\ \addlinespace
				Endurance&&0.000&\sym{***}&0.046&\sym{***}&0.044&\sym{***}&0.031&\sym{***}&0.017&\sym{***}&0.004&\sym{***}\\
&&(0.000&)&(0.001&)&(0.001&)&(0.001&)&(0.000&)&(0.000&)\\
Ability&&0.001&\sym{***}&0.130&\sym{***}&0.113&\sym{***}&0.092&\sym{***}&0.051&\sym{***}&0.016&\sym{***}\\
&&(0.000&)&(0.002&)&(0.001&)&(0.001&)&(0.001&)&(0.000&)\\
 \midrule				
				\multicolumn{14}{l}{\hspace{-1em} \textbf{Panel B. Excluding students in the bottom or top 10\% of the endurance distribution}}  \\ \addlinespace
				Endurance&&0.000&\sym{***}&0.053&\sym{***}&0.050&\sym{***}&0.035&\sym{***}&0.016&\sym{***}&0.004&\sym{***}\\
&&(0.000&)&(0.001&)&(0.001&)&(0.001&)&(0.001&)&(0.000&)\\
Ability&&0.001&\sym{***}&0.149&\sym{***}&0.131&\sym{***}&0.103&\sym{***}&0.049&\sym{***}&0.014&\sym{***}\\
&&(0.000&)&(0.001&)&(0.001&)&(0.001&)&(0.001&)&(0.000&)\\
 \midrule		
				\multicolumn{14}{l}{\hspace{-1em} \textbf{Panel C. Excluding students in the bottom or top 10\% of either distribution}}  \\ \addlinespace
				Endurance&&0.000&\sym{***}&0.044&\sym{***}&0.042&\sym{***}&0.029&\sym{***}&0.017&\sym{***}&0.004&\sym{***}\\
&&(0.000&)&(0.001&)&(0.001&)&(0.001&)&(0.001&)&(0.000&)\\
Ability&&0.001&\sym{***}&0.127&\sym{***}&0.112&\sym{***}&0.089&\sym{***}&0.050&\sym{***}&0.015&\sym{***}\\
&&(0.000&)&(0.002&)&(0.001&)&(0.001&)&(0.001&)&(0.000&)\\
 \midrule		
				\multicolumn{14}{l}{\hspace{-1em} \textbf{Panel D. Excluding students in the bottom or top 20\% of either distribution}}  \\ \addlinespace
				Endurance&&0.000&\sym{***}&0.039&\sym{***}&0.037&\sym{***}&0.025&\sym{***}&0.016&\sym{***}&0.004&\sym{***}\\
&&(0.000&)&(0.002&)&(0.001&)&(0.001&)&(0.001&)&(0.001&)\\
Ability&&0.001&\sym{***}&0.117&\sym{***}&0.105&\sym{***}&0.083&\sym{***}&0.052&\sym{***}&0.016&\sym{***}\\
&&(0.000&)&(0.002&)&(0.002&)&(0.002&)&(0.001&)&(0.001&)\\
 \midrule		
				\multicolumn{14}{l}{\hspace{-1em} \textbf{Panel E.  Excluding individuals with positive estimated endurance}}  \\ \addlinespace
				Endurance&&0.001&\sym{***}&0.054&\sym{***}&0.055&\sym{***}&0.033&\sym{***}&0.017&\sym{***}&0.003&\sym{***}\\
&&(0.000&)&(0.001&)&(0.001&)&(0.001&)&(0.001&)&(0.000&)\\
Ability&&0.002&\sym{***}&0.158&\sym{***}&0.137&\sym{***}&0.112&\sym{***}&0.049&\sym{***}&0.014&\sym{***}\\
&&(0.000&)&(0.002&)&(0.002&)&(0.002&)&(0.001&)&(0.000&)\\
 \midrule		
				
			\end{tabular}%
		\end{center}
		\begin{singlespace}  \vspace{-.5cm}
			\noindent \justify \textit{Notes:} This table shows estimates of the relationship between ability/endurance and labor-market outcomes using alternative sample restrictions. 
			
			Each column shows the result for a different dependent variable. Each panel shows the result for a different sample of students. In Panel A, I exclude students in the bottom and top deciles of the ability distribution. In Panel B, I exclude students in the bottom and top deciles of the endurance distribution. In Panel C, I exclude students in the bottom and top deciles of the distribution of either skill. In Panel D, I exclude students in the bottom and top quintiles of the distribution of either skill. In Panel E, I exclude students with a positive estimate of endurance ($\hat{\beta}>0$). I construct the deciles and quintiles using all the students in the high-school-students sample.
			
			Heteroskedasticity-robust standard errors clustered at the individual level in parentheses.$^{***}$, $^{**}$ and $^*$ denote significance at 10\%, 5\% and 1\% levels, respectively.

		\end{singlespace} 	
	}
\end{table}%

\begin{table}[H]{\footnotesize
		\begin{center}
			\caption{Robustness of the baseline regressions to accounting for measurement error: College outcomes} \label{tab:rob-prec-coll}
			\newcommand\w{1.45}
			\begin{tabular}{l@{}lR{\w cm}@{}L{0.45cm}R{\w cm}@{}L{0.45cm}R{\w cm}@{}L{0.45cm}R{\w cm}@{}L{0.45cm}R{\w cm}@{}L{0.45cm}R{\w cm}@{}L{0.45cm}}
				
				\midrule
				& \multicolumn{12}{c}{Dependent variable} \\ \cmidrule{3-14} 
				&& Enrolled    && College && Degree   && 1st-year && Grad.      && Time to    \\
				&& college     && quality && quality  && credits  && on time    && grad.      \\
				&& (1)         && (2)     && (3)      && (4)      && (5)        &&  (6)       \\ \midrule
				\multicolumn{14}{l}{\hspace{-1em} \textbf{Panel A. Weighting each observation by its precision}}  \\ \addlinespace
				
				Endurance&&0.031&\sym{***}&0.030&\sym{***}&0.051&\sym{***}&0.006&\sym{***}&0.026&\sym{***}&$-$0.048&\sym{***}\\
&&(0.000&)&(0.000&)&(0.000&)&(0.000&)&(0.000&)&(0.001&)\\
Ability&&0.100&\sym{***}&0.094&\sym{***}&0.139&\sym{***}&0.016&\sym{***}&0.073&\sym{***}&$-$0.140&\sym{***}\\
&&(0.000&)&(0.000&)&(0.000&)&(0.000&)&(0.001&)&(0.002&)\\
 \midrule
				\multicolumn{14}{l}{\hspace{-1em} \textbf{Panel B. Shrunk estimator of ability and endurance}}  \\ \addlinespace
				Endurance&&0.045&\sym{***}&0.043&\sym{***}&0.073&\sym{***}&0.012&\sym{***}&0.037&\sym{***}&$-$0.067&\sym{***}\\
&&(0.000&)&(0.000&)&(0.000&)&(0.000&)&(0.001&)&(0.002&)\\
Ability&&0.105&\sym{***}&0.099&\sym{***}&0.145&\sym{***}&0.023&\sym{***}&0.074&\sym{***}&$-$0.143&\sym{***}\\
&&(0.000&)&(0.000&)&(0.001&)&(0.000&)&(0.001&)&(0.002&)\\
 \midrule	\addlinespace Mean DV&&0.244&&3.326&&3.244&&0.158&&0.418&&3.817&\\
\(N\)&&2,501,519&&1,800,546&&1,768,707&&1,124,972&&1,472,916&&793,822&\\

				\addlinespace \midrule
				
			\end{tabular}%
		\end{center}
		\begin{singlespace}  \vspace{-.5cm}
			\noindent \justify \textit{Notes:} This table displays estimates of the relationship between ability/endurance and college outcomes accounting for measurement error in the estimates of ability and endurance.
			
			Each column shows the result for a different dependent variable. In Panel A, I weight each observation by the inverse of the standard error of the ability and endurance estimates. Specifically, the weight of each observation is $w = 1/(\text{SE}_{\hat{\alpha}_i}^2 + \text{SE}_{\hat{\beta}_i}^2)$, where $\text{SE}_{\hat{\alpha}_i}$ and $\text{SE}_{\hat{\beta}_i}^2$ are the standard errors of $\hat{\alpha}_i$ and $\hat{\beta}_i$. In Panel B, I estimate the baseline regression using a shrunk estimator of ability and endurance. I compute the shrunk estimator of endurance as $\beta^s_i = \omega_i  \hat{\beta}_i + (1 - \omega_j)\bar{\beta},$ where $\bar{\beta}$ is the average cognitive endurance in my sample. The individual-specific weight is $\omega_i = \frac{\Var[\beta_i] - \E[\text{SE}_{\hat{\beta}_i}^2]}{\Var[\beta_i] - \E[\text{SE}_{\hat{\beta}_i}^2] + \text{SE}_{\hat{\beta}_i}^2}.$ The shrunk estimator, $\beta^s_i$, puts more weight on estimates of $\beta_i$ that are more precisely estimated, as measured by a low standard error. I compute the shrunk estimator of ability analogously.
			
			Heteroskedasticity-robust standard errors clustered at the individual level in parentheses.$^{***}$, $^{**}$ and $^*$ denote significance at 10\%, 5\% and 1\% levels, respectively.

		\end{singlespace} 	
	}
\end{table}%

\begin{table}[H]{\footnotesize
		\begin{center}
			\caption{Robustness of the baseline regressions to accounting for measurement error: Labor-market outcomes} \label{tab:rob-prec-lmkt}
			\newcommand\w{1.45}
			\begin{tabular}{l@{}lR{\w cm}@{}L{0.45cm}R{\w cm}@{}L{0.45cm}R{\w cm}@{}L{0.45cm}R{\w cm}@{}L{0.45cm}R{\w cm}@{}L{0.45cm}R{\w cm}@{}L{0.45cm}}
				
				\midrule				
				& \multicolumn{12}{c}{Dependent variable} \\ \cmidrule{3-14} 
				&& Formal        && Hourly && Monthly   && Firm  && Occup.     && Industry  \\
				&& sector        && wage   && earnings  && wage  && wage       && wage      \\
				&& (1)           && (2)     && (3)      && (4)   && (5)        && (6)       \\	\midrule
				\multicolumn{14}{l}{\hspace{-1em} \textbf{Panel A. Weighting each observation by its precision}}  \\ \addlinespace
				Endurance&&0.000&\sym{***}&0.053&\sym{***}&0.051&\sym{***}&0.035&\sym{***}&0.017&\sym{***}&0.004&\sym{***}\\
&&(0.000&)&(0.001&)&(0.001&)&(0.001&)&(0.000&)&(0.000&)\\
Ability&&0.001&\sym{***}&0.151&\sym{***}&0.133&\sym{***}&0.107&\sym{***}&0.048&\sym{***}&0.014&\sym{***}\\
&&(0.000&)&(0.001&)&(0.001&)&(0.001&)&(0.001&)&(0.000&)\\
  \midrule
				\multicolumn{14}{l}{\hspace{-1em} \textbf{Panel B. Shrunk estimator of ability and endurance}}  \\ \addlinespace
				Endurance&&0.001&\sym{***}&0.076&\sym{***}&0.074&\sym{***}&0.050&\sym{***}&0.024&\sym{***}&0.005&\sym{***}\\
&&(0.000&)&(0.001&)&(0.001&)&(0.001&)&(0.001&)&(0.000&)\\
Ability&&0.002&\sym{***}&0.157&\sym{***}&0.138&\sym{***}&0.111&\sym{***}&0.050&\sym{***}&0.015&\sym{***}\\
&&(0.000&)&(0.001&)&(0.001&)&(0.001&)&(0.001&)&(0.000&)\\
\midrule			\addlinespace Mean DV&&0.326&&3.865&&7.551&&3.885&&3.886&&3.858&\\
\(N\)&&2,523,029&&818,590&&818,590&&692,880&&818,374&&818,590&\\

				\midrule
				
			\end{tabular}%
		\end{center}
		\begin{singlespace}  \vspace{-.5cm}
			\noindent \justify \textit{Notes:} This table displays estimates of the relationship between ability/endurance and labor-market outcomes accounting for measurement error in the estimates of ability and endurance.
			
			Each column shows the result for a different dependent variable. In Panel A, I weight each observation by the inverse of the standard error of the ability and endurance estimates. Specifically, the weight of each observation is $w = 1/(\text{SE}_{\hat{\alpha}_i}^2 + \text{SE}_{\hat{\beta}_i}^2)$, where $\text{SE}_{\hat{\alpha}_i}$ and $\text{SE}_{\hat{\beta}_i}^2$ are the standard errors of $\hat{\alpha}_i$ and $\hat{\beta}_i$. In Panel B, I estimate the baseline regression using a shrunk estimator of ability and endurance. I compute the shrunk estimator of endurance as $\beta^s_i = \omega_i  \hat{\beta}_i + (1 - \omega_j)\bar{\beta},$ where $\bar{\beta}$ is the average cognitive endurance in my sample. The individual-specific weight is $\omega_i = \frac{\Var[\beta_i] - \E[\text{SE}_{\hat{\beta}_i}^2]}{\Var[\beta_i] - \E[\text{SE}_{\hat{\beta}_i}^2] + \text{SE}_{\hat{\beta}_i}^2}.$ The shrunk estimator, $\beta^s_i$, puts more weight on estimates of $\beta_i$ that are more precisely estimated, as measured by a low standard error. I compute the shrunk estimator of ability analogously.
			
			Heteroskedasticity-robust standard errors clustered at the individual level in parentheses.$^{***}$, $^{**}$ and $^*$ denote significance at 10\%, 5\% and 1\% levels, respectively.

		\end{singlespace} 	
	}
\end{table}%

	\clearpage 
\section{The ENEM} \label{app:enem-exam}

\setcounter{table}{0}
\setcounter{figure}{0}
\setcounter{equation}{0}	
\renewcommand{\thetable}{C\arabic{table}}
\renewcommand{\thefigure}{C\arabic{figure}}
\renewcommand{\theequation}{C\arabic{equation}}

In this Appendix, I describe the changing role of the ENEM in the higher-education system over time, compare the ENEM to the US SAT and ACT exams, and describe the IRT grading system used by the Ministry of Education to generate ENEM test scores.

\subsection{The Role of the ENEM in the Higher-education System} \label{app:enem}

The ENEM was created in 1998 by the National Institute of Educational Studies (INEP), a unit of the Brazilian Ministry of Education, with the goal of evaluating student performance at the end of high school (Appendix Figure \ref{fig:timeline}). The ENEM is an achievement test, that is, it was designed to test for mastery of material individuals should learn by the end of high school.\footnote{Researchers often divide standardized tests into two types: reasoning tests and achievement tests. Reasoning tests measure a student's verbal reasoning, critical reading, and skills. Achievement tests measure a student's mastery of specific subjects, like biology or physics. In practice, performance on both types of tests is highly correlated \citep{soares2015sat}.}

The first ENEM contained 63 multiple-choice interdisciplinary questions and was conducted over a five-hour testing block. The test score was calculated as the fraction of correct responses. In its first edition, fewer than 200,000 individuals enrolled to take the ENEM. 

\begin{figure}[H]
	\centering
	\caption{Timeline of the ENEM} \label{fig:timeline}
	\begin{tikzpicture}[snake=zigzag, line before snake = 5mm, line after snake = 5mm, scale=1.1, every node/.style={scale=.8}]
		\draw [->] (0,0) -- (13,0);
		\foreach \x in {0,3, 6.2, 10, 12} \draw (\x cm,3pt) -- (\x cm,-3pt);
		\draw (0,0)  node[below=3pt] {1998} node[above=3pt] {\stackanchor{First edition: HS}{\stackanchor{accountability test}{(non-mandatory)}}};
		
		\draw (3,0)  node[below=3pt] {2004} node[above=3pt] {\stackanchor{PROUNI program:}{\stackanchor{scholarships to}{low-income students}}};
		
		\draw (6.2,0)  node[below=3pt] {2009} node[above=3pt] {\stackanchor{Expansion: Federal}{\stackanchor{college admission exam}{\& HS certification}}};
		
		\draw (10,0) node[below=3pt] {2017} node[above=3pt] {\stackanchor{New}{schedule}};
		
		\draw (12,0) node[below=3pt] {2020} node[above=3pt] {\stackanchor{Online}{option}};
		
		\draw [thick,decorate,decoration={brace,amplitude=5pt}] (5.5,-.9) -- +(-5.5,0)
		node [black,midway,font=\footnotesize, below=4pt] {\stackanchor{Unique test (1 day, 5 hours)}{63 interdisciplinary questions}};
		
		\draw [thick,decorate,decoration={brace,amplitude=5pt}] (9.7,-.9) -- +(-3.7,0)
		node [black,midway,font=\footnotesize, below=3pt] {};
		
		\draw [decorate,decoration={brace,amplitude=0pt}] (7.8,-.9) -- +(0,0)
		node [black,midway,font=\footnotesize, below=4pt] {\stackanchor{180 q's divided into 4 subjects:}{\stackanchor{Day 1: Soc. sci., Nat. sci. (4.5h)}{Day 2: Essay, Lang., Math (5.5h)}}};
		
		\draw [thick,decorate,decoration={brace,amplitude=5pt}] (13,-.9) -- +(-3.0,0)
		node [black,midway,font=\footnotesize, below=3pt] {};
		
		\draw [decorate,decoration={brace,amplitude=0pt}] (12,-.9) -- +(0,0)
		node [black,midway,font=\footnotesize, below=4pt] {\stackanchor{2 consecutive Sundays:}{\stackanchor{Day 1: Lang., Essay, S. sci. (5.5h)}{Day 2: Nat. sci., Math (5h)}}};

	\end{tikzpicture}
\end{figure}

In 2004, the government created a college scholarship program for low-income students called ProUni (\textit{Programa Universidade para Todos}). ProUni used ENEM scores to allocate the scholarships to applicants, with program-specific score cutoffs based on the number of seats available in each program. After ProUni was implemented, the number of individuals who signed up to take the ENEM doubled from 1.5 million in 2004 to 3.0 million in 2005.


In 2009, the Ministry of Education reformed the ENEM with the aim of encouraging colleges to use it as an admission exam. The new ENEM consists of 180 multiple-choice questions conducted over two consecutive days of testing during a weekend. The new exam contains questions in four subjects: mathematics, natural sciences (which includes biology, physics, and chemistry questions), social sciences (which includes history, geography, philosophy, and sociology questions), and language arts (which includes questions on Portuguese language, literature, foreign language, arts, physical education, and information and communication technologies). On the first day of testing, individuals had five and a half hours to take the social science test, the natural science test, and the essay. On the second day of testing, individuals had five hours to take math and language arts tests. The new ENEM is graded according to Item Response Theory (IRT), which enables colleges to compare test scores over time (see Appendix \ref{app-sub:irt-grading}). 

In 2010, the Ministry of Education introduced a centralized admission system called SISU (\textit{Sistema de Seleção Unificada}) with the goal of simplifying the college application process for federal universities. The centralized system used ENEM scores to allocate students to participating colleges. All federal universities are part of the system, but other universities (including state and municipal universities) are not mandated to be part of it. Also in 2010, the Government started using ENEM scores to allocate student loans through a program called FIES (\textit{Fundo de Financiamento ao Estudante do Ensino Superior}). In addition, starting in 2010 (and finishing in 2016), ENEM scores could be used to certify the attainment of high-school-level skills (analogously to the GED in the US). By 2010, over 4.6 million individuals enrolled to take the ENEM.

In 2017, INEP changed the schedule of the ENEM. The exam started being conducted over two consecutive Sundays. On the first Sunday, individuals have five and a half hours to answer the language arts test, the social science test, and the essay. On the second Sunday, individuals have five hours to answer the natural science and math tests. The other features of the exam remained constant.

In 2020, individuals had the option to take the ENEM through a computer without internet access. Over 5.7 million individuals enrolled to take the ENEM this year.

\subsection{ENEM Sample Questions} \label{app-sub:enem-examples}

\setlist[enumerate,1]{align=left}
\newlist{choices}{enumerate}{2}
\setlist[choices]{label=\Alph*,format=\ovalnode,align=left}
\newcommand\ovalnode[1]{%
	\tikz[baseline=(oval.base)]\node[draw,ellipse,inner ysep=0pt,inner xsep=1pt](oval){\makebox[0.8em]{#1\vphantom{A}}};}
\let\choice\item

\begin{figure}[H]
	\centering
	\caption{Social Science sample question (item \#97290)} \label{fig:ssci-ex}
	
	\textbf{Panel A. Original (in portuguese)}	
	\begin{centering}
		\fbox{\begin{minipage}{\textwidth}
				
				\centering \textbf{Parceria Transpacífica}
				
				\includegraphics[width=.35\linewidth]{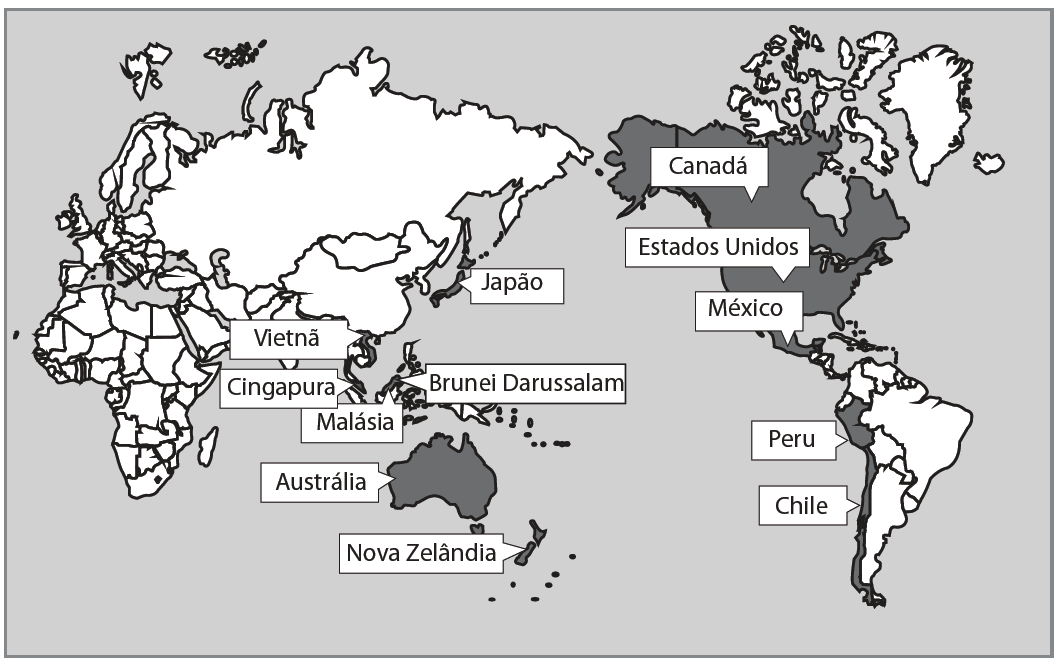}				
				
				\raggedright Dentro das atuais redes produtivas, o referido bloco apresenta composição estratégica por se tratar de um conjunto de países com
				
				\begin{choices}
					\choice Elevado padrão social.
					\choice Sistema monetário integrado.
					\choice Alto desenvolvimento tecnológico.
					\choice Identidades culturais semelhantes.
					\choice \underline{Vantagens locacionais complementares.}\vspace{.1cm}
					
				\end{choices}		
		\end{minipage}	}\\~\\
	\end{centering}
	
	\textbf{Panel B. Translation}\vspace{.2cm}
	\begin{centering}
		\fbox{\begin{minipage}{\textwidth}
				
				\centering \textbf{Trans-Pacific Partnership}
				
				\includegraphics[width=.35\linewidth]{results/ssci_item_97290}				
				
				\raggedright Within the current production networks, the aforementioned bloc has a strategic composition because it is a group of countries with:
				
				\begin{choices}
					\choice High social standard.
					\choice Integrated monetary system.
					\choice High technological development.
					\choice Similar cultural identities.
					\choice \underline{Complementary locational advantages.}\vspace{.1cm}
				\end{choices}		
			\end{minipage}	
		}
	\end{centering}	
	
	{\footnotesize	\singlespacing \justify
		
		\textit{Notes:} The correct answer is underlined.
		
	}
	
\end{figure}

Appendix Figures \ref{fig:nsci-ex}--\ref{fig:math-ex} present sample questions from the natural science, social science, language arts, and math components of the ENEM. The questions come from the 2016 ENEM. The questions are average in terms of their difficulty.

\begin{figure}[H]
	\centering
	\caption{Natural Science sample question (item \#11898)} \label{fig:nsci-ex}
	
	\textbf{Panel A. Original (in portuguese)}	
	\begin{centering}
		\fbox{\begin{minipage}{\textwidth}
				Portadores de diabetes \textit{insipidus} reclamam da confusão feita pelos profissionais da saúde quanto aos dois tipos de diabetes: \textit{mellitus} e \textit{insipidus}. Enquanto o primeiro tipo está associado aos níveis ou à ação da insulina, o segundo não está ligado à deficiência desse hormônio. O diabetes \textit{insipidus} é caracterizado por um distúrbio na produção ou no funcionamento do hormônio antidiurético (na sigla em inglés, ADH), secretado pela neuro-hipófise para controlar a reabsorção de água pelos túbulos renais.\\
				
				Tendo em vista o papel funcional do ADH, qual é um sintoma clássico de um paciente acometido por diabetes \textit{insipidus}? 
				
				\begin{choices}
					\choice Alta taxa de glicose no sangue.
					\choice Aumento da pressão arterial.
					\choice Ganho de massa corporal.
					\choice Anemia crônica.
					\choice \underline{Desidratação.} \vspace{.1cm}
				\end{choices}		
		\end{minipage}	}\\~\\
	\end{centering}
	
	\textbf{Panel B. Translation}\vspace{.2cm}
	\begin{centering}
		\fbox{\begin{minipage}{\textwidth}
				
				Patients with diabetes \textit{insipidus} complain about the confusion made by health professionals about the two types of diabetes: \textit{mellitus} and \textit{insipidus}. While the first type is associated with insulin levels or action, the second is not linked to insulin deficiency. Diabetes insipidus is characterized by a disturbance in the production or functioning of the antidiuretic hormone (ADH), secreted by the neurohypophysis to control the reabsorption of water by the renal tubules.\\
				
				In view of the functional role of ADH, what is a classic symptom of a patient with diabetes \textit{insipidus}?
				
				\begin{choices}
					\choice High blood glucose.
					\choice Increase in blood pressure.
					\choice Body mass gain.
					\choice Chronic anemia.
					\choice \underline{Dehydration.} \vspace{.1cm}
				\end{choices}		
			\end{minipage}	
		}
	\end{centering}	
	
	{\footnotesize	\singlespacing \justify
		
		\textit{Notes:} The correct answer is underlined.
		
	}
	
\end{figure}

\begin{figure}[H]
	\centering
	\caption{Language Arts sample question (item \#86509)} \label{fig:lang-ex}
	
	\textbf{Panel A. Original (in portuguese)}	
	\begin{centering}
		\fbox{\begin{minipage}{\textwidth}
				
				O último longa de Carlão acompanha a operária Silmara, que vive com o pai, um ex-presidiário, numa casa da periferia paulistana. Ciente de sua beleza, o que lhe dá certa soberba, a jovem acredita que terá um destino diferente do de suas colegas. Cruza o caminho de dois cantores por quem é apaixonada. E constata, na prática, que o romantismo dos contos de fada tem perna curta.
				
				\raggedleft \footnotesize  VOMERO, M. F. Romantismo de araque. \textbf{Vida Simples}, n. 121, ago. 2012.\\~\\
				
				\raggedright \normalsize Reconhece-se, nesse trecho, uma posição crítica aos ideais de amor e felicidade encontrados nos contos de fada. Essa crítica é traduzida
				
				\begin{choices}	
					\choice Pela descrição da dura realidade da vida das operárias.
					\choice Pelas decepções semelhantes às encontradas nos contos de fada.
					\choice Pela ilusão de que a beleza garantiria melhor sorte na vida e no amor.
					\choice \underline{Pelas fantasias existentes apenas na imaginação de pessoas apaixonadas.}
					\choice Pelos sentimentos intensos dos apaixonados enquanto vivem o romantismo. \vspace{.1cm}
				\end{choices}								
				
		\end{minipage}	}\\~\\
	\end{centering}
	
	\textbf{Panel B. Translation}\vspace{.2cm}
	\begin{centering}
		\fbox{\begin{minipage}{\textwidth}
				
				Carlão's latest feature follows the worker Silmara, who lives with her father, an ex-convict, in a house on the outskirts of São Paulo. Aware of her beauty, which gives her a certain arrogance, the young woman believes that she will have a different destiny from her colleagues. She crosses paths with two singers she is in love with. And she finds, in practice, that the romanticism of fairy tales has short legs.
				
				\raggedleft \footnotesize VOMERO, M. F. Romanticism of arak. \textbf{Simple Life}, n. 121, Aug. 2012.\\~\\
				
				\raggedright \normalsize This passage recognizes a critical position on the ideals of love and happiness found in fairy tales. This criticism is translated
				
				\begin{choices}
					\choice For the description of the harsh reality of the workers' lives.
					\choice For disappointments similar to those found in fairy tales.
					\choice For the illusion that beauty would guarantee better luck in life and in love.
					\choice \underline{For the fantasies that exist only in the imagination of people in love.}
					\choice For the intense feelings of those in love while living romanticism. \vspace{.1cm}
				\end{choices}
				
			\end{minipage}	
		}
	\end{centering}	
	
	{\footnotesize	\singlespacing \justify
		
		\textit{Notes:} The correct answer is underlined.
		
	}
	
\end{figure}

\begin{figure}[H]
	\centering
	\caption{Math sample question (item \#37515)} \label{fig:math-ex}
	
	\textbf{Panel A. Original (in portuguese)}	
	\begin{centering}
		\fbox{\begin{minipage}{\textwidth}		
				
				Para evitar uma epidemia, a Secretaria de Saúde de uma cidade dedetizou todos os bairros, de modo a evitar a proliferação do mosquito da dengue. Sabe-se que o número $f$ de infectados é dado pela função $f(t) = -2t^2 + 120t$ (em que $t$ é expresso em dia e $t=0$ é o dia anterior à primeira infecção) e que tal expressão é válida para os 60 primeiros dias da epidemia.\\
				
				A Secretaria de Saúde decidiu que uma segunda dedetização deveria ser feita no dia em que o número de infectados chegasse à marca de 1600 pessoas, e uma segunda dedetização precisou acontecer.\\ 
				
				A segunda dedetização começou no
				
				\begin{choices}				
					\choice 19° dia.
					\choice 20° dia.
					\choice 29° dia.
					\choice 30° dia.
					\choice \underline{60° dia.}
				\end{choices}

		\end{minipage}	}\\~\\
	\end{centering}
	
	\textbf{Panel B. Translation}\vspace{.2cm}
	\begin{centering}
		\fbox{\begin{minipage}{\textwidth}
				
				To prevent an epidemic, the Health Department of a city sprayed all neighborhoods, in order to prevent the proliferation of the dengue mosquito. It is known that the number $f$ of infected people is given by the function $f(t) = -2t^2 + 120t$ (where $t$ is expressed in day and $t=0$ is the day before the first infection) and that this expression is valid for the first 60 days of the epidemic.\\
				
				The Health Department decided that a second extermination should be carried out on the day when the number of infected people reached the mark of 1,600 people, and a second extermination had to take place.\\
				
				The second extermination started in
				
				\begin{choices}
					\choice 19th day.
					\choice 20th day.
					\choice 29th day.
					\choice 30th day.
					\choice \underline{60th day.}
				\end{choices}
				
			\end{minipage}	
		}
	\end{centering}	
	
	{\footnotesize	\singlespacing \justify
		
		\textit{Notes:} The correct answer is underlined.
		
	}
	
\end{figure}

\subsection{Comparison of the ENEM to the ACT and SAT exams}

Appendix Table \ref{tab:sat} compares important features of the SAT, ACT, and ENEM. The SAT contains 154 multiple-choice questions divided into three sections: reading, writing and language, and math, plus an optional essay. Including the essay, individuals have 3 hours and 50 minutes to take the test. On average across sections, test-takers have about 1 minute and 10 seconds to answer each question. Raw scores are converted into scaled scores through a score conversion table. 

The ACT contains 215 multiple-choice questions divided into four sections: English, math, reading, and science, plus an optional essay. Including the essay, individuals have 3 hours and 35 minutes to take the test. On average across sections, test-takers have less than 1 minute to answer each question. Raw scores are converted into scaled scores through a score conversion table.

There are some notable differences between the SAT/ACT and the ENEM. First, the ENEM is conducted over two days of testing. Second, individuals in the ENEM have no assigned breaks. Third, the booklet ENEM test-takers receive contains all the questions they have to answer during the testing day. Thus, they may allocate time disproportionally across sections. In contrast, in the SAT and ACT, each section has an assigned amount of time. Finally, in the ENEM, each question is associated with a different text passage or prompt (in some cases, two questions share a prompt or passage). In contrast, in the SAT and ACT, a given passage is associated with multiple questions. This partly explains why the time per question is higher in the ENEM than in the ACT/SAT.

\clearpage
\begin{sidewaystable}
	\begin{table}[H]{\footnotesize
			\begin{center}
				\caption{Comparison of the SAT, ACT, and ENEM college admission exams} \label{tab:sat}
				\begin{tabular}{llll}
					\toprule
					& SAT   & ACT  & ENEM \\ \midrule
					Cost  & $\sim$\$60 & $\sim$\$88  & $\sim$\$17 \\
					Grading & Score conversion chart using raw scores  & Score conversion chart using raw scores & Item Response Theory (IRT) \\					
					Starting time & Between 8:30 and 9am & Between 8:30 and 9am & 1pm Brasilia time \\
					Number of items & 154 questions & 215 questions & 180 questions \\
					Total length & 3 hours and 50 mins over 1 testing day & 3 hours and 35 mins over 1 testing day & 10 hours over 2 testing days \\
					Time per question  & 1 minute and 10 seconds & 50 seconds & 3 minutes \\
					Breaks & 10 mins break after reading section & 10 min break after math section & N/A  \\
					& 5 min break between math sections & 5 min break before essay & \\
					& 2 min break before the essay &       &  \\\midrule
					Sections & Reading (65 mins, 52 items) & English (45 mins, 75 items) & Social science (day 1, 45 items) \\
					& Writing and Language (35 mins, 44 items) & Math (60 mins, 60 items) & Natural science (day 1, 45 items) \\
					& Math w/o calculator (25 mins, 20 items) & Reading (35 mins, 40 items) & Language arts (day 2, 45 items) \\
					& Math w/ calculator (55 mins, 38 items) & Science (35 mins, 40 items) & Math (day 2, 45 items) \\
					& Optional essay (50 mins) & Optional essay (40 mins) & Mandatory essay (day 2) \\
					\midrule

				\end{tabular}%
			\end{center}
			\begin{singlespace}  \vspace{-.5cm}
				\noindent \justify \textit{Notes:} The SAT refers to the post-2016 version of the SAT, which includes an optional essay. This optional essay was eliminated in 2021. The ENEM refers to the 2009--2016 version of the exam (see Section \ref{app:enem} for information on the pre-2009 and post-2016 versions). The exam length was computed excluding breaks and including the essay. The time per question does not account for the essay.
				
			\end{singlespace} 	
		}
	\end{table}%
	
\end{sidewaystable}

\clearpage
\subsection{IRT Grading} \label{app-sub:irt-grading}

The Brazilian Testing Agency grades the ENEM exam based on the three-parameter item response theory (IRT). According to IRT, the probability that an individual $i$ with ability \textcolor{black}{$\theta_i$} correctly answers question $j$ is:
\begin{align}\label{eq:irt}
	\Pr(C_{ij} = 1 | \theta_i) = p_{j}({\theta_i }) = c_{j} + {\frac {1-c_{j}}{1+e^{{-a_{j}({\theta_i }-b_{i})}}}},
\end{align}
where $a_j$, $b_j$, and $c_j$ are three question-level parameters that represent, respectively, a question's ``discrimination,'' ``difficulty,'' and ``pseudo-guess.'' A question's discrimination refers to its ability to discriminate between low- and high-ability individuals; the difficulty represents the value of $\theta$ at which $p_{j}({\theta_i})$ has the maximum slope, and the pseudo-guess parameter indicates the likelihood that a student with an infinitely negative ability has to correctly respond to the question. Notice that in equation \eqref{eq:irt}, the probability of correctly answering a question does not depend on its position. Thus, the type of position effects documented above suggests that the IRT estimates of individual-level ability are biased. Modern IRT approaches \citep[e.g.,][]{debeer2013modeling} include item position into the framework.

Each question's parameters are known from pre-testing. The testing agency estimates the $\theta_i$ that maximizes the empirical likelihood of the entire sequence of responses. They do this separately for each student and academic subject. ENEM scores are normalized to have a mean of 500 and a standard deviation of 100.

\begin{figure}[H]
	\caption{Comparison of IRT-estimated ENEM score and fraction of correct responses}\label{fig:irt-pct-corr}
	\centering
	\begin{subfigure}[t]{.48\textwidth}
		\caption*{Panel A. In levels}
		\centering
		\includegraphics[width=\linewidth]{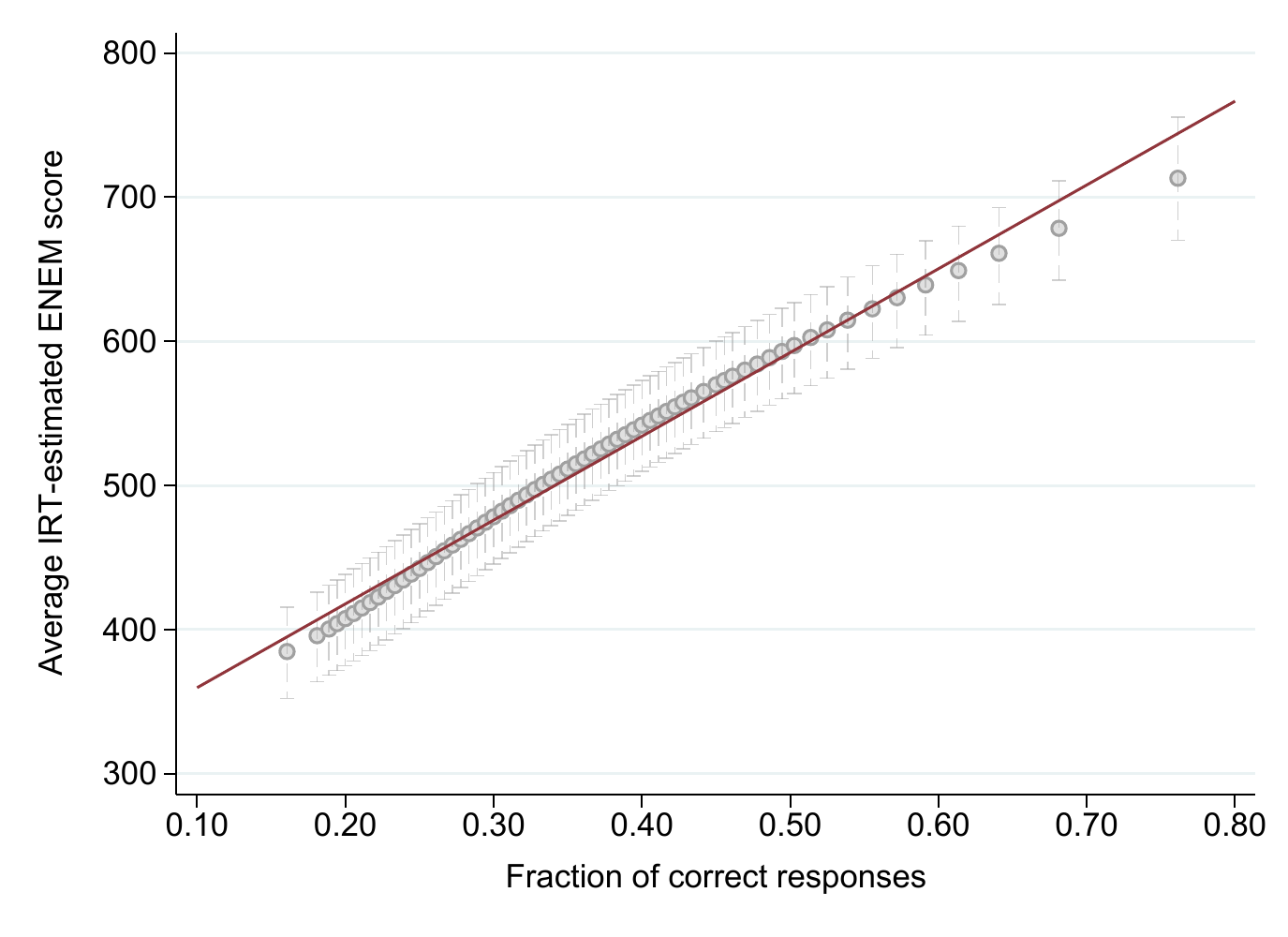}
	\end{subfigure}
	\hfill		
	\begin{subfigure}[t]{0.48\textwidth}
		\caption*{Panel B. In percentiles}
		\centering
		\includegraphics[width=\linewidth]{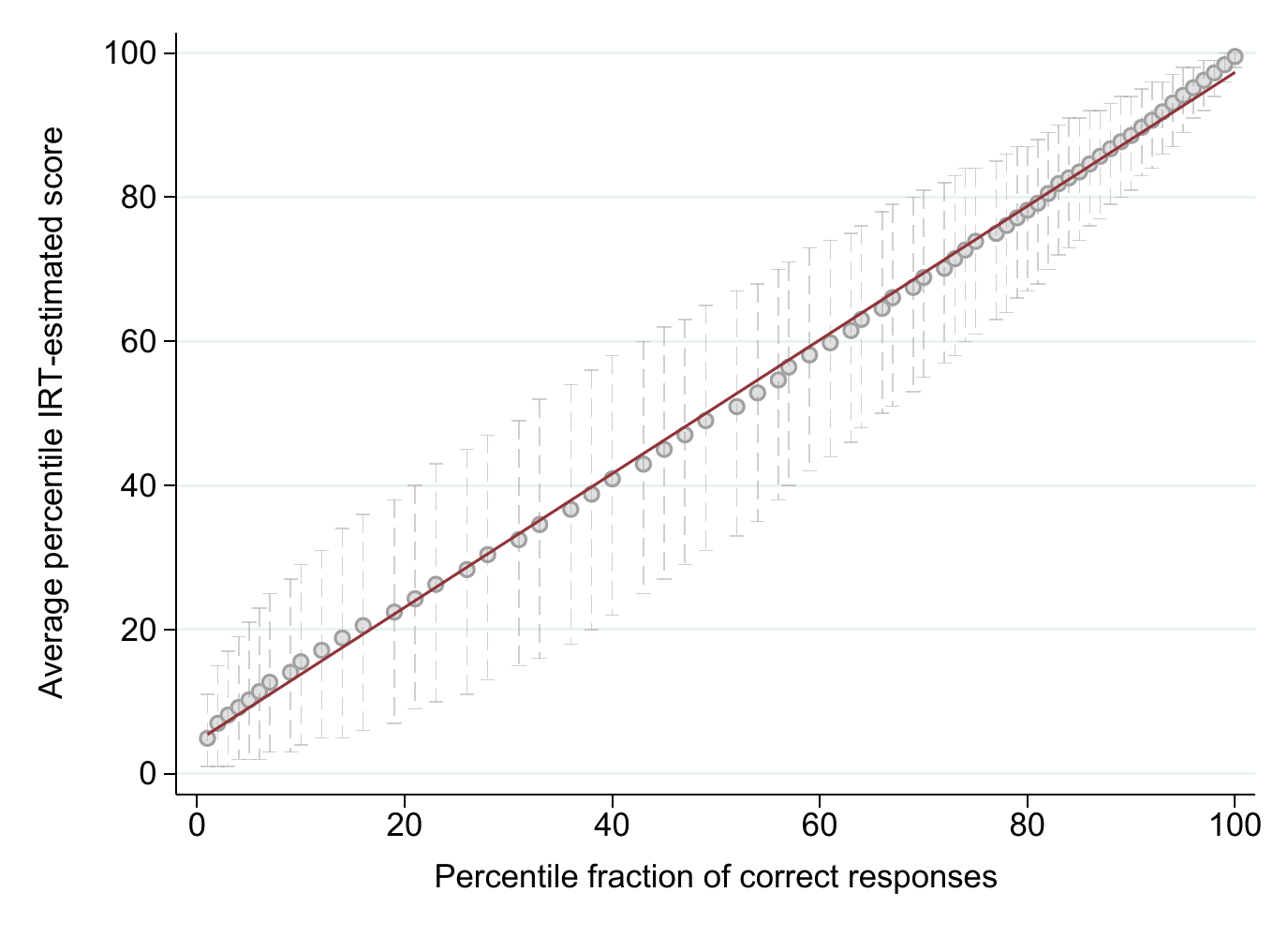}
	\end{subfigure}	
	\hfill						
	{\footnotesize
		\singlespacing \justify
		
		\textit{Notes:} This figure shows binned scatterplots plotting the average IRT-estimated ENEM score across all four academic subjects ($y$-axis) against the fraction of correct responses on the exam ($x$-axis). Panel A shows the results in levels and Panel B in percentiles.  I first group students into 100 equally-sized bins based on their fraction of correct responses. Then, I calculate the average IRT-estimated ENEM score or score percentile in each bin. The vertical lines denote the 10th and 90th percentiles of the ENEM score distribution. The solid red line shows the predicted values from a linear regression on the plotted points.	
		
	}
\end{figure}

Despite its complexity, most of the variation in IRT-estimated ENEM scores is driven by variation in the fraction of correct responses in the exam. A regression of IRT-estimated ENEM scores on the fraction of correct responses yields an R-squared of 0.88 (the rank correlation between the two variables is 0.93). Consistent with this, Appendix Figure \ref{fig:irt-pct-corr} shows that the relationship between these two variables is linear in both levels (Panel A) and percentiles (Panel B). The strong relationship between IRT-estimated scores and the fraction of correct responses holds not just for the overall score but also for the score in each academic subject (Appendix Table \ref{tab:irt-pct-corr}).

\begin{table}[H]{\footnotesize
		\begin{center}
			\caption{Correlation between IRT-estimated ENEM score and fraction of correct responses on each subject} \label{tab:irt-pct-corr}
			\newcommand\w{1.65}
			\begin{tabular}{l@{}lR{\w cm}@{}L{0.45cm}R{\w cm}@{}L{0.45cm}R{\w cm}@{}L{0.45cm}R{\w cm}@{}L{0.45cm}R{\w cm}@{}L{0.45cm}R{\w cm}@{}L{0.45cm}}
				\midrule
				& \multicolumn{8}{c}{Academic subject} \\ \cmidrule{3-10} 
				&& Social        && Natural && Language && Math  && Average \\
				&& science       && science && arts     &&       && score   \\
				&& (1)           && (2)     && (3)      && (4)   && (5)     \\			
				\midrule
				\multicolumn{12}{l}{\hspace{-1em} \textbf{Panel A. Variables measured in levels}} \\   
				Fraction correct resp.&&0.892&\sym{***}&0.880&\sym{***}&0.907&\sym{***}&0.885&\sym{***}&0.937&\sym{***}\\
&&(0.000&)&(0.000&)&(0.000&)&(0.000&)&(0.000&)\\
 
				\(N\)&&14,936,699&&14,936,699&&14,936,699&&14,936,699&&14,936,699&\\
R$-$squared&&0.79&&0.77&&0.82&&0.78&&0.88&\\
 \midrule								

				\multicolumn{12}{l}{\hspace{-1em} \textbf{Panel B. Variables measured in percentiles}} \\ 
				Fraction correct resp.&&0.904&\sym{***}&0.858&\sym{***}&0.917&\sym{***}&0.845&\sym{***}&0.931&\sym{***}\\
&&(0.000&)&(0.000&)&(0.000&)&(0.000&)&(0.000&)\\
 
				\(N\)&&14,936,699&&14,936,699&&14,936,699&&14,936,699&&14,936,699&\\
R$-$squared&&0.82&&0.74&&0.84&&0.71&&0.87&\\
 \midrule												

			\end{tabular}%
		\end{center}
		\begin{singlespace}  \vspace{-.5cm}
			\noindent \justify \textit{Notes:} This table displays the correlation between the IRT-estimated ENEM score and the fraction of correct responses. Columns 1--4 present the correlations separately for each academic subject. Column 5 presents the correlation between the average score across all subjects and the fraction of correct responses in the entire exam. Heteroskedasticity-robust standard errors clustered at the question level in parentheses. $^{***}$, $^{**}$ and $^*$ denote significance at 10\%, 5\% and 1\% levels, respectively.
			
		\end{singlespace} 	
	}
\end{table}%

	\clearpage
\section{Measuring Question Difficulty} \label{app:difficulty}

\setcounter{table}{0}
\setcounter{figure}{0}
\setcounter{equation}{0}	
\renewcommand{\thetable}{D\arabic{table}}
\renewcommand{\thefigure}{D\arabic{figure}}
\renewcommand{\theequation}{D\arabic{equation}}

In this Appendix, I describe my measures of question difficulty. Instead of taking a strong stance on what the right measure of difficulty is, I show that the results are robust to measuring question difficulty in several ways. 

\begin{figure}[H]
	\centering
	\caption{Performance on a natural science question (item \#11898)} \label{fig:item-11898}
	\includegraphics[width=.60\linewidth]{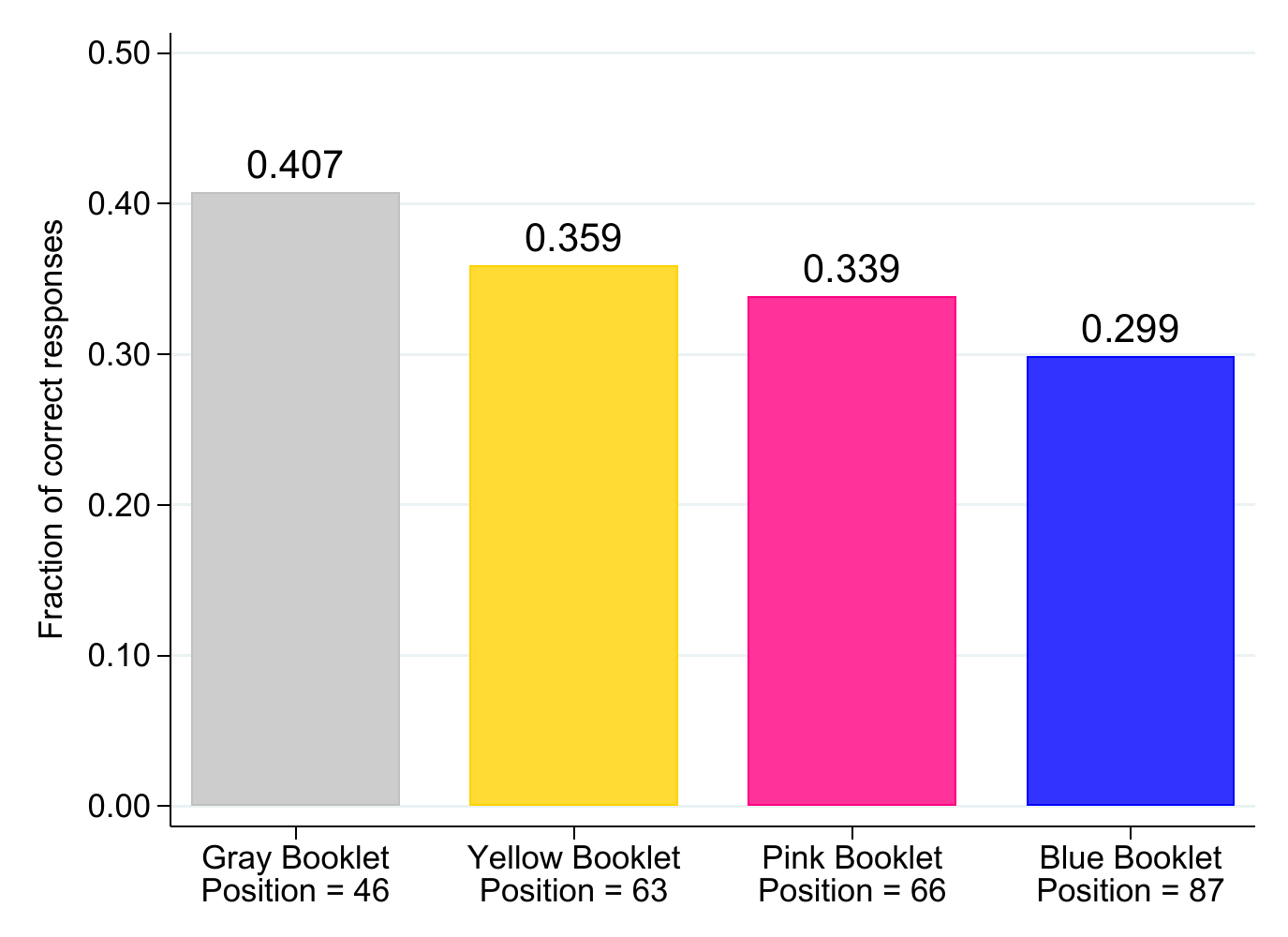}
	
	{\footnotesize
		\singlespacing \justify
		
		\textit{Notes:} This figure shows the fraction of individuals who correctly responded to item \#11898 in each of the four booklets. See Appendix Figure \ref{fig:nsci-ex} for the question's text.
		
	}
	
\end{figure}

An intuitive measure of a question's difficulty is the fraction of students who correctly answer the question. This measure is problematic in the presence of fatigue effects since a given question has a different fraction of correct responses depending on its location. To illustrate this problem, Appendix Figure \ref{fig:item-11898} plots students' performance on a natural science question in each booklet. The position of this item ranged from position 46 in the gray booklet to position 87 in the blue booklet. Correspondingly, student performance varied from 40.7\% in the gray booklet to 29.9\% in the blue booklet.

The fact that performance on a question varies according to its position raises an important challenge for measuring question difficulty. It is hard to know whether questions that appear later in the exam are less likely to be correctly answered because they test more difficult material or because students are more fatigued by the time they get to these questions.

To account for fatigue effects, I estimate measures of question difficulty that represent the fraction of students who would correctly answer a question if the question appeared in the first position of the exam. To estimate this fraction, I follow a three-step process. First, I compute the average position of each question across all booklets. Second, I estimate the effect of a one-position increase of a question position on performance on the question (``position effect''). Third, I multiply the average question position calculated in the first step by the position effect estimated in the second step and subtract this figure from the fraction of correct responses across all booklets. This yields a position-adjusted estimate of question difficulty. Appendix Table \ref{tab:example-diff} illustrates these steps in calculating the difficulty of item \#11898. 

The measures of question difficulty differ in how I estimate the position effect in the second step. My baseline measure of question difficulty uses the position effect estimated by pooling all questions (Table \ref{tab:perf-pos}, column 3). This measure assumes that the effect of a one-position increase on performance is homogeneous across questions. 

The second measure of question difficulty uses a question-specific position effect. I estimate equation \eqref{eq:pot-out-reg} separately for each question and use the intercept from the regression as the measure of difficulty. This does not assume homogeneity in position effects; however, for some questions the position effect is imprecisely estimated.

The third measure of question difficulty combines the first two by shrinking the question-specific position effect to the average effect by its signal-to-noise ratio. Specifically, let $\beta_j$ be the position effect estimating using data only from question $j$ and $\bar{\beta}$ be the average position effect across all questions. The shrunk position effect of question $j$, $\beta^s_j$, is a convex combination of $\beta_j$ and $\bar{\beta}$:
\begin{align*}
	\beta^s_j = \omega_j  \beta_j + (1 - \omega_j)\bar{\beta},
\end{align*}
where the question-specific weight, $\omega_j$, is
\begin{align*}
	\omega_j = \frac{\Var[\hat{\beta}_j] - \E[\text{SE}_{\hat{\beta}_j}^2]}{\Var[\hat{\beta}_j] - \E[\text{SE}_{\hat{\beta}_j}^2] + \text{SE}_{\hat{\beta}_j}^2}.
\end{align*}

The shrunk estimator puts more weight on position effects that are more precisely estimated, as measured by a low standard error of $\hat{\beta}_j$, $\text{SE}_{\hat{\beta}_j}^2$.

The fourth measure estimates the position effect separately for questions with a below/above median fraction of correct responses. The fifth measure estimates the effect separately for each academic subject. These measures assume that the effect of a one-position increase on performance is homogeneous within a type of question.

\begin{table}[H]{\footnotesize
		\begin{center}
			\caption{\centering Alternative measures of the difficulty of item \#11898} \label{tab:example-diff}
			\begin{tabular}{ccccc}
				\midrule
				Position effect         & Average fraction  & Fatigue effect (in pp) & Question      \\ 
				estimation method      & correct responses & $\times$  average position  & difficulty         \\  
				(1)                    &   (2)             &    (3)                     & (4)                \\ \midrule
				None                   &	0.36&	0 $\times$ 64 = 0&	0.36	\\
				Pooling all items      &	0.36&	-0.08 $\times$ 64 = -5.1&	0.41	\\
				Item-specific effect   &	0.36&	-0.24 $\times$ 64 = -15.3&	0.51	\\
				Shrinkage estimator    &	0.36&	-0.24 $\times$ 64 = -15.3&	0.51	\\
				By fraction corr. resp.&	0.36&	-0.15 $\times$ 64 = -9.6&	0.45	\\
				By academic subject    &	0.36&	-0.03 $\times$ 64 = -1.6&	0.37	\\
				
				\midrule
			\end{tabular}%
		\end{center}
		\begin{singlespace} \vspace{-.5cm}
			\noindent \justify \textit{Notes:} This table illustrates how the six measures of a question's difficulty are calculated. The average fraction of correct responses and the average question position are calculated using the number of students with each booklet as weights.
		\end{singlespace}		
	}
\end{table}

\begin{figure}[H]
	\centering
	\caption{Cross-question correlation matrix of item difficulty measures} \label{fig:matrix-diff-measures}
	\includegraphics[width=.6\linewidth]{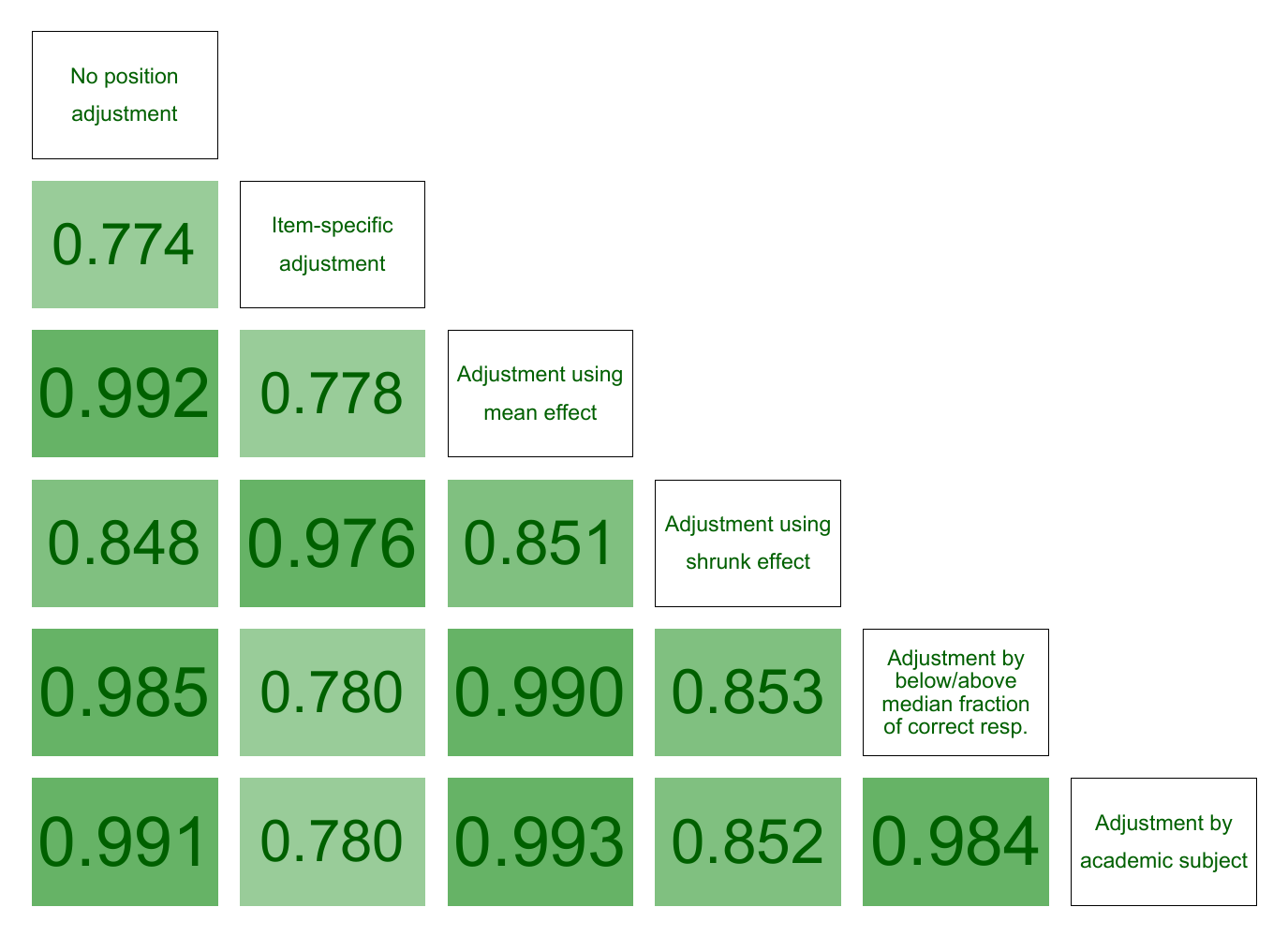}
	
	{\footnotesize
		\singlespacing \justify
		
		\textit{Notes:} This figure shows the relationship between the different measures of question difficulty. Each cell shows the cross-question linear correlation between two measures of question difficulty. The sample size is $N = 1,842$ across all cells.
		
	}
	
\end{figure}

Appendix Figure \ref{fig:matrix-diff-measures} shows the cross-question correlation between the measures of question difficulty. Reassuringly, all difficulty measures are highly correlated, with coefficients ranging from 0.77 to 0.99.

	\clearpage 
\section{Exam Content and Cognitive Endurance} \label{app:het-question}

\setcounter{table}{0}
\setcounter{figure}{0}
\setcounter{equation}{0}	
\renewcommand{\thetable}{E\arabic{table}}
\renewcommand{\thefigure}{E\arabic{figure}}
\renewcommand{\theequation}{E\arabic{equation}}

In this Appendix, I assess whether the type of questions of an exam can influence the effect of limited endurance on performance. For this, I estimate heterogeneity in the effect of limited endurance on student performance based on two question characteristics: difficulty and length.   

First, I explore heterogeneity based on question difficulty. Previous research has identified task difficulty as a moderator of cognitive fatigue \citep{ackerman2011cognitive}. We might expect cognitive endurance to matter only for questions in a certain difficulty range. Students should be able to answer very easy questions regardless of how tired they are. Similarly, some questions might be too difficult for students to answer regardless of how rested they are. The ENEM contains very difficult questions. On average, students only answer 34.3\% of questions correctly (random chance would imply 20\%). Individuals who answer 50\% of questions correctly are in the top 10\% of the score distribution. Thus, we might expect limited endurance to affect performance in the relatively-easier questions. 

\begin{figure}[H]
	\caption{The effect of cognitive endurance on performance by question characteristics}\label{fig:chg-pos-char}
	\centering
	\begin{subfigure}[t]{.48\textwidth}
		\caption*{Panel A. Change in performance vs. change in question position by question difficulty \\ }
		\centering
		\includegraphics[width=\linewidth]{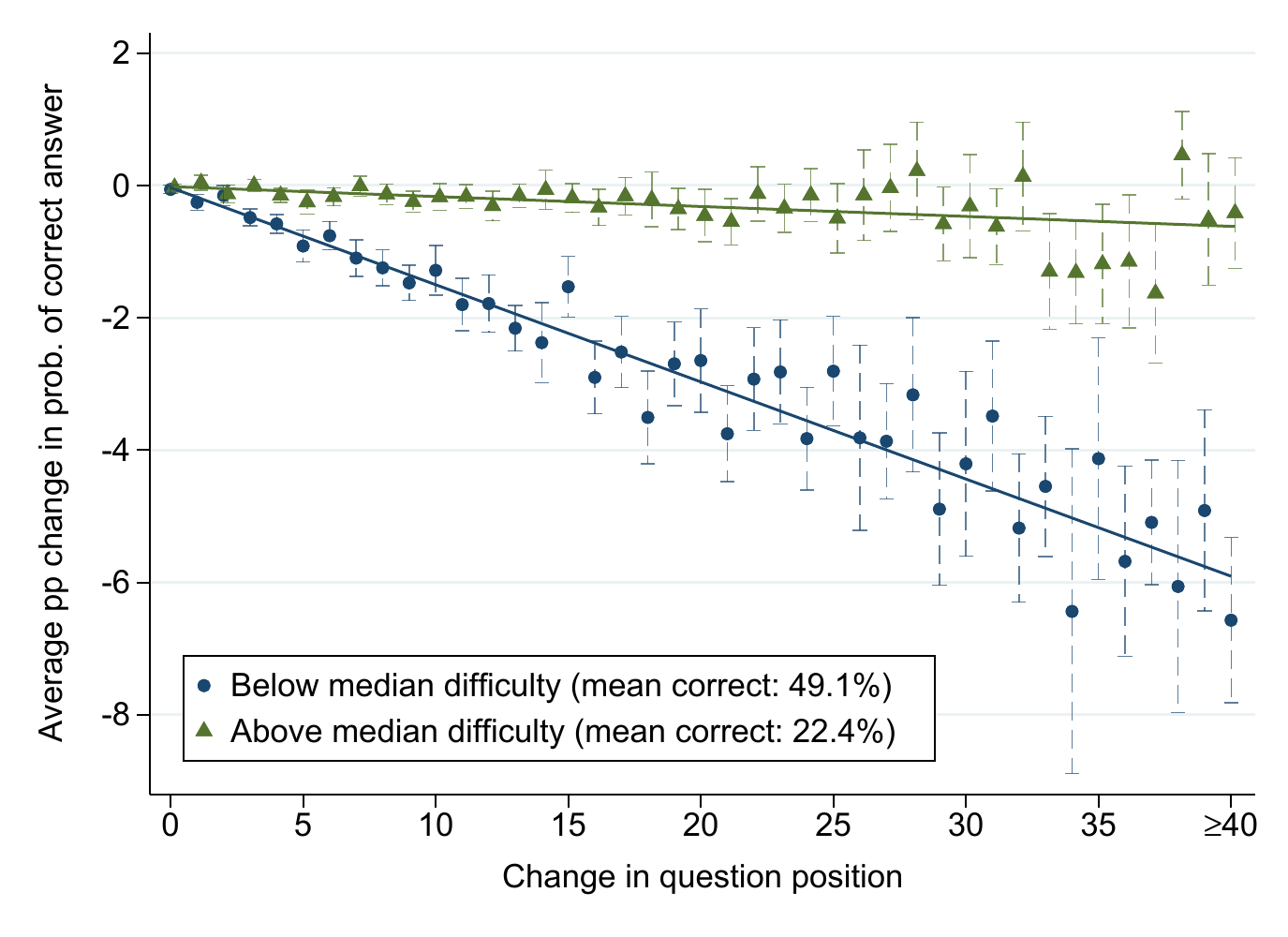}
	\end{subfigure}
	\hfill		
	\begin{subfigure}[t]{0.48\textwidth}
		\caption*{Panel B. Binned scatterplot: \\ Endurance effect vs. question difficulty}
		\centering
		\includegraphics[width=\linewidth]{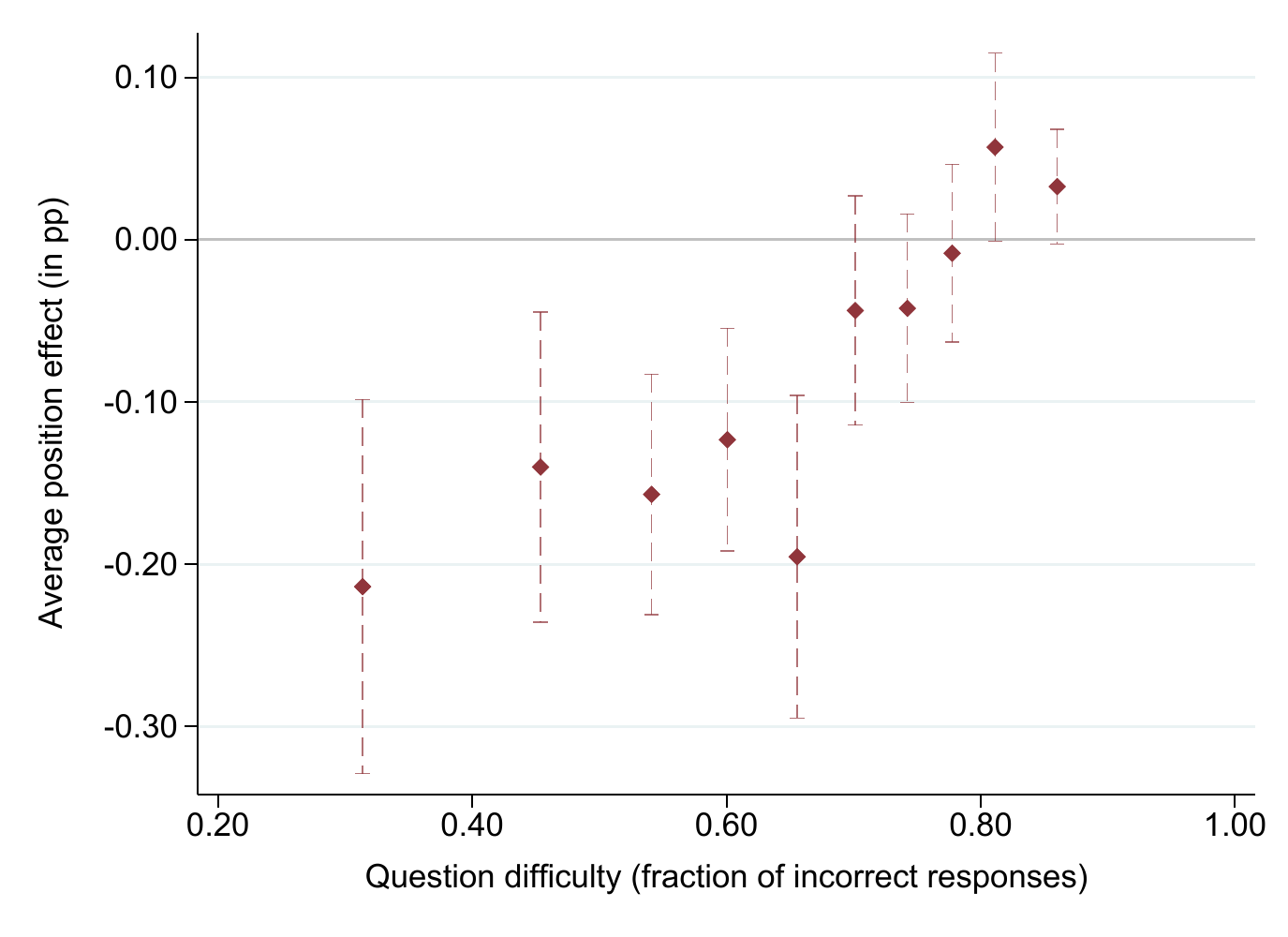}
	\end{subfigure}	
	\hfill		
	\begin{subfigure}[t]{.48\textwidth}
		\caption*{Panel C. Change in performance vs. change in question position by question length\\}
		\centering
		\includegraphics[width=\linewidth]{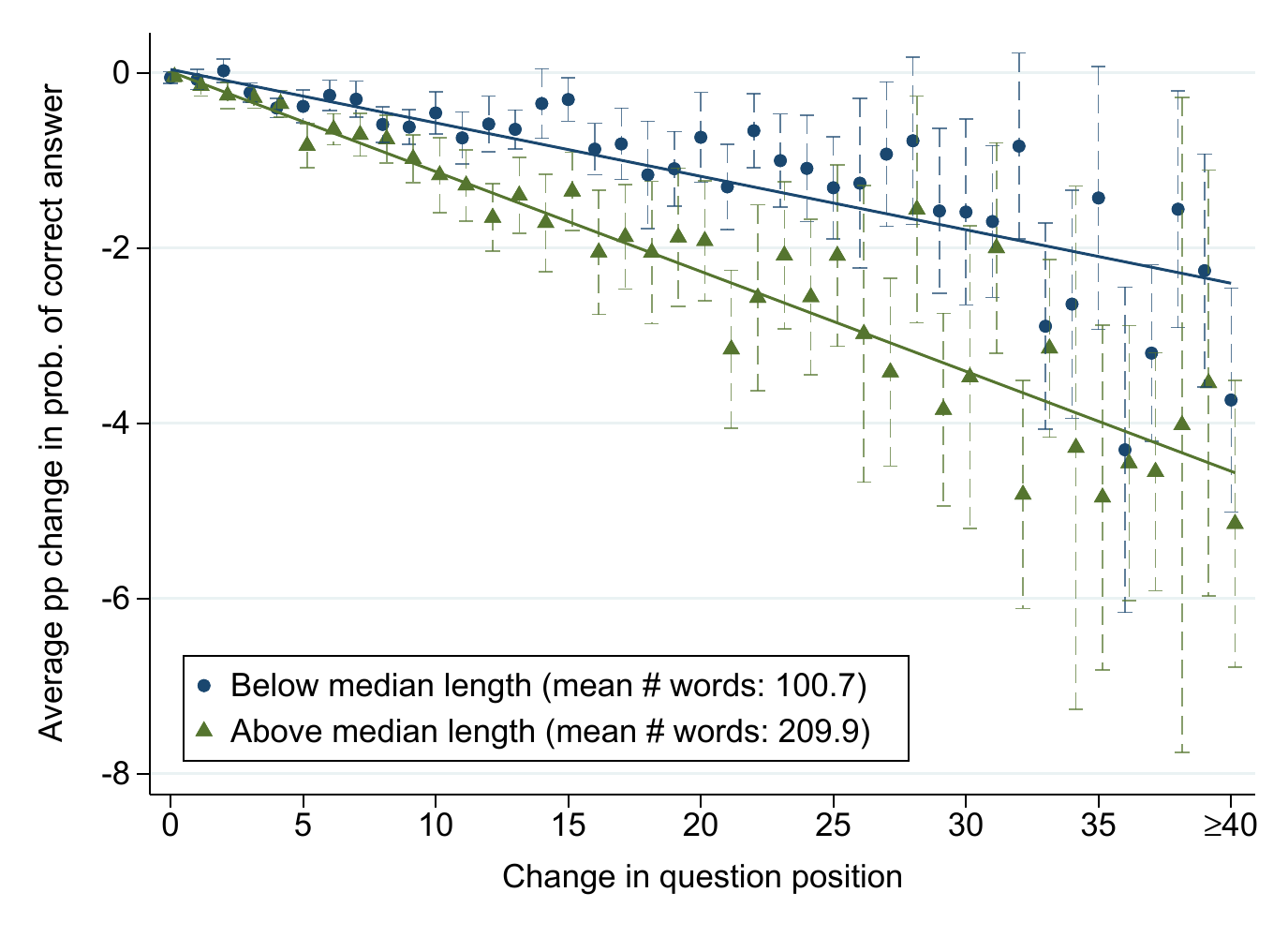}
	\end{subfigure}
	\hfill		
	\begin{subfigure}[t]{0.48\textwidth}
		\caption*{Panel D. Binned scatterplot: \\ Endurance effect vs. question length}
		\centering
		\includegraphics[width=\linewidth]{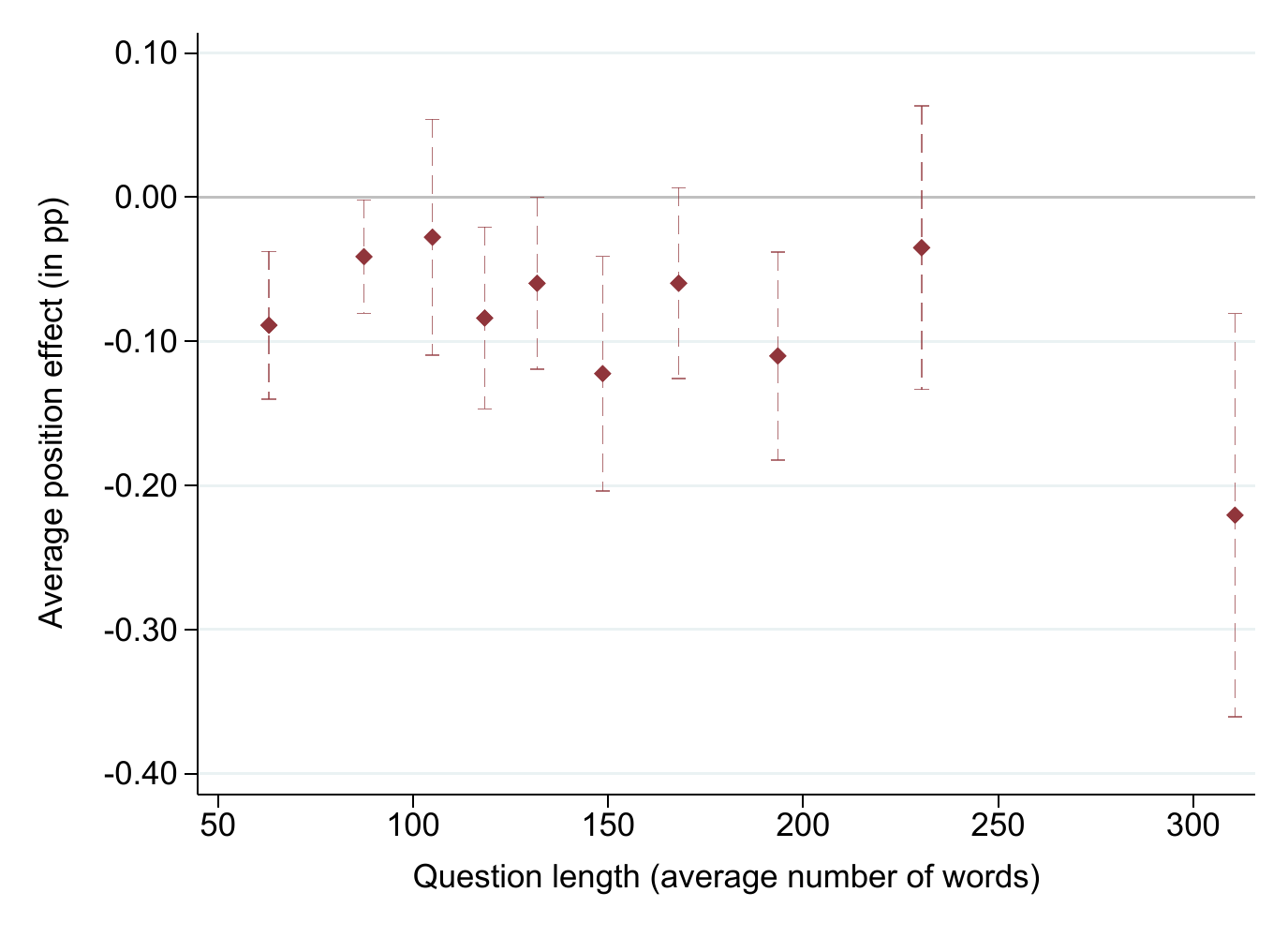}
	\end{subfigure}
	\hfill		
	{\footnotesize
		\singlespacing \justify
		
		\textit{Notes:} This figure shows heterogeneity in the effect of limited endurance on student performance by question difficulty and length. Panels A and C are analogous to Figure \ref{fig:chg-pos}, but the fatigue effect is estimated separately for below/above median difficulty questions (Panel A) and below/above median length questions (Panel C). The $y$-axis shows the average change (in percentage points) in the fraction of students who correctly respond to a question. The $x$-axis plots the change in the question position between each possible booklet pair. The dashed line denotes predicted values from a linear regression estimated on the plotted points, using the number of questions used to estimate each point as weights. Panels B and D show a series of binned scatterplots plotting the average endurance effect among questions in a given difficulty bin (Panel B) or length bin (Panel B). To construct this figure, I divide questions into ten equally-sized bins based on their difficulty or length. Then, I calculate the effect of limited endurance on performance on questions in each bin.
		
	}
\end{figure}

In Appendix Table \ref{tab:perf-pos-het-quest}, I estimate equation \eqref{eq:reg-spec-fe} separately for below/above median-difficulty questions.\footnote{Above [below] median difficulty questions are responded correctly 22\% [48\%] of the time.} Appendix Figure \ref{fig:chg-pos-char}, Panel A, plots corresponding binned scatterplots. The effect of limited endurance is driven by relatively easier questions (columns 3--4). The estimated coefficient is thirteen times larger for below-median-difficulty questions than for above-median-difficulty questions (-13.2 vs. -1.0 percentage points, respectively). More broadly, there is a negative---although non-monotonic---relationship between a question's difficulty and the magnitude of the limited endurance effect (Appendix Figure \ref{fig:chg-pos-char}, Panel B). The performance of students only declines when responding to questions below a certain difficulty, possibly because they do not have the preparation required to respond to the hardest questions regardless of their location.

Second, I explore heterogeneity based on question length. Previous research has shown that time-on-task is one of the main predictors of cognitive fatigue \citep{ackerman2011cognitive}. We might expect fatigue effects to be larger for lengthy questions if students are more likely to have an attentional lapse in long questions. To measure question length, I compute the number of words in each question using text-scraped data. Appendix Table \ref{tab:perf-pos-het-quest} estimates equation \eqref{eq:reg-spec-fe} separately for relatively long questions (above-median number of words) and short questions (below-median number of words).\footnote{Above [below] median length questions have 209 [100] words.} I find that limited endurance effects are about twice as large for longer questions (-10.2 vs. -5.3 percentage points, respectively). Overall, there seems to be a negative relationship between a question's length and the size of the limited endurance effect (Appendix Figure \ref{fig:chg-pos-char}, Panels C--D).

\begin{sidewaystable}
	\begin{table}[H]{\footnotesize
			\begin{center}
				\caption{The heterogeneous effect of limited cognitive endurance on performance by question characteristics} \label{tab:perf-pos-het-quest}
				\newcommand\w{1.5}			     
				\begin{tabular}{l@{}lR{\w cm}@{}L{0.5cm}R{\w cm}@{}L{0.5cm}R{\w cm}@{}L{0.5cm}R{\w cm}@{}L{0.5cm}R{\w cm}@{}L{0.5cm}R{\w cm}@{}L{0.5cm}}
					
					\midrule
					&& \multicolumn{12}{c}{Outcome: Fraction of correct responses} \\ \cmidrule{3-14}
					&& \multicolumn{4}{c}{Question position} & \multicolumn{4}{c}{\hspace{-.5cm} Question difficulty} & \multicolumn{4}{c}{Question length}  \\ \cmidrule{3-5}  \cmidrule{7-9} \cmidrule{11-13} 
					&& 1st half && 2nd half && Below  && Above  && Below   && Above   \\ 
					&& each day && each day && median && median && median  && median  \\ 
					&& (1)      && (2)      && (3)    && (4)    && (5)     && (6)        \\
					\midrule
					Position (normalized)&&$-$0.097&\sym{***}&$-$0.051&\sym{***}&$-$0.132&\sym{***}&$-$0.010&\sym{***}&$-$0.053&\sym{***}&$-$0.102&\sym{***}\\
&&(0.006&)&(0.006&)&(0.006&)&(0.003&)&(0.005&)&(0.007&)\\
    \addlinespace
					Question fixed effects && Yes      && Yes      && Yes    && Yes    && Yes     && Yes   \\ \addlinespace
					\(N\) (Item$-$Booklets)&&3,016&&2,880&&2,935&&2,911&&2,940&&2,956&\\
  \midrule  \addlinespace  
					
				\end{tabular}
			\end{center}

			\begin{singlespace}  \vspace{-.5cm}
				\noindent \justify \textit{Notes:} This table shows the heterogeneous effect of limited cognitive endurance on daily student performance based on question characteristics.
				
				Each column displays the estimate of $\beta$ in equation \eqref{eq:reg-spec-fe} estimated on the sample listed in the column header. I normalize question position such that the first question in each testing day is equal to zero and the last question is equal to one.
				
				Heteroskedasticity-robust standard errors clustered at the question level in parentheses. $^{***}$, $^{**}$ and $^*$ denote significance at 10\%, 5\% and 1\% levels, respectively.
				
			\end{singlespace} 	
		}
	\end{table}
\end{sidewaystable}

\end{document}